# LISA

## Laser Interferometer Space Antenna

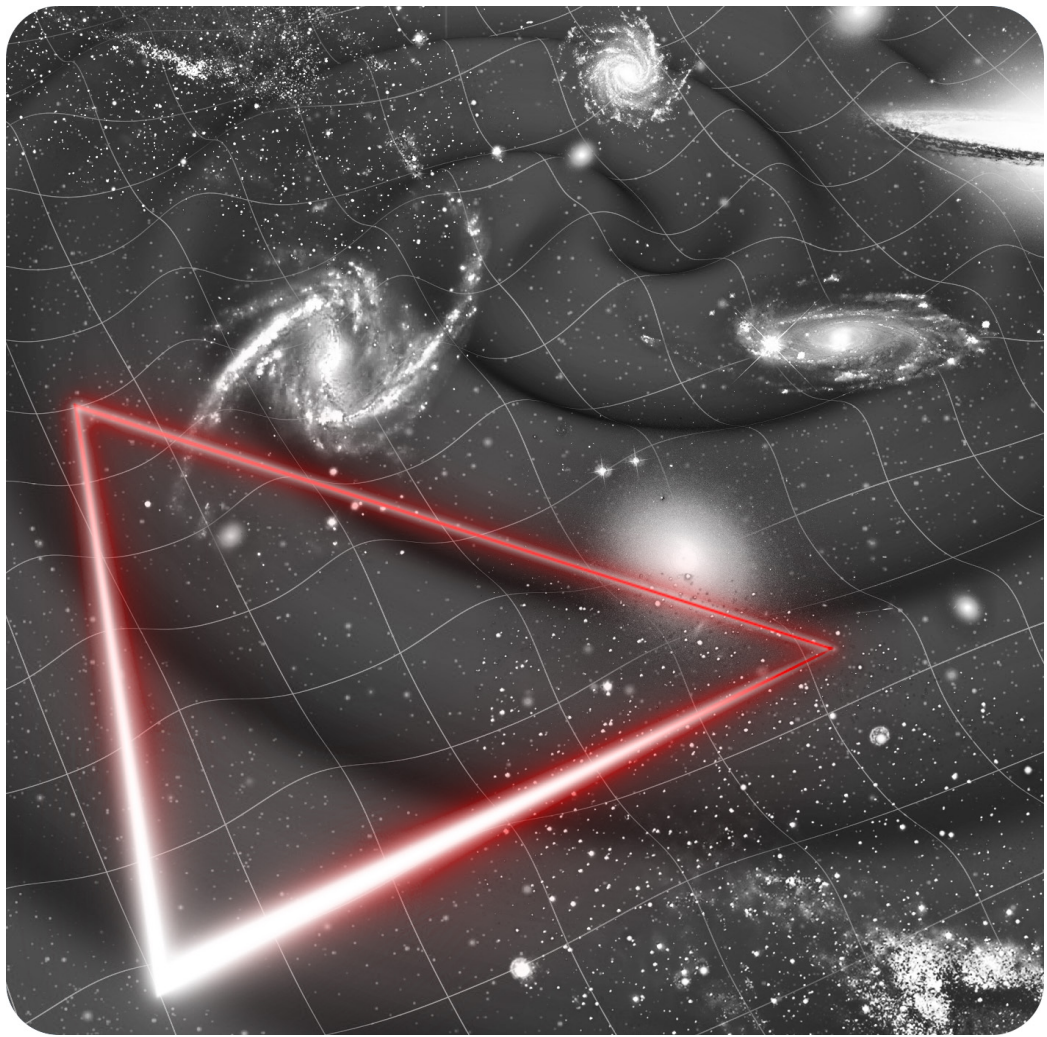

Definition Study Report

→ THE EUROPEAN SPACE AGENCY

# LISA MISSION SUMMARY

## Science Objectives

- Study the formation and evolution of **compact binary stars** and the structure of the Milky Way Galaxy

- Trace the origins, growth and merger histories of **massive Black Holes** across cosmic epochs

- Probe the properties and immediate environments of Black Holes in the local Universe using **extreme mass-ratio inspirals** and **intermediate mass-ratio inspirals**

- Understand the astrophysics of **stellar-mass Black Holes**

- Explore the **fundamental nature of gravity** and Black Holes

- Probe the rate of **expansion of the Universe** with standard sirens

- Understand **stochastic gravitational wave backgrounds** and their implications for the early Universe and TeV-scale particle physics

- Search for gravitational wave bursts and **unforeseen sources**

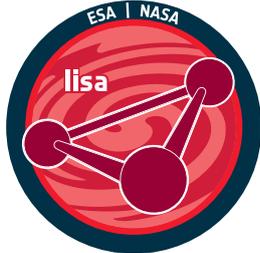

## Measurement

Gravitational waves (GWs) in the Frequency Band of 0.1 mHz - 1.0 Hz with a GW Strain Spectral Density: $10^{-21}$–$10^{-23}$

## Payload

| | |
|---|---|
| Lasers | 2 per spacecraft ● 2 W output power ● wavelength 1064 nm ● frequency stability 300 Hz/$\sqrt{\text{Hz}}$ |
| Optical Bench | 2 per spacecraft ● double-sided use ● high thermal stability (Zerodur) |
| Interferometry | heterodyne interferometry ● 15 pm/$\sqrt{\text{Hz}}$ precision ● Inter-spacecraft ranging to ~1 m |
| Telescope | 2 per spacecraft ● 30 cm off-axis telescope ● high thermal stability |
| Gravitational Reference System | 2 per spacecraft ● acceleration noise <3 fm/(s$^2$ $\sqrt{\text{Hz}}$) ● 46 mm cubic AuPt test mass ● Faraday cage housing ● electrostatic actuation in 5 degree of freedom |

## Mission

| | |
|---|---|
| Duration | 4.5 years science orbit ● >82 % duty cycle ● ~6.25 years including transfer and commissioning |
| Constellation | Three drag-free satellites forming an equilateral triangle ● $2.5 \times 10^6$ km separation ● trailing/leading Earth by ~20° ● inclined by 60° with respect to the ecliptic |
| Orbits | Heliocentric orbits ● semimajor axis ~1 AU ● eccentricity $e \approx 0.0096$ ● inclination $i \approx 0.96°$ |

## Data Analysis

| | | |
|---|---|---|
| Noise Reductions | Laser noise suppression with time-delay interferometry ● Ranging processing and delay estimation ● Spacecraft jitter suppression and reduction to 3 lasers ● Tilt-to-length effect correction ● Clock noise suppression ● Clock synchronisation | |
| Data Levels | Level 0 | Raw data, de-multiplexed, time-ordered, corruption removed |
| | Level 0.5 | Primary science telemetry, decommutated, time-stamped, unit-level calibrations applied |
| | Level 1 | Time-Delay Interferometry (TDI) variables (GW strain) |
| | Level 2 | Output from a global fit pipeline, statistical evidence for candidate sources |
| | Level 3 | Catalogue of GW sources (detection confidence, estimated astrophysical parameters) |

# FOREWORD

The first mission concept studies for a space-borne gravitational wave observatory can be traced back to activities by Peter Bender in the 1970s at the Joint Institute For Laboratory Astrophysics (JILA, Boulder, USA) leading to a first full description of a mission comprising three drag-free spacecraft in a heliocentric orbit, then named Laser Antenna for Gravitational-Radiation Observation in Space (LAGOS). In the early 1990s, LISA (Laser Interferometer Space Antenna) was proposed to ESA, first to the then M3-cycle, later as a cornerstone to the "Horizon 2000 Plus" programme. At that time LISA consisted of six spacecraft, but showed already the key features of today's LISA: Interferometric measurement of distances, long baselines, drag-free spacecraft based on inertial sensors, and the familiar LISA "cartwheel" orbits. The number of spacecraft was reduced to the current three in 1997. Also in 1997, the study team and ESA's Fundamental Physics Advisory Group recommended to carry out LISA in collaboration with NASA, laying the grounds for the present day collaboration.

The first industrial study on LISA published its final report in July 2000, proposing a mission design for LISA that persists – three spacecraft, single launch, and a measurement approach that has been refined, but is unchanged in the underlying principles. In 2003, LISA underwent a first of a series of US reviews aimed at technology readiness that culminated with LISA being identified as the mission with the highest readiness in NASA's Beyond Einstein programme. When ESA formulated the Cosmic Vision 2015–2025 programme in 2005, and started the assessment phase in 2007, LISA was identified early on as one of the potential candidates for the L1 launch slot due to the relative maturity of the concept and the proceeding industrial study.

In 2011, however, financial complications in the US led to an ESA-only mission which was being reduced to a new concept called the New Gravitational Observatory (NGO) which consisted of only two arms with room for contributions. The twin successes of the first gravitational wave (GW) discovery by LIGO/Virgo and the demonstration of LISA-level requirements in free-falling test masses by LISA Pathfinder (LPF) bolstered both excitement and confidence in the LISA concept. A new proposal was submitted in 2017 by the LISA Consortium with three spacecrafts and armlengths of $2.5 \times 10^6$ km. This was followed by a short Phase 0 and a 3 year Phase A for competitive industrial studies carried out by Airbus Defence & Space and Thales Alenia Space concluding in a successful Mission Formulation Review at end of 2021. From 2021-2023 LISA was in Phase B1 for a preliminary definition of the mission conducted by the two prime contractors which leads to the Mission Adoption Review at the end of 2023 and the reason for this document.

This definition study report presents a summary of more than fifty years of refinement and optimisation. It is the result of the recent industrial study together with many contributions from technology development activities, either sponsored by ESA or undertaken by laboratories and institutes in the US and Europe on national funding. Over the last several decades, the scientific objectives and requirements for LISA were developed by the LISA community and are described in this report. Today, computational tools are available to assess the impact of changes in the sensitivity on the science output, leading to mission requirements that are traceable to the scientific objectives. Furthermore, the scientific community rose to the challenge of demonstrating the feasibility of the LISA Data Analysis, closing the loop between science objectives and science output.

# AUTHORSHIP AND ACKNOWLEDGEMENTS

| LISA Science Study Team | | |
|---|---|---|
| *Name* | *Affiliation* | *City, Country* |
| Monica Colpi | University of Milano Bicocca | Milan, Italy |
| Karsten Danzmann | Albert Einstein Institute | Hannover, Germany |
| Martin Hewitson | Albert Einstein Institute | Hannover, Germany |
| Kelly Holley-Bockelmann | Vanderbilt University | Nashville, United States |
| Philippe Jetzer | University of Zurich | Zurich, Switzerland |
| Gijs Nelemans | Radboud University | Nijmegen, The Netherlands |
| Antoine Petiteau | IRFU, CEA | Gif-sur-Yvette, France |
| David Shoemaker | MIT Kavli Institute | Cambridge, United States |
| Carlos Sopuerta | Institute of Space Sciences, CSIC | Barcelona, Spain |
| Robin Stebbins | JILA/University of Colorado | Boulder, United States |
| Nial Tanvir | University of Leicester | Leicester, United Kingdom |
| Henry Ward | University of Glasgow | Glasgow, United Kingdom |
| William Joseph Weber | University of Trento | Trento, Italy |
| Ira Thorpe (NASA Study Scientist) | NASA GSFC | Greenbelt, United States |

| ESA Study Team | | |
|---|---|---|
| *Name* | *Affiliation* | *City, Country* |
| Martin Gehler (Study Manager) | ESTEC | Noordwijk, The Netherlands |
| Oliver Jennrich (Study Scientist) | | |
| Nora Lützgendorf (Study Scientist) | | |
| Linda Mondin (Payload Manager) | | |
| Jonan Larrañaga (Mission System Engineer) | | |
| Eric Joffre (System Engineer) | | |
| Ignacio Fernández Núñez (Payload System Engineer) | | |
| César García Marirrodriga (LISA Coordinator) | | |
| Maike Lieser (PRODEX) | | |
| Jean-Philippe Halain (PRODEX) | | |
| Anna Daurskikh (PRODEX) | | |
| Atul Deep (PRODEX) | | |
| Uwe Lammers (SOC) | ESAC | Madrid, Spain |
| Ana Piris Niño (MOC) | ESOC | Darmstadt, Germany |
| Waldemar Martens (Mission Analysis) | | |

→ THE EUROPEAN SPACE AGENCY

# Additional Authors

Adam Pound, University of Southampton; Alberto Mangiagli, APC; Alberto Sesana, University of Milano-Bicocca, INFN; Alberto Vecchio, University of Birmingham; Alejandro Torres-Orjuela, The University of Hong Kong; Alexandre Toubiana, AEI; Alvin Chua, National University of Singapore; Andrea Caputo, CERN; Andrea Derdzinski, Fisk University, Vanderbilt University; Andrea Maselli, GSSI; Andrea Sartirana, APC; Angelo Ricciardone, University of Pisa; Anna Heffernan, University of the Balearic Islands; Antoine Klein, Birmingham University; Astrid Lamberts, CNRS; Aurelien Hees, SYRTE; Chiara Caprini, University of Geneva; Christopher Berry, University of Glasgow; Danièle Steer, Paris Diderot University; Daniela Doneva, University of Tübingen; Danny Laghi, L2IT-IN2P3; David Izquierdo, University of Milano-Bicocca; Deirdre Shoemaker, University of Texas at Austin; Elena Maria Rossi, Leiden Observatory; Eleonora Castelli, NASA GSFC; Emanuele Berti, Johns Hopkins University; Eric Madge, Weizmann Institute Rehovot Israel; Etienne Savalle, CEA; Gareth Davies, University of Portsmouth; Germano Nardini, Stavanger University; Gianluca Calcagni, Fundación General CSIC; Gianmassimo Tasinato, Swansea University; Guillaume Boileau, Universiteit Antwerpen; Hippolyte Quelquejay-Leclere, APC; Hsin-Yu Chen, The University of Texas at Austin; Ian Harry, University of Portsmouth; Irna Dvorkin, IAP; Ivan Martin Vilchez, Institute of Space Sciences (ICE, CSIC and IEEC); Jacob Slutsky, NASA GSFC; Jean-Baptiste Bayle, University of Glasgow; Jesus Torrado, University of Padua, INFN; Jonathan Gair, AEI; Jose María Ezquiaga, Niels Bohr International Academy; Josh Mathews, National University of Singapore; Kent Yagi, University of Virginia; Laura Sberna, AEI; Lorenz Zwick, Niels Bohr Institute; Lorenzo Speri, AEI; Maarten van de Meent, Niels Bohr Institute; Manuel Arca Sedda, GSSI; Marco Peloso, Padua University; Marek Lewicki, Warsaw University; Marta Volonteri, Sorbonne University; Martin Staab, AEI; Martina Muratore, AEI; Matteo Bonetti, University of Milano-Bicocca; Maude Le Jeune, APC; Mauro Pieroni, CERN; Michele Vallisneri, NASA JPL; Natalia Korsakova, CNRS; Nicola Tamanini, L2IT-IN2P3; Nikolaos Karnesis, Aristotle University of Thessaloniki; Olaf Hartwig, SYRTE; Paolo Pani, University of Roma La Sapienza; Pau Amaro-Seoane, Universitat Politécnica de València; Pedro R. Capelo, University of Zurich; Pierre Auclair, Louvain University; Quentin Baghi, IRFU CEA; Riccardo Buscicchio, University of Milano-Bicocca, INFN Milano; Richard Brito, Instituto Superior Técnico; Sascha Husa, University of the Balearic Islands; Shane Larson, Northwestern University; Stanislav Babak, APC; Sylvain Marsat, L2IT-IN2P3; Tessa Baker, Queen Mary University of London; Thomas Kupfer, Texas Tech University; Thomas Sotiriou, University of Nottingham; Tyson Littenberg, NASA MSFC; Valerie Fiona Domcke, LPPC, EPFL, Lausanne, CERN; Valeriya Korol, MPA; Vishal Baibhav, CIERA; Xian Chen, Kavli Institute; Zoltan Haiman, Columbia University;

→ THE EUROPEAN SPACE AGENCY

## Acknowledgements

Bernard Kelly, NASA GSFC; Carlo Zanoni, INFN / TIFPA; Constantino Pacilio, University of Rome; Curt Cutler, NASA JPL; Daniele Vetrugno, University of Trento; Daryl Haggard, McGill University; Edward Porter, APC; Elisa Bortolas, University of Milano-Bicocca; Enrico Barausse, SISSA; Erik Arends, SRON; Ewan Fitzsimons, STFC; Gareth Davies, Portsmouth University; Gerhard Heinzel, AEI; Gudrun Wanner, AEI; Guido Müller, AEI; Hannah Middleton, University of Birmingham; Jeffrey Livas, NASA GSFC; Joey Shapiro Key, University of Washington Bothell; John Baker, NASA GSFC; John Conklin, University of Florida; Joseph Martino, APC; Juan Jose Esteban Delgado, AEI; Kayhan Gultekin, University of Michigan; Kenji Numata, NASA GSFC; Luigi Ferraioli, ETH Zurich; Neil Cornish, Montana State University; Peter Wass, University of Florida; Priya Natarajan, Yale University; Rita Dolesi, University of Trento; Rosa Valiante, INAF-Italy; Ryan DeRosa, NASA GSFC; Sara Paczkowski, AEI; Sascha Rieger, Milde International Science Communication, AEI; Susanne Milde, Milde International Science Communication, AEI; Tamara Bogdanović, Giorgia Institute of Technology; Valerio Ferroni, University of Trento;

## Red Team Reviewers

Katie Breivik, Flatiron Institute; Anthony Brown, Leiden University; Jan Harms, Gran Sasso Science Institute; Scott Noble, NASA GSFC; Stefano Vitale, University of Trento; Graham Woan, University of Glasgow;

The LISA Science Study Team acknowledges all members of the LISA Consortium and the science community at large for their support on the LISA study.

→ THE EUROPEAN SPACE AGENCY

**Table of Contents:**

→ THE EUROPEAN SPACE AGENCY

**List of Figures:**

→ THE EUROPEAN SPACE AGENCY

→ THE EUROPEAN SPACE AGENCY

## List of Tables:

→ THE EUROPEAN SPACE AGENCY

# 1    EXECUTIVE SUMMARY

One of the revolutionary predictions of Einstein's theory of General Relativity is that accelerated masses produce ripples in spacetime itself, known as gravitational waves. While electromagnetic radiation is caused by acceleration of electric charges, gravitational radiation is caused by acceleration of mass or energy, but due to momentum conservation the leading order is the quadrupole radiation rather than dipole radiation. Gravitational waves interact very weakly and thus are not absorbed, perturbing spacetime far from the source with a dimensionless metric-strain amplitude $h \approx \Delta L/L$ transverse to their propagation direction, which can be observed by the change of distances ($L$) between objects. This strain even for the most powerful sources of gravitational waves is exceedingly small, of order $10^{-20}$ or smaller. It was therefore not until 2015 that the LIGO/Virgo collaboration made the ground-breaking discovery of the first gravitational wave ever directly measured [5] and paved the way for a new era in science. In June 2023, evidence for an ultra-low frequency gravitational wave background was published by the various Pulsar Timing Array collaborations around the world [24, 354, 437]. Different sources emit in different gravitational wave frequency bands, and the mHz band, in between the LIGO/Virgo and Pulsar Timing Array bands, is richly populated with strong sources, but is as yet unexplored.

Unlocking this mHz gravitational wave band is the objective of the LISA (Laser Interferometer Space Antenna) space mission as proposed by the LISA Consortium in 2017 [153]. The LISA mission is an international collaboration between ESA, its member states, and NASA. It will consist of three identical spacecraft forming a near equilateral triangle that exchange laser beams over the $2.5 \times 10^6$ km long arms. LISA will use precision laser interferometry to compare separations between test masses that are protected by the spacecraft from non-gravitational disturbances. LISA will coherently measure spacetime strain variations as a function of time, including frequency, phase, and polarisation, forming a continuous, all-sky observatory sensitive to all sources simultaneously.

LISA's science objectives cover a wide range of outstanding questions in astrophysics, fundamental physics and cosmology including ESA's Cosmic Vision [145] questions such as "What are the fundamental laws of the Universe?" and "How did the Universe originate and of what is it made?". LISA is unique in that it will be sensitive to Massive Black Hole mergers across the Universe, directly probing the origin of Massive Black Holes and their growth and important role in galaxy evolution. Some of these sources will have signal-to-noise ratios as large as 1000, providing exquisite tests of General Relativity and the nature of event horizons. Multi-messenger observations will provide additional information about matter surrounding the mergers and a test of the propagation of gravitational waves. Stellar mass compact objects surrounding Massive Black Holes give rise to extreme mass-ratio inspiral signals, unique for LISA, that probe the stellar population in the central regions of galaxies and measure the mass and spin of the Massive Black Hole. They are also outstanding sources to map the detailed structure of the spacetime metric of the central Massive Black Hole, allowing tests of the no-hair theorem, the existence of additional fields and other alternative Gravity theories. LISA is guaranteed to detect a large number of Galactic sources, consisting of binary and multiple compact objects, which will give detailed insight into the final stages of stellar binary evolution, into the physics of tides and mass transfer, and can be used as independent probes of Galactic structure. Stellar-mass Black Hole binaries can be detected in the nearby Universe and can evolve from the LISA measurement band into that of the ground-based detectors, potentially yielding multi-band gravitational wave sources. Because gravitational waves provide a direct distance measurement to the source, LISA will determine the expansion of the Universe in a complementary way to Euclid and other cosmological probes. Finally, there are several potential stochastic signals from collective phenomena in the Early Universe, such as first-order phase transitions, that, if detected, would provide the unique direct measurement of the first seconds after the Big Bang.

In order to achieve these science goals, the LISA mission will leverage a 4.5 year observation period via an orbit that preserves the constellation without the need for continuous maintenance. Sufficient consumables will allow a mission extension up to ten years. ESA leads the mission and will provide the spacecraft, launch, mission operations and data handling. The ultra-stable lasers, the 30 cm telescopes, and the UV light source for test mass discharge will be provided by NASA. Key instrumental elements are the free-falling test masses, shielded from external forces in the Gravitational Reference System (provided by Italy and Switzerland with contributions from NASA), the picometre accuracy interferometric detection systems (provided by Germany, the UK, France, the Netherlands, Belgium, Denmark and the Czech Republic), the Science Diagnostics Subsystem (provided by Spain), and supporting ground test equipment (provided by France). These instruments have now reached mature designs, carry a strong heritage from the successful LISA Pathfinder mission, and all are at, or on a solid path to, Technology Readiness Level 6. Several elements not tested by LISA Pathfinder (e. g. the phasemeter post-processing techniques to remove the residual laser phase noise and the mechanism to compensate for the angle between send and receive beam) have been demonstrated experimentally to be able to fully meet the requirements of LISA. The classical distinction between spacecraft and payload does not fit LISA very well, as the spacecraft must be co-designed (and then co-operated) with the payload that controls the position of the spacecraft during science operations, rendering the spacecraft effectively a part of the constellation-level instrument.

As outlined in Figure 1.1 the LISA observing will start with the three spacecraft measuring the Doppler shifts between the emitted and received laser beams exchanged between the spacecraft, as caused by gravitational wave tidal accelerations superimposed with slow frequency changes that arise from the spacecraft orbits. Those measurements will be transmitted to earth where the ESA Science Operation Centre will transform the data from Level 0 data (raw data) to Level 0.5 data that are synchronised, in physical units and in directly usable format. Initial post-processing by the Science Operation Centre will remove the major noise sources from different components (such as the laser frequency fluctuations using time delay interferometry), producing Level 1 data. These Level 1 data will then be dominated by overlayed gravitational wave signals and will be distributed to the Distributed Data Processing Centre (provided by the member states, led by France) and to the NASA Science Ground Segment. Both entities will make the computationally demanding, global fits to the data that consists of a linear combination of waveform models of a large number of sources. Based on decades of work in the community, the waveforms of many signals can be accurately calculated, and LISA Consortium data challenges have demonstrated the feasibility of the data analysis task. The global fit output (Level 2 data) will provide a collection of posterior parameter distributions for identified sources. In a final step those distributions from both the Distributed Data Processing Centre and the NASA Science Ground Segment will be transformed into a single catalogue of gravitational wave source candidates (Level 3 data) with detection confidence and estimated astrophysical parameters by a joint working group. In a parallel "low latency" analysis process, transient events will be detected and alerts will be issued for relevant sources as soon as they become evident to allow electro-magnetic observations of the event.

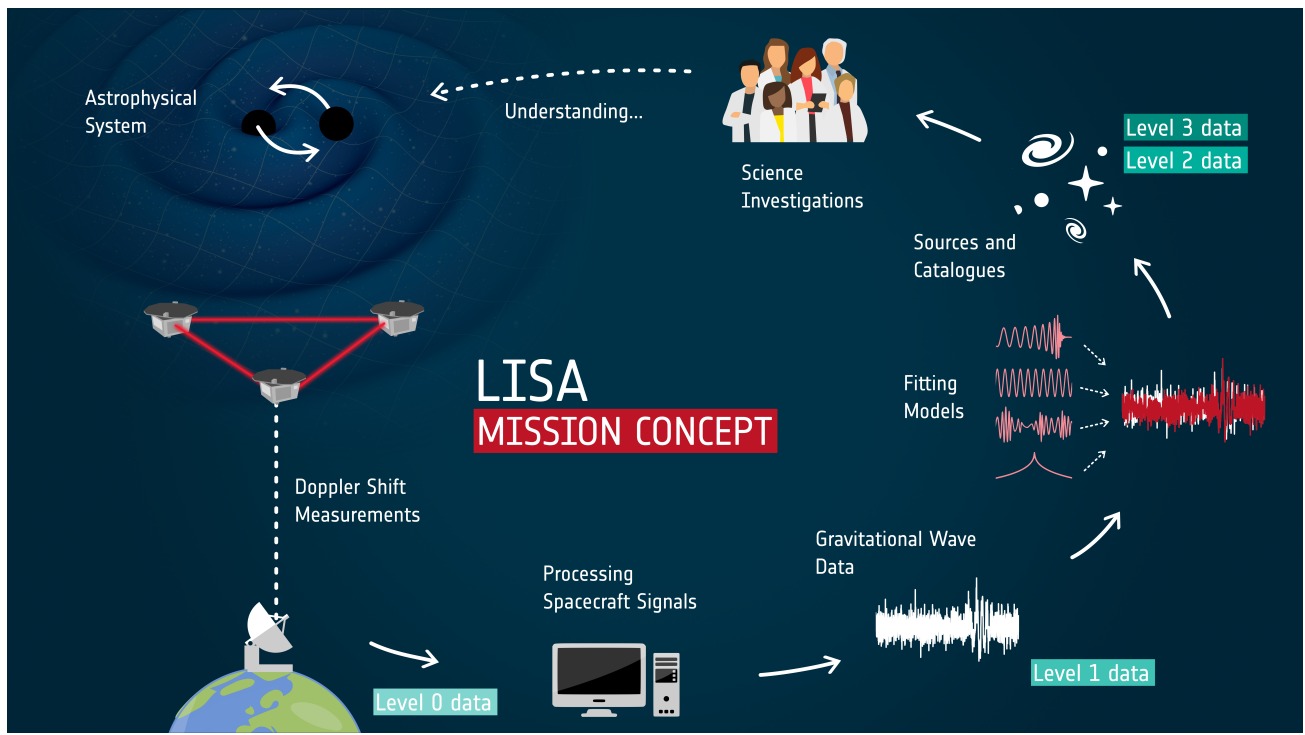

**Figure 1.1:** Overview of the LISA data analysis concept.

Level 0.5 to Level 3 data will be released to the scientific community in regular data releases at planned intervals with a first interval (Early Release Science Time, ERST) that also is used to validate the data and provide initial results from science topical teams that are formed via an open call. The LISA mission is managed by ESA in collaboration with NASA and the member states and in close contact with the LISA Consortium and the wider community ensuring maximal science return from the mission and recognition of those who have made LISA happen. The results from this completely new chapter in space science and the challenges encountered on the path towards it, will be shared with the public, in close collaboration between the ESA, NASA and LISA Consortium outreach efforts. The LISA mission is well on its way to write science history and it puts Europe in a world-leading position in this completely new and groundbreaking science field.

# 2   MISSION CONCEPT OVERVIEW

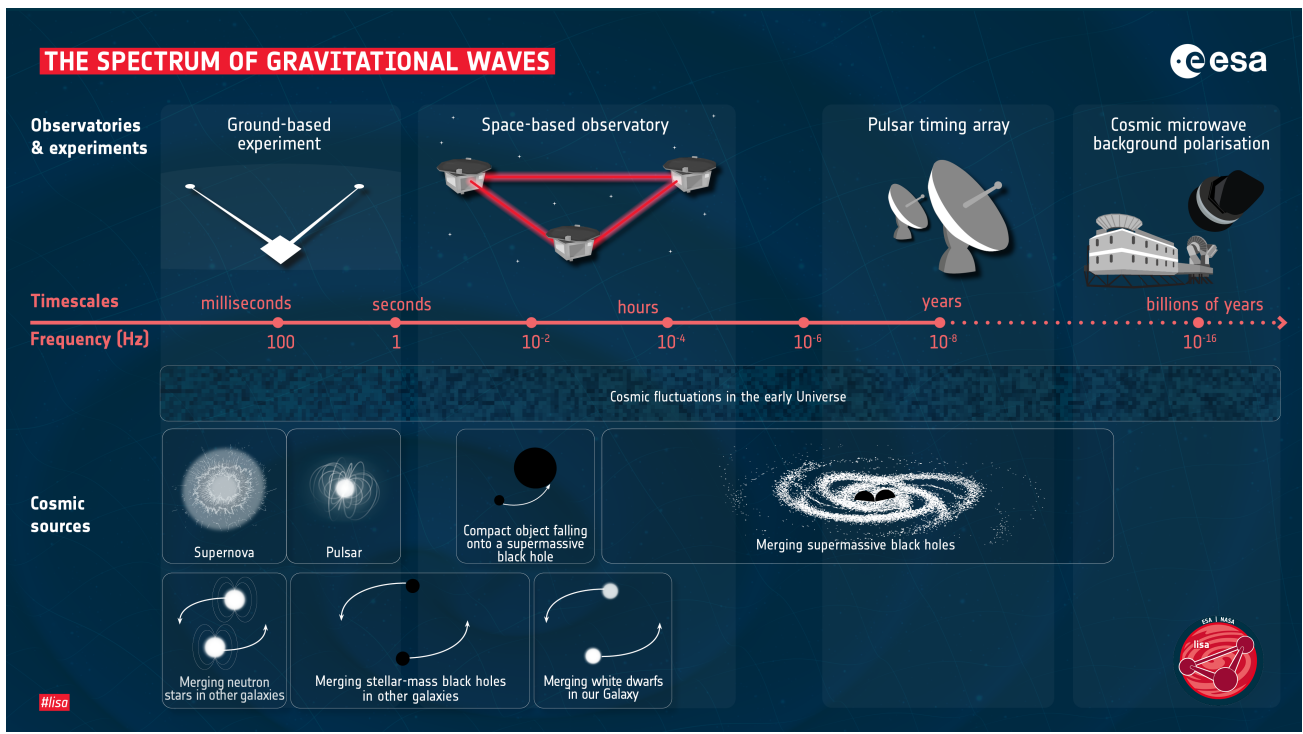

**Figure 2.1:** LISA targets the millihertz band of gravitational waves, lying between the nanohertz regime probed by pulsar timing arrays and the decahertz regime accessible to ground-based detectors. Several types of astrophysical sources produce gravitational waves (GWs) in this band.

## 2.1   Gravitational waves and sources

Gravitational waves (GWs) are perturbations of spacetime which provide a unique and powerful probe of the exotic astrophysical and cosmological phenomena that produce them. Terrestrial detectors (LIGO, Virgo), which made the first direct detection of GWs less than a decade ago [1], have demonstrated the scientific impact of the GW observations through various groundbreaking results including the discovery, and subsequent population study, of a previously unknown class of heavy Black Holes [10]; the first GW-electro-magnetic (EM) multimessenger observation, providing confirmation of the connection between neutron star mergers and short gamma ray bursts [11]; and placement of some of the most stringent limits to date on several possible violations of General Relativity [12]. On the other end of the spectrum, in June 2023, evidence for an ultra-low frequency gravitational wave background most likely originating from a population of merging Massive Black Holes (MBHs) was published by the NANOGrav, EPTA/InPTA [24], PPTA [354], and CPTA [437] collaborations.

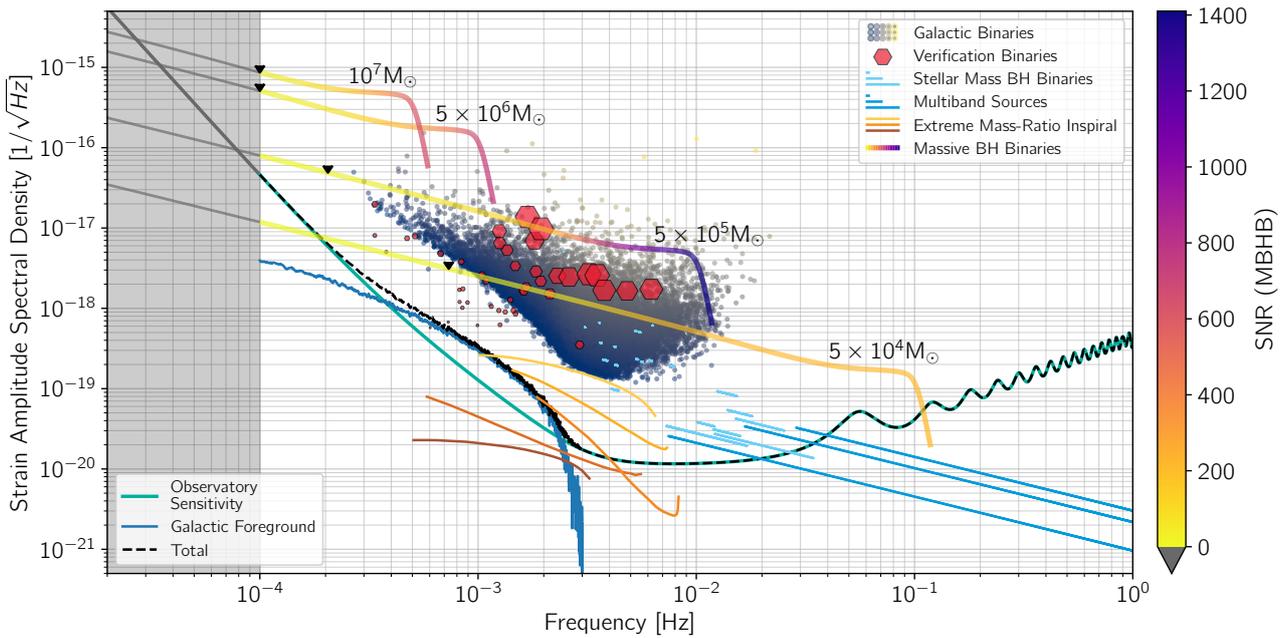

**Figure 2.2:** Illustration of the primary LISA source classes in the gravitational wave (GW) frequency-amplitude plane. Included are merging massive Black Hole binaries (MBHBs) and an extreme mass-ratio inspiral (EMRI) at moderate redshift; stellar-mass Black Holes (sBHs), including potential multiband sources, at low redshift; and Galactic binaries (GBs), including verification binaries (VBs), in the Milky Way. Chapter 3 presents each of these sources and their science opportunities in detail. Solid teal, solid blue and dashed black lines denote sensitivity limits from instrumental noise alone, the unresolved GW foreground, and their sum, respectively. The displacement of the cloud of resolvable sources above the noise is due to the detection threshold being set to signal-to-noise ratio (SNR)=7. The grey shaded area is the extrapolation of LISA's instrumental noise below 0.1 mHz. All quantities are expressed as Strain Amplitude Spectral Densities (ASDs) in order to facilitate a unified plot. For deterministic signals, the ASD is not formally defined but can be approximated as $A_f \sqrt{f}$ where $A_f$ is the Fourier amplitude and $f$ is the Fourier frequency.

---

### Spectral Densities

Spectral densities, which describe the distribution of signal energy as a function of frequency, are a useful tool for expressing LISA's instrument performance. Formally, the Power Spectral Density is defined as the Fourier transform of the autocorrelation function. For a stochastic signal $x(t)$ with units $[\cdot]$, the Power Spectral Density (PSD) gives the expectation value for the variance of the Fourier transform, $S_x(f) \propto |\langle \bar{x}(f) \rangle|^2$ and has units $[\cdot]^2$/Hz. The PSD is useful as it allows the strength of a potential GW signal to be compared to instrument noise only over the relevant portion of the measurement band. In most of this document, the Amplitude Spectral Density (ASD), $\sqrt{S_x(f)}$, with units $[\cdot]/\sqrt{\text{Hz}}$, is used.

---

As with electromagnetic radiation, different science opportunities reside in different bands of the gravitational wave spectrum but require distinct approaches to realise sufficiently sensitive instruments. Figure 2.1 presents a schematic representation of the GW spectrum, spanning more than ten decades in frequency. The millihertz frequency band targeted by LISA sits between the higher frequencies covered by ground-based detectors and the lower frequencies observed by pulsar timing arrays.

LISA's measurement band is expected to have a rich and diverse population of astrophysical – and potentially cosmological – sources, and thus provides an extremely broad science case for GW astronomy. Figure 2.2 provides an illustration of a selection of LISA sources in the GW frequency vs. amplitude plane as well as a representation of the instrumental noise assuming nominal mission lifetime. A complete discussion of each source class and the associated science case is found in Chapter 3 while a breakdown of the major noise sources is found in Chapter 7.

Working upwards from the bottom of Figure 2.2, the teal line shows LISA instrumental noise requirements expressed as equivalent GW strain. The dashed black trace adds the unresolved GW foreground (blue line) from persistent sources in the Milky Way which are not individually identifiable by LISA. This foreground limits LISA's sensitivity to resolved sources in the frequency band $0.3\,\mathrm{mHz} \lesssim f \lesssim 3\,\mathrm{mHz}$, but is a probe of Galactic sources.

Near the bottom of LISA's sensitivity "bucket", the extreme mass-ratio inspiral (EMRI) sources radiate between $1\,\mathrm{mHz} \lesssim f \lesssim 10\,\mathrm{mHz}$. EMRIs, which result from the orbit of a stellar-mass Black Hole (sBH) around a much more massive companion Black Hole (BH), produce complex GW waveforms with multiple harmonics of the fundamental GW frequency simultaneously detectable by LISA. EMRIs are long-lasting sources with timescales of several years. The example EMRI signal in Figure 2.2 shows the fundamental and first four harmonics from the inspiral of a $35\,\mathrm{M_\odot}$ BH into a $2 \times 10^5\,\mathrm{M_\odot}$ BH at a redshift of $z = 2$.

Binary systems of stellar-mass BHs, such as those that are now routinely detected by ground-based GW observatories, are also detectable by LISA during earlier phases in their evolution in the nearby Universe. These are represented by the downward-sloping cyan traces located at higher frequencies than the EMRIs. Note that the downward slope is a consequence of representing the signals with their Amplitude Spectral Density (ASD, see infobox); the instantaneous signal amplitude *increases* with frequency, which in turn increases with time. Some of the sources (lighter lines) do not merge during the observations while others (darker lines) do reach the merger phase, at which point the frequency evolves through the upper sensitivity limit of LISA and into the sensitivity band of ground-based GW instruments, providing the opportunity for *multiband* GW astronomy: the ability to detect the same system in multiple frequency bands with different GW instruments.

Near the centre of the figure reside the Galactic binaries, the most numerous class of LISA sources. These represent binary systems containing two compact objects in the Milky Way with periods of hours to minutes. More than ten thousand individual sources are expected to be detected by LISA during the nominal science phase. For the most part, these sources do not evolve appreciably in frequency and so they are represented by points in the figure. For some systems, the slow evolution of their frequency can be measured over the lifetime of the LISA observations, which allows for a direct estimate of their luminosity distance via the "standard siren" technique [373]. The red hexagons indicate a subset of the Galactic binaries which have already been identified through electromagnetic observations. These guaranteed multimessenger sources provide an opportunity to perform additional science investigations and can serve as a *verification* of LISA's on-orbit performance. The size of the hexagons illustrates the SNR of the systems. In the years between the writing of this report and the launch of LISA, many more such systems are expected to be discovered by Gaia [190], the Vera Rubin Observatory [243], and other facilities.

At the upper part of Figure 2.2, and consequently the loudest individual signals in LISA, lie the massive Black Hole binaries. These represent the late inspiral, merger, and ringdown of Black Holes roughly a million times the mass of our Sun across the Universe. Four representative systems are shown with different masses at $z = 3$. As discussed in Chapter 3, LISA will be sensitive to systems with masses ranging from $10^3\,\mathrm{M_\odot} \lesssim M \lesssim 10^7\,\mathrm{M_\odot}$ at redshifts $z \lesssim 15$.

## 2.2 The LISA mission and instrument concept

Gravitational waves can be observed by measuring variations in spacetime curvature, for instance by establishing a pair of freely-falling (inertial or geodesic) reference objects separated by some distance and monitoring the proper distance between them using beams of light. This is the approach successfully employed by ground-based GW detectors with pendulum-suspended mirrors providing the (horizontally) inertial reference objects [45], and laser interferometry providing the proper distance measurement.

The amplitude of the GW signal is typically quantified as spacetime *strain*, denoted by $h(t)$, a dimensionless quantity that describes the fractional change in curvature. For example, a light beam traversing between two freely-falling objects separated by a nominal proper distance $L$ will accumulate an additional phase shift proportional to the GW strain:

$$\Delta\phi = \frac{2\pi}{\lambda}\frac{c}{2}\int_{t-T}^{t} h(t')\,\mathrm{d}t' \approx \frac{2\pi}{\lambda}\frac{L}{2}h(t) \qquad (2.1)$$

where $\lambda$ is the optical wavelength, $c$ is the speed of light, and $T = L/c$ is the nominal light travel time between the reference objects. Here we assume for simplicity an "optimally" oriented and polarised GW. A more general expression can be easily derived [e. g. 273] which includes coupling constants that depend on polarisation of the GW and the orientation of the baseline relative to the direction in which the GW is travelling. The final approximation in Equation (2.1) is valid in the low frequency limit $f \ll c/L$, which is applicable for the lowest portions of LISA's measurement band.

In addition to dimensionless strain, GW amplitudes can also be expressed in terms of other physical quantities that are useful when comparing with limiting noise sources in various subsystems of the LISA observatory. The most common set of such quantities are:

**Displacement** The phase shift in Equation (2.1) is equivalent to a displacement of a test object by $\Delta x = (\lambda/2\pi)\Delta\phi \approx (L/2)h$. Typical values for the displacement induced in the LISA arms by a GW signal are tens to hundreds of picometers.

**Doppler shifts** Differentiating $\Delta\phi$ results in a frequency which physically represents an optical Doppler shift from the effective relative motion between the two freely-falling test masses. The frequency change is $\Delta\nu(t) \approx (h(t) - h(t - T))\nu/2$ with typical values of order μHz.

**Acceleration** Differentiating the displacement twice in time yields a signal that is an effective differential or *tidal* acceleration between the freely-falling test masses. This can be compared with any non-gravitational forces acting on the test masses, which limit LISA's sensitivity at low frequencies. Typical values of differential accelerations produced by GW signals acting on the constellation are of the order of femto-g (or $10^{-14}$ m/s$^2$).

The foundational step in space-based GW observatory design is the selection of the orbits, which determine the size and dynamics of the constellation, and thus plays a determining role in the science measurement band, mission performance, instrument requirements and key mission design parameters. LISA utilises three spacecraft (S/C) in Earth-like heliocentric orbits, which permits an approximately equilateral triangular constellation to be maintained for roughly a decade with no active station keeping (Figure 2.3). A minimum of two measurement baselines is required for a "common-mode" rejection of laser frequency noise as in a Michelson interferometer – the implementation of this for LISA is discussed in the next section. The fully-symmetric three-arm constellation used for LISA has benefits beyond the minimal two-arm configuration, allowing direct and instantaneous measurement of GW polarisation information, increasing mission robustness, and enabling an interferometric channel with substantially reduced sensitivity to GW and thus a probe of instrument noise.

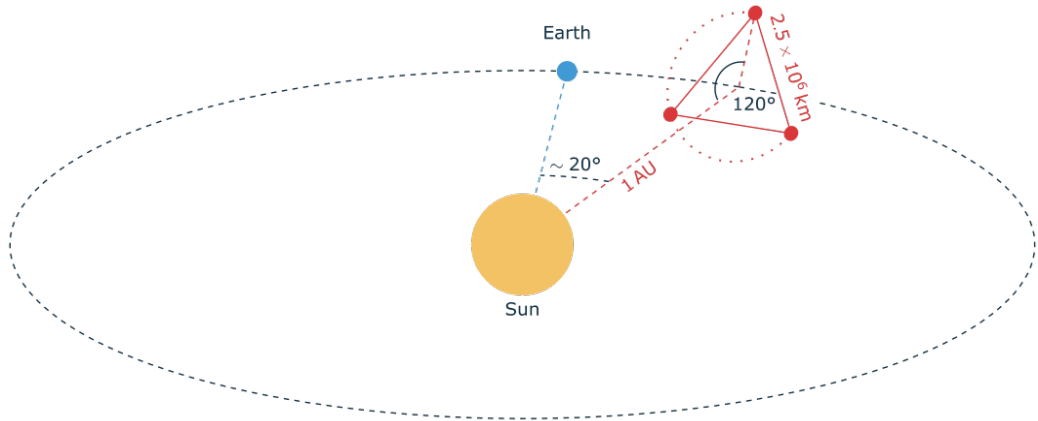

**Figure 2.3:** Schematic depiction of the LISA orbit, not to scale. The three satellites are arranged in a equilateral triangle, the constellation barycentre follows a heliocentric orbit lagging or leading approximately 20°, or about $50 \times 10^6$ km, behind Earth. The plane of the constellation (marked with the dotted line) is inclined at 60° with respect to the Ecliptic and the triangular array undergoes an annual rotation within the plane. See Chapter 6 for further details.

The 2.5 million km baselines (or "arms") between the three S/C are chosen to optimise science return while minimising costs associated with orbital transfer and communications. Shorter arms reduce the variation of the angles between the S/C as well as the variation of the armlengths whereas longer arms increase the effect that GW have on the constellation, and hence the signals. A more detailed description of the orbits can be found in Section 6.2.1.

LISA employs a constellation of geodesic reference objects, comprising two free-falling test masses, shielded from external solar radiation pressure and micrometeoroids disturbances, inside each of the three S/C. The test masses are 46 mm cubes of a Au-Pt alloy, each free-falling in vacuum inside its own cubic conducting housing that also serves as a capacitive position sensor and electrostatic force actuator. This Gravitational Reference System (GRS) is described in detail in Section 5.2. The technique employed by LISA to realise free-fall of the test masses at sufficient fidelity is known as 'drag-free control'. The position and orientation of the test masses relative to the S/C is measured using a combination of capacitive and optical sensors. These same quantities can be controlled by applying forces and torques either to the test masses via a capacitive actuation system or the S/C via the micropropulsion system. For science operations, the position along the measurement axis defined by the inter-S/C optical link is controlled exclusively through actuating the spacecraft, thereby avoiding disturbances on the test masses and allowing them to trace purely geodesic trajectories defined by the underlying spacetime (see Figure 2.4 for a schematic description). The major limits to the performance of this system are internally-generated force disturbances on the test masses, including residual gas pressure, thermal radiation pressure, electrostatic, and magnetic forces. The stray test mass acceleration noise sets the low frequency observatory sensitivity, with the requirement at the *femto-g* level, or $1 \times 10^{-14}$ m/(s$^2 \sqrt{\text{Hz}}$) at 0.1 mHz. Demonstrating free-fall at this level, using drag-free S/C and the LISA GRS hardware, was the primary objective, and the successful outcome, of LISA Pathfinder [32].

LISA's $2.5 \times 10^6$ km baselines between pairs of S/C provide efficient coupling to GW signals in the target measurement band. However, the large diffraction losses associated with these baselines prohibit the use of the closed Michelson-like configurations found in ground-based GW observatories. LISA employs a transponder scheme, in which each Moving Optical Sub-Assembly (MOSA) acts as an interferometric transceiver at the end of each arm, simultaneously transmitting high-power laser beams toward the distant MOSA while receiving a small portion of the beam transmitted from that MOSA. The relative phase between transmitted and received beams is measured interferometrically at each MOSA, with the resulting signals continuously measured and sent to the ground. These individual phase measurements contain a superposition of the desired GW signals as well as much larger instrument noise and a secular phase accumulation resulting from relative motion between the S/C. On-ground processing of the combined set of phasemeter signals from across the LISA constellation allows the secular phase accumulations and instrument noise to be suppressed, revealing the GW signals. This processing is described further in the following section. The bi-directional transponder technique has been successfully employed by the Laser-Ranging Interferometer (LRI) on the US-German GRACE-FO mission to track the ~300 km distance between a pair of satellites in low-Earth orbit to the ~nm precision required to monitor changes in the Earth's geoid [15]. This performance is limited by the single-baseline configuration of Laser-Ranging Interferometer (LRI), which does not permit the instrument noise to be rejected through on-ground processing. With the on-ground processing technique applied, LISA's displacement sensitivity is primarily limited by photon counting statistics, with other contributors held at a fraction of this limit. A detailed description of the optical metrology system can be found in Section 5.3.

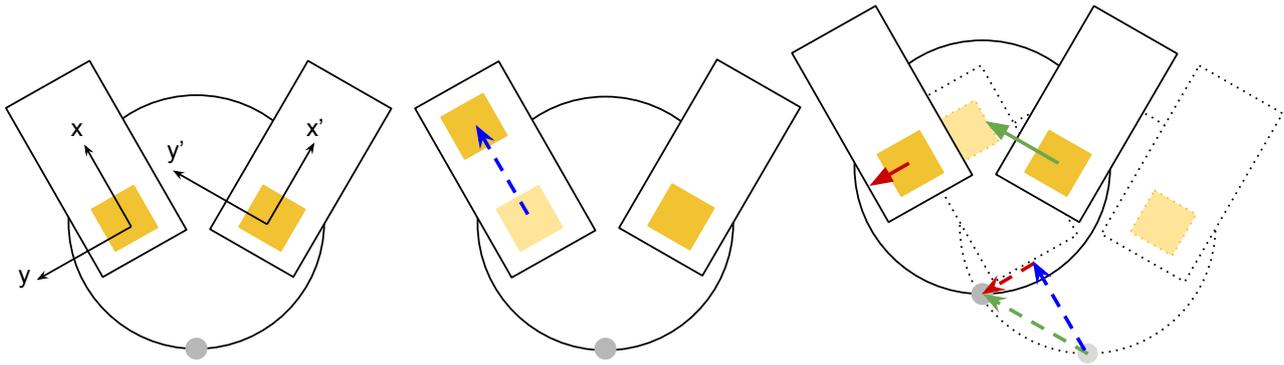

**Figure 2.4:** Schematic description of the approach employed by the Drag-Free Attitude Control System (DFACS) to enable free-fall of the two test masses in each LISA spacecraft (S/C) along the respective sensitive axes. Note that all movements are overemphasised for clarity. First panel: The LISA S/C is represented by the circle with the grey dot serving as a visual reference for displacements and rotations. The two Moving Optical Sub-Assemblies (MOSAs) are indicated by the black rectangles with the gold cubes representing the test masses and the sensitive axes are denoted as $x$ and $x'$. Second panel: the left test mass follows a different orbit and moves along the sensitive axis $x$ (blue dashed vector) with respect to the S/C as the movement along the $y$ axis is restricted by the test mass-control. Third panel: the DFACS algorithm compensates by moving the spacecraft in the direction of the insensitive axis of the other test mass ($y'$, green dashed vector) so that the displacement of the first test mass is compensated (blue dashed vector), simultaneously moving in the direction of the sensitive axis $x$ of the first test mass, applying forces to the test masses only along the insensitive axes $y'$ and $y$ (green and red solid vectors, respectively). This results in each test mass remaining in free-fall along its corresponding sensitive axis.

## 2.3 Constellation measurements

While each pair of LISA arms is superficially similar to a Michelson interferometer, the configuration is actually quite different. In a Michelson interferometer light from a single source is divided in a beamsplitter, sent along each of the arms, reflected back, and recombined optically to form the differential measurement. The coupling of frequency noise in the light source to the interferometer output is determined by the difference in the optical path length in the arms. If the arms are exactly equal, the interferometer can exhibit perfect rejection of frequency noise, as it is common to both arms and occurs at the same time so that the difference in optical phases is free of the frequency noise ("common-mode rejection").

In LISA, the large distances involved make a fully optical chain of division, reflection, and recombination infeasible. Instead, each one-way link between pairs of spacecraft has its own light source and the interferometric measurement that is made is between the outgoing light and the light received from the distant spacecraft, similar to a coherent radio transponder. As illustrated in Figure 2.5, the recorded phase of each of these measurements contains three primary elements: the phase of the *local* beam that is being transmitted to the distant spacecraft; the phase of the *received* beam, which is delayed due to the light travel time from the far spacecraft; and a (much smaller) signal that arises from the interaction between the travelling optical beam and GWs present in the constellation. Because the two light sources are independent, there is no cancellation of their phase noise at any given interferometric measurement. However, since each pair of lasers is compared in two places – each end of a given arm – there is an opportunity to remove the noise by taking differences in the measurements. In practice, measurements from all three spacecraft are needed to fully cancel the frequency noise in the constellation. The relative time delays of the phase signals in the various measurements must be accounted for in order to ensure the cancellation is sufficient. Because the GW signals induce different correlations between the one-way measurements that are present in the laser noise, the combinations of measurements which suppress laser frequency noise retain GW sensitivity. These quantities are referred to as *Time Delay Interferometry* or TDI variables and form the basic observables for LISA. Time-Delay Interferometry (TDI) has been extensively studied analytically [e. g. 405, and references therein], through high-fidelity simulations [82, 413], and through experimental analogues [211, 308, 309, 449]. The fundamental limits in the capabilities of TDI to suppress laser frequency noise are the linearity of the phase measurements as well as the ability to place measurements from different spacecraft on a common time grid.

## 2.4    Operations and Data Analysis

Once commissioning activities to establish the working constellation are complete, operations of the LISA constellation are relatively simple. The instrument is sensitive to GW signals from all directions and requires no active pointing or targeting. Aside from short, periodic maintenance intervals, the observatory produces continuous time series of TDI observables, which together are sensitive to GWs across the LISA measurement band and in all areas of the sky – somewhat like an omnidirectional microphone for spacetime. Tens of thousands of individual signals from distinct astrophysical sources are expected to be present in these observables. Extracting each of these sources and characterising their astrophysical parameters requires a simultaneous fit that produces a catalogue of sources with probability distributions of parameters. Having removed the detected sources, the residuals can be used to search for unmodelled astrophysical or cosmological signals as well as to characterise observatory performance. Development of the sophisticated algorithms required for this source identification and extraction has been a major effort of the LISA research community over the last two decades, including through LISA Consortium-led data challenge efforts.

While a catalogue of millihertz GW sources represents the immediate objective of the LISA measure-ment effort, it is only the beginning of the scientific inference that motivates building the observatory. LISA's catalogues will give scientists direct access to information including masses, spins, distances, and sky positions for systems containing compact objects across a huge range of mass and distance, providing information to constrain a broad range of theoretical models and to provide opportunities for multimessenger astrophysics, fundamental physics, and cosmology. For some of the loudest signals, LISA will also provide time-domain information in the form of alerts to facilitate contemporaneous multimessenger observations of astrophysical objects such as accretion around merging massive Black Holes.

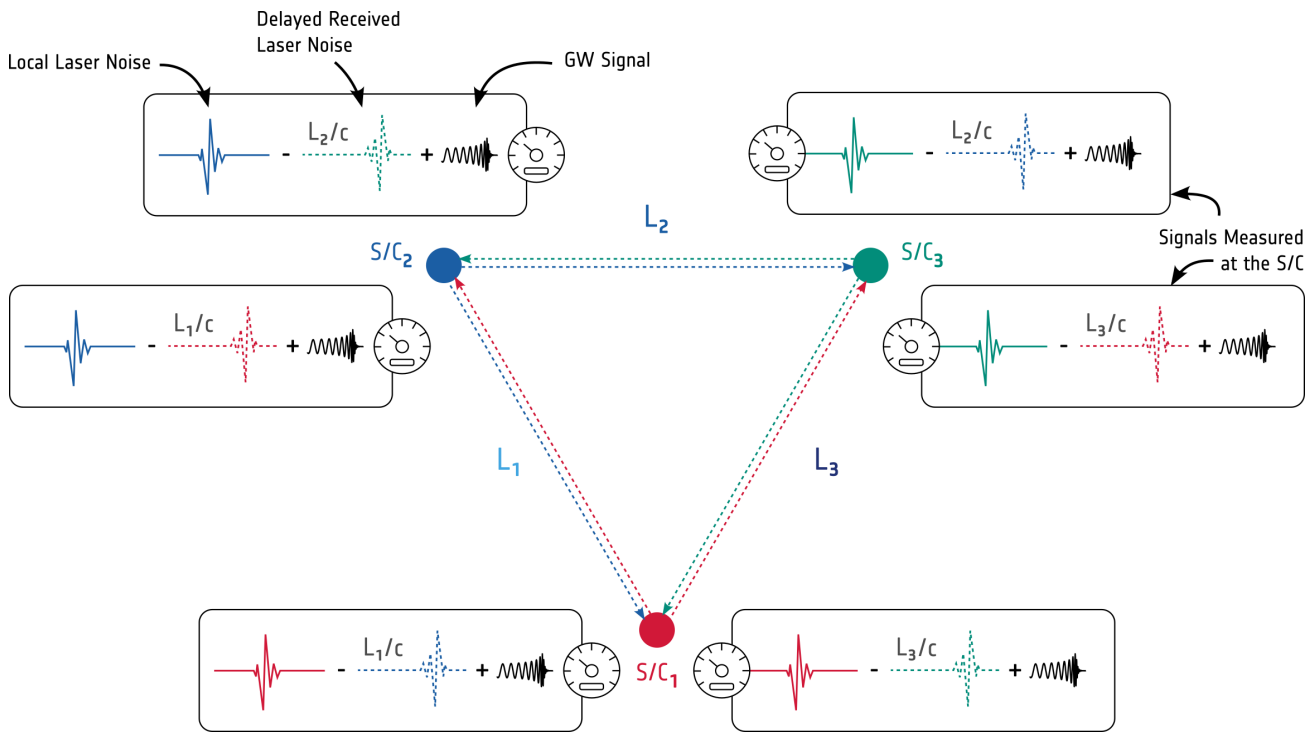

**Figure 2.5:** Schematic representation of LISA's Time-Delay Interferometry (TDI) measurement scheme. Each spacecraft records the phase of the interference between its local laser (solid signal) and the received signal, which contains a time-delayed copy of the distant spacecraft's laser noise (dashed signal) plus a GW signal. Combining measurements from across the constellation with appropriate time delays allow the laser noise to be suppressed while retaining the GW signal.

## 2.5 Evolution of the LISA Concept

Concepts for a space-based GW observatory employing laser interferometry were first proposed in the mid-1970s and were developed in academic institutions through the 1980s and 1990s. The first mission concept studies were carried out by ESA from 1993-2000 [263] and by NASA in 1997 [187], resulting in the establishment of the joint NASA/ESA LISA study which ran from 2001-2011 and made substantial progress in refining the mission concept and developing of critical technologies [207, 265, 266].

The origin of today's LISA is the proposal made to ESA [153] by the LISA Consortium which was the basis for Mission Selection in 2017. That proposal built on decades of prior work, experience with LISA Pathfinder, and progress in understanding of the science landscape to yield a mission concept that delivered transformational science while minimising cost, complexity, and risk. The parameters of that mission have remained largely stable during the subsequent study phase. Here we highlight a few important changes:

- Change in high frequency optical metrology performance. The original mission proposal was based on an optical metrology performance of $10 \, \text{pm}/\sqrt{\text{Hz}}$. During the study phase, establishing the full performance budget and setting allocations to all the key contributors resulted in the need to relax this requirement to remain within the feasibility of the current technology (largely due to laser power and photoreceiver performance). As such, the metrology performance was relaxed to $15 \, \text{pm}/\sqrt{\text{Hz}}$.

- Change in total accumulated science data. The mission proposal of 2017 assumed a total observing time for the nominal mission of 4 years with no interruptions. In establishing a realistic scenario for the mission operations and the observatory availability, this number was adjusted to be 82 % availability over a total nominal science operations phase of 4.5 years, resulting in 3.69 years of accumulated science data during this phase. The mission is also designed, specifically in terms of fuel capacity and orbital configurations, to accommodate an extended science phase which would yield additional science return. While the specific targets for a mission extension proposal are beyond the scope of this document, opportunities include improved parameter estimation for persistent sources, increased population statistics, and increased probability of detecting rare events of exceptional significance.

- Assessment of the lower limit of the measurement band. The mission proposal from 2017 proposed a goal of $20 \, \mu\text{Hz}$ for the lower limit of the measurement band. This was motivated by the strong science return for sensitivity in this band. During the study phase, it became clear that carrying this lower limit as a requirement would substantially drive costs associated with pre-flight verification as well as place constraints on certain material choices in the optical chain. The decision was taken to remain with the proposed requirement of 0.1 mHz, which fully supports the science case presented here. However, the project will maintain an analysis of the expected performance in below the formal requirement so that the science team can prepare for science opportunities obtained by the realised performance. For context, it is notable that LISA Pathfinder (LPF) achieved excellent performance down to $20 \, \mu\text{Hz}$ despite having a formal lower bandwidth requirement of 1 mHz.

- Change in the upper limit of the measurement band. The mission proposal defined the upper limit of the measurement band at 100 mHz, with a goal of extending to 1 Hz. That goal is now promoted to a requirement. This supports the science objectives of multiband GW astronomy and discovery space.

These changes have some impact on LISA's science capabilities, although the original Science Objectives (SOs) described in [153] are still met. In addition, current and foreseen advances in both traditional and GW astronomy have yielded some adaptations of LISA's scientific impact and are included in detail in Chapter 3. In particular, there has been significant evolution in the understanding of the impact of LISA on studies of the structure of the Milky Way (Section 3.1.2); on the impact of LISA on the study of Intermediate-Mass Black Holes (IMBHs) (now described in Section 3.3 rather than Section 3.2), on EMRIs (Section 3.3) and on the impact of LISA on the stellar-mass Black Hole binaries (sBHBs) (Section 3.4). For the latter, the number of expected sources is lower than originally estimated and a new SO is added for very heavy BHs, as found in GW190521 (Section 3.4.2). The study of anisotropies in stochastic gravitational-wave backgrounds (SGWBs) is also a new SO (Section 3.7.3), while the study of new light fields around BHs is now discussed jointly with the study of the multipolar structure of BHs in Section 3.5.2.

→ THE EUROPEAN SPACE AGENCY

# 3    SCIENCE OBJECTIVES

The science of the LISA mission is extremely broad and diverse and has been studied and described in a series of white papers [63, 85, 91, 97, 116, 138, 143, 149, 283, 302, 318]. The following sections briefly describe the science objectives of the LISA mission, following the Science Objectives (SOs) introduced in the LISA L3 proposal [153].

---

**Gravitational Wave signals, science observables and notation: a short guide**

LISA sources are of two kinds. The first are isolated binaries, observed either far from coalescence as (quasi) monochromatic sources, such as Galactic binaries (GBs) of stellar compact objects, or when in-spiralling as extreme mass-ratio inspirals (EMRIs), and/or merging as massive Black Hole binaries (MBHBs). A binary is characterised in its rest-frame (source-frame, sf) by the component masses $m_1$ and $m_2$ or its mass ratio, $q \equiv m_2/m_1 < 1$, the total mass $M$ and chirp mass $M_c = \nu^{3/5}M$, where $\nu \equiv m_1 m_2/M^2$ is the symmetric mass ratio. To leading order, the instantaneous frequency $f_{sf}$ of the gravitational wave (GW) emitted by a circular binary of two point masses is twice the Keplerian frequency. As GWs carry away energy, linear momentum and angular momentum, the frequency $f_{sf}$ increases with time, producing a *chirping* signal. Its time derivative $\dot{f}_{sf}$ depends on $M_c$ and $f_{sf}$, providing a direct measurement of $M_c$ (in presence of non-GW torques $\dot{f}$ can be either positive or negative). The evolution in time of the GW strain amplitude, frequency and phase allows the measurement of the source's luminosity distance $d_L$, and the inclination angle $\iota$ between the binary orbital angular momentum and the line of sight. Internal degrees of freedom, such as the dimensionless Black Hole spin amplitudes ($\chi_{1,2}$), the polarisation and sky position angles can be extracted from the GW signal. GWs are redshifted and are observed at a frequency $f = f_{sf}(1+z)^{-1}$. Likewise, the chirp mass (time) in the observer-frame has a larger value $\mathcal{M} = M_c(1+z)$ $(t = t_{sf}(1+z))$. [a] Astrophysical and cosmological stochastic backgrounds comprise the second kind. They are characterised by specifying how the energy is distributed in (observed) frequency, using the dimensionless quantity $\Omega_{GW}(f) \equiv (d\rho_{GW}/d \ln f)/\rho_{crit}$, where $\rho_{GW}$ is the energy density of GWs and $\rho_{crit}$ the critical energy density to close the Universe.

[a] When presenting results of parameter estimation campaigns we fix the total source-frame mass and distance/redshift, while randomly varying the binary mass ratio (unless otherwise specified), Black Hole's spin amplitudes, and the extrinsic parameters (i.e. inclination, polarisation angle, phase, and sky-position extracted from uniform distributions). We provide in tabular form the median fractional error and associated dispersion at 90 % (unless otherwise specified) confidence level for a selection of parameters. The error on the redshift is obtained by propagating the uncertainty on the luminosity distance, using standard cosmology [339]. Waveforms, computed and used in the analysis, are an essential tool for LISA science recovery.

---

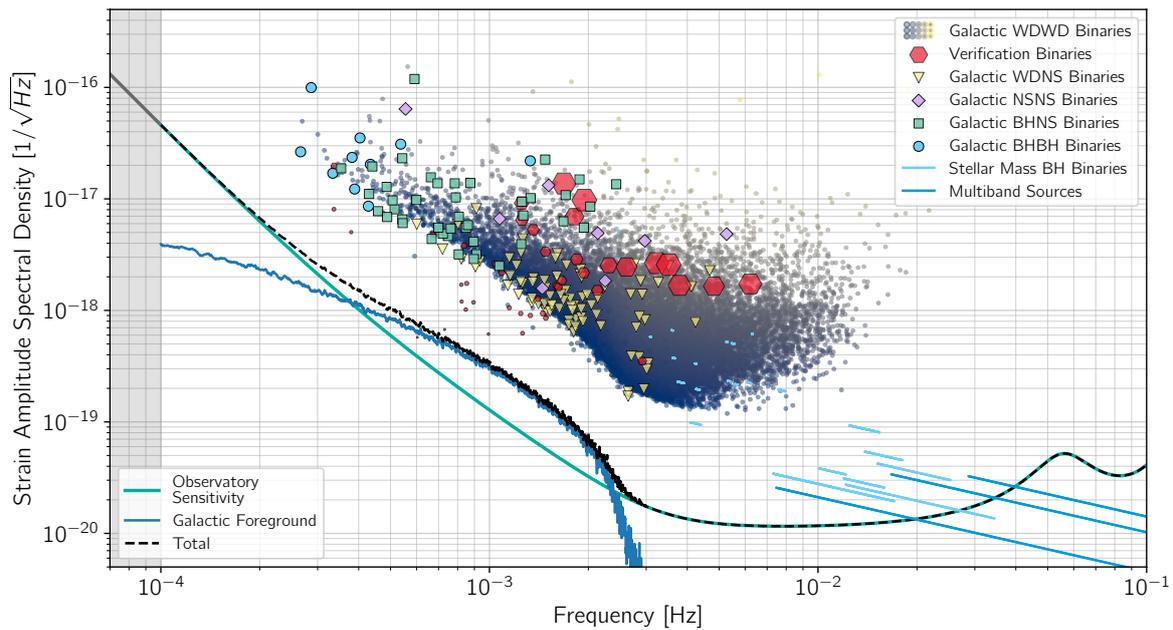

**Figure 3.1:** Simulations of different classes of resolved Galactic binaries with symbols as in the legend based on [259, 406, 421]. Hexagons denote the verification binaries (VBs) with size indicating the signal-to-noise ratio (SNR). Solid teal, solid blue and dashed black lines denote the LISA's instrumental strain amplitude, the stochastic Galactic foreground, and their sum, respectively. The displacement of the cloud of resolvable sources above the noise is due to the detection threshold being set to SNR=7. High-$f$, light- dark- blue lines show tracks of chirping extragalactic stellar-mass Black Hole binaries (sBHBs) described in Section 3.4
.

## 3.1 SO1: Study the formation and evolution of compact binary stars and the structure of the Milky Way Galaxy

### Binary and multiple systems in the Milky Way

Binary stars often follow a complex evolution which is responsible for many high-energy and transient phenomena. Stellar compact binaries will be the most numerous GW sources in the LISA band; many are dark, invisible in electro-magnetic radiation, while some will be multimessenger sources. The bulk of the population in the Milky Way will form a detectable stochastic Galactic foreground. The LISA survey of compact binaries will improve our understanding of binary interactions, currently poorly understood, and will probe the structure of the Milky Way in a complementary way to electro-magnetic (EM) probes.

Observations show that virtually all massive stars and a significant fraction of low-mass stars are multiples (binaries, triples etc). Many of these systems have orbits tight enough that the stars interact at some point during their evolution. These interactions change the course of evolution of the stars and lead to important phenomena, such as X-ray binaries, type Ia supernovae, novae, gamma-ray bursts, and stellar and compact-object mergers [e. g. 235]. Through these processes, binaries inject energy into the host galaxy, and contribute to the synthesis of heavy nuclei and metal pollution, important for the chemical evolution of the Universe. At the same time, binaries are a laboratory for extreme physics, such as the structure of compact objects, the production of relativistic outflows and of measurable GWs [e. g. 261, 343].

Based on the current observational knowledge, coupled with best understanding of binary evolution and Galactic structure, a large number (about $10^7$) of binary and multiple sources is expected to emit GWs in the LISA band [e. g. 112, 269, 320, 365] (see Figure 3.1). Their orbital periods range from a few hours to several minutes. Due to the relatively weak GW emission, the vast majority contributes to a stochastic Galactic foreground. But, about $10^4$ are detectable as continuous, nearly monochromatic individual sources in the Milky Way. A small fraction of these (up to a few percent) will come from the Magellanic Clouds, and a handful from other nearby satellite galaxies [258]. This will allow new measurements of the Galactic structure and increase our knowledge of this population [Section 3.1.2, 259].

The expected types of stellar multiples detected by LISA are listed below; more details are presented in [20]:

- Detached and interacting double White Dwarfs (WDs) are the remnants of low- or intermediate-mass stars (95 % of all stars). From the millions of expected systems in the Milky Way, about $10^4$ will be detected individually by LISA but are dark, several hundreds have EM counterparts that can be used for joint GW and EM studies [378].
- There are a few tens of LISA sources already known, the so called verification binaries (VBs) [182, 262]. Some of them can be detected within several months and will play an important role in the science verification of the LISA data.
- More massive stars form Neutron Stars (NSs) or Black Holes (BHs). Several tens to hundreds of systems are expected to be found in the Milky Way by LISA, including binaries of Neutron Stars or stellar-mass Black Holes with White Dwarfs.
- Any type of system with two objects in a tight enough orbit is a potential LISA source, such as helium burning stars and low-mass main sequence stars. In addition, triple stars, or even higher multiples will be found, most likely in a hierarchical structure. The presence of exoplanets around double WD systems can be detected if they exist, providing a complementary way to study planets, typically at large distances from the Earth [391].

The combined signals of unresolved binaries will form a stochastic Galactic foreground signal, that is sometimes referred to as confusion noise. It is concentrated towards the Galactic centre and thus is non-isotropic, leading to a signal that is modulated over the year due to the LISA motion. It will be detected, and its amplitude characterised at the per cent level, and subtracted from the data [171, see also Section 3.7].

## 3.1.1 Formation and evolution pathways of dark compact binary stars in the Milky Way and in neighbouring galaxies

**Science Investigation – SI 1.1** – aims to probe the evolutionary pathways leading to the observed binaries and answer the following questions:

> - **How do binary stars evolve into close double compact systems?**
> - **What is their current merger rate in the Milky Way?**

LISA will measure the GW frequency and amplitude, and for a fraction of sources the chirp mass and distance of systems mostly invisible to EM. After some months of observations the first set of new sources will be detected, together with a number of VBs, increasing to about $10^4$ at the end of the LISA nominal mission.

Mass transfer is the main driver of early binary evolution, happening typically when one of the stars expands [e. g. 428]. The mass transferred to the companion star cannot always be accreted and material and angular momentum are lost from the system, changing the orbit, which will increase or decrease the mass transfer. Increased mass transfer likely results in a common-envelope configuration: the companion is engulfed in the envelope of the giant which leads to strong orbital shrinkage. This complex phase is poorly understood but the existence of close double compact objects and mergers is a strong indication of its existence [see 242]. For massive stars that form NSs or Black Holes (BHs), there are additional uncertainties: whether a star forms a NS or a BH is largely unknown [e. g. 421]. Even more, VLBI measurements of magnetised NSs observed as radio pulsars show that some NSs acquire high natal velocities, which implies that at least some NSs form in an asymmetric collapse that imparts a kick at birth [238]. The kick velocity distribution, and whether (some) BHs also receive a kick, are not well known and can have a large influence on the formation of tight compact NS and BH binaries, since large kicks tend to disrupt the binary [e. g. 199]. Survival of mass transfer, common envelope and kicks can lead to the formation of close binaries consisting of WDs, NSs or BHs depending on the mass, which emit GWs in the LISA band [e. g. 320].

Subsequent energy and angular momentum loss due to the emission of GWs will bring the compact objects closer and closer together until they merge, leading to explosive phenomena such as type Ia supernovae or kilonovae [e. g. 20, 428]. If at least one of the stars is a WD, merger occasionally is avoided leading to an ultra-compact interacting binary [20] that evolves back to longer periods and lower GW frequencies. The fate of binary systems is firmly anchored in the measured properties of the detected population of compact binary systems in the Milky Way.

> **SI 1.1** aims to detect GWs from the population of compact binaries in the Milky Way to study the evolution leading up to the formation of close double compact objects, including common envelopes, NS and BH formation and kicks. The measurements can be used to directly infer the merger rates of WDs, NSs and BHs in our Galaxy and thus better constrain the explosive events associated with these mergers.

**Nature of the measurements** – LISA will detect the Galactic foreground and a very large number of continuous sources in the Milky Way: several (tens of) thousands of double WDs and several tens to hundreds of NS and BH binaries [see 20]. Table 3.1 shows some typical final uncertainties of sources at a distance of the Galactic centre, at different frequencies. Detection and initial parameter estimation takes at least some months after which the uncertainties decrease over time.

**Diagnostics on the evolution of binaries** – The merger rates of the different source classes can be inferred by LISA, since the number of systems as function of frequency is directly related to the rate at which they merge, since the observed GW frequency provides the time to merger. For WD binaries this will be determined at the 1 % level, for NSs and BHs binaries at the several tens of percent level. Table 3.1 shows that for frequencies above ~2 mHz (orbital periods below 17 min) the parameters are well determined. The number of detections and the distribution of source properties for the different classes can be compared to sophisticated models to reconstruct the previous evolution and put constraints on the uncertain physics [e. g. 112, 269, 320, 365]. The level and shape of the unresolved foreground can be compared to the same models.

**Table 3.1:** Parameter estimation accuracies from LISA observations of Galactic stellar sources at the Galactic centre distance of ~8 kpc, with random inclination, polarisation and initial phase. Columns report the median and the 68 % ranges of the fractional uncertainty on the observed frequency $f$, the amplitude $\mathcal{A}$, luminosity distance $d_L$, observed chirp mass $\mathcal{M}$ and the sky error at different frequencies. We assume WD+WD has masses $0.4\,\mathrm{M_\odot} + 0.3\,\mathrm{M_\odot}$, NS+NS: $1.35\,\mathrm{M_\odot} + 1.3\,\mathrm{M_\odot}$ and BH+BH: $15\,\mathrm{M_\odot} + 25\,\mathrm{M_\odot}$. Above ~2 mHz the parameters are well determined.

| $f$ (mHz) | Source | $\mathcal{M}$ ($\mathrm{M_\odot}$) | $\dfrac{\Delta f}{f}$ | $\dfrac{\Delta \mathcal{A}}{\mathcal{A}}$ | $\dfrac{\Delta d_L}{d_L}$ | $\dfrac{\Delta \mathcal{M}}{\mathcal{M}}$ | $\Delta\Omega$ ($deg^2$) |
|---|---|---|---|---|---|---|---|
| 1 | WD + WD | 0.30 | $2.2^{+1.1}_{-0.7}\,10^{-5}$ | $>1$ | $>1$ | $>1$ | $2.0^{+2.5}_{-1.1}\,10^{4}$ |
|  | NS + NS | 1.15 | $2.4^{+1.1}_{-0.8}\,10^{-6}$ | $0.37^{+0.63}_{-0.12}$ | $>1$ | $>1$ | $2.3^{+2.8}_{-1.3}\,10^{2}$ |
|  | BH + BH | 16.7 | $2.8^{+1.3}_{-0.9}\,10^{-8}$ | $4.1^{+8.9}_{-1.6}\,10^{-3}$ | $4.2^{+2.5}_{-1.1}\,10^{-3}$ | $3.8 \pm 1.2\,10^{-4}$ | $3.2^{+1.9}_{-0.2}\,10^{-2}$ |
| 3 | WD + WD | 0.30 | $2.5^{+1.3}_{-0.8}\,10^{-7}$ | $1.1^{+2.4}_{-0.3}\,10^{-1}$ | $3.2^{+0.8}_{-0.6}\,10^{-1}$ | $0.15 \pm 0.05$ | $6.9^{+9.1}_{-3.8}$ |
|  | NS + NS | 1.15 | $2.7^{+1.3}_{-0.9}\,10^{-8}$ | $1.2^{+2.2}_{-0.4}\,10^{-2}$ | $1.2^{+2.2}_{-0.3}\,10^{-2}$ | $1.7 \pm 0.5\,10^{-3}$ | $8.1^{+9.9}_{-4.5}\,10^{-2}$ |
|  | BH + BH | 16.7 | $1.1^{+0.4}_{-0.4}\,10^{-9}$ | $1.4^{+2.6}_{-0.4}\,10^{-4}$ | $1.4^{+2.6}_{-0.4}\,10^{-4}$ | $4.4 \pm 1.3\,10^{-7}$ | $1.3^{+1.4}_{-0.7}\,10^{-5}$ |
| 5 | WD + WD | 0.30 | $7.7^{3.3}_{-2.6}\,10^{-8}$ | $5.4^{+11}_{-1.5}\,10^{-2}$ | $5.7^{+11}_{-1.0}\,10^{-2}$ | $0.012 \pm 0.004$ | $8.4^{+10}_{-4.7}\,10^{-1}$ |
|  | NS + NS | 1.15 | $8.0^{+4.0}_{-2.6}\,10^{-9}$ | $6.0^{+12}_{-1.8}\,10^{-3}$ | $6.0^{+12}_{-1.8}\,10^{-3}$ | $1.3 \pm 0.4\,10^{-5}$ | $9.0^{+11}_{-4.9}\,10^{-3}$ |
|  | BH + BH | 16.7 | $4.2^{+2.0}_{-1.4}\,10^{-10}$ | $7.0^{+15}_{-2.2}\,10^{-5}$ | $7.0^{+15}_{-2.2}\,10^{-5}$ | $6.7 \pm 2.1\,10^{-7}$ | $1.2^{+1.3}_{-0.7}\,10^{-6}$ |

> LISA will uniquely probe the population of binary systems near the end of their evolution, detecting (tens of) thousands of systems with very short periods that are invisible to EM instruments. The parameters of many hundreds of these systems are determined with high accuracy, allowing tests of binary evolution models and improved determination of the merger rate of compact object binaries.

## 3.1.2     The Milky Way mass distribution

Being intrinsically dim and physically small, WDs are difficult to find at large distances with EM telescopes. **Science Investigation SI 1.2** aims at exploiting LISA's measurements, which are expected to be almost complete for orbital periods of less than ~15 min up to distances of ~50 kpc. This will enable the investigation of the following key questions:

> ▪ **What is the spatial distribution of ultra-compact binaries detected by LISA that are too dim for detection with EM telescopes?**
> ▪ **What can we learn from double WDs about the structure of the Milky Way as a whole?**

The Milky Way is the only galaxy whose formation history can be studied using detailed distributions of different types of stars and with this aim, large surveys have been carried out across the entire sky [100, 190]. Still, large parts of our Galaxy remain uncharted as dust extinction, stellar crowding, and severe selection effects with the distance hamper our ability to map the Milky Way at low Galactic latitudes and large distances (e. g. Galactic centre and beyond). By measuring the amplitude of a GW, which scales as $1/d_L$, LISA is sensitive to double WDs in the entire Milky Way and, thus, will provide a unique opportunity for mapping our Galaxy, complementing EM maps [262], as shown in Figure 3.2.

Similarly to other spiral galaxies, the Milky Way has recognisable stellar components: central bulge/bar, thin and thick disks and halo. These can be described by density profiles which are useful when comparing our Galaxy with other galaxies and with cosmological simulations. The profile parameters can be derived from the density distribution of individually resolved LISA detections. Moreover, building upon the analogy with stellar population models used to infer stellar masses of galaxies based on their total light, the total stellar mass of the Milky Way can be inferred from the number of LISA sources and the level of unresolved Galactic foreground based on binary population models [194, 260].

**Nature of the measurements** – LISA's measurements of 3D positions (sky location and distance) of resolved double WDs will be the key for characterising the shape of the Milky Way.

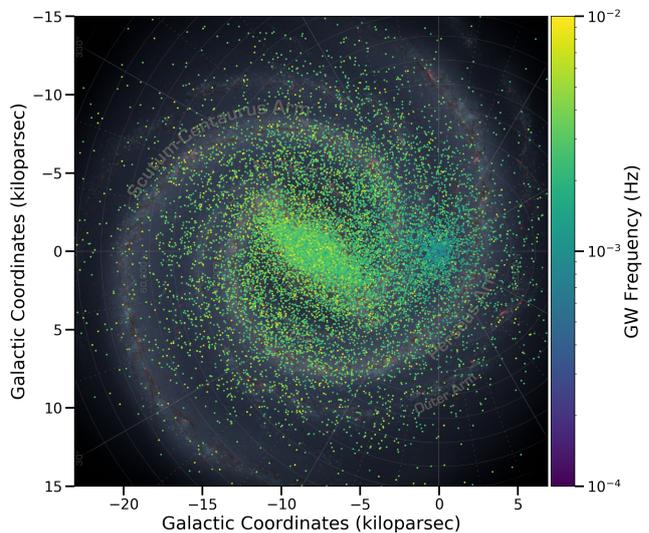

**Figure 3.2:** Spatial distribution of numerous individual double WDs (dots, coloured according to GW frequency) projected onto the Galactic plane, from [20]. LISA will enable mapping our Galaxy. LISA's position is at (0, 0). In the background an artist impression of the Milky Way.

Several thousands of these measurements will have distance accuracy better than 30 % and sky localisation of less than a few square degree, allowing determination of the underlying stellar density. More than a thousand of these are located beyond the Galactic centre (Figure 3.2). Scale parameters encoding the density profile of the bulge/bar and disk can be derived to an accuracy of ~10 % without additional modelling [259, 431].

> LISA will survey WD binaries with orbital periods of less than ~15 minutes across the entire Milky Way volume, allowing us to reconstruct the stellar mass distribution of our Galaxy using evolved stars invisible to EM telescopes.

### 3.1.3 The interplay between gravitational waves and tidal dissipation

To study the interplay between GW, mass transfer and tides, **Science Investigation – SI 1.3** – aims at detailed examination of the shortest period systems in order to answer the following questions:

> - **What fraction of detached ultra-compact binaries evolve into interacting binaries and avoid merger?**
> - **What is the role of mass transfer, tides and GWs in the merger of systems, and what does it tell us about the explosion mechanism of type Ia supernovae?**

When double WD binaries spiral-in due to GW radiation, they reach a point where a WD starts transferring mass to its companion. At that moment, the radius of the WD is comparable to the binary separation and tides become important. The interplay between the mass transfer, GW emission and tides, which are poorly understood, determine the final fate of the system: merger or transition into a stable interacting binary [e. g. 296]. This outcome is currently highly uncertain and it is crucial for our understanding of several transient phenomena including type Ia supernovae.

Joint GW and EM observations are important in this final phase. Note that here the EM observations do not have to be simultaneously with the GW measurements. The tidal interaction influences both the GW evolution (by changing the phase evolution) as well as the EM brightness (by deforming and heating WDs). GWs will allow us to measure the chirp mass, whereas EM observations can provide the mass ratio, yielding a measurement of the individual masses. This enables a direct comparison between the rate of orbital decay observed in GWs or EM radiation (for eclipsing and/or tidally distorted systems) and the respective rate from GWs only, allowing to disentangle the contribution of GWs and tidal effects [378]. A merger would lead to transient EM emission, in the most extreme case, a type Ia supernova. Therefore, GW and EM observations have a unique potential to provide insight into tides and measure the branching ratio between mergers and transition into stable mass transfer [295]. The latter can also be derived from the measurement of the ratio of the number of inspiralling to interacting systems.

> **SI 1.3** aims to determine the branching ratio between mergers and transition into stable mass transfer and at providing insight into the physics of tides in white dwarfs.

**Nature of the measurements** – LISA will measure the number and frequency evolution of those binaries evolving towards mass transfer and those evolving away, and thus directly determine the branching ratio between merger and transition into stable mass transfer. With expected numbers of both at least in the hundreds, this ratio can be determined to better than 10%. In addition, a number of binaries will be detected jointly with GW and EM instruments, allowing complementary measurements. Finally, GW results can be combined with statistics of EM bright transients originating from compact object mergers.

**Diagnostics on merger physics, mass transfer stability and tides** – The measured population can be compared to models to constrain these processes. In addition, joint GW and EM measurements can be directly compared to models with different assumptions about the effectiveness of tides. Finally, the combination of GW measurements with statistics of EM bright transients can be compared to simulated populations constructed with different assumptions.

> LISA measurements of systems at high frequencies ($f \gtrsim 2\,\text{mHz}$) will uniquely probe the uncertain physics of tides, mass transfer and WD mergers, in particular when combined with EM measurements.

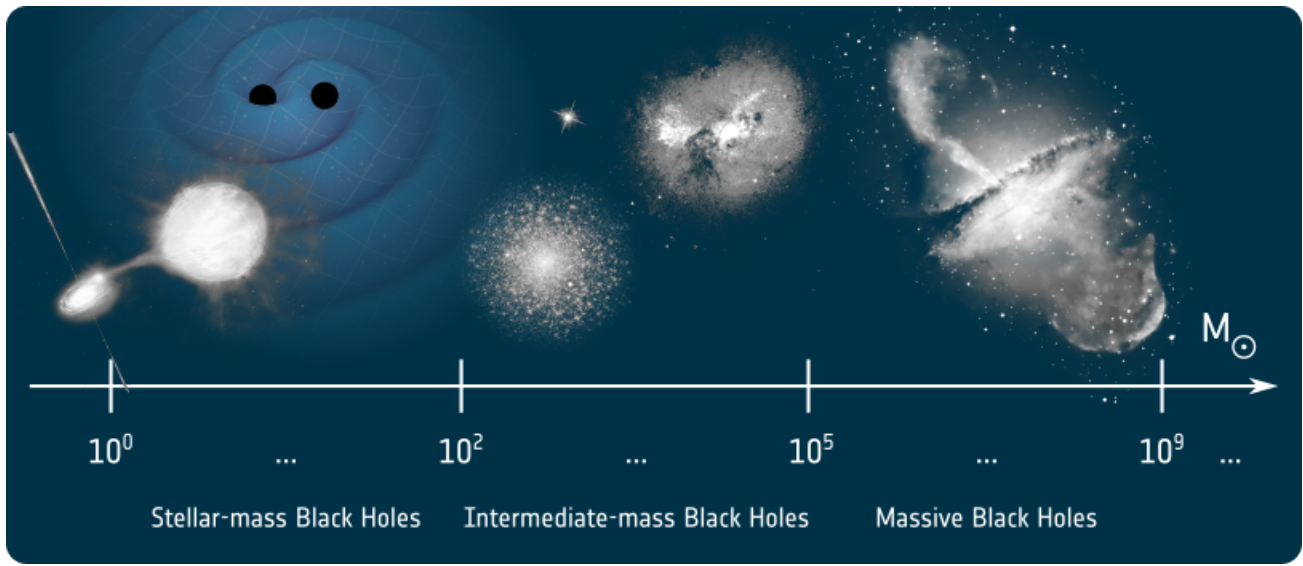

**Figure 3.3:** Black-hole types, their mass ranges and possible environments (not to scale).

## 3.2 SO2: Trace the origins, growth and merger histories of massive Black Holes

> **Massive Black Hole binaries in the gravitational universe**
>
> LISA will detect the gravitational wave (GW) signals from merging massive Black Hole binaries (MBHBs) in the largely unexplored segment of the mass spectrum, between about $10^4\,M_\odot$ and $10^7\,M_\odot$ and out to very high redshifts, when the first galaxies are forming. Unveiling the origin and evolution of Massive Black Holes (MBHs) along cosmic history is a key objective of contemporary astrophysics. Knowledge is acquired through the measurement of the masses and spins of the merging MBHBs, and the luminosity distance imprinted in the GW signal.

**Astrophysical Black Holes in the cosmic landscape** – Astrophysical BHs are currently known to come in two distinct families: stellar-mass Black Holes (sBHs), linked to the life-cycle of massive stars, widespread in all galaxies, and Massive Black Holes (MBHs) at the centres of galaxies with masses from $10^6\,M_\odot$ to $10^9\,M_\odot$ or even larger. MBHs can be observed indirectly by tracking the motion of stars and/or gas in close orbits around a massive dark object, or by detecting the EM emission, released close to the event horizon when matter dissipates the energy stored in the highly curved space-time through magnetic and viscous processes. Chief examples are the MBH at the Galactic centre [176, 193, 197] and that in the galaxy M87 [175]. In this phase they are called quasars or less-luminous Active Galactic Nuclei (AGN). A further class mostly elusive to EM observations is that of Intermediate-Mass Black Holes (IMBHs) with nominal masses around $10^2$–$10^5\,M_\odot$ [208, 420]. IMBHs bridge the gap between the high-mass end of the sBHs and the low-mass end of MBHs (see Figure 3.3). Their existence, origin, formation redshift and evolution is largely unknown, and GW observatories will shed new light on this population. LISA in particular is sensitive to IMBHs between $10^3\,M_\odot$ and $10^5\,M_\odot$ (depending on the source's redshift; see Figure 3.5).

Most of what we know about MBHs thus far has been informed by observations of quiescent (single) MBHs, which allowed to discover their co-evolution with the galaxy they inhabit [257], and by AGN, present at *cosmic noon*, around redshift $z \sim 2$, when the star formation rate and AGN activity reach their peak [288]. EM observations further reveal the existence of rare, distant quasars at redshift $z \sim 6 - 7$ [439], shining when the Universe was only 700–800 Myrs old and the neutral baryons of the intergalactic medium were still turning into a tenuous plasma, completing *cosmic reionisation*. They have masses around $10^9$ M$_\odot$ that can not form directly from the gravitational collapse of gigantic gas clouds of such mass or any other object, and thus pose questions on their rapid mass growth from *seeds* of about $10^2$ to $10^4$ M$_\odot$, on time-scales of less than a billion years. MBHs are expected all to be born from *seeds*, and current EM observations indicate that their present-day mass is acquired by accretion and to a lesser extent by mergers. Early Black Hole growth is now being unveiled by JWST [372] with the discovery of a vigorously accreting $10^6$ M$_\odot$ MBH at $z \sim 10$ [289]. In addition JWST is now observing a rich population of AGN with masses around $10^6$–$10^7$ M$_\odot$ in the redshift interval $4 < z < 11$, with evidence of candidate merging MBHs around $z \sim 4$ [290]. LISA will be sensitive to mergers of systems in this mass and redshift range.

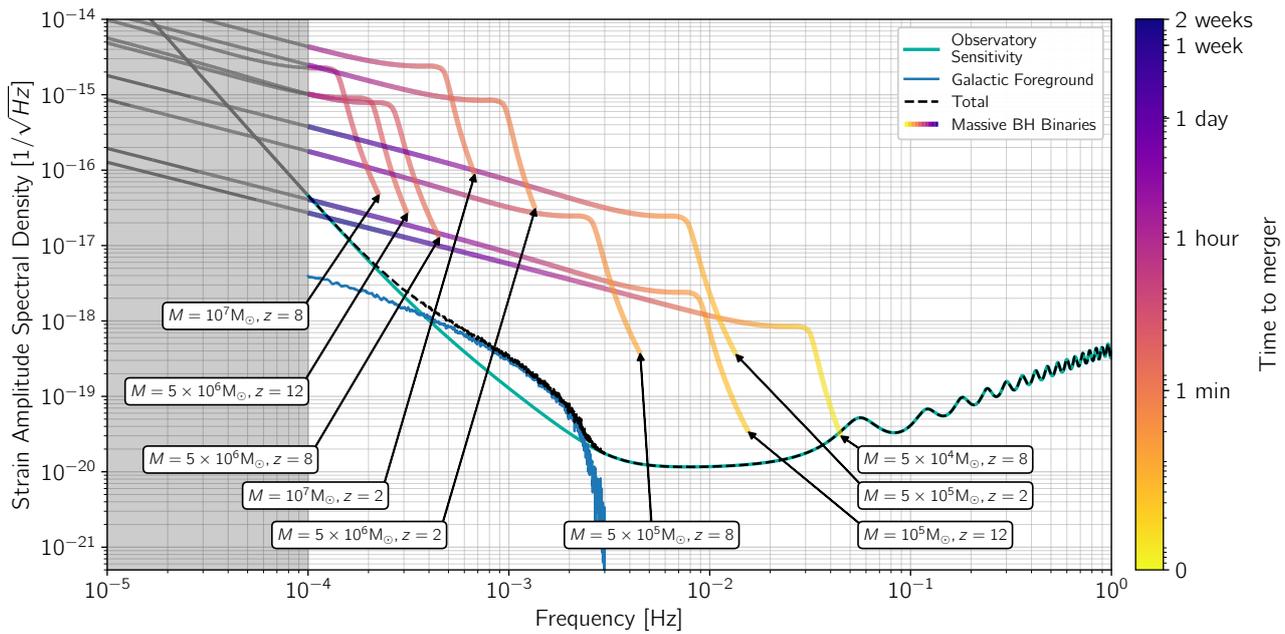

**Figure 3.4:** Strain amplitude spectral density of the GW signal of merging MBHBs versus the observer-frame frequency, for systems studied in Tables 3.2, 3.3, 3.4, with mass ratio 0.5. Source-frame masses and redshifts are indicated in round boxes (coloured traces). Lines are colour coded according to the time to merger at each frequency. Solid teal, solid blue and dashed black lines denote LISA's instrumental strain amplitude, the Galactic foreground, and their sum, respectively. The grey shaded area is the extrapolation of LISA's instrumental noise below 0.1 mHz.

**MBHBs as LISA transient sources and the nature of the measurements** – LISA will detect the Inspiral-Merger-Ringdown (IMR) signal of MBHBs with total mass from $10^5$ M$_\odot$ up to about $10^7$ M$_\odot$, and the inspiral-only of binaries of a few $10^3$–$10^4$ M$_\odot$. These sources become detectable by LISA weeks/days/hours before the merger, depending on mass and redshift, and become quiescence shortly after coalescence, forming a class of *transient* sources, in contrast to the continuous sources such as Galactic binaries and EMRIs. Figure 3.4 shows IMR tracks of merging MBHB as a function of the GW frequency $f$ as measured in the LISA-frame. Figure 3.5 shows LISA's cosmic horizon, which extends even beyond $z \sim 12$. Thus, LISA will let us pierce deep into the epoch of *cosmic reionisation*, when the first stars, Black Holes and galaxies are forming.

**The cosmic history of massive Black Hole binaries**

LISA science objectives for MBHs are rooted in our current model of cosmic structure. Participating in the formation, growth, and assembly of galaxies, MBHs pair in colliding galaxies, forming *binaries* which then merge, becoming strong sources of GWs. With masses between about $10^4\,M_\odot$ to a few $10^7\,M_\odot$ MBHBs invariably cross the LISA frequency bandwidth (see Figure 3.4 and Figure 3.5). They are the least studied population in terms of basic demographics, birth, growth, dynamics, and connection to their galaxy host. We also do not know how their mass function extends and evolves with redshift, how their spins shaped by mergers and accretion episodes evolve, and whether there is a physical and evolutionary connection with the population of stellar-mass Black Holes (sBHs). GW observations will be key to unveiling the existence of, and the physics governing, MBHs throughout cosmic history. LISA will detect for the first time GWs from merging MBHBs, some with extremely high signal-to-noise ratios (SNRs), carrying out the first census of this new population.

### 3.2.1 Discover seed Black Holes at cosmic reionisation

MBHs are not massive at birth, but grow by several orders of magnitude from *seed Black Holes* [215, 420]. This **Science Investigation – SI 2.1 –** aims to investigate the origins of MBHs to answer the following questions:

> ▪ **How were MBHs born and how did they grow?**
> ▪ **What is the nature of the seed masses and how and when did they form?**
> ▪ **Are sBHs the only elementary building blocks of MBHs?**

Theoretical investigations indicate that *seed Black Holes* may have been born with masses from a few $100\,M_\odot$ to about $10^4\,M_\odot$, straddling the two observed families of sBHs and MBHs, in what is currently an observational "desert". Seeds have been hypothesised to belong to the class of sBHs and/or IMBHs. They formed at early cosmic times and grew into MBHs by accretion of stars and gas as well as by mergers with other seeds, participating in the assembly of cosmic structures [385]. Some seeds evolved in "starvation" with very little growth, and could be observed as accreting or merging IMBHs in local galaxies today [31, Section 3.3.2]. The observation of accreting seeds with masses below $10^5\,M_\odot$ at $z \gtrsim 10$ is challenging with current EM facilities [206, 319, 412]. Thus, GW observations are a way to discover seeds and their early growth.

Various paths of seed formation and growth have been proposed, and multiple processes may have operated [240, 420]. Seeds can have *primordial origin* [125], resulting from the collapse of high-contrast density perturbations during phase transitions in the very early Universe, and form before galaxies do. In these models, their initial mass has no lower bound and can be as large as $\sim\!10^5\,M_\odot$. *Light seeds* form through mechanisms rooted in known stellar physics, but extrapolated to untested regimes inside the earliest-forming dark matter halos of $\sim\!10^6\,M_\odot$. Light seeds resulting from the gravitational collapse of the first metal-free stars at $z \sim 30-15$ are expected to have masses between $10\,M_\odot$ and a few times $10^2\,M_\odot$. *Heavy seeds* result from the direct collapse of single supermassive stars of $10^4$–$10^5\,M_\odot$ forming around $z \sim 10-15$ in more massive halos ($10^8\,M_\odot$ or larger). Additional pathways have been proposed, which call for *stellar runaway collisions* in dense metal-poor star clusters [385]. Typical masses of these *runaway seeds* are up to several $10^3\,M_\odot$, possibly up to $10^4$–$10^5\,M_\odot$ [240]. Likewise *hierarchical Black Hole mergers* in dense nuclear star clusters with high escape velocities can lead to the formation of Black Holes of $10^2$–$10^5\,M_\odot$ [27].

*Binary seeds*, fated to become GW sources, may form *in situ* via fission of rapidly rotating supermassive stars [357] or via dynamical interactions in fragmenting proto-stellar discs [215], and for the heaviest seed via halo-halo mergers [20, 412]. Binary formation introduces the concept of *delayed coalescence* as a finite cosmic time elapses between formation and merger, depending on the formation process, initial separation and interaction of the binary with the environment [20, 418]. The net effect is to reduce merger rates, shifting detectable events to lower redshifts [254].

> **SI 2.1** aims to detect the GW signals from the earliest MBHBs in the mass interval between about $5 \times 10^3\,M_\odot$ and $5 \times 10^6\,M_\odot$, as measured in the source-frame, and at formation redshifts between $10 \lesssim z \lesssim 15$, to inform us about the physics producing the seeds and their early growth and assembly. This will provide knowledge of this pristine population that will anchor the initial conditions of MBH cosmic evolution.

In Table 3.2 we collect the results of the parameter estimation carried out on a family of binaries with fixed source-frame total mass consistent with the astrophysically-informed model by [254]. We let the mass ratio vary to account for the uncertainties on this parameter related to the way MBHs pair in close binaries during galaxy collisions [20, 70, 106]. Table 3.2 contains mostly "threshold" sources with median errors on the masses in the source frame between 1 % and 100 %. Binaries with masses equal and below $10^4\,M_\odot$ have an SNR near the detection threshold.

**Table 3.2:** Parameter estimation accuracies from LISA observations of non-spinning binaries at redshift $z = 12$ ($d_L = 130\,\mathrm{Gpc}$ − age of the Universe of 0.372 Gyr) with source-frame total masses given in the first column. The mass ratio $q$ in the range $q \in [0.3, 1]$ is sampled uniformly in $1/q$, and polarisation and sky-position angles are sampled uniformly from the sphere. The median fractional errors and dispersions refer to the source-frame individual masses $m_1$ and $m_2$, total initial mass $M$, mass ratio $q$, luminosity distance $d_L$, and redshift $z$. The chirp masses in the observer frame are determined with fractional errors at sub-percent level, with the exception of the heaviest sources having uncertainties up to a few %. Fractional errors exceeding 100% are associated to weaker sources with low mass ratio or/and placed in unfavourable positions in the LISA sky.

| Total Mass $M_\odot$ | $\left(\dfrac{\Delta m_1}{m_1}\right)_{\mathrm{sf}}$ | $\left(\dfrac{\Delta m_2}{m_2}\right)_{\mathrm{sf}}$ | $\left(\dfrac{\Delta M}{M}\right)_{\mathrm{sf}}$ | $\dfrac{\Delta q}{q}$ | $\dfrac{\Delta d_L}{d_L}$ | $\dfrac{\Delta z}{z}$ | SNR |
|---|---|---|---|---|---|---|---|
| $5 \times 10^3$ | $0.6^{+1.4}_{-0.4}$ | $0.6^{+1.6}_{-0.4}$ | $0.33^{+0.72}_{-0.26}$ | $0.5^{+0.8}_{-0.3}$ | $0.6^{+1.0}_{-0.4}$ | $0.7^{+0.9}_{-0.5}$ | $13.28^{+14.79}_{-7.02}$ |
| $1 \times 10^4$ | $0.4^{+1.0}_{-0.2}$ | $0.4^{+1.1}_{-0.3}$ | $0.4^{+0.6}_{-0.3}$ | $0.3^{+0.5}_{-0.2}$ | $0.4^{+0.8}_{-0.3}$ | $0.43^{+1.01}_{-0.33}$ | $23.06^{+25.29}_{-12.16}$ |
| $1 \times 10^5$ | $0.08^{+0.84}_{-0.06}$ | $0.09^{+0.85}_{-0.07}$ | $0.05^{+0.08}_{-0.04}$ | $0.03^{+0.02}_{-0.01}$ | $0.10^{+0.82}_{-0.07}$ | $0.09^{+1.11}_{-0.08}$ | $131.876^{+147.008}_{-69.544}$ |
| $1 \times 10^6$ | $0.80^{+4.61}_{-0.75}$ | $0.80^{+4.66}_{-0.75}$ | $0.2^{+0.6}_{-0.1}$ | $0.04^{+0.10}_{-0.02}$ | $0.82^{+10.92}_{-0.77}$ | $1.00^{+0.74}_{-0.95}$ | $65.04^{+61.93}_{-32.67}$ |

> The BH binaries that LISA can identify before cosmic reionisation are those with total masses centred around $10^5\,M_\odot$ (see Figure 3.5 and Figure 3.4). The number of detections and source-frame properties are our diagnostic tool to infer statistically the physical model of seed formation.

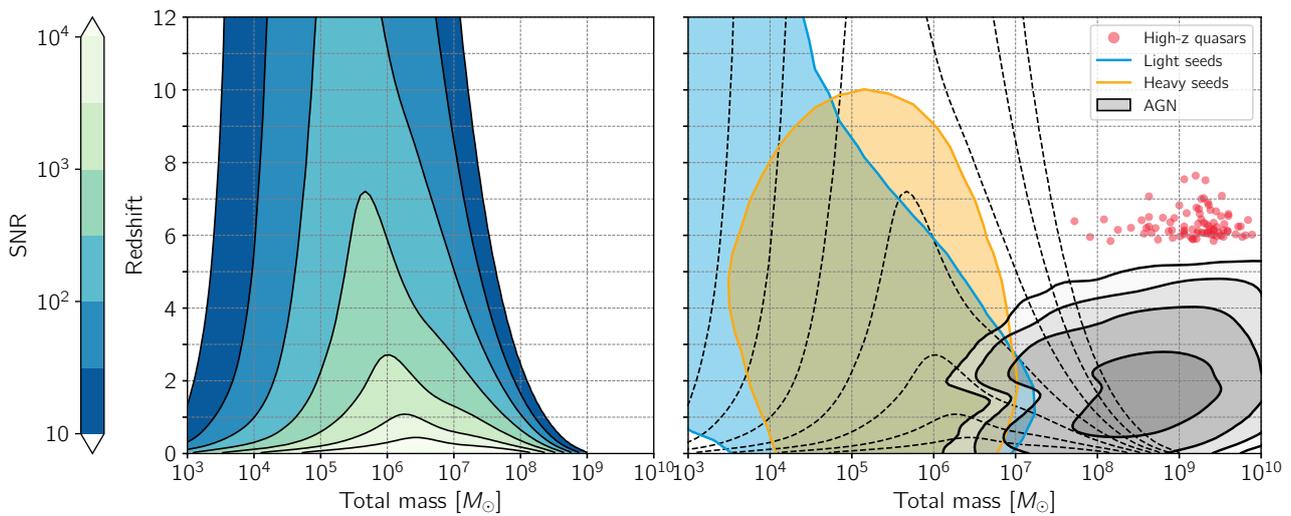

**Figure 3.5:** Left panel: contour lines of constant SNR from MBHBs. Binaries have mass ratio $q = M_1/M_2 = 0.5$ and aligned spins of amplitude $\chi_{1,2} = 0.2$. The SNR, shown in the colour bar is the result of sky, polarisation and inclination averaged for the dominant quadrupole mode. Right panel: Overlaid on the dashed contour lines of constant SNR is the distribution of the two populations of MBHB mergers resulting from *light seeds* (light blue) and *heavy seeds* (yellow) as simulated in [106]. The grey shaded areas show the current distribution of the AGN population [436], and red dots indicate the quasars observed at $6.3 < z < 7.5$ [439]. For AGN and quasars, the mass $M$ on the $x$-axis is that of a single MBH.

### 3.2.2 Study the growth mechanism and merger history of massive Black Holes from the epoch of the earliest quasars

Once the seeds are established, an extended epoch of MBH growth takes place from the time of *cosmic reionisation* around $z \sim 8$ down to the present, crossing the epoch of *cosmic noon* at $z \sim 2$. Figure 3.5 (right panel) depicts redshift and mass loci where mergers are expected to occur according to an example model [106]. Accretion [253] and mergers [71] influence in different ways the evolution of the BH mass and the spin in modulus and direction [92], informing whether accretion is chaotic, leading to randomly oriented spins, or coherent, leading to aligned spins, parallel to the binary orbital angular momentum [164].

The **Science Investigation SI 2.2** aims to investigate the growth mechanisms of MBHs, from the epoch of cosmic reionisation to the present, to answer the following questions:

> - **How do MBHs grow in mass?**
> - **How do MBH spins evolve?**
> - **How do MBHs assemble inside the cosmic web?**
> - **How efficiently do MBHs merge and when?**

> **SI 2.2** aims to detect the IMR signal from MBHB between a few $10^4\,M_\odot$ up to $\sim\!10^7\,M_\odot$, at redshift $z \lesssim 8$.

This investigation is split into two parts: **SI 2.2a** to show LISA's capability of detecting and astrophysically interpreting merger signals from MBHBs around the epoch of *cosmic reionisation*, and **SI**

**2.2b,** GW signals from *cosmic noon.* In Table 3.3 and Table 3.4 we show the results of the parameter estimation, which also highlights the nature of the measurements.

**Table 3.3:** Parameter estimation accuracies for **SI 2.2a**: binaries are at redshift $z = 8$ ($d_L = 10$ Gpc − age of the Universe of 0.648 Gyr) and have dimensionless spins aligned to the orbital angular momentum with amplitudes randomly drawn in the range [0,1]. Mass ratios, sky and polarisation angles are extracted as in Table 3.1. Uncertainties in the observed chirp masses are at sub-percent level, except for the heaviest binaries. Last column: absolute error on $\chi_{\rm eff}$, defined as the sum of the mass-weighted spins close to coalescence. SNR distributions by rows are: $85.7^{+109.9}_{-47.0}$, $467.09^{+722.70}_{-280.18}$, $169.4^{+245.7}_{-107.2}$, and $102.3^{+157.6}_{-65.3}$, respectively.

| Total Mass $M_\odot$ | $\left(\dfrac{\Delta m_1}{m_1}\right)_{\rm sf}$ | $\left(\dfrac{\Delta m_2}{m_2}\right)_{\rm sf}$ | $\left(\dfrac{\Delta M}{M}\right)_{\rm sf}$ | $\dfrac{\Delta q}{q}$ | $\dfrac{\Delta d_L}{d_L}$ | $\dfrac{\Delta z}{z}$ | $\Delta\chi_{\rm eff}$ |
|---|---|---|---|---|---|---|---|
| $5 \times 10^4$ | $0.08^{+0.37}_{-0.06}$ | $0.09^{+0.37}_{-0.07}$ | $0.10^{+0.10}_{-0.09}$ | $0.04^{+0.05}_{-0.02}$ | $0.11^{+0.43}_{-0.08}$ | $0.09^{+0.47}_{-0.07}$ | $0.005^{+0.011}_{-0.004}$ |
| $5 \times 10^5$ | $0.033^{+0.484}_{-0.030}$ | $0.033^{+0.487}_{-0.029}$ | $0.010^{+0.022}_{-0.006}$ | $0.003^{+0.007}_{-0.002}$ | $0.042^{+0.582}_{-0.038}$ | $0.04^{+0.65}_{-0.03}$ | $0.002^{+0.004}_{-0.001}$ |
| $5 \times 10^6$ | $0.2^{+2.9}_{-0.1}$ | $0.2^{+2.9}_{-0.1}$ | $0.05^{+0.12}_{-0.03}$ | $0.02^{+0.06}_{-0.01}$ | $0.20^{+3.07}_{-0.18}$ | $0.17^{+2.22}_{-0.16}$ | $0.010^{+0.039}_{-0.007}$ |
| $10^7$ | $0.34^{+3.20}_{-0.32}$ | $0.34^{+3.24}_{-0.32}$ | $0.09^{+0.29}_{-0.07}$ | $0.04^{+0.23}_{-0.03}$ | $0.43^{+4.90}_{-0.40}$ | $0.42^{+2.06}_{-0.39}$ | $0.021^{+0.095}_{-0.016}$ |

**Table 3.4:** Parameter estimation accuracies for **SI 2.2b** for families of binaries at $z = 2$ ($d_L = 16$ Gpc − age of the Universe of 3.3 Gyr) with source-frame total mass as indicated in the first column and parameter settings and fractional errors as in Table 3.3 with the addition of the accuracies on the spin of the primary MBH and binary inclination angle $\iota$. Observed chirp masses are estimated with fractional errors of a few $10^{-4}$. SNR distributions by rows are: $1133.9^{+1581.4}_{-635.6}$, $1936.5^{+2812.3}_{-1136.6}$, $1514.7^{+2116.8}_{-972.4}$, and $1040.3^{+1492.4}_{-654.8}$, respectively. For these strong GW sources merging at *cosmic noon*, parameters are determined with exquisite precision, and in particular the spin of the primary MBH. For this reason these binaries are the best sources to carry on tests of the General Theory of Relativity (GR) (see Section 3.5). Being well localised in the LISA sky at merger, they are in addition *bright standard sirens* (see Section 3.2.3).

| Total Mass $M_\odot$ | $\left(\dfrac{\Delta m_1}{m_1}\right)_{\rm sf}$ | $\left(\dfrac{\Delta m_2}{m_2}\right)_{\rm sf}$ | $\left(\dfrac{\Delta M}{M}\right)_{\rm sf}$ | $\dfrac{\Delta d_L}{d_L}$ | $\dfrac{\Delta z}{z}$ | $\Delta\chi_1$ | $\Delta\iota$ |
|---|---|---|---|---|---|---|---|
| $5 \times 10^5$ | $0.006^{+0.049}_{-0.004}$ | $0.006^{+0.049}_{-0.004}$ | $0.002^{+0.003}_{-0.001}$ | $0.010^{+0.088}_{-0.008}$ | $0.008^{+0.073}_{-0.007}$ | $0.0013^{+0.0056}_{-0.0011}$ | $0.003^{+0.013}_{-0.002}$ |
| $10^6$ | $0.005^{+0.057}_{-0.004}$ | $0.005^{+0.057}_{-0.004}$ | $0.0007^{+0.0018}_{-0.0005}$ | $0.008^{+0.103}_{-0.007}$ | $0.007^{+0.084}_{-0.006}$ | $0.0009^{+0.0037}_{-0.0007}$ | $0.002^{+0.009}_{-0.001}$ |
| $5 \times 10^6$ | $0.009^{+0.127}_{-0.008}$ | $0.009^{+0.128}_{-0.008}$ | $0.0012^{+0.0030}_{-0.0007}$ | $0.02^{+0.22}_{-0.01}$ | $0.013^{+0.196}_{-0.012}$ | $0.002^{+0.007}_{-0.001}$ | $0.0014^{+0.0099}_{-0.0010}$ |
| $10^7$ | $0.02^{+0.25}_{-0.01}$ | $0.02^{+0.25}_{-0.01}$ | $0.002^{+0.005}_{-0.001}$ | $0.028^{+0.442}_{-0.026}$ | $0.023^{+0.420}_{-0.021}$ | $0.003^{+0.011}_{-0.002}$ | $0.003^{+0.017}_{-0.002}$ |

> The analysis exemplified in Table 3.3 and Table 3.4 shows the large improvement in the parameter estimation of sources near the epoch of *cosmic reionisation* and *cosmic noon*. The source-frame total mass will be determined to a precision between 0.1 % and 30 % (for the heaviest binaries). Redshift determinations are uncertain only for the lightest and heaviest binaries at high redshift. Spins are determined with an absolute error of $\sim 10^{-3}$.

Many *dual* AGNs in merging galaxies at $10^3$–100 pc separation have been observed, but discovering *binaries* along their track to coalescence down to micro-parsec scales when GWs drive their evolution is challenging, and only candidate MBHBs have been claimed [103, 156]. But they might have been discovered through the emission of a stochastic GW background.

> The evidence of a cosmic, stochastic background of GWs in the nano-Hertz band by Pulsar Timing Array (PTA) experiments has the potential to shed new light on the formation and evolution of MBHBs with masses about $10^8$–$10^9$ M$_\odot$. The PTA data are consistent with a population of slowing inspiralling MBHBs, present at redshift $z \sim 1$, that will merge efficiently on timescales of only thousands of years or even less [25]. If this observation and the MBHBs origin will be confirmed soon, as expected, this discovery brings first, direct evidence that MBHBs merge in Nature, adding an important observational piece to the puzzle of structure formation and galaxy-BH co-evolution.

**Rates and diagnostics on the MBHB cosmic evolution** – The process of MBHB formation is believed to start with the assembly of dark matter halos, followed by galaxy mergers, which all occur on cosmological scales and across cosmic epochs. Forming a MBHB after a galaxy collision requires dissipation of orbital energy and efficient transport of angular momentum from the Galactic scale of kiloparsecs down to the micro-parsec scale, when the merger gives birth to a new, single MBH. The *coalescence time delay* between the merger of the galaxies and the merger of the two MBHs can be as short as a few 100 Myr or as long as the age of the Universe, and sometimes the MBHs never merge [20, 212]. The formation and shrinking of the binary via stellar and gaseous torques of MBHBs down to coalescence is a complex dynamical problem involving highly-coupled non-linear physical processes [137]. Given the overwhelmingly large dynamical range involved in this problem, numerical simulations and semi-analytical models are precious tools to model the avenues of MBHBs formation and final coalescence, but they are affected by large uncertainties. In most models, the merging rate peaks at $z \sim 2-4$ (dominated by sources with high SNR), shows a slow decay at lower redshifts and a tail extending up to $z \sim 10-15$ dominated by the seed/evolved population. LISA signal rates integrated over masses and redshift are expected to be between a few to about 100 per year, with redshift distribution depending on the seeding model and the time-delayed distribution [20, 70, 106, 154, 254].

> Masses, mass ratios, spin magnitudes, and luminosity distance which are encoded in the GW signal from MBHBs can be connected to the physical processes leading to coalescence, in different astrophysical settings. Through Bayesian model selection, LISA will shed light into the avenue of MBH and MBHB formation and evolution.

## 3.2.3    Identify the electromagnetic counterparts of massive Black Hole binary coalescences

MBHB coalescences may occur in gas-rich environments and the GW signal may be accompanied by EM emission [103]. This **Science Investigation – SI 2.3** – aims to answer the following questions:

> - **Can we identify the host galaxies of GW-identified coalescence events?**
> - **How does accretion proceed in the violently changing spacetime of a merger?**
> - **Which are the EM signatures of the precursor and post-merger emission?**

Linking masses, spins and luminosity distance determined by the GW signal with a joint EM detection will provide an unprecedented and unique opportunity to understand the physics of accretion and jet launching in a time-varying spacetime. If the two MBHs are surrounded by a circumbinary disc, accretion is a continuous process down to coalescence and further on [109, 180, 392]. In the *pre-merger* phase, as the MBHs in-spiral, high energy EM emission [446] is expected to rise from the mini-discs, which surround the MBHs, modulated in time with periodicies linking the MBH orbital motion with non-axisymmetric patterns present in the fluid flow [128, 139, 216, 392]. In the *post-merger* phase, from radio to X-ray emission can reveal the luminous consequences of a shock-driven readjustment of the circumbinary disc around the recoiling new MBH, witnessing the turn-on of an AGN [144, 362] or the multi-wavelength emission from a collimated jet [443]. Alerts will be issued, in order to stimulate multi-messenger observations with facilities like Vera Rubin Observatory, Nancy Grace Roman Space Telescope, Athena, and SKA.

> **SI 2.3** aims to detect joint GW and EM signals from Black Hole binary (BHB) mergers with source-frame masses between $10^5\,M_\odot$ and $\lesssim 10^7\,M_\odot$ below $z \lesssim 3$.

**Nature of the measurement** - LISA is sensitive to sources at all points on the sky, albeit with a sensitivity that varies depending on the position of the source with respect to the constellation. To build localisation information, LISA can exploit its motion around the Sun and its own rotation (see Figure 6.3) to triangulate long-lived (weeks) transient events. Its instrumental response at high frequencies also informs us about the source's location [294].

Recent studies [294] suggest modifications to the observational requirements for **SI 2.3** with respect to the LISA Proposal and the Science Requirements Document (SciRD) [153, 267], which is now split into two parts: **SI 2.3a** addresses the issue of alerts to EM observatories during the inspiral phase to search for *precursor* emission; **SI 2.3b** focus on alerts in the *post-merger* phase.

For **SI 2.3a**, evidence has been recently found that LISA localisation pattern may show multi-modality in the sky [60, 292, 294, 347], complicating the preparation of alerts during the late inspiral. Investigations [284, 338] indicate that BHBs with total masses between $10^5\,M_\odot$ and $10^6\,M_\odot$, at $z \lesssim 0.3$ can be localised with an uncertainty $\Delta\Omega \lesssim 10\,\mathrm{deg}^2$ a few days before merger and an uncertainty $\Delta\Omega \lesssim 0.4\,\mathrm{deg}^2$ a few hours before merger, allowing to issue alerts on these timescales. Post merger, these events can be localised within $0.1\,\mathrm{deg}^2$ or even less ($0.01\,\mathrm{deg}^2$ for best oriented sources). While the number of galaxies in these fields can still be very large (between 100 to 1000), selection in the X-rays helps discover the true counterpart in a field with only few AGNs present [284, 292, 338]. Due to their low redshift, these events could be rare.

> Joint GW and EM observations of BHB coalescences during the *inspiral* phase, to detect *precursor* emission and a potential EM *chirp*, are limited to the strongest, best oriented and longest-lived GW signals at redshift $z \lesssim 0.3$ with masses between $10^5\,M_\odot$ and $10^6\,M_\odot$.

Source sky localisation improves significantly in the immediate vicinity of the merger, when GW multipoles (beyond the quadrupolar) can be detected, which extend the signal to higher frequencies and enhance the SNR and LISA response. These so called high-order modes help in breaking degeneracies in the parameter estimation, preventing the appearance of multi-modality patterns in the sky for most of the sources [292, 294]. To address **SI 2.3b** in Table 3.5 we show the sky-position

→ **THE EUROPEAN SPACE AGENCY**

uncertainties $\Delta\Omega$ for binaries at $z = 1$ before and at merger. In [347] it is further shown that binaries of masses around $10^6$–$10^7$ M$_\odot$ at $z = 3$ can be localised with $0.1 \deg^2 < \Delta\Omega < 0.4 \deg^2$, if located in a favourable portion of the sky.

**Table 3.5:** LISA sky localisation uncertainties $\Delta\Omega$ (in $\deg^2$) at three different times to merger, for three families of binaries at $z = 1$ ($d_L = 6.7$ Gpc – age of the Universe of 5.9 Gyr) with source-frame mass as indicated in the first column, with parameters extracted with the same procedure as in Table 3.4.

| Total Mass | | $\Delta\Omega$ | |
| --- | --- | --- | --- |
| M$_\odot$ | 1 week | at 4 hours | at merger |
| $3 \times 10^5$ | $86.9^{+253}_{-78.3}$ | $28.5^{+137}_{-28.1}$ | $1.60^{+13.3}_{-1.59}$ |
| $3 \times 10^6$ | $3.69 \times 10^{3}{}^{+4.93\times10^4}_{-3.58\times10^3}$ | $189^{+3.79\times10^3}_{-189}$ | $0.238^{+36.8}_{-0.237}$ |
| $10^7$ | $2.53 \times 10^{5}{}^{+9.88\times10^6}_{-2.49\times10^5}$ | $632^{+4.50\times10^4}_{-630}$ | $0.942^{+481}_{-0.939}$ |

> Joint GW and EM observations after MBHB coalescence are promising below $10^7$ M$_\odot$ and below about $z \sim 3$ [347]. At higher masses GW signals are shorter-lived and sky localisation is poor. About $7 - 20$ such joint observations at, and after merger, are estimated in 4 years, enabling follow-up observations in radio and X-rays [163, 292].

## 3.3    SO3: Probe the properties and immediate environments of Black Holes in the local Universe using EMRIs and IMRIs

**Single massive Black Holes in the gravitational universe**

LISA will observe single, quiescent Massive Black Holes (MBHs), residing in the centres of galaxies and, possibly, in massive and dense star clusters. LISA will detect the gravitational wave (GW) signal emitted by stellar-mass compact objects swirling around the MBH in generic, highly relativistic, mildly eccentric orbits. Depending on the MBH-to-companion mass ratio, these sources are called either extreme mass-ratio inspirals (EMRIs), extremely mass-ratio inspirals (XMRIs), or intermediate mass-ratio inspirals (IMRIs). Their observation will provide constraints on the origins and evolution of the MBH population, including Intermediate-Mass Black Holes (IMBHs, $10^2$–$10^5$ M$_\odot$). Insights are obtained by precisely measuring the masses of the two objects, the spin of the primary BH, the orbit inclination, and eccentricity and the luminosity distance of the source. Additional insight comes from searching for the imprints of the environment in the observed signals.

The title and scope of this science objective has expanded since the LISA Science Requirements Document (SciRD) was written [153, 267], due to advances in our understanding of these systems. Furthermore, this section includes the analysis on the detectability of IMBHs, omitted in Section 3.2.

**Extreme- and Intermediate- Mass Ratio Inspirals** – MBHs of millions to billions of solar masses are found in the centres of most present-day galaxies [257]. IMBH candidates have been recently discovered in dwarf galaxies with masses around $10^4$–$10^5\,M_\odot$ [208]. There have also been claims of IMBHs residing in the centres of stellar clusters, based on measurements of cusps in the velocity dispersion profiles [286, 325], but those have been heavily disputed [270, 453]. Identifying these IMBHs is challenging, owing to the small dynamical effect that they imprint on the surrounding stars [304]. The observation of GWs emitted by a binary that resulted in a BH with mass of about $150\,M_\odot$ [8], and the detection of multi-wavelength EM emission driven by a star disrupted by an IMBH candidate in a distant galaxy [280, 281] constitute two of the few robust indications of the existence of IMBHs in the nominal interval between $10^2$ and $10^5\,M_\odot$.

MBHs and IMBHs do not live in isolation, but occupy dense stellar environments in which they are regularly interacting with stellar objects in their immediate neighbourhood. It was the observations of stars orbiting the weak radio source Sgr A* at the centre of the Milky Way Galaxy that provided compelling evidence that it is an MBH, and allowed us to determine its mass as $10^6\,M_\odot$ [102, 202], which was recently confirmed using the image of the MBH made with the Event Horizon Telescope [176].

There is abundant EM evidence for energetic dynamical encounters between MBHs and stellar objects. Main sequence stars are observed being torn apart during tidal disruption events [196], while hyper-velocity stars [113] are thought to be formed when MBHs disrupt a stellar binary. LISA will offer a new perspective into MBH–star interactions, by directly observing GWs produced by compact objects (sBHs and perhaps neutron stars and white dwarfs) and brown dwarfs in the last phase of inspiral before they are consumed.

Depending on the mass-ratio $q$, we define extreme mass-ratio inspirals (EMRIs) those systems with $10^{-6} < q < 10^{-4}$; extremely mass-ratio inspirals (XMRIs) with $q < 10^{-6}$; and intermediate mass-ratio inspirals (IMRIs) with $10^{-4} < q < 10^{-3}$. The latter generally involves an IMBH either as primary component of the binary (defined in this case as *light* IMRI), or as secondary component (defined as *heavy* IMRI). *Light* IMRIs possibly form in the centre of dense clusters, whilst EMRIs and *heavy* IMRIs occur in the centres of galaxies. The detection of GWs from these classes of sources will shed light on the properties of MBHs and IMBHs and their environments.

EMRIs and IMRIs radiate in the LISA band for a long time and undergo complex, fully relativistic orbits that depend on the properties of the under-lying spacetime and are encoded in the complex gravitational waveforms emitted (see Section 8.3.1). A typical EMRI system with $10^6\,M_\odot$ as primary and $10\,M_\odot$ as secondary BH takes approximately one year to inspiral from a starting GW frequency of 3 mHz down to plunge at around 10 mHz, during which time it generates approximately $10^5$ waveform cycles [183]. The number of cycles is high as it scales as $M^2$ (the primary mass), as $m^{-1}$ (the secondary mass) and as $f_{\min}^{-5/3}$, with $f_{\min}$ the starting frequency.

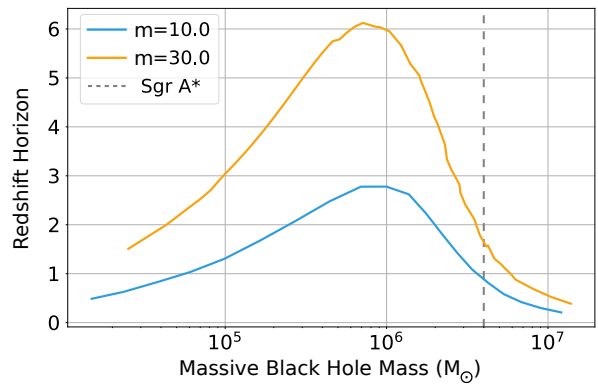

**Figure 3.6:** Sky-averaged horizon distance for LISA EMRI observations, assuming that an SNR of 20 is required for a confident EMRI detection. Orange (blue) line refers to a secondary of mass $m$ equal to $10\,M_\odot$ ($30\,M_\odot$). The horizon is computed for observations of EMRIs plunging at the end of the nominal mission lifetime. SNR is averaged over orientation, sky location and orbital inclination. Vertical line indicates Sgr A*

In the same way that observations of the dynamics of the S-stars near Sgr A* have provided a ~10 % measurement of the mass of the Milky Way MBH, LISA observations of the dynamics of stellar objects falling into extragalactic MBHs, imprinted in these long GW signals, can provide measurements of the BH properties at redshifts $z \lesssim 3$. Figure 3.6 shows the *horizon redshift* for EMRI observations with LISA, as a function of the central object mass $M$. LISA is most sensitive to EMRI systems with mass comparable to, but a factor of a few to ten times lower than Sgr A*. Since these waves originate much closer to the MBH than the S-stars orbiting Sgr A*, the resulting measurements will be orders of magnitude more precise. EMRIs and IMRIs thus offer a unique opportunity to test fundamental physics (see Section 3.5), to measure cosmological parameters (see Section 3.6), and to probe the astrophysical environments of quiescent MBHs/IMBHs, which is the focus of this Section.

**Extremely mass-ratio inspirals (XMRIs)** form when the smaller object is a brown dwarf. In this case the mass ratio relative to the MBH becomes extremely small (~$10^{-8}$). XMRIs with MBH mass in the range $10^4$–$10^7$ M$_\odot$ generate GWs in the LISA frequency range, but are considerably weaker than regular EMRIs and they stay in band for a longer time. This means that XMRIs are only detectable in the Milky Way or nearby galaxies, but there could be several XMRIs in band [415, 440]. A few dozen of galaxies are at distances of around 10 to 100 times the distance of Sgr A*, and any of these galaxies could contain detectable XMRIs. In the early stages of inspiral, EMRIs and XMRIs can be on extremely eccentric orbits. During this phase there is significant GW emission only when the secondary object is making its closest approach to the primary. As these periapse passages are well separated in time, the emission is burst-like [218, 407]. As many as a few tens of these bursts could be detected from the Milky Way [88] and galaxies within ~100 Mpc with a suitable central MBH [89], providing weak constraints on MBH masses and spins.

**Rates** – Predictions for the number of EMRIs that LISA will observe in the final phase of inspiral and plunge into the MBH vary between a few and several thousand [53]. Predictions for IMRI rates rely on assumptions about the IMBH population that is very poorly known, but could be as high as a few hundred [31]. The XMRI event rate is uncertain as the existence of these systems has only been recently proposed. These rate uncertainties reflect current astrophysical uncertainties and are an indication of LISA's great discovery potential.

---

### Probing the closest environment of massive Black Holes with LISA

Our understanding of the astrophysical environment of MBHs is almost entirely based on EM observations of the nearby Sgr A*. LISA science objectives for EMRIs extend our understanding of Sgr A* as well as to a larger, more distant population of MBHs. Interactions between MBHs and stars and gas in their vicinity play a crucial role in the co-evolution of MBHs with their host galaxies, as well as explain a number of astrophysical phenomena including tidal disruption events, hypervelocity stars and the enigmatic stellar population at the Galactic centre. LISA EMRI observations will provide a direct probe of MBH–star interactions in the low redshift Universe, and provide constraints on the gaseous environment in these systems.

### 3.3.1 Study the properties and immediate environment of Milky Way-like MBHs using EMRIs

There are several mechanisms that can lead to the formation of an EMRI near a MBH. Diffusion of orbital angular momentum through stellar encounters can put compact objects on orbits that pass sufficiently close to the central MBH that they are captured via emission of a burst of GWs [19, 21, 232]. Binary tidal disruption not only produces a hypervelocity star, but delivers the other component of the binary sufficiently close to the MBH that it becomes an EMRI through GW emission [231]. Similarly, giant stars can be tidally stripped by the MBH if they pass too close, leaving the dense core on an orbit that eventually becomes an EMRI [21, 422]. Finally, star formation in an accretion disc can create compact objects, which become EMRIs [277]. This **Science Investigation – SI 3.1** – aims to use EMRIs to investigate MBHs to answer the following questions:

> - **What are the mass and spin distributions of MBHs?**
> - **What stellar and gaseous environments do these MBHs live in?**
> - **Which physical process dominate stellar dynamics near to these MBHs?**

> **SI 3.1** aims to detect the GW signals from EMRIs with MBH masses between a few times $10^4\,M_\odot$ and a few times $10^6\,M_\odot$ at redshifts up to $z \sim 3$, for inspiralling objects of $m \gtrsim 10\,M_\odot$.

**EMRIs as probes of massive Black Holes** – EMRI observations will provide precise measurements of the masses and spins of the quiescent population of MBHs that are neither merging nor necessarily accreting (a requirement for EM measurements of BH spins). The mass and spin distributions of these two populations together encode the formation and assembly history of MBHs [417, 419]. In Table 3.6 we report the expected accuracy with which LISA observations of a typical EMRI will determine the properties of the MBH in the system. Measurements of (redshifted) masses and spins are expected to be exceptionally precise, driven by the large number of orbital cycles that LISA can observe. Although results are shown only for one EMRI, these values are typical of results obtained when the MBH mass or spin and small object mass are varied away from the reference values.

> LISA EMRIs are a unique probe of the properties of MBHs that are in a quiescent state, neither undergoing major mergers nor accreting. For most observed EMRI systems, the MBH mass in the source rest frame and the spin amplitude will be determined to a precision of $2\,\%$ and $10^{-5}$ respectively.

**EMRIs as probes of stellar cluster dynamics** – LISA observations of EMRIs will also provide high precision measurements (given in Table 3.7) of the inspiralling object's mass and orbital properties (primarily eccentricity and inclination). These measurements will provide the exciting prospect to distinguish EMRI formation channels. EMRIs generated via two-body relaxation require initially high eccentricities ($e \gtrsim 0.9999$ at $r \gtrsim 0.01\,pc$), leading to a moderate eccentricity by the time they reach the LISA band ($0.1 \lesssim e \lesssim 0.8$ [234], if not higher [19]), and a nonzero eccentricity at the last stable orbit ($e \lesssim 0.25$ [53]). Conversely, EMRIs resulting from binary disruption form at lower eccentricities ($e \sim 0.99$ at a few Astronomical Unit (AU)), leading to near-zero eccentricity in the LISA band ($e \lesssim 0.01$, [306]). EMRIs formed in accretion disks around active MBHs are also expected to have low eccentricity ($e \sim 0$), while their inclinations should be close to equatorial. In other channels the

**→ THE EUROPEAN SPACE AGENCY**

**Table 3.6:** Parameter estimation accuracies from LISA EMRI (Section 3.3.1) and IMRI observations (Section 3.3.2). The EMRI has (source frame) MBH mass $M = 10^6 \, M_\odot$, BHs mass $m = 10 \, M_\odot$, MBH spin $\chi_1 = 0.9$ and eccentricity at plunge $e_{pl} = 0.3$. The source is at $z = 1$ and has an SNR of 25. The light IMRI has (source frame) IMBH mass $M = 5 \times 10^4 \, M_\odot$, BHs mass $m = 10 \, M_\odot$, IMBH spin $\chi_1 = 0$ and eccentricity at plunge $e_{pl} = 0$. The source is at $z = 0.5$, at which the SNR is 30. The first heavy IMRI system has MBH mass $M = 10^6 \, M_\odot$, IMBHs mass $m = 1000 \, M_\odot$, MBH spin $\chi_1 = 0.9$ and eccentricity at plunge $e_{pl} = 0$. For the second heavy IMRI system, the MBH mass is $M = 10^7 \, M_\odot$ and the IMBH mass $m = 1000 \, M_\odot$. For the third heavy IMRI, the MBH mass is $M = 10^6 \, M_\odot$ and the IMBH mass is $500 \, M_\odot$. All other parameters are unchanged and all heavy IMRIs are at redshift $z = 1$, with SNRs of 40, 20, and 35, respectively. The parameter estimation uncertainties are quoted as medians and offsets to min/max values of a distribution over inclinations, sky locations and orientations of the EMRI and IMRI systems that plunge at the end of the mission duration. Sky location uncertainties are quoted in square degrees. Uncertainties on the MBH mass are given in both the observer (denoted "obs") and source frames (denoted "sf"). All waveforms were computed using the analytic kludge model [53, 67].

| System | $\left(\frac{\Delta M}{M}\right)_{obs}$ | $\Delta \chi_1$ | $\frac{\Delta d_L}{d_L}$ | $\Delta \Omega$ | $\left(\frac{\Delta M}{M}\right)_{sf}$ |
|---|---|---|---|---|---|
| EMRI | $3.5 \times 10^{-5 \, +6 \times 10^{-5}}_{\phantom{-5} -1.9 \times 10^{-5}}$ | $3.1 \times 10^{-5 \, +1.5 \times 10^{-4}}_{\phantom{-5} -1.1 \times 10^{-5}}$ | $0.07^{+0.1}_{-0.04}$ | $10^{+8}_{-8}$ | $0.02^{+0.04}_{-0.005}$ |
| Light IMRI | $2.2 \times 10^{-5 \, +2.8 \times 10^{-5}}_{\phantom{-5} -1.3 \times 10^{-5}}$ | $2.8 \times 10^{-5 \, +4.6 \times 10^{-5}}_{\phantom{-5} -1.4 \times 10^{-5}}$ | $0.14^{+3.3}_{-0.08}$ | $7.0^{+113}_{-5.7}$ | $0.02^{+0.35}_{-0.01}$ |
| Heavy IMRI 1 | $7.9 \times 10^{-6 \, +1.7 \times 10^{-5}}_{\phantom{-6} -6.7 \times 10^{-6}}$ | $2.0 \times 10^{-5 \, +5.4 \times 10^{-5}}_{\phantom{-5} -1.5 \times 10^{-5}}$ | $0.002^{+0.02}_{-0.001}$ | $0.028^{+4}_{-0.026}$ | $0.0033^{+0.016}_{-0.0009}$ |
| Heavy IMRI 2 | $4.4 \times 10^{-4 \, +2 \times 10^{-4}}_{\phantom{-4} -2.2 \times 10^{-4}}$ | $0.004^{+0.003}_{-0.003}$ | $0.07^{+0.08}_{-0.03}$ | $50^{+80}_{-40}$ | $0.03^{+0.03}_{-0.01}$ |
| Heavy IMRI 3 | $8.7 \times 10^{-6 \, +2.4 \times 10^{-5}}_{\phantom{-6} -6.1 \times 10^{-6}}$ | $2.6 \times 10^{-5 \, +1.1 \times 10^{-4}}_{\phantom{-5} -2.0 \times 10^{-5}}$ | $0.003^{+0.026}_{-0.002}$ | $0.094^{+7}_{-0.088}$ | $0.005^{+0.02}_{-0.001}$ |

inclination should be uncorrelated with the direction of the MBH spin. LISA will be able to distinguish between a circular EMRI and one with eccentricity greater than $6.3 \times 10^{-6}$.

**Table 3.7:** Parameter estimation accuracies from LISA EMRI and IMRI observations. Systems are as in Table 3.6, but we now show accuracies on the measurement of properties of the sBH (IMBH in heavy IMRI) orbit, including mass $m$ in both observer (denoted "obs") and source frames (denoted "sf"), the eccentricity at plunge, $e_{pl}$, and orbital inclination, $\iota$. Results are computed using the same procedure as Table 3.6.

| System | $\left(\frac{\Delta m}{m}\right)_{obs}$ | $\Delta e_{pl}$ | $\Delta \iota$ | $\left(\frac{\Delta m}{m}\right)_{sf}$ |
|---|---|---|---|---|
| EMRI | $2.1 \times 10^{-5 \, +1.8 \times 10^{-5}}_{\phantom{-5} -1.2 \times 10^{-5}}$ | $6.3 \times 10^{-6 \, +2.4 \times 10^{-5}}_{\phantom{-6} -5.8 \times 10^{-6}}$ | $1.2 \times 10^{-4 \, +0.01}_{\phantom{-4} -7.2 \times 10^{-5}}$ | $0.02^{+0.04}_{-0.005}$ |
| Light IMRI | $1.0 \times 10^{-5 \, +1.4 \times 10^{-5}}_{\phantom{-5} -6.1 \times 10^{-6}}$ | $8.5 \times 10^{-8 \, +1.5 \times 10^{-7}}_{\phantom{-8} -5.0 \times 10^{-8}}$ | $1.9 \times 10^{-2 \, +0.5}_{\phantom{-2} -0.013}$ | $0.02^{+0.35}_{-0.01}$ |
| Heavy IMRI 1 | $4 \times 10^{-6 \, +5.8 \times 10^{-6}}_{\phantom{-6} -3.3 \times 10^{-6}}$ | $3.3 \times 10^{-7 \, +1.5 \times 10^{-6}}_{\phantom{-7} -2.8 \times 10^{-7}}$ | $3.6 \times 10^{-5 \, +0.002}_{\phantom{-5} -2.8 \times 10^{-5}}$ | $0.0033^{+0.016}_{-0.0009}$ |
| Heavy IMRI 2 | $3.5 \times 10^{-4 \, +5.6 \times 10^{-4}}_{\phantom{-4} -2.3 \times 10^{-4}}$ | $0.0011^{+0.0004}_{-0.0009}$ | $0.006^{+0.04}_{-0.005}$ | $0.03^{+0.03}_{-0.01}$ |
| Heavy IMRI 3 | $4.9 \times 10^{-6 \, +6.1 \times 10^{-6}}_{\phantom{-6} -3.0 \times 10^{-6}}$ | $7.2 \times 10^{-7 \, +2.7 \times 10^{-6}}_{\phantom{-7} -6.4 \times 10^{-7}}$ | $3.6 \times 10^{-5 \, +0.003}_{\phantom{-5} -2.9 \times 10^{-5}}$ | $0.005^{+0.02}_{-0.001}$ |

**EMRI constraints on their astrophysical environment** – Interaction with matter backgrounds near the MBH, most notably the presence of gas and dark matter [69, 132, 256], induces slight deviations in the EMRI's orbital evolution, which in turn can produce a shift in the expected GW phase.

The amount of phase shift is directly determined by the properties of the environment in question and can provide a way to measure crucial unknown quantities such as the density of accretion discs, dark matter spikes or the gravitational influence of a third body. LISA can detect small fractional phase shifts, due to the large number of cycles observed for a typical EMRI, and hence potentially identify material surrounding a MBH.

LISA's ability to constrain environmental effects has been examined both using complete numerical simulations, e. g., of gas embedded EMRIs [157, 445] or EMRIs in the presence of dark matter candidates [123, 278], and using phenomenological models focused on parameter reconstruction and bias estimation. Recent studies have shown that a LISA observation of an EMRI embedded in an accretion disc described by an $\alpha$-disc model would provide a constraint on the accretion rate at the level of a few to a few tens of percent [383]. Occasional interactions with nearby stars could also leave a detectable imprint in the EMRI waveform [22].

> EMRIs are a novel probe of stellar populations in the close vicinity of MBHs. A typical EMRI observation will provide a measurement of the sBH mass to $\sim 1\,\%$ and the eccentricity and inclination of the sBH orbit to $10^{-5}$ and $10^{-4}$ respectively. The emerging picture is that environmental effects will be detectable in a variety of realistic astrophysical scenarios. Even a single successful measurement would provide invaluable information on the presence of matter in the form of stars, gas or dark matter, only a few Schwarzschild radii from the MBH horizon.

## 3.3.2    Study the IMBH population using IMRI

This **Science Investigation – SI 3.2 –** will study the properties of any IMRIs that LISA observes to answer the following questions:

> - **How readily do IMBHs form in stellar clusters and Galactic nuclei?**
> - **How do these IMBHs subsequently grow and what are their properties?**
> - **How often do these IMBHs merge with MBHs?**

> **SI 3.2** aims at detecting gravitational waves from IMRIs at low redshift, $z \lesssim 2$, in which the IMBH has mass in the range $10^3$–$10^4\,M_\odot$.

***Light*-intermediate-mass ratio inspirals** There are multiple channels that could form IMBHs. They could be failed MBH seeds, but a much more efficient way is to form IMBHs via collision and merger of objects in dense star clusters or Galactic nuclei. For these dynamically assembled IMBHs, we can distinguish three main sub-channels: (i) a series of stellar collisions occurring in the earliest cluster phase lead to the formation of a Very Massive Star (VMS) with a mass in the 200–500 $M_\odot$ range, that directly collapse to an IMBH [342]; (ii) a VMS that accretes part of its matter onto a sBH [201]; and (iii) repeated (hierarchical) mergers between stellar BHs [31, 293]. Recoil kicks of the mergers will lead to IMBHs being ejected from isolated clusters [359]. However, clusters with deeper potential wells such as nuclear star clusters in the centres of galaxies are able to retain more massive objects [189]. Determining the IMBH population will constrain which formation channels contribute.

After formation, the IMBH will continue to merge with compact objects in the dense stellar environment

→ THE EUROPEAN SPACE AGENCY

in its vicinity as mass segregation causes the most compact objects to sink to the centre of the cluster [31, 188, 189, 275]. These IMBH-compact object mergers are light IMRIs. LISA could detect light IMRIs with IMBH mass in the $10^3$–$10^4\,M_\odot$ range in the nearby Universe. The rate is very uncertain, but could be as high as a few tens of events per year [31]. Table 3.6 and Table 3.7 show the precise measurements that LISA would be able to make of such systems. Combining LISA observations with those of future ground-based detectors would provide even more powerful constraints on the IMBH population with mass < $10^3\,M_\odot$ [18].

**Heavy-intermediate-mass ratio inspirals** – IMBHs may also generate GWs detectable by LISA when they merge with MBHs. These *heavy* IMRIs can form during the early stages of the build-up of Galactic nuclei, or via the inspiral and merger of massive star clusters formed in the inner Galactic regions [411] which can deliver IMBHs to the Galactic centre [30, 169, 341]. A *heavy* IMRI could even be lurking in the centre of the Milky Way [213]. The expected parameter estimation precision from LISA *heavy*-IMRI observations was also shown in both Table 3.6 and Table 3.7.

> LISA observations of *light* IMRIs in star clusters would enable us to place constraints on the IMBH mass and spin distributions in the mass range between about $10^3\,M_\odot$ and a few $10^4\,M_\odot$. This, in turn, will help us to infer the most likely formation channel for IMBHs. LISA's observations of *heavy* IMRIs could help us to reconstruct the currently unconstrained rate, mass and spin distributions of heavy IMRIs, providing vital clues to the role of stellar clusters in galaxy evolution.

## 3.4    SO4: Understand the astrophysics of stellar-mass Black Holes

### Stellar-mass Black Hole Binaries in the gravitational Universe

> LISA will observe individual stellar-mass Black Hole binaries (sBHBs) up to several hundreds of years before coalescence, enabling measurements that are complementary to those made with ground-based detectors. This will help to pin down their nature, including formation channels and interactions with the environment.

LIGO, Virgo, and KAGRA (LVK) observations have unveiled a population of stellar-mass Black Hole binaries (sBHBs) with masses as large as ~140 $M_\odot$ [395]. These ground-based observatories will continue to discover new sBHB in the coming years, likely accumulating about 10 000 detections by the time LISA will fly. This will greatly improve the empirical measurements of their merger rate (up to $z \gtrsim 1$), component masses and spin distributions.

By observing extragalactic sBHBs at frequencies around 10 mHz, much lower than those accessible with ground-based observatories (but higher than the Galactic sBHBs discussed in Section 3.1), LISA will enable the measurement of properties that are harder to access from the ground, such as binary orbital eccentricity, or environmental effects due to embedding in a dense gaseous disk, or the proximity to a MBH. Some of these systems will be caught by LISA in the last few years of their inspiral and will be later observed by ground-based detectors, opening the novel opportunity of multi-band GW astronomy.

Given recent progress in this field, the Science Investigations follow a different ordering compared to the LISA Proposal [153] and a new investigation is added at the end of this section (**SI 4.2**).

The numbers presented below are based on the sBHB population model which accounts for the LVK GWTC3 constraints [396]. Strain amplitudes from a typical population of sBHBs are shown in Figure 3.1.

## 3.4.1    Study the properties of sBHs far from merger

This Science Investigation aims to probe the origins of sBHBs by answering the questions:

> - **How are stellar-mass Black Hole binaries born?**
> - **Are there multiple formation channels? To what extent does each channel contribute to the overall population?**

The fact that most massive stars live in binaries and associations [369] suggests that sBHBs might be the natural endpoint of stellar binary evolution. At the same time, sBHBs can efficiently form in dense star clusters (due to dynamical capture [361] or secular dynamics involving triplets and multiplets [26]), or in active Galactic nuclei (where sBHs can form, grow and pair in binaries in fragmentation episodes occurring in the outer regions of AGN disks [80]), or might have a primordial origin [370].

Each formation scenario leaves distinctive signatures in the properties of the sBHB population. Field binary evolution should mostly result in circular sBHBs possibly favouring aligned spins, while dynamical capture should favour binaries with high eccentricities and random spin orientations [368]. Evolution in dense clusters and AGN disks opens the possibility of multiple generation of mergers and gas accretion, thus favouring the existence of more massive binaries [195, 388]. While masses and spins can be reasonably well measured by ground-based detectors, GW circularisation implies that eccentricity can be measured much better far from coalescence, when sBHBs are emitting GWs in the mHz frequency range accessible to LISA.

> **SI 4.1** aims at the individual detection of several sBHBs. The measurement of their eccentricity with sub-percent accuracy will inform us about the physical mechanisms producing them, thus constraining the physics of binary stellar evolution and dynamics in dense environments.

LISA will measure the eccentricity of systems detected with $SNR > 8$ to a precision of about $10^{-3}$ [323]. We expect detection of up to ten sBHBs with $SNR > 8$ (see Table 3.8), and up to a comparable number of sBHBs with $5 < SNR < 8$ and coalescence time $t_c < 15$ years, which can be extracted via archival searches informed by later ground-based observations. This will allow us to discriminate dynamical formation scenarios, and to separate the standard stellar binary evolution formation scenario from dynamical capture in star clusters at $99\%$ confidence [322].

> By allowing a precise measurement of orbital binary eccentricity, LISA is a unique probe of sBHB formation channels.

## 3.4.2    Detecting high mass sBHBs and probing their environment

The observation of GW190521 [8] suggests the existence of BHs within the expected pair instability mass gap. This Science Investigation – **SI 4.2** – aims to understand their formation by answering the following questions:

**Table 3.8:** Number of expected sBHB detections in support of the different SIs calculated over 1000 realisations of sBHB cosmic populations drawn from the posterior, based on GWTC-3 sBHB mergers, for the fiducial Power-Law+Peak model used by LVK [396]. For each defined category of sources, we report the average number of detections $\langle N \rangle$, the 90 % credible interval and the percentage of realisations with zero objects of that category.

|        | sBHB type | definition | $\langle N \rangle$ | 90 % confidence | no sBHB (%) |
|--------|-----------|------------|---------------------|-----------------|-------------|
| SI 4.1 | detected  | $SNR > 8$  | 4.9 | $0.4 - 9.8$ | 2.2 |
|        | archival  | $5 < SNR < 8$ & $t_c < 15\,\mathrm{yr}$ | 5.6 | $0.8 - 10.0$ | 1.4 |
| SI 4.2 | massive   | $SNR > 8$ & $m_1 > 50\,\mathrm{M_\odot}$ | 1.3 | $0 - 3.6$ | 34.1 |
| SI 4.3 | multiband | $SNR > 8$ & $t_c < 15\,\mathrm{yr}$ | 1.5 | $0 - 3.8$ | 26.7 |
|        |           | $SNR > 8$ & $t_c < 4.5\,\mathrm{yr}$ | 0.4 | $0 - 1.4$ | 67.7 |

> - **How do BHs within and beyond the pair instability gap form?**
> - **Do sBHBs efficiently form in AGN disks?**
> - **Do hierarchical mergers of sBHBs occur in nature?**

According to our current understanding of stellar evolution, stars cannot produce BHs with masses in the range between $50\,\mathrm{M_\odot}$ and $120\,\mathrm{M_\odot}$, the so-called pair instability gap [223]. However, the more massive component of the LVK event GW190521 falls within this mass range with 90 % confidence [8], although there are caveats to this statement [435]. BHs with masses within the mass gap might be the product of a hierarchy of coalescences of lighter BHs [195], the result of accretion on sBHs in gas rich environment, such as the accretion disk of an AGN [388], or they might have primordial origin [155]. Each of these has particular signatures, such as Center of Mass (CoM) acceleration and dynamical friction in the AGN scenario [239].

> **SI 4.2** aims at the detection of GW190521-like binaries. The measurement of their Center of Mass (CoM) acceleration and other environmental effects will help assessing their origin.

As soon as a sBHB can be detected with $SNR > 8$, several environmental effects can be measured in the phase evolution of the source when the GW190521-like binary is embedded in an AGN disc around a MBH, as shown in [408]. The acceleration of the binary CoM is measurable for orbits as wide as $\sim 0.5\,\mathrm{pc}$ around a $10^8\,\mathrm{M_\odot}$ MBH. Both the mass of the central MBH and semimajor axis of the sBHB CoM orbit can be measured at a percent level. Dynamical friction exerted by the gas onto the sBHB allows a percent precision level measurement of the density of the disk, whereas accretion onto the sBHB can be measured only if around the Eddington limit or larger. The measurement (or lack thereof) of these effects will therefore enable us to pin down the environment of mass gap BHs. For example, if no acceleration is detected, it is likely that the binary formed via hierarchical mergers in dense star clusters, whereas detection of a cumulative dephasing will allow us to probe the AGN disk formation channel and will provide invaluable information about the density and structure of AGN disks.

LISA's capability of probing mass gap objects is determined by their detectability. We therefore extract from our models the number of observable mass gap objects, defined as those having at least one component mass larger than $50\,M_\odot$. LISA will allow the identification of at most a handful of such sBHBs (see Table 3.8 second row).

> By allowing the measurement of environmental effects on their slow inspiral, LISA is a unique probe of the nature of sBHBs falling within the pair instability mass gap.

### 3.4.3   Enabling multiband and multimessenger observations at the time of coalescence

This Science Investigation – **SI 4.3** – aims to use LISA forewarnings to perform multimessenger observations of sBHBs, and answer the question:

> - **Are there any associated EM counterparts to merging sBHBs?**

If LISA had been operating at the time, GW150914 would have been detectable with SNR > 5 many years prior to coalescence [376]. In fact, every LIGO/Virgo sBHB coalescence is the product of a slow inspiral that crosses through the LISA band. Therefore, sBHBs are appealing multiband GW sources. In particular, LISA detection of sBHBs in the last few years of their slow inspiral provides forewarnings about their sky position and time of coalescence. This will enable joint observations of these systems at merger with ground-based GW detectors such as Einstein Telescope and Cosmic Explorer. Although most sBHBs are not expected to coalesce in gas rich environments, several non-vacuum scenarios leading to EM counterparts have been proposed [333, 388, 447], and could be tested by coincident GW and EM observations.

> **SI 4.3** aims to detect GW150914-like binaries a few years prior to their coalescence. The inferred sky position and time to coalescence will inform ground-based GW detectors and EM observatories to enable the first GW *multiband* and *multimessenger* observations in astronomy.

LISA will measure the sky position of a sBHB 5.5 years prior to the merger at a level of a few deg$^2$ and its time of coalescence within several hours. Prediction of the merger time is reduced to minutes a few months before the merger takes place in the band of the ground-based detectors. This will be sufficient for pointing instruments of the class of Vera Rubin Observatory, Athena, and SKA for deep observations to look for a coincident signal across the EM spectrum. Even in the absence of EM counterparts, multi-band sBHBs are exquisite (astro)physical and cosmological probes *per se*; they can be used as dark sirens [316], and probe non-standard gravitational physics [65].

LISA will identify at most a handful of sBHBs with SNR > 8 and coalescence time $t_c < 15$ years, including an ~33 % chance of detecting at least one source merging within the mission lifetime (see Table 3.8), thus offering the first opportunity of performing GW *multiband* and *multimessenger* observations.

> LISA will enable the first *multiband* observation of a GW source, which has applications in cosmography and tests of gravity, and will facilitate *multimessenger* observations of sBHBs.

## 3.5     SO5: Explore the fundamental nature of gravity and Black Holes

> **The nature of Black Holes, strong gravity, and physics beyond the standard model**
>
> LISA is likely to detect loud, transient sources such as coalescing massive Black Hole binaries (MBHBs) or sources with thousands of cycles or more in band such as extreme mass-ratio inspirals (EMRIs) and intermediate mass-ratio inspirals (IMRIs). These sources allow us to test with high precision the nature of BHs in the range of $10^5$–$10^7\,M_\odot$ and the validity of the General Theory of Relativity (GR) in the strong-field regime. LISA may also constrain extensions of GR and the standard model of particle physics, probe the properties of a variety of dark matter models, and allow us to detect new fundamental fields.

GWs provide a new and unique probe of fundamental physics, ranging from the nature of BHs to the validity of GR and alternative theories of gravity. LISA is an extraordinary laboratory, instrumental to answer some of the key questions that link fundamental physics to astronomy [43, 66, 72, 96].

LISA has the ability to detect the merger of MBHBs with large SNRs as shown in Section 3.2. The high SNRs, in the hundreds up to a few thousands, for MBHBs at redshift $z \lesssim 3$, allow for tests of the strong-field generation properties of GWs in unprecedented ways; their great distance will probe the propagation properties of GWs, and possibly how they are affected by gravitational lensing. Furthermore, LISA can detect the signal from EMRIs and IMRIs, which can spend hundreds of thousands of cycles in band (see Section 3.3). These allow us to produce an exquisite map of the spacetime around the massive central object, and hence test if it is indeed described by the Kerr metric. Their long, intricate inspiral is also ideal for searches of new polarisations and their effect on their highly relativistic orbital dynamics. As such LISA promises to detect these sources to constrain new fundamental fields. Changes in the orbital dynamics could also reveal elusive properties of dark matter. Tests of the nature of BHs and searches for new fields and polarisations can be performed in several complementary ways, which we discuss below.

### 3.5.1     Use ringdown characteristics observed in MBHB coalescences to test whether the post-merger objects are the MBHs predicted by GR

This **Science Investigation – SI 5.1 –** aims to test the nature of BHs by answering the following questions:

> - **Are the massive objects that merge and their remnants consistent with being rotating MBHs described by the Kerr solution?**
> - **If not, are they horizonless ultracompact objects?**

Black hole spectroscopy is effectively a null test of the Kerr nature of BHs. In GR, perturbation theory predicts that all of the damped (hence complex) oscillation frequencies of the remnant MBH formed in the aftermath of a merger will depend only on the remnant's mass and dimensionless spin. The measurement of one those *quasi-normal modes* yields an estimate of the mass and spin [170]. The measurement of additional frequencies yields better estimates of the mass and spin, as well as tests of the Kerr nature of the source and of the underlying theory of gravity [93–95, 158, 166]. The details of this test depend on which oscillation modes, characterized by angular indices $(\ell, m)$ and by the so-called overtone index $n$, are excited in the merger.

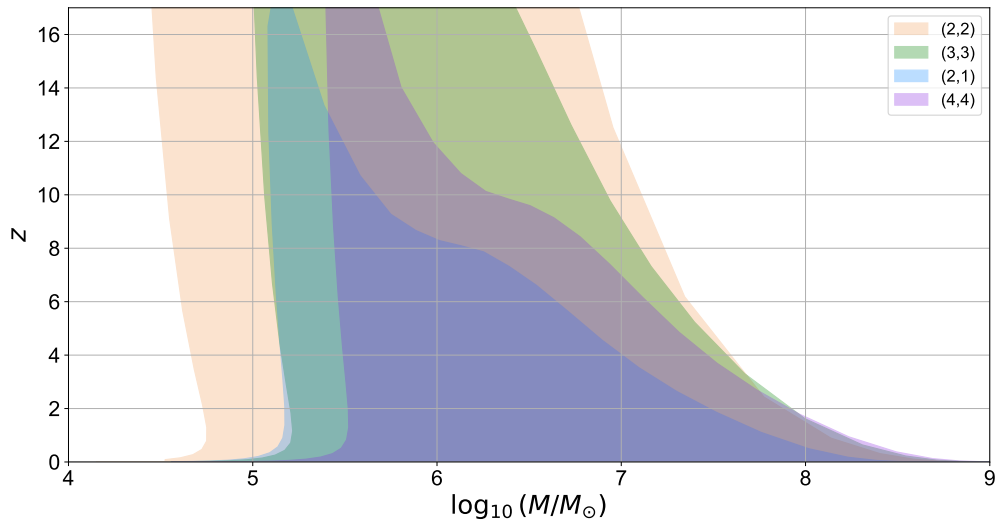

**Figure 3.7:** Horizon redshift $z$ for the detection (with SNR = 8) of the fundamental $(\ell, m) = (2, 2), (3, 3), (2, 1), (4, 4)$ ringdown quasi-normal modes as a function of the source-frame remnant mass $M$ for non-spinning binaries with mass ratio $q = 0.5$.

Drastic modifications of the oscillation spectrum can be possible if the central object has no horizon. If horizonless objects with compactness comparable to Black Holes do exist – as may be possible e.g. for "firewalls", in fuzzball/microstate scenarios, and even within GR (e.g., because of superradiant instabilities or the formation of boson stars) – other exciting possibilities arise. For example, the oscillation spectrum could show signatures of Planck-scale physics through the presence of "echoes" of the original ringdown signal [14, 327].

> **SI 5.1** aims to detect multiple ringdown "spectral lines" in the post-merger signal of MBHBs and put limits on GW echoes.

Numerical simulations in GR show that the dominant multipoles of the radiation, depending on the masses and spins of the remnant progenitors, are expected to be $(\ell, m) = (2, 2), (3, 3), (2, 1), (4, 4)$. In order to establish the Kerr nature of astrophysical BHs *at least three* ringdown modes need to be detectable, each with SNR larger than 8.

Figure 3.7 shows the horizon redshift for each multipole as a function of the total mass of the binary. Three or more modes can be detected up to high redshifts for low-mass MBH; up to redshift $z \sim 6$ for a source with total mass $> 1 \times 10^7\,M_\odot$; and at moderate redshifts $z \lesssim 0.5$ when the total mass of the remnant is large enough ($> 5 \times 10^8\,M_\odot$), despite the inspiral-merger part of the signal is out of band. Using information from the whole waveform, including harmonics, deviations from GR in the ring-down signal can be constrained to within 10 % and in the best cases to 1 % [409].

New physics may prevent the formation of an event horizon (a fully absorbing surface) in gravitational collapse. Horizonless objects can form in quantum-gravity motivated extensions of GR, or even in GR in the presence of bosonic fields or dark matter [124, 279]. These objects, if ultra-compact, can mimic a BH. However a BH has exactly zero reflectivity to GWs, while a non-zero reflectivity can be used as strong evidence for an object different from a BH. Very high values of the ringdown SNR are needed in order to place constraints on the existence of horizonless massive compact objects such as boson stars, or to detect the hypothetical echoes produced by quantum gravity modifications at the horizon scale. Assuming a reference horizonless remnant with mass $M = 10^6 \, M_\odot$, dimensionless spin 0.7, and an SNR in the ringdown of $\sim 500$, the minimum reflectivity $R$ of gravitational waves on the horizonless object that can be measured with an accuracy of $\sim 1\,\%$ is $R \sim 0.01$ [124].

> Using the ringdown signal from MBHB mergers, LISA will establish if the merger remnant is Kerr or not and test the no-hair theorem. Deviations from GR can be measured to within 1to10 %. Furthermore LISA has the potential to discover horizonless objects, which would provide strong evidence physics beyond General Relativity.

## 3.5.2 Use EMRIs to explore the multipolar structure of MBHs and search for the presence of new light fields

The **Science Investigations – SI 5.2 and SI 5.5 –** [153] aim to use EMRIs to answer the following questions:

> - **Are the massive objects observed at the centres of galaxies consistent with the rotating Kerr MBHs predicted by GR?**
> - **Are there new fundamental fields leading to hairy BHs?**

During the late inspiral of an EMRI, and in particular during the last year before plunging into the MBH, a compact object such as a sBH can produce up to $10^5$ GW cycles in the LISA band as it moves at relativistic velocities around the MBH. This makes EMRIs an ideal source to probe the geometry of spacetime around the MBH and, at the same time, search for new fundamental fields beyond the standard model or beyond GR. In particular, the so-called *golden* EMRIs, on prograde orbits with SNR > 50 and high dimensionless spins of the central object, $\chi_1 > 0.9$, allow for exquisite accuracy in the estimation of the source parameters, as illustrated in Section 3.3.

> **SI 5.2** LISA aims to observe stellar and intermediate mass objects spiralling into putative MBHs for $10^3$ to $10^5$ cycles, with SNR in excess of 50, thus testing the structure of the spacetime around these objects, probing the presence of dark matter, and potentially measuring charges on the orbiting body associated with new fundamental fields.

The GW emission contains a detailed and precise map of the central object's geometry, and hence of its nature [53]. EMRI signals allow us to measure the mass of the primary to a relative accuracy of about $10^{-5}$, and its dimensionless spin with an absolute error of $10^{-4}$: see Table 3.6. They also allow us to measure higher multipole moments of the MBH: for example, as shown in Table 3.9, different EMRI systems can constrain deviations of the spacetime's quadrupole moment from Kerr down to a part in $10^4$ [66, 367]. Such measurements will provide observational test of the no-hair theorem that are not achievable by other techniques [229, 380].

**Table 3.9:** Parameter estimation accuracies for the quadrupole moment $Q$ of LISA EMRI (Section 3.3.1) and IMRI observations (Section 3.3.2). The basic parameters of these systems are described in Table 3.6.

| System: | EMRI | Light IMRI | Heavy IMRI 1 | Heavy IMRI 2 | Heavy IMRI 3 |
|---------|------|------------|--------------|--------------|--------------|
| $\frac{\Delta Q}{Q}$ | $2.47 \times 10^{-5}$ | $4.53 \times 10^{-4}$ | $1.01 \times 10^{-4}$ | $1.73 \times 10^{-2}$ | $1.39 \times 10^{-4}$ |

> **SI 5.5** EMRIs have the potential to reveal new fundamental physics that affects the primary: a deviation from GR that can introduce Black Hole hair (see, e.g. [43, 162, 229, 380]), dark matter in the form of a light boson cloud [44], or the fact that the primary or secondary are not BHs, but rather something more exotic [76, 203, 297, 298].

If the secondary in an EMRI is endowed with some new, light fundamental field, and hence carries some new charge ("hair") it can emit, e.g., scalar radiation as it orbits the primary, affecting the inspiral waveform. The additional loss of energy will in turn affect the orbital dynamics, and hence the emission pattern of conventional (tensor) polarisations. This effect persists even when the primary is adequately described by the Kerr metric to leading order in the binary mass ratio – a valid approximation in a variety of theoretical scenarios – and it may allow LISA to measure the charge of the secondary with high precision [76, 297, 298].

> EMRIs provide an exquisite map of the geometry of spacetime of MBHs and can test the no-hair conjecture with high accuracy. They also allow for the detection of new fundamental fields.

### 3.5.3 Test the presence of beyond-GR emission channels

The **Science Investigation – SI 5.3** – aims to answer the following questions:

> - **Are there GW emission channels beyond GR?**
> - **Are there new physical degrees of freedom and extra GW polarisations, as predicted by some extensions of the standard model and of GR?**

LISA will allow for new types of GW tests of GR using observations of MBHBs and EMRIs. When additional fields are present in a theory beyond GR, MBHs may acquire "hair", i.e. they will be characterised by quantities other than mass and spin, which can lead to additional radiation channels through scalar and vector radiation to which LISA is sensitive. For example, the sensitivity to scalar-longitudinal and vector polarisation modes can be much higher than that to the tensor modes at relatively high frequency [401].

Another interesting possibility is presented by multi-detector observations of a single event, discussed in Section 3.4.3. If a sBHB is detected by LISA in the inspiral phase and later by ground-based detectors in the merging phase, the time at which the binary re-enters the ground-based sensitivity band will be distinct from GR if additional emission channels are present. LISA may be able to detect a few such events [310], and their value for testing modifications of GR is tremendous [73, 126, 127, 204, 332]. The large frequency range spanned by these events implies high precision in detecting or disproving beyond-GR emission channels, because the test spans a large range of binary separations that would not be accessible otherwise. Furthermore, joint observations with space- and ground-based detectors allow us to break some of the degeneracies between GR and non-GR parameters, improving the measurability of the latter.

> **SI 5.3** LISA aims to probe the existence of dynamical fields by searching for additional radiation channels and polarisations that would provide strong evidence for non-GR theories.

The presence of additional radiation channels can be probed within the framework of model-independent tests of GR. One such test is the Parameterized Post-Einsteinian (PPE) formalism [444]. In this framework, one can include the leading post-Newtonian non-GR correction to the waveform phase. Then, modified-gravity effects are parameterised by a PPE parameter $\beta$ that determines the overall magnitude of the non-GR correction term, and by an exponent $n$ that refers to the dominant post-Newtonian (PN) order at which the PPE correction enters the waveform phasing in the frequency domain. A known mapping exists between $(\beta, n)$ and theoretical parameters in specific modified theories of gravity [332, 389]. When BHs possess scalar and vector charges, BH binaries typically generate scalar/vector dipole radiation that modifies the waveform phase at $-1$ post-Newtonian order (namely $n = -1$). Therefore, by studying how well one can probe the PPE parameter $\beta$ at $n = -1$, one can predict how well LISA can probe the presence of the beyond-GR emission channels.

We can study the bound on the PPE parameter $\beta$ by comparing it to the one of GW150914, $\beta_{\mathrm{GW150914}}$. For example, for an EMRI with $(10^6, 10)\,M_\odot$ at $z \sim 0.5$, the bound on $\beta$ at $n = -1$ over a 4-year LISA observation is $\beta < 6.5 \times 10^{-10} \equiv \beta_{\mathrm{EMRI}}$. Normalising this bound with the existing bound from GW150914, $\beta_{\mathrm{GW150914}}$, we find $\beta_{\mathrm{EMRI}}/\beta_{\mathrm{GW150914}} = 4 \times 10^{-6}$. This means that LISA can improve the GW150914 bound on additional radiation channels by more than five orders of magnitude.

> MBHBs and EMRIs will allow to test beyond GR theories by looking into possible effects of new radiation channels in the gravitational emission.

## 3.5.4    Test the propagation properties of GWs

The **Science Investigation – SI 5.4 –** aims to answer the following questions:

> - **Does gravity respect Lorentz symmetry and parity invariance?**
> - **How do GWs propagate over cosmological scales?**

Changes to the way GWs propagate from the source to the observer are a generic signature of departures from GR, motivated by dark energy or dark matter models: see e. g. [247] for a review. LISA will test deviations from GR at mHz frequencies, outside the waveband of ground-based GW observatories (see e. g.  [133]).

Modifications in the GW propagation properties affect both the phase and the amplitude of gravitational waveforms. While the speed of GWs is strongly constrained at the frequencies of ground-based detectors [4], the possible frequency dependence of the GW speed provides a theoretical motivation to re-consider this bound at the smaller (mHz) LISA frequencies [448]. The strongest consequence of a change in the GW propagation speed is a shift in the coalescence time of a binary, though such shifts are likely to be detectable for multiband systems and MBHBs (see Section 3.4.3 and Section 3.2). In such cases, bounds on the deviations from luminal propagation speed comparable with the GW170817 bound may be achieved [65, 219, 410]. For a GW150914-like system, the fractional change in the GW propagation speed can be constrained to order $10^{-15}$ or better [65]. In a standard MBHB merger without a counterpart the measurement must rely on more subtle changes to the waveform phase and amplitude [64], resulting in a much weaker constraint.

Theories of massive gravity, or more generally scenarios with non-standard dispersion relations or Lorentz violating effects in the gravity sector, can also be constrained through their effects on the phase of GWs [307, 432]. For example, the combination of current LIGO/Virgo sources yields a constraint on the graviton mass $m_g \leq 4.7 \times 10^{-23}$ eV/c$^2$ [2]. More distant and more massive LISA sources, as well as the lower frequency LISA waveband, can be more effective in constraining GW dispersion relations: LISA will be able to tighten the bounds on $m_g$ by 2-3 orders of magnitude with respect to ground-based GW experiments [133, 237, 332]. Certain theories predicting Lorentz-violating effects in the GW sector, such as Einstein-Aether and chronometric theories, can also be better tested with LISA at low frequencies, through a study of the waveform phase [133].

> **SI 5.4** aims to detect GWs from golden MBHB coalescences or/and from EMRIs, all with SNR > 200, to probe the propagation of GWs over very large distances, imposing new stringent constraints on dark energy models, modified graviton dispersion relations, and theories of gravity beyond GR.

In several dark energy scenarios, a damping in the GW amplitude is predicted as the GW travels through large distances in the cosmic medium (see e. g. [84]). LISA can detect such effects by measuring the GW luminosity distance from merging MBHBs and comparing its value with the luminosity distance measured from electromagnetic counterparts or other techniques (see Section 3.6). The phenomenon of GW oscillations is predicted in theories with more than one spin-2 field, such as bigravity or vector-tensor theories of gravity, and can be analysed with theoretical techniques similar to the ones developed in studying neutrino oscillations. Recent studies proved the capabilities of LISA in constraining these theories by studying their effects on the GW amplitude [84], although the full potential of LISA to test their consequences for the waveform phase is still a subject of investigation.

> The observation of MBHB coalescences and EMRIs will provide tests of fundamental symmetries like Lorentz and parity invariance. We can also infer the propagation properties of GWs with high accuracy over cosmological distances.

## 3.6    SO6: Probe the rate of expansion of the Universe with standard sirens

**Mapping the expansion of the Universe at high redshift**

As a cosmological probe, LISA will map the expansion of the Universe in still poorly charted high-redshift epochs by providing self-calibrated, i.e. not relying on local Universe rulers, distance measurements across multiple cosmic ages, between approximately $0.01 \lesssim z \lesssim 7$.

The luminosity distance of a GW source can be measured directly from the amplitude of the signal, without the need for calibration with local distance rulers. With a measurement of the luminosity distance and redshift, GW sources become "standard sirens", allowing to probe the expansion rate of the Universe and infer the cosmological parameters [373]. However, GWs do not provide information on the redshift of the source and thus, depending on the nature of the GW source, the approach to estimate the redshift is different. Massive Black Hole binary (MBHB) mergers are expected to be surrounded by gas and an electro-magnetic (EM) signal can emerge at different stages before, after, and during the merger. The redshift of the source is then inferred after identification of the host galaxy. The system becomes a "bright siren" [284, 292, 390]. extreme mass-ratio inspiralS (EMRIs) are not expected to produce detectable EM counterparts. With EMRIs however, it will still be possible to obtain the redshift statistically from the inhomogeneous galaxy distribution, an approach known as *dark sirens* [268, 316]. Population analyses of the spectrum of masses may also help in cosmological inference, a method usually named "spectral sirens" [7, 177].

## 3.6.1     Cosmology from bright sirens: massive Black Hole binaries

MBHB mergers with an associated EM counterpart can be used to measure the expansion of the Universe to answer the following questions:

> - **How fast did the Universe expand beyond $z \sim 2$?**
> - **Is there any deviation from ΛCDM at high redshift?**

> **SI 6.1** aims at constraining the expansion rate of the Universe by combining multimessenger GW and EM observations from MBHB mergers in the $10^5$–$10^6\,M_\odot$ mass range at $z \lesssim 7$.

The most recent estimates [292] predict between $\sim 7$ and $\sim 20$ MBH bright sirens combining $\sim 4$ years of LISA observations with future EM facilities such as Vera Rubin Observatory, SKA, ELT, and Athena. Although these sources will not provide constraints on $H_0$ better than a few percent, they will constrain the expansion rate of the Universe at redshifts between $1.5 \lesssim z \lesssim 3$ where the majority of joint observations can still be well localised (see Section 3.2.3).

> LISA will be able to constrain the rate of expansion at $z \sim 2$ with an accuracy better than 10 %, competing with results expected from future high-redshift surveys. This will provide an unique opportunity to test the evolution of the Universe at early epochs which are still poorly traced by EM observations and may thus hide novel deviations from ΛCDM [118, 384].

## 3.6.2     Cosmology from dark sirens: extreme mass ratio inspirals

GW events without EM counterpart can be used to measure the expansion of the Universe to answer the following questions:

> - **What is the value of the Hubble constant $H_0$ as determined by GWs?**
> - **What is the equation of state of dark energy?**
> - **How does the expansion of the Universe behave around the matter to dark energy transition ($z \sim 0.7$)?**

**Table 3.10:** Accuracy of key cosmological parameter measurements (68 % confidence level) obtained by different LISA sources in 4 years of observations. Upper (lower) bounds represent pessimistic (optimistic) population scenarios.

| Source | $H_0$ | $\Omega_M$ | $w_0$ |
|--------|-------|-----------|-------|
| EMRIs  | $[1-6]\%$ | $[25-60]\%$ | $[7-12]\%$ |
| MBHBs  | $[3-7]\%$ | $[4-9]\%$ | - |

> **SI 6.2** aims to constrain the expansion rate of the Universe with EMRIs out to $z \lesssim 1$.

Predictions for the number of extreme mass-ratio inspirals (EMRIs) that LISA will detect vary between a few to several thousands per year [53] out to $z \sim 3$, with large uncertainties on their population due to yet poorly understood astrophysics. Since EM counterparts from these events are not expected to be bright enough to be observed, EMRIs can only be used as dark sirens. Assuming galaxy catalogues complete up to $z = 1$, a recent investigation [268] showed that a few high-SNR EMRIs, observed up to $z \sim 0.7$, can safely be used to probe $\Lambda$CDM with a joint estimate of $H_0$ at few % but large uncertainties on $\Omega_M$ (see Table 3.10). The same high-SNR events may allow us to get some information also on the dark energy equation-of-state parameter $w_0$, assuming $H_0$ and $\Omega_M$ known a priori. Constraints on $H_0$ by a few percent will provide useful complementary insights to solve the Hubble tension, assuming a solution will not be found before, while a gravitational-wave measurement of dark energy will yield information not accessible by EM observations [84].

> LISA aims to determine the rate of expansion of the Universe at low and medium redshifts by combining GW observations and galaxy redshift surveys, with the further goal of testing $\Lambda$CDM and obtaining insights on the nature of dark energy.

### 3.6.3 Cosmology at all redshift: combining local and high-redshift LISA standard sirens measurements

LISA will be a unique observatory measuring the expansion rate of the Universe across cosmic history. The joint-inference on $H_0$ and $\Omega_M$ resulting from the combined analyses of the standard siren populations described above, in particular of EMRIs and MBHBs, is expected to provide compelling constraints, possibly reaching the sub-percent level on $H_0$ and a few percent on $\Omega_M$. These LISA sources probe different redshift regimes, allowing to partly break correlations between the cosmological parameters.

LISA cosmological measurements could also be extended in different ways, via e.g., MBHB merger events at $z \gtrsim 2$ could be strongly lensed gravitationally. GWs emitted by these events may travel close enough to a galaxy so that multiple signals from the same event are detected with their time delays depending on the cosmological parameters, which could then be measured by LISA [375].

> By mapping the nearby Universe with dark sirens and the distant Universe with bright sirens, LISA will open unprecedented opportunities to unveil the mysteries of dark energy and test the foundations of the standard cosmological model.

## 3.7     SO7: Understand stochastic GW backgrounds and their implications for the early Universe and TeV-scale particle physics

**Signal or noise?**

Stochastic gravitational-wave backgrounds (SGWBs) of both astrophysical and cosmological origin are predicted in the LISA band. Detecting them can lead to groundbreaking scientific progress. However, these signals appear in the detector as additional sources of noise. The loudest SGWBs can mask other GW sources, while the identification of the faintest ones constitutes a challenging data analysis task.

The superposition of low SNR, overlapping GW signals from astrophysical sources can appear as a SGWB in the detector. GW sources, present in the early Universe, also give rise to SGWBs as they are homogeneous and isotropic over the sky, and/or correlated on scales much smaller than the detector resolution. The SGWBs are characterised by their energy density per logarithmic frequency normalised to the critical energy density in the universe $h^2\Omega_{\mathrm{GW}}$, where $h$ is the dimensionless Hubble constant (not to be confused with the GW characteristic strain $h_c$ [47]).

### *Astrophysical GW backgrounds in the LISA band*

Since the astrophysical sources contributing to the SGWB are those with low masses and/or high redshift (i.e. low SNR), the SGWB conveys information on their population complementary to the one gathered by detecting individually resolved, high SNR binaries [13, 54]. All classes of LISA sources can potentially generate a SGWB: for example, MBHBs (see Section 3.2), if they have formed from low-mass ($\sim 10^3\,\mathrm{M_\odot}$) remnants of Population III stars [70, 106]; or EMRIs (see Section 3.3), if they are numerous enough [105]. However, the astrophysical SGWBs which are guaranteed in the LISA band, better characterized, and have the highest amplitude, are:

*Stochastic Galactic foreground from Galactic binaries (GBs)* (see Section 3.1): while a fraction of the GBs in their slow inspiral phase can be resolved, the majority gives rise to a yearly modulated stochastic signal (displayed in Figure 3.8) [104, 251, 324]. It is crucial to accurately model it, as it will be the first guaranteed stochastic signal to emerge in the LISA band. Its detection will inform us on the population properties of double WDs (see Section 3.1.1).

*SGWB from extragalactic stellar mass binaries*: slowly inspiralling sBHBs from across the Universe give rise to a power-law SGWB (displayed in Figure 3.8). This signal is dominated by the contribution of sBHBs located at redshift $z \lesssim 5$, and its precise measurement by LISA will improve knowledge on the redshift-dependence of the sBHBs merger rate [54]. Similarly, extragalactic WD binaries form a SGWB expected to be comparable in amplitude [179].

### *Cosmological GW backgrounds in the LISA band*

GW sources in the early Universe form a fossil radiation, which carries information on the first instants of the Universe's evolution and thereby on high energy physics. Remarkably, LISA's frequency band encompasses the TeV energy scale, constituing the frontier of our knowledge of the theory of fundamental interactions, tested at the Large Hadron Collider (LHC). LISA has therefore discovery potential in complementarity with the LHC and future colliders, which are scheduled on a longer time-scale than LISA.

Futhermore, LISA is sensitive to many possible GW sources operating at a variety of energy scales in the early Universe, with the potential of clarifying, in complementarity with cosmological observations such as the Cosmic Microwave Background (CMB) and large scale structure, some outstanding problems in cosmology: for example the nature of Dark Matter and of the matter-antimatter asymmetry, the scale of Grand Unification, Inflation and its particle content [43, 47, 48, 79, 119, 121]. Even an upper limit on the cosmological SGWB amplitude will have far-reaching consequences in constraining cosmological scenarios.

Among the many proposals leading to observable cosmological SGWBs, we identify the following state-of-the-art science drivers (displayed in Figure 3.8):

*SGWB from First-Order Phase Transitions (FOPTs)*: well-motivated scenarios in Physics beyond the Standard Model (BSM) predict strong First-Order Phase Transitions (FOPTs) at the Electroweak (EW) scale and beyond, which give rise to SGWBs peaking in the LISA band [119, 121].

*SGWB from Cosmic Strings (CSs)*: CSs leftover from Phase Transitions at very high energy, such as the Grand Unified Theory (GUT) one, can source large SGWBs, allowing to probe BSM beyond any collider capability. LISA will constrain the string tension down to $G\mu \sim 10^{-17}$ [48], i. e. seven orders of magnitude smaller than the present constraints (or detection? [25]) achievable at PTA frequencies ($1 \times 10^{-9}$–$1 \times 10^{-6}$ Hz).

*SGWB from Primordial Black Holes (PBHs)*: Earth-based observations of GWs from sBHBs, combined with the lack of experimental evidence of Dark Matter (DM), have renewed interest in PBHs as DM candidates. By probing the mass range $10^{-16}$–$10^{-10}$ M$_\odot$, LISA will be able to support or rule out the hypothesis that the totality of DM is made of PBHs [47, 77].

### SGWB detection challenges in relation to the Science Investigations

Detecting the SGWB requires overcoming the crucial challenge of distinguishing the signal from the instrument noise. Some difficulties can be mitigated in the case of astrophysical SGWBs, for which the expected amplitudes and spectral shapes can be predicted based on observational knowledge of sources' populations. On the other hand, theoretical forecasts of cosmological SGWB sources operating in the early Universe, albeit numerous and promising [47], are based on BSM, which so far lacks experimental confirmation. While this opens up the discovery potential, it also prevents any certitude on the characteristics of the expected signal.

If the SGWB signal is strong (see e. g. the scientifically motivated examples of Figure 3.8), the detection could be claimed rather convincingly, leading to a groundbreaking discovery. If it is weak, it will be challenging to distinguish it from the instrument noise, lacking prior knowledge of any of the two. This is especially true since the efficiency of "null" channels, i. e. Time-Delay Interferometry (TDI) combinations which suppress the GW signal and allow thereby to characterise the instrument noise [16, 315, 349], is compromised in realistic detector configurations [313]. Noise and SGWB separation via null channels could be complemented with other techniques: for example, by exploiting statistical properties, such as stationarity and Gaussianity.[1] The presence of a dipole anisotropy due to the movement of LISA with respect to the cosmological frame might also help to identify what

---

[1]Only the quasi-stationary (e. g. periodic over a period of 1 year), and quasi-Gaussian component of both the noise and the residuals from the iterative source subtraction procedure, if large, bias the SGWB reconstruction.

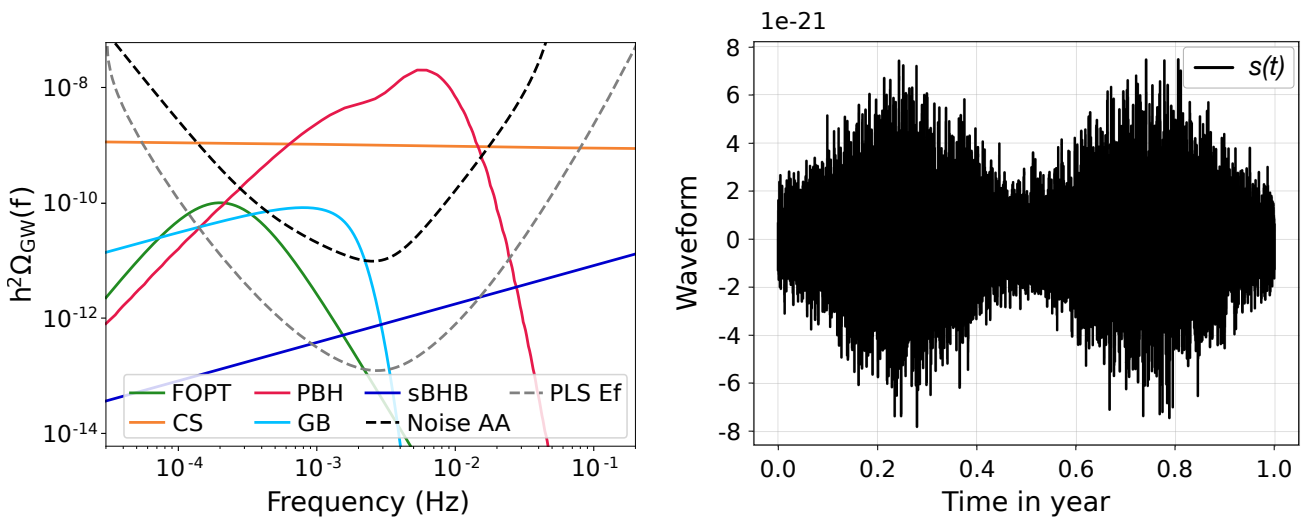

**Figure 3.8: Left panel:** Examples of SGWBs in the LISA band, together with the instrument sensitivity in the *A*-channel (*black, dashed*) and the effective Power Law Sensitivity [399] (*grey, dashed*). The cosmological SGWBs are: in *red*, the SGWB from Primordial Black Holes (PBHs) in a mass range for which they could constitute the totality of the Dark Matter today [77]; in *orange*, the SGWB from Cosmic Strings (CSs) with tension providing a signal that would account for the SGWB detection by PTAs [25]; in *green*, the SGWB from a primordial First-Order Phase Transition (FOPT) at the Electroweak (EW) scale, in the context of a singlet extension of the Standard Model of particle physics, testable at particle colliders. The astrophysical SGWB from unresolved stellar-mass Black Hole binaries (sBHBs), taken from [54] assuming GWTC-3 population constraints [396] is shown in *dark blue*. The Galactic foreground is shown in *light blue*, taken from [251], averaged over time. **Right panel:** The stochastic Galactic foreground in the time domain, where the periodic time variability of the signal amplitude is apparent (figure taken from [104]).

appears as extra noise components with proper SGWB signals (see Section 3.7.3).

In the following, we detail the **Science Investigations** aiming at detecting and characterising both astrophysical signals (Section 3.7.1), and one benchmark example of cosmological SGWB (Section 3.7.2). We place ourselves in the oversimplified situation of perfectly triangular detector configurations, for which the TDI *T*-channel can be used to gather information on the instrument noise. Note that the development of SGWB detection techniques is the subject of forefront, ongoing research: for these **SIs**, we have developed yet unpublished analyses. The **SI** in Section 3.7.3 highlights the scientific importance of probing the SGWB anisotropic component.

---

### Unveiling the early Universe beyond standard cosmological observations

Detecting a cosmological SGWB can have groundbreaking consequences for our understanding of the Universe (and thereby of fundamental high energy physics) over energy scales ranging from Big Bang Nucleosynthesis (~0.1 MeV) up to Inflation (≲$10^{16}$ GeV). Several puzzling phenomena are expected at these energies, such as the spontaneous breaking of the fundamental symmetries of the theory of particle interactions, the generation of the matter-antimatter asymmetry, the production of Dark Matter candidates of diverse nature. LISA has the potential to probe these phenomena in complementarity with both particle colliders such as the LHC, and cosmological observations such as the CMB and Large-Scale Structure.

---

## 3.7.1    Characterise the astrophysical SGWB amplitude

This **Science Investigation – SI 7.1 –** focuses on the SGWB from sBHBs and the ability of LISA to characterise it, to answer the following questions:

> - **What is the amplitude of the astrophysical SGWB?**
> - **What can this signal tell us about the source population?**

At LISA frequencies, the astrophysical SGWB is expected to be a power-law $h^2\Omega_{\mathrm{AGWB}}(f) \simeq A_{\mathrm{AGWB}}[f/f_{\mathrm{ref}}]^{n_{\mathrm{AGWB}}}$, with $n_{\mathrm{AGWB}} = 2/3$ [335]. For sBHBs the amplitude can be inferred from observation-based models of the sBHB mass distribution and local merger rate [13]: for this **SI**, we set $A_{\mathrm{sBHB}} = 7.87 \times 10^{-13}$ at reference frequency $f_{\mathrm{ref}} = 0.003\,\mathrm{Hz}$, coherent with GWTC-3 population constraints [54, 115]. Available information on the population of GBs allows to infer the associated foreground [324]: in this **SI**, we adopt the model of [251] (specifically, with parameters set to the last column of their Table II).

Prior knowledge on the expected spectral shape of both astrophysical signals can therefore be used to search for them in the simulated data. On the other hand, it is instructive to adopt a template-free approach to search for the instrument noise: building on [59], one can model it with a spline function of frequency.

> **SI 7.1** Aims at characterising the SGWB from sBHBs, if their population obeys current LVK models, together with the GB foreground and the instrument noise.

It can be appreciated from Figure 3.8 that the sBHB SGWB emerges at high frequency, where the GB foreground declines. This separation in frequency, together with the template-based search for both astrophysical signals, and with the use of the signal-orthogonal $T$-channel within the idealised case of an equal-arm constellation, lead to the identification of the sBHB SGWB and an unbiased estimation of its amplitude. Table 3.11 summarises the results of a parallel-tempered Markov-Chain Monte Carlo (MCMC) search over a set of parameters comprising the frequencies and amplitudes of the 6 noise Power Spectral Density (PSD) nodes, the GB foreground amplitude[2] $A_{\mathrm{GB}}$, and the sBHB SGWB amplitude $A_{\mathrm{sBHB}}$, fixing its spectral index to $n_{\mathrm{sBHB}} = 2/3$. Table 3.11 reports the relative $2\sigma$ errors on the parameters of the astrophysical signals and on the frequency localisation (in logarithmic scale $\log f_{\mathrm{n}}$) and on the log-amplitude ($\log A_{\mathrm{n}}$) of the noise PSD node for which these errors are the worst.

> The SGWB from sBHBs is expected to be strong enough in the LISA band to be detected. LISA will thus either discover this signal, or provide a confirmation in a new frequency band with respect to the one probed by Earth-based interferometers. In both cases, the detection by LISA will improve the characterisation of the sBHB population model.

---

[2]The amplitude $A_{\mathrm{GB}}$ is the only parameter characterising the foreground which is exclusively connected to the GBs population; it is therefore left free in the analysis, while the others are kept fixed to their default values taken from the Table II of [251]

**Table 3.11:** Ability of LISA to characterise the sBHB SGWB. From left to right: relative $2\sigma$-errors on the reconstructed amplitudes of the GB foreground and of the sBHB SGWB, and on the parameters of the noise PSD node for which the errors are the worst.

| | SI 7.1 | | |
|---|---|---|---|
| $\dfrac{\Delta A_{\mathrm{GB}}}{A_{\mathrm{GB}}}$ | $\dfrac{\Delta A_{\mathrm{sBHB}}}{A_{\mathrm{sBHB}}}$ | $\dfrac{\Delta \log f_{\mathrm{n}}}{\log f_{\mathrm{n}}}$ | $\dfrac{\Delta \log A_{\mathrm{n}}}{\log A_{\mathrm{n}}}$ |
| 1.2 % | 20 % | 11 % | 2 % |

## 3.7.2    Measure, or set upper limits on, the spectral shape of the cosmological SGWB

This **Science Investigation – SI 7.2 –** aims at illustrating that a cosmological SGWB can be identified in the LISA data in order to answer the following questions:

> - **Is there an SGWB of cosmological origin in LISA data?**
> - **If so, can we reconstruct the cosmological SGWB spectral shape to gather information about the process generating it?**

Similarly to **SI 7.1**, for **SI 7.2** we also set ourselves in the idealised case of an equal-arm detector, allowing to exploit the signal-orthogonal $T$-channel. Because of the inherent uncertainty on the spectral features of cosmological SGWBs, they must be searched for by minimising any a priori assumption on their spectrum, while still remaining as agnostic as possible about the instrument noise. Within **SI 7.2**, we therefore abandon the template-based inference for the signal, and adopt a different perspective. **SI 7.2** is divided in two parts: **SI 7.2a** aims at identifying the presence of a cosmological signal in the data; **SI 7.2b** aims at gathering information on its spectral shape via a piece-wise power-law reconstruction. Determining the spectral shape of the cosmological SGWB is the simplest resource available to assess the early Universe process that generated it.

> **SI 7.2** aims at illustrating that cosmological SGWBs can be identified in LISA data assuming a hard prior on the instrument noise (**SI 7.2a**), thereby allowing the application of agnostic searches aimed at reconstructing the signal spectral shape (**SI 7.2b**).

One example of physically motivated cosmological signal is the SGWB generated by a FOPT at the EW scale, within a singlet extension of the Standard Model testable at particle colliders [119]. In this context, the production of sound waves in the cosmic fluid by the broken phase bubbles leads to a SGWB whose spectral shape can be approximated by $h^2\Omega_{\mathrm{PT}}(f) = h^2\Omega_p \left(f/f_p\right)^3 \left\{7 / \left[4 + 3\left(f/f_p\right)^2\right]\right\}^{7/2}$ [119, 121]. For **SI 7.2**, we choose this benchmark signal, setting $h^2\Omega_p = 10^{-10}$ and $f_p = 2 \times 10^{-4}$ Hz

(displayed in Figure 3.8).[3]

**SI 7.2a**: Following the procedure of [250], one can show that the presence of the above benchmark cosmological signal can be identified in simulated LISA data containing i) the signal itself, ii) the GB foreground and the sBHB SGWB, iii) the instrument noise. The LISA frequency range is binned and the presence of an "extra noise component" (i.e. the cosmological SGWB) is identified bin by bin, assuming as input a hard prior on the combination of the instrument noise PSD + the astrophysical signals, at each binned frequency. Two values for the relative uncertainty of the noise PSD + astrophysical signals are tested: $\varepsilon = 5\%$ and $\varepsilon = 30\%$. Given $\varepsilon$, at each binned frequency, one evaluates the Bayes factor between the model containing noise + astrophysical signals + FOPT SGWB, and the model containing only noise + astrophysical signals: if $\log \mathcal{B}_{\text{FOPT}} > 50$ for at least one bin, the model including the FOPT SGWB is robustly preferred, and it is considered that the presence of the cosmological SGWB can be identified. The result is summarised in the left panel of Figure 3.9: the range of frequencies for which the FOPT SGWB can be identified is highlighted in yellow for $\varepsilon = 5\%$ and in orange for $\varepsilon = 30\%$, over the benchmark signal (green curve). The FOPT SGWB can be identified in the data over a sizeable range of frequencies for both levels of uncertainty tested; naturally, the smaller the noise uncertainty $\varepsilon$, the weaker the accessible signal over a broader frequency range.

**SI 7.2b**: With the methodology of the SGWBinner code [120], one can piece-wise reconstruct the FOPT SGWB spectral shape (**SI 7.2b**). The LISA frequency range is divided in macroscopic bins, in each of which the SGWB is searched for in the form of a power law, characterised by two parameters: its spectral index $n_i$ and its amplitude $\Omega_i$ at a pivot frequency (given by the geometrical mean of the bin $i$). Simultaneously, a template-based search is performed both for the astrophysical signals (sBHBs SGWB and GB foreground) and for the instrument noise. The noise template has two parameters: A (mass acceleration noise) and P (optical metrology system noise) [186]. The prior for the noise parameters is established using the $\mathcal{T}$-channel. The posterior for the $A$-channel noise and signals is minimised independently in each bin, over the full set of parameters: the two noise parameters $(A, P)$, the two astrophysical signals' parameters $(A_{\text{GB}}, A_{\text{sBHB}})$ (the spectral index of the sBHBs SGWB is fixed to 2/3), and the two FOPT SGWB parameters per bin $i$ $(\log \Omega_i, n_i)$. For all pairs of neighbouring bins, the code iteratively checks whether merging them is statistically favoured. The result is shown in the right panel of Figure 3.9: the merging procedure gives two final bins (horizontal dashed grey lines), allowing to recognise the presence of a break in the FOPT SGWB spectral shape (solid green lines). In Table 3.12 we report the SNR and the marginalised $2\sigma$ errors on the FOPT SGWB parameter set for each bin, as well as on the other parameters of the search.

> Determining the cosmological SGWB spectral shape constitutes the first step for the identification of the early Universe process that generated it. LISA can reconstruct basic features in the spectral shape of physically motivated cosmological SGWBs.

---

[3]Note that SGWBs with much higher SNR are predicted for the same BSM scenario in the corners of the model parameter space, or arising from different cosmological sources; however, we focus on this benchmark signal precisely because it is not very special.

→ THE EUROPEAN SPACE AGENCY

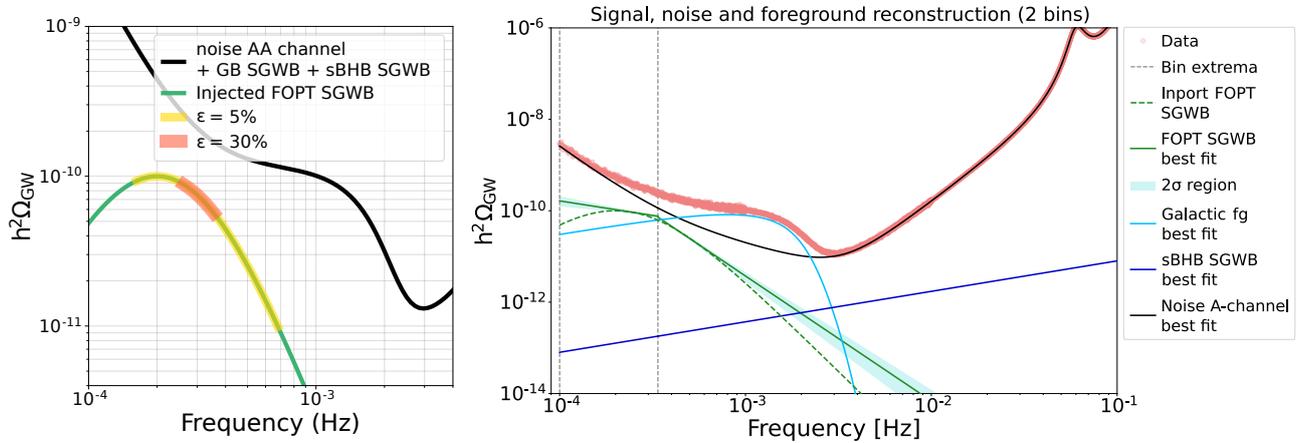

**Figure 3.9: Left panel, SI 7.2a:** the detection of a cosmological SGWB. Here the presence of the injected FOPT SGWB (green line) can be identified in the simulated LISA data imposing a hard prior on the instrument noise, for two tested values of the relative uncertainty on the noise + astrophysical signals: $\varepsilon = 5\%$ and $30\%$. The frequency region over which the signal is identified (Bayes factor $\log \mathcal{B}_{FOPT} > 50$) is shown as the coloured regions: orange, $\varepsilon = 30\%$ and yellow, $\varepsilon = 5\%$. **Right panel, SI 7.2b:** The outcome of a signal reconstructing procedure based on searching for the FOPT SGWB (green, dashed line: injected signal) as a series of power laws in frequency bins. The SGWBinner code has been run on simulated data (red dots) containing the FOPT SGWB, the astrophysical signals (sBHBs SGWB in dark blue and Galactic foreground in light blue), and the instrument noise (black). The code has iteratively merged several initial frequency bins in two final ones. The presence of a break in the FOPT SGWB can be reconstructed piece-wise, with two power laws (green, solid). The $2\sigma$ error on the reconstructed signals and noise are shown as shaded areas.

**Table 3.12:** Relative $2\sigma$ errors on the parameter set of the reconstruction of the FOPT SGWB, performed dividing the LISA frequency range in bins and fitting piece-wise power-laws in each bin. The analysis results in two final bins: the second column gives the SNR in each bin $i$, the third and fourth the errors on the SGWB parameters: amplitude at a pivot frequency and slope ($\log \Omega_i, n_i$). Also shown are the errors on the two noise parameters ($A, P$), and on the two astrophysical signals' amplitudes ($A_{GB}, A_{sBHB}$) (the spectral index of the sBHBs SGWB is fixed to 2/3). The errors on these latter are smaller than those given in Table 3.11 because for SI 7.1 the noise is reconstructed together with the astrophysical signals, while for SI 7.2b a model for the noise is assumed.

| SI 7.2b | SNR | $\dfrac{\Delta \log \Omega_i}{\log \Omega_i}$ | $\dfrac{\Delta n_i}{n_i}$ | $\dfrac{\Delta A}{A}$ | $\dfrac{\Delta P}{P}$ | $\dfrac{\Delta A_{GB}}{A_{GB}}$ | $\dfrac{\Delta A_{sBHB}}{A_{sBHB}}$ |
|---|---|---|---|---|---|---|---|
| Bin 1 | 67.72 | 0.45 % | 7.08 % | 0.85 % | 0.014 % | 0.045 % | 0.27 % |
| Bin 2 | 67.39 | 1.94 % | 10.18 % | | | | |

### 3.7.3      Characterise the large-scale anisotropy of the SGWB

Similarly to the CMB, the SGWB is expected to be statistically homogeneous and isotropic in the cosmological frame, and therefore to feature a dipole fluctuation [184], induced by the motion of the detector. This guaranteed SGWB dipole anisotropy is potentially detectable by LISA [78], if the product of the detector velocity with the amplitude of the SGWB isotropic component satisfies $(v/c)\,\Omega_{\mathrm{GW}}(f) \gtrsim 10^{-11}$. This requires the presence of a SGWB with high enough amplitude (see Figure 3.8). Assuming that the SGWB rest frame coincides with the cosmological one, $v/c \sim 10^{-3}$; however, this might not be the case. The observation of the kinematic dipole with LISA will allow a test of this hypothesis [47], and thereby the cosmological concordance model [192].

Furthermore, the GW propagation through the matter structure of the Universe, as well as the clustered distribution of the binaries in the sky (e.g. the Galactic binaries), can induce a directional dependence of the SGWB amplitude [147]. While subdominant with respect to the isotropic component, this variation can be exploited to disentangle SGWBs of different origin, or to probe the properties of the large-scale matter perturbations through the cross-correlation of the SGWB with the CMB [358] and with galaxies, tracers of the Large-Scale Structure.

Detecting the large-scale anisotropy of the SGWB with LISA has thus the potential to answer the following questions:

> ■ **Is the SGWB frame the same as the CMB one?**
> ■ **What are the host galaxies of sBHBs?**

## 3.8      SO8: Search for GW bursts and unforeseen sources

LISA will lead us into uncharted territories with the potential of discovering unforeseen sources with unknown features and signatures in their GW signal or cosmological sources that do not form a SGWB signal.

### 3.8.1      Search for cusps and kinks of cosmic strings

Apart from their SGWB signal as described above in Section 3.7, cosmic strings may produce time-wise well-separated, GW bursts that can be observed individually.

The relativistic nature of the strings typically leads them to form *cusps*, corresponding to points on the string momentarily moving at the speed of light. Furthermore, intersections of strings will also generate discontinuities in their tangent vector known as *kinks* [47]. *Cusps* and *kinks* can generate GW burst signals.

On a loop of length $\ell$ at redshift $z$, the corresponding linearly polarised wave form is given by [416] $h(\ell, z, f) = A_s(\ell, z) f^{-s}$ where $A_s$ is the wave's amplitude, and $s = 4/3$ for *cusps* and $s = 5/3$ for *kinks*. The emitted radiation from cusps (resp. kinks) is non-isotropic and concentrated inside a cone (resp. ring) of beaming angle [151, 152] $\theta_{\mathrm{cutoff}}(\ell, z, f) \sim [f(1+z)\ell]^{-1/3}$.

This inverse scaling with $f$ means that at LISA frequencies, the opening angle is larger than at LIGO/Virgo frequencies, thus enhancing the probability of burst detection at LISA frequencies and answering the following questions:

> ▪ **Are cosmic strings present in the Universe? If so, what is their string tension?**

Using the LISA sensitivity curve, we can determine the SNR of a burst as a function of $A_s$, depending on its sky localisation and polarisation. Under the assumption that bursts with SNR > 20 can be detected, and convolved with the predicted burst rate $\partial^2 R / \partial A_s \partial z$, which depends on the loop distribution $n(\ell, t)$ (see [7] for the details), one can then determine the expected burst rate for LISA as a function of $G\mu$, where $\mu$ is the energy per unit length of a string and $G$ the Gravitational constant, ([46], see Figure 3.10).

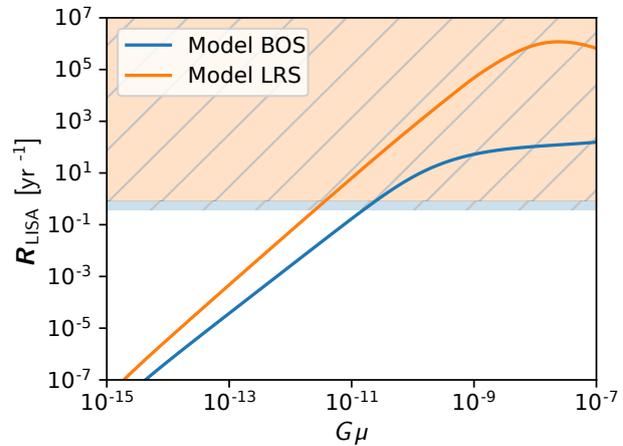

**Figure 3.10:** Expected rate of cusps as function of the string energy $\mu$ and LISA upper limits for both cosmic string models on the market (BOS [99] and LRS [285]). In case LISA does not detect bursts from cosmic string cusps, the orange hatched region is excluded after $T_{obs} = 82\,\% \times 4.5$ years and the blue hatched region is excluded after $T_{obs} = 82\,\% \times 10$ years.

> The non-detection of cosmic string bursts in LISA would constrain the string tension down to $G\mu \lesssim 10^{-11}$. Alternatively, if $G\mu \approx 5 \times 10^{-11}$ (compatible with the PTA signal), LISA would observe between $4 - 30$ cosmic string bursts per year.

### 3.8.2 Search for unmodelled sources

When opening a new observational band, one of the most intriguing possibilities is the discovery of yet completely unknown and unexpected types of sources. This has happened in the past, e.g. with new EM bands [110, 430]. Although these unknown sources are difficult to plan for, the mission is designed in such a way that non-stationary noise and artefacts are minimised and adequate calibration measures are in place in order to enable the discovery of unknown sources if strong enough. In the data analysis pipelines, in particular the global fits to the data, the possibility of unmodelled sources will be explicitly kept in mind (see Chapter 8).

> Detection of yet unknown types of sources is an exciting prospect for LISA and mission and data analysis pipeline design takes this into account.

# 4 MISSION REQUIREMENTS

In this chapter, we present the key performance requirements for the LISA mission and their relationship to the Science Objectives outlined in Chapter 3. The presentation here is not intended to replace the formal flowdown from science requirements to requirements on the mission, the space segment, and the instruments, but rather to highlight the key characteristics of the LISA design and trace their connection to specific science objectives. Readers new to LISA should review Chapter 2 to familiarise themselves with the measurement concept.

**Table 4.1:** Summary of LISA's key design characteristics and their corresponding mission and instrument performance requirements.

| Characteristic | Performance Metric | Requirement |
|---|---|---|
| Free-fall purity | Test mass residual acceleration | $\delta_a(1\,\text{mHz}) \lesssim 3 \times 10^{-15}\,\text{m}\,\text{s}^{-2}/\sqrt{\text{Hz}}$ |
| Displacement sensitivity | Interferometer displacement noise | $\delta_x(1\,\text{mHz}) \lesssim 15 \times 10^{-12}\,\text{m}/\sqrt{\text{Hz}}$ |
| GW measurement band | $100\,\mu\text{Hz} \le f_{\text{GW}} \le 1\,\text{Hz}$ | $L_{\text{arm}} \approx 2.5 \times 10^6\,\text{km}$ |
| Quantity of Science Observations | Mission duration and availability | $T_{\text{data}} \ge 3.7\,\text{yrs}$ |
| Polarisation & null channel | Constellation configuration | 3-arm symmetric triangle |
| Observatory knowledge | Constellation stability | Ephemeris knowledge |

Table 4.1 lists several key characteristics of the LISA mission along with their primary performance metrics and performance requirements. Meeting this set of performance requirements is the primary objective of the mission architecture as described in Chapter 5 and Chapter 6. The combination of free-fall purity and displacement sensitivity determine the observatory's gravitational wave (GW) strain sensitivity, which is the noise of the instrument expressed as equivalent GW strain (see Figure 7.1). The GW measurement band is determined by the interferometer baseline. The colours used in Table 4.1 are also used in the Performance Model in Section 7.1. LISA's ability to detect and characterise particular GW sources is determined by the strain sensitivity, the source's GW waveform, and the observation duration as described in detail in Section 8.3.

In the remainder of this chapter, we describe how each of LISA's design characteristics enable specific Science Objectives already described in Chapter 3. The presentation is organised as follows. Each of the design characteristics enables one or more *capabilities* for the observatory. These capabilities in turn enable *observations* of various types of sources. By meeting a certain level of *performance*, observations will enable scientific *studies* that support LISA's Science Objectives as defined in Chapter 3. The tables below present the linkage from Characteristics to Science Objectives via Capabilities, Observations, Performance, and Studies.

## 4.1    Free-fall purity

A fundamental limit on the performance of a GW observatory are disturbances on the fiducial references used to track local inertial frames at the vertices of the observatory. For LISA, this is quantified by the acceleration due to disturbances of the test masses relative to ideal free-fall. Table 4.2 presents the relationship between free-fall purity and science objectives. This directly limits LISA's GW strain sensitivity at the lower end of the measurement band ($f \lesssim 3\,\mathrm{mHz}$). This sensitivity in turn defines LISA's observational capabilities for sources in this band, including massive Black Hole binary (MBHB) mergers, Galactic binary systems, as well as cosmological and stochastic astrophysical sources. In addition, the low-frequency performance determines how early before merger LISA can detect MBHB sources, which has implications for multi-messenger astronomy. LISA's approach to meeting the free-fall purity requirements is discussed in Section 5.2.

**Table 4.2:** Science studies and performance requirements supported by the free-fall purity design characteristics.

| Characteristic | Capability | Observations | Performance | Studies | Science Objective |
|---|---|---|---|---|---|
| Free-fall purity | Strain sensitivity between 0.1 mHz and 3 mHz | Observation of post-merger phase of GW signal from MBHBs with $M \sim 10^7\,M_\odot$ around $z \sim 4$. | Detect $\geq 3$ ringdown modes with SNR $\geq 8$. | Test the "no-hair" conjecture of GR. | SO5 FoM-SO5a |
| | | Observation of MBHB mergers with mass $\sim 10^5$–$\sim 10^7\,M_\odot$ at $z < 2$ and measurement of masses and spins. | • $\delta m_i / m_i < 2\%$<br>• $\delta \chi_i / \chi_i < 0.5\%$ | Investigate growth mechanism & merger history of Massive Black Holes (MBHs). | SO2 FoM-SO2a |
| | | Observation of resolvable Galactic binary systems with sky localisation and measurement of their orbital frequency evolution. | For $\geq 3\%$ of the population, measure:<br>• Sky position $\leq 1\,\mathrm{deg}^2$<br>• $\dot{f}_f / \dot{f} \leq 30\%$ | Study the interplay between gravitational damping, tidal heating, and perform tests of GR. | SO1 FoM-SO1b |
| | | Detect or set upper limits on stochastic gravitational-wave background (SGWB) at low frequencies. | Residual GW strain measurement after subtraction of bright sources and instrument effects. | Detect or set upper limits on SGWB, probe a broken power-law stochastic background from the early Universe from first order phase transitions, cosmic strings, or inflation. | SO7 FoM-SO7 |
| | | Multimessenger observations of coalescing MBHBs. | $3 \times 10^5\,M_\odot$ binary at $z = 1$ with:<br>• Detection at $t_{merge} < 3$ days<br>• Merger time to 1 hr at $t_{merge} < 3$ days<br>• Sky location to $200\,\mathrm{deg}^2$ at $t_{merge} < 4$ hrs | Study the environment of MBHB mergers and the physics of electro-magnetic (EM) emission. | SO2 FoM-SO2a |

## 4.2    Interferometer Performance

LISA's performance in the mid-band, $3\,\mathrm{mHz} \lesssim f \lesssim 30\,\mathrm{mHz}$, is largely determined by the performance of the interferometer that measures the relative displacement of the test masses at either end of a LISA arm. This subsystem is described in Section 5.3. Table 4.3 demonstrates the linkage between the interferometer mid-band performance and several science objectives. Numerous sources are anticipated in this band, with a correspondingly wide array of scientific studies enabled through their observations. The most stringent requirements come from observations of MBHBs in the early universe (SO2) and the detection and characterisation of extreme mass-ratio inspiral (EMRI) systems (SO3 and SO5).

**Table 4.3:** Science studies and performance requirements supported by the Optical Metrology System design characteristics.

| Characteristic | Capability | Observations | Performance | Studies | Science Objective |
|---|---|---|---|---|---|
| Interferometer Performance | Strain sensitivity between 3 mHz and 30 mHz | Observation of resolvable Galactic binaries, sky localisation and frequency evolution. | For 30 % of the population, determine:<br>• Sky position $\leq 1\,\mathrm{deg}^2$<br>• $\delta_f/f \leq 30\%$ | Elucidate the formation and evolution of Galactic Binaries. | SO1 FoM-SO1b |
| | | Observation of Black Hole binaries ($10^3\,M_\odot - 10^5\,M_\odot$) in early Universe ($z$ of $10-15$). | Detect $M = 5 \times 10^3\,M_\odot$ at $z = 12$ (SNR > 15) and measure distance $\delta D_L/D_L < 50\%$, individual masses $\delta m_i/m_i < 50\%$ and spin $\delta \chi_1 < 0.3$. | Study seed Black Holes in the early Universe and distinguish seed formation channels. | SO2 FoM-SO2a |
| | | Post-merger MBHB localisation. | For $3 \times 10^5$–$5 \times 10^6\,M_\odot$ at $z < 3$, localise better than $3\,\mathrm{deg}^2$. | Multimessenger observations of MBHBs. | SO2 FoM-SO2a-1 |
| | | Observation of nearly equal mass IMBHs binaries of total intrinsic mass between $10^3\,M_\odot$ and $10^4\,M_\odot$ at $z < 1$. | Detect $M = 1200\,M_\odot$ at $z = 0.5$ with SNR > 20 and individual masses to $\delta m_i/m_i < 5\%$. | Test the existence of IMBHs. | SO2 FoM-SO2a-2 |
| | | . | • Detect $10 - 3 \times 10^5\,M_\odot$ to $z = 2$<br>• Detect $30 - 1 \times 10^5\,M_\odot$ to $z = 5$ | Study demographics of EMRIs | SO3 FoM-SO3 |
| | | Observation of EMRIs | For $\sim 10$–$\sim 30\,M_\odot$ merging into $1 \times 10^6\,M_\odot$ at $z = 1$, measure:<br>• $\delta M/M < 5 \times 10^{-5}$<br>• $\delta m/m < 2 \times 10^{-5}$<br>• $\delta S < 5 \times 10^{-5}$<br>• $\delta Q < 3 \times 10^{-4}$ | Detailed study of EMRIs systems and perform precision General Theory of Relativity (GR) tests. | • SO3<br>• SO5<br>FoM-SO3a |
| | | Multiband GW astronomy for GW150914-like events. | Detect GW150914 at optimal orientation with:<br>• coalescence time $\delta t_{merge} < 5.5\,\mathrm{hr}$<br>• sky position $\delta \Omega < 5\,\mathrm{deg}^2$ | Precision GR and cosmology tests via GW astronomy. | SO4 FoM-150914 |
| | | Observation of inspiral signal from stellar-mass Black Hole binaries (sBHBs) events with potential for multiband GW astronomy. | From a catalogue based on LVK O3 with $\geq 30\%$ probability of having $\geq 2$ multiband sources:<br>• Detect 15 sBHBs with SNR > 7<br>• Detect 2 multiband sBHBs ($T_c < 10\,\mathrm{years}$) | Study demographics of sBHBs, enabling multiband observations. | SO4 FoM-SO4 |
| | | Observation of heavy sBHBs ($m_i > 50\,M_\odot$). | Detect GW190521-like max-mean $D_L \geq 600\,\mathrm{Mpc}$. | Study heavy sBHBs, which could be embedded in gaseous environment. | SO4 FoM-190621 |
| | | Observation of astrophysical and cosmological SGWBs. | Detect and characterise SGWBs from:<br>• Galactic binaries<br>• extragalactic population of coalescing compact binaries | Study SGWBs. | SO7 FoM-SO7-2 |

## 4.3  Observatory Response

The sensitivity of LISA at the highest frequencies is set by the interferometry measurement noise and an observatory response that decreases with the observatory armlength $L$. As with electromagnetic antennas, GW antennas couple most efficiently when they are of similar size to the wavelength of the incoming signal. Wavelengths shorter than the antenna will suffer from an averaging effect which degrades the response as the wavelength decreases. For LISA's $2.5 \times 10^6$ km arms, this limits the sensitivity above $\sim 30$ mHz. Table 4.4 presents the relationship between the high-frequency observatory response and science objectives. The anticipated sources in this frequency range are stellar-mass Black Hole (sBH) inspirals, with rates now constrained from ground-based GW observations as well as potential mergers of Intermediate-Mass Black Hole (IMBH) systems with much more uncertain event rates. The detectability of multi-band sBH signals, as described in SO6, is particularly sensitive to LISA's performance in this band.

**Table 4.4:** Science studies and performance requirements supported by the observatory response design characteristics.

| Characteristic | Capability | Observations | Performance | Studies | Science Objective |
|---|---|---|---|---|---|
| Response to GWs | Strain sensitivity between 30 mHz and 1 Hz | Observation of the inspiral signal from sBHBs events with potential for multiband GW astronomy. | Detect $\geq 15$ events with 30% chance of more than 2 multiband events based on current rate estimates. | Study demographics of sBHBs, enabling multiband observations. | SO4 FoM-SO4 |
| | | Multiband GW astronomy for GW150914-like events. | Detect GW150914 at optimal orientation with: <br>• coalescence time $\delta t_{merge} < 2.5$ hours<br>• sky position $\delta\Omega < 5\,deg^2$ | Precision GR and cosmology tests via multiband GW astronomy. | SO6 FoM-150914 |
| | | Observation of EMRIs. | For $\sim 10 - 30\,M_\odot$ merging into $1 \times 10^6\,M_\odot$ at $z = 1$, measure:<br>• $\delta M/M < 5 \times 10^{-5}$<br>• $\delta m/m < 2 \times 10^{-5}$<br>• $\delta S < 5 \times 10^{-5}$<br>• $\delta Q < 3 \times 10^{-4}$ | Demographics of EMRIs systems, precision GR tests. | SO3 FoM-SO3a |
| | | Discovery potential for IMBH mergers. | Detect $M = 1.2 \times 10^3\,M_\odot$ at $z = 0.5$ with:<br>• SNR > 20<br>• $\delta m/m < 5\%$ | Probe the Black Hole mass gap. | SO2 FoM-SO2a-2 |

## 4.4 Mission Duration & Availability

The quantity of LISA observations, determined by the product of mission lifetime and observatory availability, is a strong driver of the scientific return. The impact differs for the various LISA source classes. Persistent sources benefit from additional data through an accumulation of SNR whereas transient sources benefit through an increase in the detected population. For rare sources, many of which enable important scientific investigations, an adequate duration of collected data dramatically decreases the risk that certain science objectives are not met. The detailed impact on each source class and their resulting scientific investigations is treated in [23]. The scaling of various figures of merit with integrated observation time ranges from $T^{1/2} \sim T^3$ depending on the particular investigation under consideration. Table 4.5 presents a simplified relationship between data quantity and science objectives. LISA's nominal mission duration of 4.5 years with an anticipated duty cycle of >82 % is predicted to yield a total integrated observation time of at least ~3.7 yrs, which is sufficient to meet all of the Scientific Objectives with high confidence.

**Table 4.5:** Science studies and performance requirements supported by the duration & availability design characteristics.

| Characteristic | Capability | Observations | Performance | Studies | Science Objective |
|---|---|---|---|---|---|
| Mission duration and observatory availability | Total integrated observation time $\geq 3.7\,\mathrm{yr}$ | Catalogue of frequency-resolved GBs with measured GW period. | a few thousand systems detected with $dP/P \sim 1 \times 10^{-6}$ | Study formation and evolution via survey of their period distribution. | SO1 |
| | | Catalogue of sky-resolved GBs. | ~20 % of systems with sky location <1 deg$^2$. | Identification of possible EM counterparts. | SO1 FoM-SO1-1a |
| | | Catalogue of 3D spatially-resolved Galactic binaries (GBs). | ~10 % of systems with:<br>• sky location $\delta\Omega < 1\,\mathrm{deg}^2$<br>• distance error $\delta_D/D < 10\%$ | Study the co-evolution of GBs and galaxy by surveying spatial distribution of GBs. | SO1 FoM-SO1-1a |
| | | Catalogue of spatially-resolved, chirping GBs. | a few percent of systems with<br>• sky location $\delta\Omega < 1\,\mathrm{deg}^2$<br>• chirp error $\delta\dot{f}/\dot{f} < 10\%$ | Study the interplay between gravitational damping, tidal heating, and to perform tests of GR. | SO1 FoM-SO1b |
| | | Catalogue of Massive Black Holes with measured masses. | at least $30 - 200$ systems with:<br>• $\delta_{m_1}/m_1 < 30\%$<br>• $\delta_{m_2}/m_2 < 50\%$ | Study origin and evolution of MBHB. | SO2 FoM-SO6-1 |
| | | Massive Black Holes standard sirens. | measure $\Delta H_0 < 20\%$ in a combined astro model. | Probe of cosmological parameters. | SO6 FoM-SO6-3 |
| | | Catalogue of EMRI systems. | a few to a few 10s of systems (depending on event rates) with SNR > 8. | • Study the immediate environment of Milky Way like MBHs at low redshift<br>• Perform precision tests of GR using EMRIs | • SO3<br>• SO5 |
| | | Multiband sBH systems with LISA alerts for GW merger in ground-based detection band. | 30% chance of >1 system with SNR > 8 and $t_c < 10\,\mathrm{yr}$. | Precision tests of GR (e.g., GW dispersion). | SO4 FoM-SO4 |
| Communications plan and ground segment performance | Latency for transient alerts | Enable searches for EM counterparts of MBHBs mergers. | Alert distribution, including relevant GW signal parameters within 1 hr of constellation measurement. | • Accretion environments of merging MBHBs<br>• Cosmology with bright standard sirens | • SO2<br>• SO6 |

# 4.5     Constellation Geometry

The three-arm configuration of LISA optimises the balance between the capabilities and the complexity of the observatory, as indicated in Table 4.6. While a single arm is in principle sensitive to GW, in practice a two-arm configuration is required to mitigate the otherwise-overwhelming frequency noise in the interferometer via common-mode rejection. A third arm provides additional scientific benefits through the ability to simultaneously measure both polarisation components of a GW signal, which is important in constraining the luminosity distance to GW sources through breaking the amplitude/inclination degeneracy. In addition, a third arm provides the possibility to construct *null-streams* which, in the long-wavelength limit, can be insensitive to external GW signals and can be important diagnostic tools for searches of unmodelled GW signals. From an engineering perspective, a three-arm design provides graceful degradation by enabling a two-arm design with significant science performance in the event of a catastrophic failure in a one arm. LISA's scientific performance is assessed assuming the full three-arm constellation is available. In the event that a significant failure impacts the operation of one arm, much of the science capability is preserved, with an approximate reduction in sensitivity of $\sqrt{2}$.

**Table 4.6:** Science studies and performance requirements supported by the constellation geometry design characteristics.

| Characteristic | Capability | Observations | Performance | Studies | Science Objective |
|---|---|---|---|---|---|
| Three-arm constellation | Simultaneous measurement of both GW polarisations | High-SNR MBHB mergers. | Improved measurement of distance and inclination through breaking of polarisation degeneracies. | Cosmology with MBHB standard sirens. | SO6 |
| | | GW stochastic background. | Measure polarisation. | Investigate non-GR metric theories which permit additional polarisations. | SO7 |
| | Null-stream observable | Simultaneous measurement with three-arm/six-link constellation. | Generation of TDI null-stream observables (suppressed GW response for long wavelengths). | Distinguish between part of instrumental noise and potential stochastic signals of astrophysical or cosmological origin. | SO7 |

# 4.6      Observatory Knowledge

Knowledge of the observatory parameters impacts science through two mechanisms. The first is in quantifying the response of the observatory to GW signals, what is typically referred to as *calibration* for astronomical instruments. For LISA, the quantity which most directly impacts the GW amplitude response is the wavelength of the laser. Laser wavelengths are known to ~0.1 nm, or $10^{-5}$ fractionally, in order to ensure that heterodyne frequencies can be brought into the phasemeter measurement bandwidth. As a result, LISA measurements of GW amplitudes are calibrated to approximately this level, which is sufficient for anticipated science investigations.

An additional aspect of calibration is associated with relating LISA's observations to standard astronomical reference frames. This places constraints on the knowledge of the constellation ephemeris and the time-transfer chain between the LISA spacecraft and Earth-based time standards.

A different category of Observatory Knowledge that is especially relevant for certain science objectives is the accuracy with which the instrument noise can be modelled as equivalent GW strain. A performance model informed by on-orbit measurements will be critical for distinguishing between instrumental artefacts and potential unmodelled signals, especially stochastic and isotropic signals which are not easily identified through matched filtering techniques. Table 4.7 links both signal calibration and noise knowledge capabilities to their corresponding science objectives.

**Table 4.7:** Science studies and performance requirements supported by the observatory knowledge design characteristics.

| Characteristic | Capability | Observations | Performance | Studies | Science Objective |
|---|---|---|---|---|---|
| Observatory Knowledge | Instrumental glitch identification/mitigation | Distinguish transient events of cosmological and/or astrophysical origin from instrumental sources. | Unvetoed instrumental glitch rate compatible with the bounds on cosmic string cusp rates. | Gravitational wave bursts from cusps and kinks of cosmic strings. | SO8 |
| | Observatory response | Determination of phase and amplitude of induced strain in the observatory. | Knowledge of response to 0.1 % and 0.1 rad across the measurement band. | • Calibrate luminosity distance for Hubble constant measurements <br> • Explore deviations from GR through multiband GW astronomy | • SO6 <br> • SO5 |
| | Observatory Location and orientation | Absolute sky position of sources in standard reference frames. | Absolute spacecraft (S/C) positions to ≤200 km, resulting in sky angle error ≤20 arcsec. | EM counterparts and multimessenger astronomy. | • SO2 <br> • SO6 |
| | Observatory Time Reference | Absolute time stamping of LISA data products. | Post-processed time stamping to 1 ms relative to UTC. | EM counterparts and multimessenger astronomy. | • SO2 <br> • SO6 |

# 5 THE LISA INSTRUMENT

## 5.1 Measurement principle of LISA

The basic measurement in LISA, the interferometric measurement of the change of the distance between pairs of free-falling test masses within separated spacecraft (see also Chapter 2), is for practical reasons broken up into three different measurements in each of the three long arms of LISA. The *long-arm interferometer*, or inter-satellite interferometer (ISI), measures the distance between the optical benches that are rigidly attached to the two spacecraft. In addition, in each of the two spacecraft, the *short-arm interferometers*, or test mass interferometers (TMIs), measure the position of the respective test masses with respect to their local optical bench (see Figure 5.1 and Figure 5.2).

This partition of the measurement couples the measurements on-board pairs of spacecraft, and makes the measurement a *constellation* function. However, while it is not possible to separate the measurement system functionally, the various payload hardware items are housed on individual spacecraft. The partition of the measurement also follows roughly the main contributors to the limits of the mission sensitivity. In the low frequency regime, the sensitivity is limited by force noise on the test mass. In the high frequency regime, the main noise contributor is the readout noise of the interferometric measurement, including the shot-noise (see Figure 7.1).

The performance of the constellation is described by a detailed performance model [230] that contains of order 100 individual noise contributors (see Chapter 7 and also Appendix A), in part based on measurements and e.g. on-orbit performance of LISA Pathfinder (LPF), in part based on analysis. The resulting current best estimate (CBE) for the mission performance is about 20 % better than the science requirements (see Chapter 7).

Each spacecraft contains two Moving Optical Sub-Assemblies (MOSAs), the structural units that contain the elements that support the optical measurement and define the inertial reference: the telescope, the optical bench (OB) and the Gravitational Reference System (GRS) with free-falling test mass (TM). There is an optical interface between the two Moving Optical Sub-Assemblies (MOSAs) in a single spacecraft, the *backlink fibre* that passes laser light between the two optical benches for the local oscillator in the reference interferometer (RFI) interferometer on each of the two optical benches.

The "moving" attribute refers to the fact that the two MOSAs in each spacecraft define optical axes that are nominally separated by 60° but must be dynamically adjusted to accomodate the change of that angle due to orbital mechanics. Over the course of the mission duration, this angle changes by ±1.5° (see Table 6.1 in Section 6.2.1).

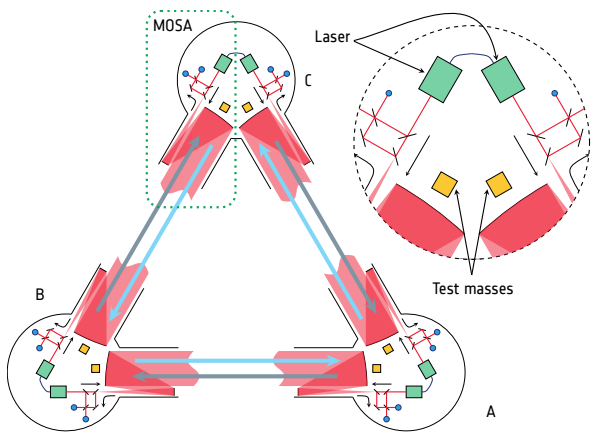

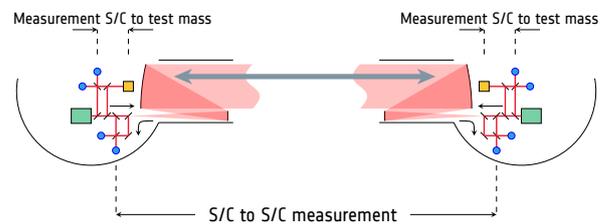

**Figure 5.1:** Schematic view of the LISA constellation and the S/C-S/C measurement. Blue dots show the location of interferometric measurements. For clarity, the redundancy options are left out and the interferometric scheme is simplified, leaving out the TMI. A single Moving Optical Sub-Assembly (MOSA) is highlighted with the teal dotted box.

**Figure 5.2:** Simplified scheme for the full measurement of one arm. The measurement is broken up in three parts, test mass to S/C, S/C to S/C, and finally S/C to test mass. Blue dots indicate interferometric measurements taking place. For clarity, redundancy options are left out and the interferometry is simplified.

## 5.2     Free-falling test masses and the gravitational reference system

The constellation of free-falling test masses traces the gravitational wave tidal acceleration. The test mass itself, plus surrounding hardware and avionics needed to hold, release, shield, sense, force, and discharge the test mass, are known collectively as the Gravitational Reference System (GRS).

The two main roles of the GRS in science operations are to provide geodesic reference at the level of $3\,\text{fm}/(\text{s}^2\,\sqrt{\text{Hz}})$ free-fall at mHz frequencies and provide an end mirror that allows its tracking as part of the $15\,\text{pm}/\sqrt{\text{Hz}}$ composite interferometric measurement between two distant test masses.

These two top requirements are *system* level, involving the measurement architecture, dynamical control, and spacecraft environment. While the GRS has only a limited role in optical metrology, as the local "short arm" end mirror test mass, the GRS hardware plays a dominant role in defining the test mass free-fall environment, and thus limiting stray force noise is the main GRS design driver.

To guarantee the dynamical stability to reach both of the above requirements, the GRS also provides nm-level (100 nrad-level, respectively) capacitive position sensing as well as electrostatic forcing of the test mass. The applied forces allow accelerations of order $\text{nm}/\text{s}^2$ (1–10 $\text{nrad}/\text{s}^2$), to align the test mass to the optical measurement system, on all degrees of freedom except the "science" *x* interferometry axis, along which there is no test mass forcing (see Figure 2.4 for description of the coordinate systems and control scheme).

To reach and maintain free-fall science operations, the GRS also provides a high-load test mass launch lock device, a low-load grabbing and release mechanism accompanied by a higher electrostatic force mode (order $\mu\text{m}/\text{s}^2$ accelerations) to release and electrostatically stop the test mass in orbit, and a UV illumination system, to photoelectrically discharge the test mass.

**Table 5.1:** Key design parameters for the LISA Gravitational Reference System. Ranges of values given for different translational and rotational degrees of freedom.

| | **Parameter** | **Values** |
|---|---|---|
| **test mass** | size, shape, material | 46 mm, 1.93 kg Au-Pt cube |
| | magnetic susceptibility | $|\chi| < 3 \times 10^{-5}$ |
| **electrode housing** | electrodes | 12 sensing, 6 injection |
| | TM-electrode gap | 4 mm ($X$), 2.9 mm ($Y$), 3.5 mm ($Z$) |
| | EH material | Au-coated Mo and Sapphire |
| | residual pressure | $p < 2\,\mu$Pa |
| **position sensing** | sensing bias | 4.9 V (injection), 0.6 V (TM) at 98 304 Hz |
| | sensing noise | 1.8-3 nm/Hz$^{1/2}$, 120-200 nrad/Hz$^{1/2}$ |
| **force/torque actuation** | high resolution authority | 1.7–3 nm/s$^2$, 20–40 nrad/s$^2$ |
| | wide range authority | 400–700 nm/s$^2$ and 17–30 $\mu$rad/s$^2$ |

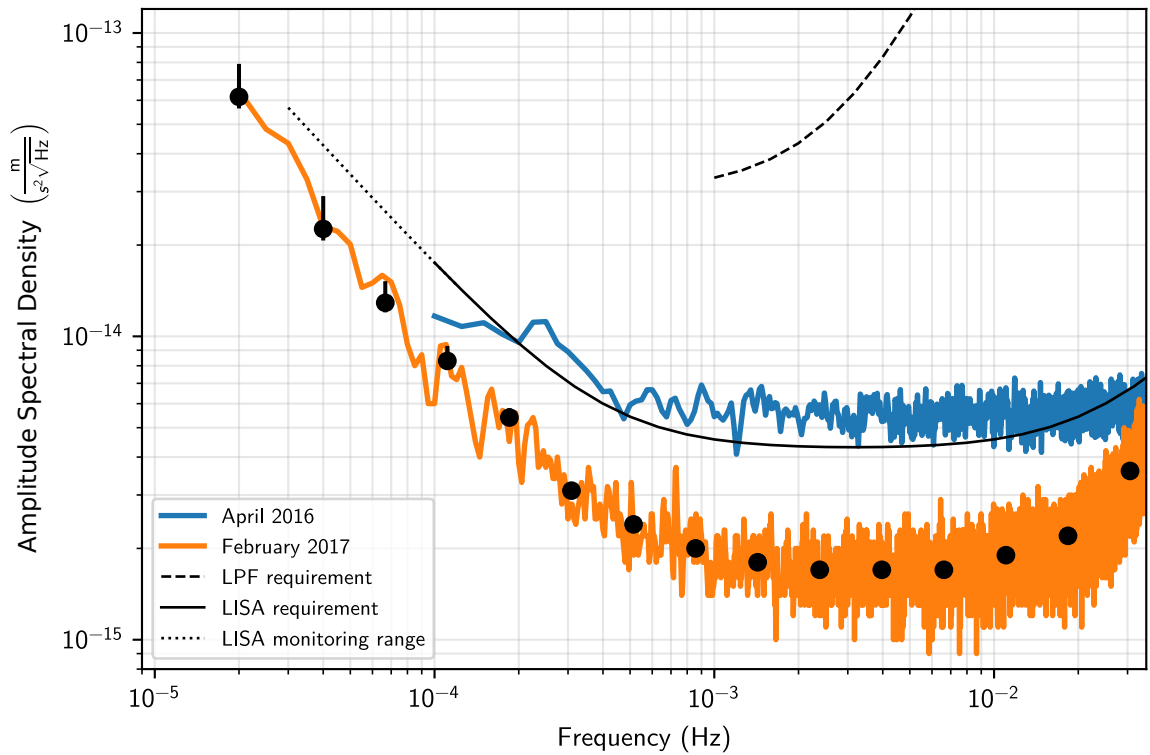

**Figure 5.3:** LISA Pathfinder data for the differential acceleration between two test masses, each inside a LISA Gravitational Reference System (GRS), compared with the equivalent two-test mass requirement for LISA. The improved performance over the year between measurements in the mHz band is due to a lower Brownian noise floor because of a decrease in the residual gas pressure. Data reprinted from [32].

All of these functionalities, with the full LISA performance − most critically the purity of free-fall as quantified by the measured residual test mass acceleration noise across the full LISA band, as illustrated in Figure 5.3 − were demonstrated in the LPF mission [32, 41]. The LISA GRS is a near rebuild of the LPF-heritage GRS [161], with small modifications developed for the slightly different LISA measurement configuration and mission profile and to improve the mission robustness.

The following sections outline the GRS hardware and operations, through the major subsystems − the mechanical GRS head on board the Moving Optical Sub-Assembly (MOSA) and two major avionics units, Front-End Electronics (FEE) and UV light source charge management device − highlighting design changes relative to the LPF GRS.

## 5.2.1 GRS head

### Test mass and electrode housing

> **Use of symbols**
>
> We are using capital letters $\pm X, \pm Y, \pm Z$ to identify *surfaces* of the test mass and lower case letters $x, y, z$ to identify *directions* in this section.

The geometry, dimensions and materials of the test mass and electrode housing are driven by the extreme test mass acceleration noise requirements. Compared with geodesy accelerometers with similar functionality such as those on, e. g., CHAMP [355], GRACE [393], and GOCE [131], the LISA design increases the test mass and the gap between test mass and electrodes to limit acceleration noise from hard-to-model, short-range electrostatic and molecular collision forces [129, 161, 429]. The allowable test mass and gap sizes are however limited by mass / volume constraints, loads on the caging and release mechanisms, and the need to achieve nm-level capacitive sensing. Additionally, we eliminate the discharge wire used to keep a neutral test mass in other space accelerometers, instead using a truly free test mass with a UV discharge system [386] (charge management device or CMD, discussed later in this section). The GRS design is thus a trade-off, fine-tuned during the LPF development [33] and the LISA design follows the LPF design closely.

The test mass is a 46 mm cube of Au-Pt (see Figure 5.5), chosen for high density and magnetic purity [40], with a total mass of 1.92 kg. The test mass is gold coated to create a photoemissive and electrostatically stable and homogeneous surface, to minimise stray potential variations [29, 34], and polished to a mirror finish for the local interferometric test mass readout. The test mass $Z$ faces include features for mating to the launch lock and release mechanisms.

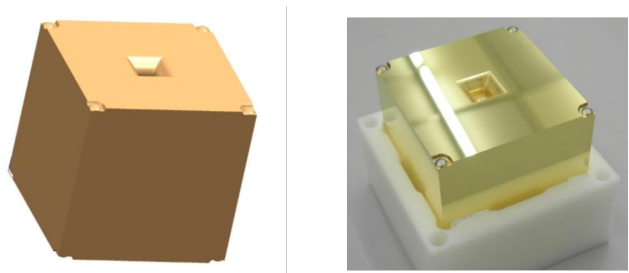

**Figure 5.4:** Left: CAD rendering of the LISA test mass. Right: Photo from the flight model of the test mass in LPF, including $Z$ axis features for mating to the launch lock (corners) and release (central indent) mechanisms.

The electrode housing forms a Faraday cage electrostatic shield including 18 electrodes used for 6 degrees of freedom capacitive sensing and electrostatic forcing of the test mass. Each face has two adjacent sensing electrodes, combined to allow force / torque actuation and translation / rotation sensing. The test mass-electrode gap ranges from 2.9 mm on the $Y$ faces to 4 mm on the performance-critical $X$ face. Injection electrodes on the $Y$ and $Z$ faces provide contact-free AC (roughly 100 kHz) polarisation of the test mass for use in the capacitive sensing readout.

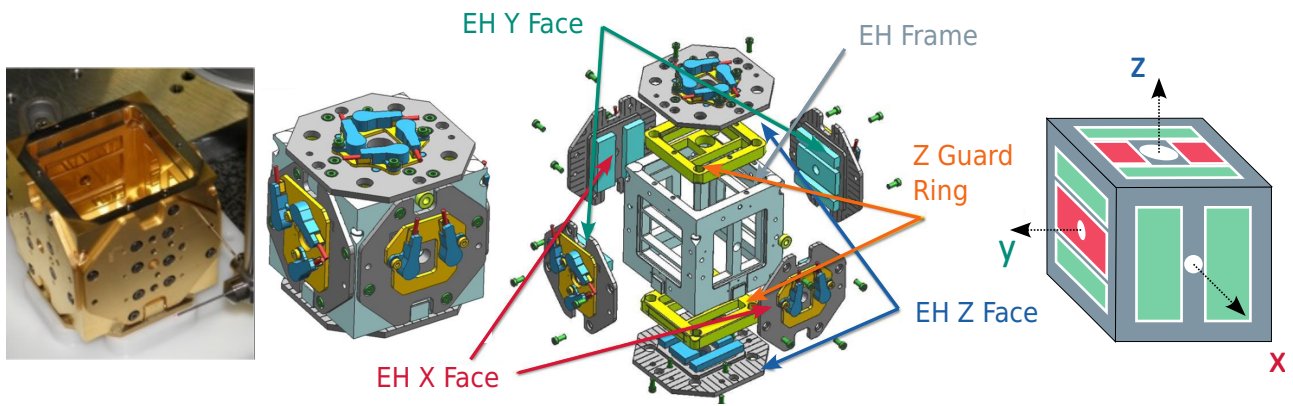

**Figure 5.5:** From left to right: Photo of the test mass electrode housing taken during integration for LISA Pathfinder (LPF); CAD rendering of the mechanical implementation of the electrode housing, in molybdenum (Mo) with sapphire insulating spacers, with shielded coaxial cable connectors; conceptual design with sensing (green), injection (red), and guard ring (grey) surfaces. CAD and photo images courtesy of OHB-Italia.

The electrode housing cage structure and electrodes are bulk molybdenum (Mo) with molybdenum electrodes isolated by sapphire dielectric spacers, chosen for high thermal conductivity, to reduce thermal gradients. For electrostatics and photoemissivity the test mass-facing surfaces are Au-coated. Interruptions in the Au-coated conducting shield are kept to a minimum, with the largest holes on the $Z$ faces to allow access for mechanisms and smaller holes on $X$ and $Y$ allowing venting and $(+X$ only) access for the interferometer laser beam. Smaller holes near the $-Z$ test mass corners allow UV illumination of test mass and electrode housing surfaces for bipolar photoelectric test mass discharge.

### GRS mechanisms

The test mass is firmly held by 8 fingers (see Figure 5.6) mating with dedicated spherical bumps at the test mass $\pm Z$ face corners with roughly 1.2 kN preload. The test mass is "decaged" in orbit by retraction of the fingers by a paraffin-actuated, single-shot *caging and venting mechanism* (CVM), simultaneously vents the entire GRS head to space.

Upon decaging, the test mass is held, with force 1–70 N, by the grabbing, positioning, and release mechanism (GPRM) using opposing piezo-walker-driven *plungers* that mate with the pyramidal recess at the centre of the $\pm Z$ test mass faces. This multiple use mechanism can grab the test mass from any position inside the electrode housing and then position it at the centre of the electrode housing before releasing it into free-fall. The post-release test mass velocity must not exceed 15 µm/s to allow capture, without hitting the electrode housing walls, by the limited-authority electrostatic actuation force system. The delicate release functionality is provided by rapid retraction by a limited-range (tens of µm) piezo-stack pins at the centre of the plunger.

Limiting the residual momentum transfer to the test mass upon release, due to a variety of effects – adhesion, asymmetric elastic forces, secondary collisions from misalignments – is one of the key technologies needed for LISA science with a free-falling test mass. LPF successfully demonstrated test mass release (see [108] for in-flight results for LPF), but typically with velocities exceeding the allowed envelope, causing collisions of the test mass with the electrode housing, requiring additional time and occasional operator intervention to control the test mass. The excess velocities have been traced mainly to misalignments at the test mass-plunger contact and resulting secondary test mass-plunger collisions. Various mitigation strategies are under development for LISA – including improved mechanism control sensors, increased piezo pin stroke, and improved alignment tolerances – while preserving the critical heritage of the mechanism architecture and interfaces.

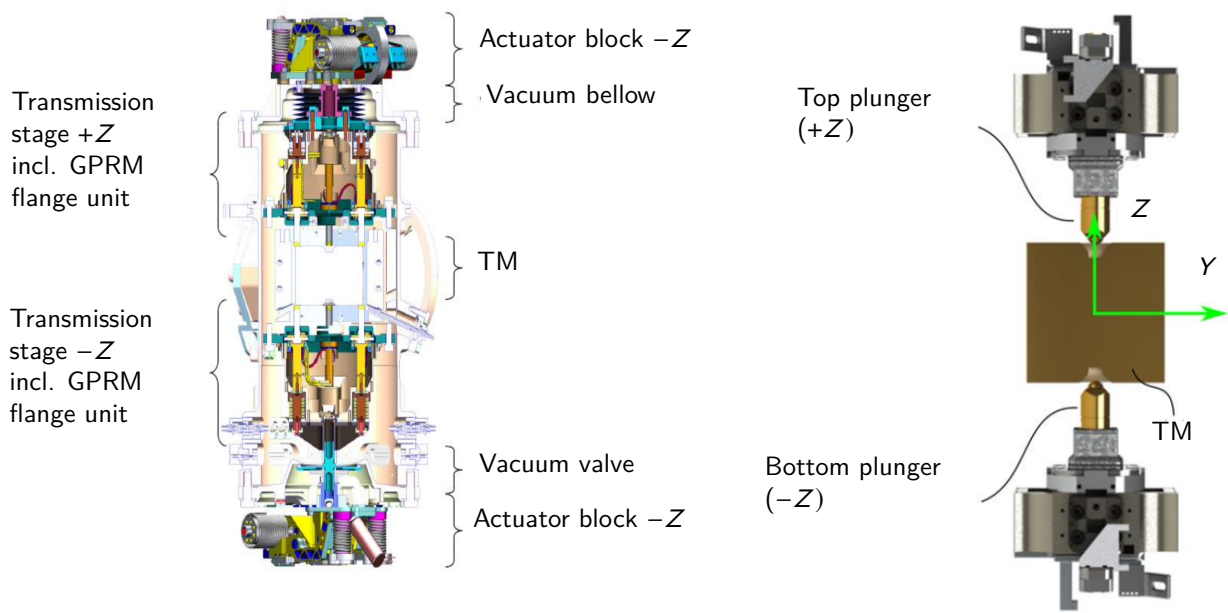

**Figure 5.6:** Illustration of key GRS mechanisms. At left, CVM elements, for the test mass launch lock (see fingers interfacing to test mass corners and associated actuator and transmission stages) and for the venting valve. At right, GPRM mechanism, with the plunger/tip interface to the test mass and piezo-actuator and alignment stages.

### Vacuum chamber

The allowed residual gas pressure around the test mass is $2\,\mu\text{Pa}$, driven by both molecular impact Brownian noise and thermal gradient radiometer effects [39], achieved over the long mission with a dedicated vacuum chamber vented to space. The cylindrical titanium vacuum chamber, illustrated in Figure 5.7 and Viton-sealed valve as part of the caging and venting mechanism (CVM). Both the vacuum chamber and venting valve are LPF heritage, as is the delivery of a sealed GRS with test mass caged until release and venting on orbit.

The vacuum chamber is then connected to the vacuum of space through a tube, whose limited molecular conductance, roughly $15\,\text{l/s}$, limits the allowable outgassing inside the vacuum chamber. The baseline design, roughly $600\,\text{mm}$ long with an L-shaped routing (see Figure 5.7) will be adapted to the final spacecraft configuration, reaching outside the spacecraft to limit possible outgassing contamination. Finally, the end terminal of the vent duct will be equipped with a fixture allowing interface, on ground, to a valve and possible pumping system, in order to allow keeping the vent duct inner surface but also harness and actuation elements for the CVM – in a clean vacuum environment that can be checked and when needed re-evacuated in the several years of shelf life on ground.

The vacuum chamber houses standard electrical feedthroughs, fibre optic feedthroughs for UV light,

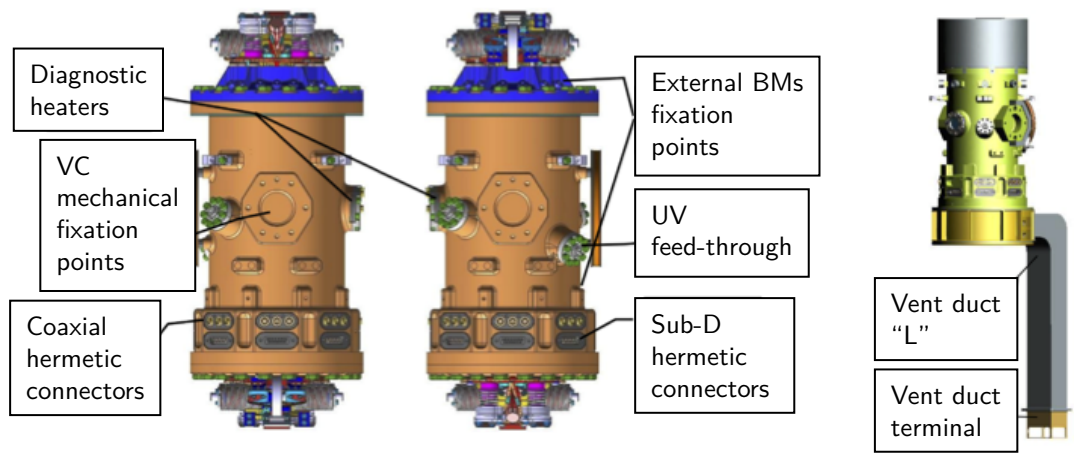

**Figure 5.7:** GRS head vacuum chamber, including (at right) the new vent duct for LISA (images courtesy of OHB-Italia).

and an optical window to allow the local interferometer test mass displacement readout.

The vacuum chamber also houses gravitational balance masses, used to balance the DC forces acting on the test mass, dominated by gravity from the spacecraft mass distribution, in particular from nearby MOSA components. Limiting DC forces and torques is critical for limiting the applied test mass actuation forces (and force noise), to reduce couplings to the spacecraft and MOSA rotational jitter, and to preserve the LISA orbits. LISA will require balancing to within an accuracy of several hundred $pm/s^2$ and to better than 1 $nrad/s^2$ angular acceleration around the test mass $z$ axis. The GRS will provide a balancing authority of up to 35 $nm/s^2$ – mainly pulling along the test mass $-x$ axis to compensate the gravitational pull of the optical bench and telescope systems – and will need to do this with roughly 100 $pm/s^2$ precision.

Multi-axis balancing with the required precision, both in translation and rotation, has been demonstrated with LPF [36, 37], with balancing at 25 $nm/s^2$ level done with a nearly 2 kg tungsten balance mass inside the vacuum chamber behind the $-X$ electrode face. Reaching the 35 $nm/s^2$ authority requested for LISA will be done with LPF-like internal balancing mass and an additional, roughly 1 kg disk on the $-X$ vacuum chamber exterior (far right in Figure 5.7). Additional external mass options exist for balancing on other axes or fine tuning.

## 5.2.2 GRS Front-end electronics

The Front-End Electronics (FEE), conceptually illustrated in Figure 5.8, primarily allows measurement of the test mass displacement with respect to the electrode housing, via an AC-excited capacitance bridge, and application of low-bandwidth test mass control forces with applied audio-frequency carrier voltages.

The interface of the sensing and actuation unit to the electrode housing is a harness of coaxial lines to each sensing/actuation electrode, simultaneously carrying sensing currents and actuation voltages, and coaxial lines to the electrode housing AC injection electrodes with a 98.304 kHz injection voltage to capacitively polarise the test mass for position sensing. A single dedicated FEE board connects to four sensing electrodes – for instance on the electrode housing (EH) $Y$ faces – with capacitive sensing readouts for the two pairs of opposing electrodes, combined to allow translation and rotation – $y$ and $\theta$, in this case – and with drive voltage circuits for actuating $y$ forces and $\theta$ torques. Identical boards are used for the $X$ electrodes ($x/\phi$ degree of freedom - DOF) and for the $Z$ electrodes ($z/\eta$ DOF).

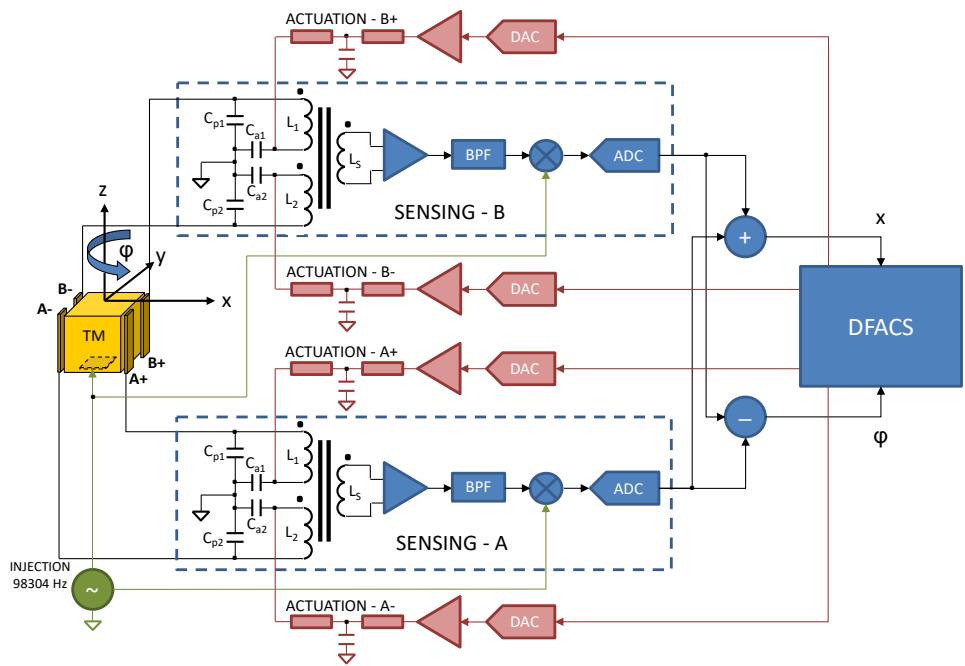

**Figure 5.8:** Conceptual illustration of Gravitational Reference System (GRS) Front-End Electronics (FEE) sensing and actuation circuitry, integrated with the GRS electrode housing/test mass for the $X$ face electrodes serving the $x$ and $\phi$ degrees of freedom. Image courtesy of ETH-Zurich and adapted from [291].

The sensing and actuation unit operates in both a *high resolution* mode, with relatively low actuation voltages for nN-level forces and low noise/small range sensing for use during LISA science operations, and a *wide-range* mode, with larger voltages for μN-level forcing and nearly "wall-to-wall" sensing of the test mass in any position inside the electrode housing, used during *accelerometer mode* following the test mass release, constellation acquisition and other non-science operations.

Sensing and actuation architecture and functionalities are LPF heritage, with LISA performance requirements based on as-measured noise levels in LPF [33]. We provide some details of the key sensing and actuation functionalities, as achieved when the FEE is integrated with the Gravitational Reference Sensor Head (GRSH), in the next two subsections.

### Capacitive Position sensing

Capacitive position sensing employs AC voltages applied in phase to the 6 injection electrodes at $98\,304\,\mathrm{Hz} = 2^{15} \times 3\,\mathrm{Hz}$ with roughly $4.9\,\mathrm{V}$ amplitude, to provide a *contact-free* AC test mass potential ($0.6\,\mathrm{V}$ amplitude). Test mass displacement from the centre of the electrode housing then results in a difference in the AC currents flowing towards pairs of sensing electrodes, a differential current sensed by a transformer bridge readout circuit, with a transimpedance amplifier, demodulation for homodyne detection, and finally AD conversion. The on-board computer combines the output voltages from the bridge readouts of adjacent electrode pairs, to produce translation and rotation outputs.

The required sensitivity of these readouts, six identical bridge circuits per test mass, is roughly

$1\,\mathrm{aF}/\sqrt{\mathrm{Hz}}$ for the 0.6 V test mass bias, which was demonstrated in LPF[33]. A low dissipation, highly symmetric differential transformer is a key component, setting the thermal-noise limit. Low-frequency sensitivity depends on the stability of the transformer imbalance and of the test mass bias voltage. The bridge capacitance, dominated by the coaxial cable harness, is tuned for resonance, which limits the impact of amplifier noise. The "virtual ground" of the sensing electrodes eliminates, to first order, AC currents towards the coax shields and thus makes the bridge readout nearly insensitive to this parasitic cable capacitance.

The science mode saturation range of motion is roughly 100 µm and 10 mrad depending on the degree of freedom, with some tuning allowed with the choice of bias amplitude. For wide-range operation after test mass release, the saturation range is increased to cover displacements up to roughly half the gap, up to several mm, with a gain switch on the readout circuit amplification chain.

### Electrostatic force and torque actuation

The science mode electrostatic force actuation is used on all degrees of freedom except the sensitive $x$ interferometer axis in science mode. It must provide limited forces and torques – order several nN and tens of pNm – needed to compensate $\mathrm{nm/s^2}$ and $\mathrm{nrad/s^2}$ level residual DC accelerations, requiring voltages of order several Volt. To avoid mixing with DC or low-frequency voltages from any test mass charge or from stray electrostatic fields, the actuation employs audio frequency carriers in the 50-300 Hz range, exploiting the $F \propto V^2$ dependence to obtain DC or slowly varying control forces.

The actuation scheme employs six different audio carrier frequencies, one per available actuation degree of freedom. Forces are applied by simultaneously pulling on both sides of the test mass, keeping the sum of the absolute value of the forces constant – holding constant the sum of the mean square voltages on the four electrodes comprising two opposing electrode housing faces – and then imbalancing the voltages on opposing sides of the test mass to obtain a net positive or negative force. This allows for a *constant stiffness* actuation scheme, where a range of applied test mass accelerations – between $\pm g_{0y}$ on the $y$ axis for instance – can be compensated without changing the force (or torque) gradients. A similar scheme is used also for the actuation torques, with most sensing electrodes thus carrying the sum of two audio-frequency voltages for both rotational and translational control. The voltage phases are chosen such that the sum of the applied voltages from a four electrode set is also zero, which keeps the test mass electrostatic potential unchanged, at least for a centred test mass. The constant stiffness scheme, with net neutral voltages and dedicated separate carrier frequencies for each degree of freedom also reduces various crosstalk effects between different degrees of freedom.

While audio carrier frequencies are used for force and torque control, the electrostatic actuation system also can provide small DC or slowly varying offset voltages to any electrode, which are used to compensate stray electrostatic surface potentials and, for charge management (see next subsection), to both measure test mass charge and to bias the flux of photoelectrons between the test mass and electrode housing surfaces.

The overall actuation scheme is full heritage from LPF and indeed is one of the key elements that allowed the LPF acceleration noise benchmark [32, 303]. The use of audio frequency carriers was an essential LPF design innovation, shifting critical electrostatic force disturbances out of the LISA band and thus limiting a potentially dominant noise from stray electrostatic fields. The LPF in-flight performance is compatible with LISA requirements, perhaps most notably for the stability of the actuation gain – *multiplicative noise* at the roughly $50 \times 10^{-6}$ $1/\sqrt{\text{Hz}}$ level at 0.1 mHz – and of the in-band *additive noise* – better than $70\,\mu\text{V}/\sqrt{\text{Hz}}$ at 0.1 mHz [34].

For test mass release, larger forces are needed to electrostatically control the test mass in all DOFs. This is done with a switch to a higher voltage output, allowing audio frequency carriers up to roughly 150 V, allowing applied accelerations up to of order $500\,\text{nm/s}^2$ and $20\,\mu\text{rad/s}^2$.

### Changes to the GRS FEE design

The overall GRS FEE design and performance enjoy extensive LPF heritage, tested down to the 100 μHz band limit and even below, though several changes in design are under development.

Some are related to the LISA era, to handle obsolescence of certain electrical components, and to meet specific LISA interfaces – such as the clock frequency and bus voltages – and finally for the 6 units needed for the LISA configuration, with each GRS head matched to a dedicated FEE (while a single FEE served both GRS in LPF). The larger LISA spacecraft will likely result in a longer coaxial harness, with some slight modifications to accommodate resonance with a larger capacitance. Other design changes include details of the wide range actuation interface, with a dedicated high-voltage amplifier replacing the step-up transformer employed in LPF for improved sensing/actuation isolation, and some flexibility introduced to the waveform commanding interface. On the high resolution side, a subtle digitisation-linearity issue [303] is being addressed.

## 5.2.3    GRS charge management device

UV light introduced into the electrode housing and illuminating preferentially either the test mass or the electrode housing surfaces (see illustration in Figure 5.9) is used to photo-electrically charge the test mass, respectively, positively or negatively, allowing neutralisation of the test mass from cosmic ray and solar particle charging and, after test mass release, tribolelectric charge transfer from contact with the GPRM. Keeping the test mass nearly neutral, to within ±70 mV potential, is a performance requirement to avoid excessive force noise from fluctuating electrostatic fields, mainly from actuation noise.

The general scheme of bipolar photoelectric test mass discharge using UV light and the test mass / electrode housing gold surfaces was tested successfully in LPF [39], with the same electrode housing and test mass illumination geometry, materials, and processing envisioned for LISA. LISA however will take advantage of UV LED technology for the light source, replacing Hg-vapour lamps, for smaller size, lower power, longer life, and the ability to rapidly pulse the LED illumination to synchronise with applied AC electrostatic fields. This allows the application of light in sync with the applied fields, most importantly the roughly 100 kHz test mass bias, so that the polarity causes the photoelectrons to be transported in the right direction to neutralise the test mass. Compared to the discharge scheme employed in LPF [39], the test mass discharge becomes more easily tuneable and thus more robust against test mass photoemissivity variations.

The Charge Management Deceive (CMD) is composed of UV light unit (ULU) connected to the GRS head through a fibre optic harness. The ULU is a redundant single unit serving both test mass aboard a LISA spacecraft. The development of a UV LED light source has proceeded rapidly [276], with successful testing of a TRL 5 unit in addition to extensive performance and lifetime characterisation of the LEDs themselves, which provide a central wavelength around 250 nm, compared to the Hg lamps used in LPF, which had a wavelength of 254 nm.

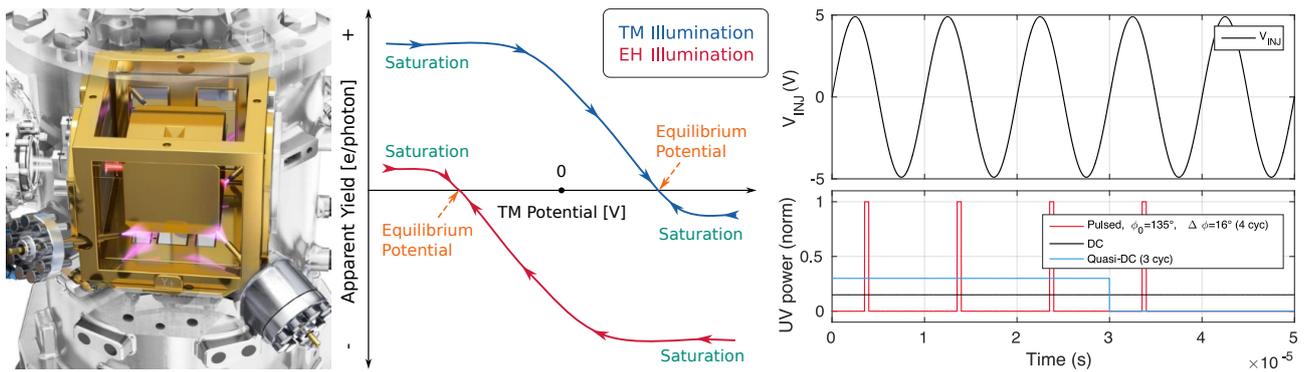

**Figure 5.9:** Illumination scheme for the heritage LPF GRS, showing Inertial Sensor UV Kit (ISUK) illumination of the electrode housing and test mass. At centre illustration of the effective yield ($Y$) governing test mass discharge, photoelectric current in charges per UV photon, as a function of the test mass potential. The test mass equilibrium voltage, where $Y = 0$ and measured in the $\pm 100$ mV range for LPF, can be tuned, moving the apparent yield curve right or left, by illuminating the test mass only at specific phases of the test mass 100 kHz bias. At right is an illustration of pulsed UV light synchronised to the injection bias, as well as DC and quasi-DC modes where light is applied for multiples of the injection period.

The light output from a single LED is digitally programmed to provide pulses that are synchronised to the 98 304 Hz injection voltage, with the pulse height (LED drive current), average pulse phase – relative to the injection voltage – and pulse width, as well as the number of ON pulses per repetition cycle (16 Hz). The pulse width can also be set to cover the full injection period, roughly 10 µs, the illumination is constant over the injection period [241]. This "quasi-DC" illumination can even be applied to all injection periods to gives a true DC illumination and thus a fallback to LPF-heritage.

Two main operation modes are foreseen for charge management, *continuous discharge*, and *fast discharge*, described in the following.

### Continuous discharge

In continuous discharge, the test mass is kept nearly neutral by compensating the environmental charging, likely to be in the 10–50 e/s (electrons/second) range over the course of the solar cycle, based on both simulations and LPF measurements [35]. Discharge to an equilibrium test mass potential, near zero for a desired neutral test mass, is passively stable: the photocurrent to the test mass becoming more negative (positive) with increased (decreased) test mass potential (see effective yield data in Figure 5.9). The relative phase of UV light pulses with respect to the 98 304 Hz test mass sensing bias field is adjusted to fine tune the test mass equilibrium potential to zero, while the UV power sets the time to reach equilibrium, likely of order several hours to days in LISA.

Continuous discharge will require initial in-flight calibration and periodic monitoring of the test mass charge, but not any active closed loop on-board control of the UV illumination in typical science operations. It has the advantage of not requiring interruptions to the LISA science, in addition to allowing reduction of the force noise caused by non-zero and linearly increasing test mass charge.

### Fast discharge

A fast discharge mode of operations is needed for quick neutralisation of a highly charged test mass following release and as an alternate science discharge scheme. This allows the test mass charge to freely accumulate while science data are acquired, until a threshold – in charge or, more simply, in elapsed time – is reached. Then the test mass is rapidly discharged, in a brief intervention during which performance requirements could be breached, before starting another period of science measurement and environmental charging.

Typical time between discharges would be several days to several weeks, set by the environmental charging and the requirement to keep the test mass potential in the ±70 mV range. duration of the fast discharge is about 10 minutes with a maximal disruption of science mode of 1 hour.

The fast discharge can be operated from ground based on timeline-scheduled illuminations. It was routinely used in LPF and is even more robust than the continuous discharge, allowing if needed even larger DC bias voltages in addition to pulse phase tuning.

## 5.3   Optical metrology interferometry instrumentation

The optical metrology instrumentation comprises the subsystems that are required to make the interferometric measurement. The telescope is needed to send the light to the other spacecraft as well as to receive it, the optical bench houses all the optical components required to direct the light to the telescope and to the detectors, the phase measurement system receives the light on the photodetectors and extracts the changing phase, providing the core of the interferometric measurement, and the laser system provides the laser light for the various interferometers and reference systems on board a spacecraft.

### 5.3.1   Telescope

Each spacecraft contains two telescopes giving a total of six telescopes in the constellation. Each telescope functions as an afocal beam expander and operates simultaneously in transmit and receive modes. Its primary purpose is to facilitate a precision length measurement between the optical benches (and ultimately the test masses) on widely separated spacecraft. For transmission, the telescope takes a collimated beam from the optical bench, with a diameter of approximately 2.24 mm, and transforms it into a collimated beam, with a diameter of about 300 mm, and a profile optimised to deliver power efficiently on-axis in the far field. For reception, the telescope collects light from the central portion of the beam sent by the far spacecraft and reduces it to a collimated beam which is delivered to the optical bench. Different polarisations of the light ensure that the transmit and receive beams can be separated on the optical bench.

An oversizing of the telescope primary mirror ensures that the small lateral beam movements produced by the beam alignment mechanism (BAM) will not lead to vignetting of the transmitted beam. Note that the transmitted beam does not have a Gaussian beam profile but a top-hat profile to minimise scattered light at the edges of the telescope mirror.

The telescope also supports the Constellation Acquisition Sensor (CAS), providing a larger field-of-view for the acquisition of the constellation than for the science optical chain.

### Telescope design

The primary design parameter has been chosen of 300 mm balances optical considerations, with larger diameters leading to a higher efficiency of both transmission and reception, against mass and size. The dimensional stability of the telescope, the quality of its output wavefront, the quality of the mapping between conjugate pupils, and the coherent backscatter of the transmitted beam are the other critical design drivers. While many design features would benefit from an on-axis (axisymmetric) layout, the coherent back-reflection (the so-called Narcissus reflection) from an on-axis secondary mirror can easily dominate the very low power being received from the far spacecraft by many orders of magnitude. To avoid that, an unobstructed off-axis design has been chosen [173].

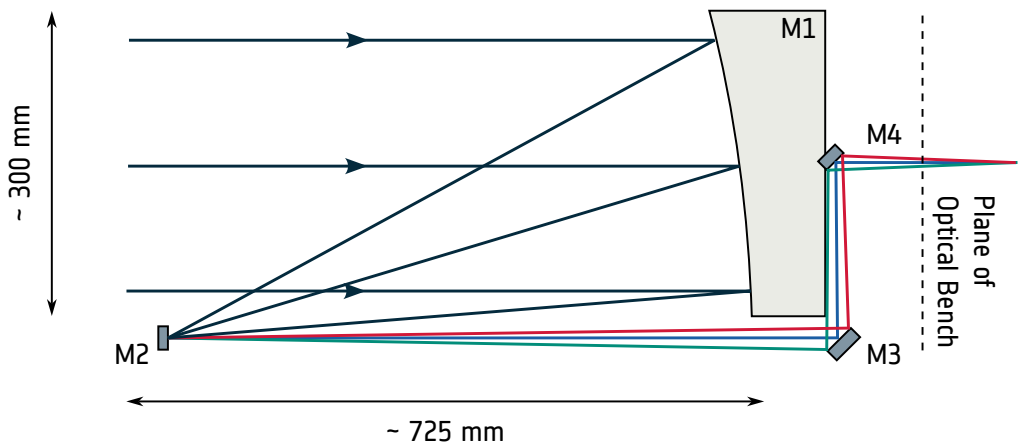

**Figure 5.10:** LISA Telescope Design with 4 mirrors with 134× magnification. The perimeter of the optical bench is only shown schematically, to indicate the location of its surfaces with respect to the beam's propagation path. In reality its body extends down and catches the telescope output and deflects it to the bench optics.

As scattered light recombining with the received beam can produce a spurious noise signal at the science heterodyne frequency, the phase jitter of the scattered light can contribute to noise in the measurement band. This drives the requirements on dimensional stability of the telescope and the associated structures in the presence of temperature fluctuations and gradients [217, 321].

The required system magnification is determined by the ratio of the size of the input aperture of the telescope and the input aperture of the quadrant photo receiver (QPR). Part of the overall magnification is provided by the telescope, the remainder is taken up by beam-matching optics in front of the QPR. The requirements on scattered light lead to reducing the number of mirrors, leading to a design similar to the 4-mirror prescription in Figure 5.10 and places strict tolerances on the position and orientation of each optic. In addition, it sets requirements on the allowable change in mirror position, both on orbit and between ground and space.

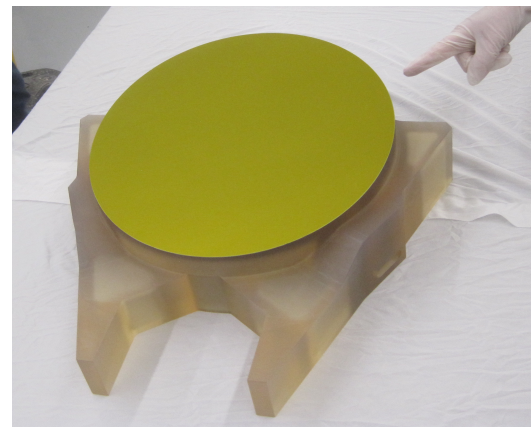

**Figure 5.11:** Integrated primary mirror and strongback structure for an Engineering Development Unit of the LISA Telescope. The NASA telescope development activity will demonstrate key performance and environmental requirements with full-scale telescope models. Image Credit: L3 Harris Corp./NASA

The baseline is a telescope layout which consists of a well-corrected front end (M1/M2) (see Figure 5.11 for an engineering model of the M1), in a Cassegrain format, with aft optics (M3/M4) which correct the pupil size, location, and boresighting. Freeform surfaces in the aft optics minimise the change in optical pathlength for light traversing the telescope due to variations in field angle. While the telescope is unobstructed and not on-axis in the usual sense, the beam is brought back to the centre axis when it is exchanged with the optical bench, in order to minimise the coupling of angular jitter at the telescope/optical bench interface into a pathlength signal. The focal planes of the two mirror-pairs are overlapped to provide a low wavefront error image of the aperture stop of the right size and location, and the position of the telescope with respect to the aperture stop on the optical bench is used to align the measurement axes.

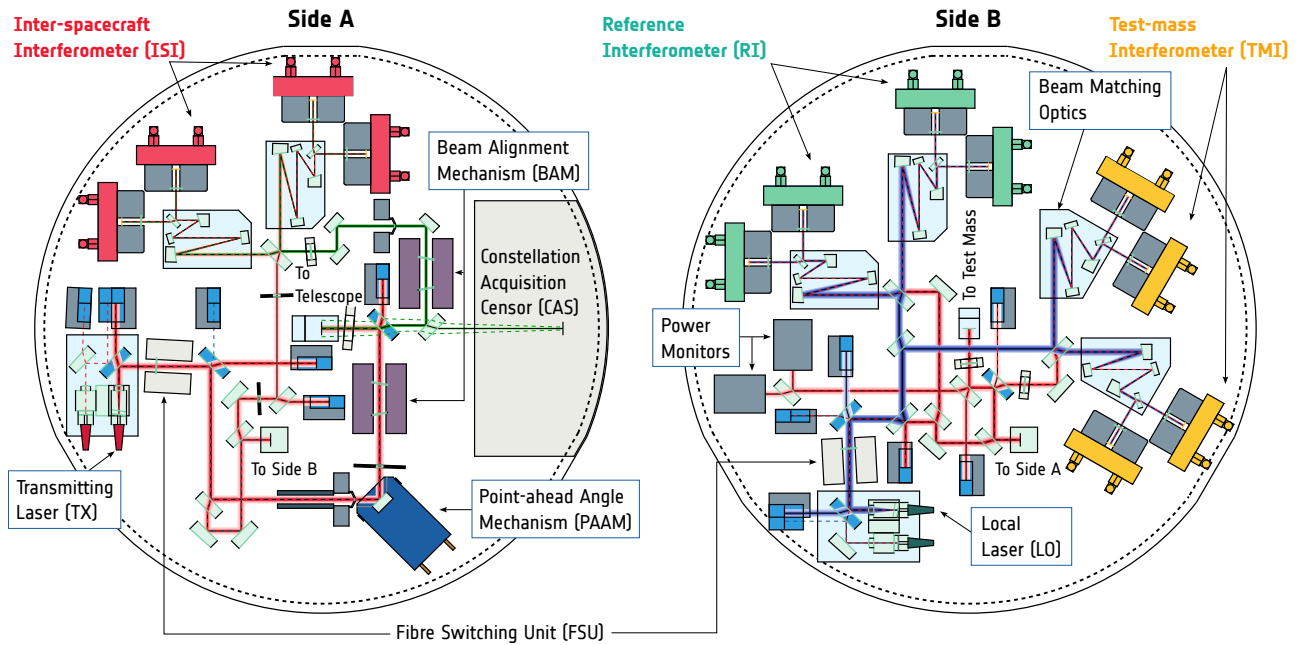

**Figure 5.12:** Schematics of the optical bench A and B-Sides. Shown are all light paths with the interferometric readout (quadrant photo receiver and Beam Matching Optics) of the inter-satellite interferometer (ISI), reference interferometer (RFI), and test mass interferometer (TMI) as well as the point-ahead angle mechanism (PAAM), fibre switching units (FSUs), beam alignment mechanism (BAM), Local (LO) and transmitting (TX) laser, power monitors (PMONs) and various interfaces between the two sides, the telescope and the Constellation Acquisition Sensor (CAS).

The telescope will be thermally isolated from in-band spacecraft temperature fluctuations by a series of passive radiative shields. A quiet thermal environment in combination with a structure composed of an ultra-low coefficient of thermal expansion material, such as Zerodur or ULE, provides a means to reach the target displacement sensitivity. A heating system of modest power consumption, to replace the heat which the telescope radiates to cold space, allows room temperature operation in orbit. This simplifies ground testing, and allows ultra-low thermal expansion coefficient (CTE) glasses to be used near their typical designed operating temperatures.

## 5.3.2 Optical Bench

The optical bench provides three interferometers that are needed in each MOSA to support the overall mission level measurement concept. One interferometer, the inter-satellite interferometer (ISI), measures the relative motion of the optical bench (and by implication, the MOSA) along its *x*-axis with respect to the incoming wavefront received from the far spacecraft; a second interferometer, the test mass interferometer (TMI), measures the relative longitudinal motion of the optical bench (and by implication, the MOSA) with respect to the free-falling test mass of that MOSA; and a third interferometer, the reference interferometer (RFI), that measures the relative phase fluctuations between the local laser used on one MOSA and the laser used on the adjacent MOSA within the same spacecraft. The exchange of laser beams between the MOSAs is enabled by the so-called backlink fibre, that delivers light from one MOSA to the other and is built from a pair of polarisation maintaining fibres, one of which is kept in cold redundancy.

The interferometers on the optical bench also measure the angles (tip and tilt around the beam axis) of both the test mass with respect to the optical bench, and of the optical bench with respect to the incoming wavefront. [130, 227, 312]. These angle measurements are used to derive attitude control signals for the Drag-Free Attitude Control System (DFACS).

The conversion of the optical signals into electrical signals is performed by the QPR (see Section 5.3.2.5).

### Optical bench, A-side

To effectively use the optical bench it is populated on both sides, the A-side and the B-side that are connected through a periscope. The A-side contains the recombining optics for the ISI, the optics and optomechanics to launch the transmit (TX) beam and optics to monitor the laser power (see left panel in Figure 5.12).

The TX beam (red) enters the optical bench via a fibre launcher and a polariser and is directed through a beam-shaping aperture (TX aperture) to the point-ahead angle mechanism (PAAM) and the BAM and then to the telescope interface where the beam leaves the optical bench (out of the drawing plane) after rotating its polarisation by 90° by a half-wave plate. A beamsplitter splits of a small amount of light to serve as local oscillator (or local laser) for the ISI and part of which is also sent to the B-side to serve as measurement beam for the TMI and the RFI.

The detector assembly has two sets of two detectors each, providing redundant readout for each of the two output ports of the recombining beamsplitter. Both output ports are used to reduce the effect of the relative intensity noise (RIN) on phase noise of the interferometric signal [434].

The receive (RX) beam (green) enters the A-side through the telescope interface and has its polarisation rotated by 90° by a half-wave plate so that it is transmitted through the immediately following polarising beamsplitter. It then passes through the RX BAM and the RX aperture and has its polarisation rotated again by 90° to enable interference with the local oscillator at the recombining beamsplitter. A fraction of the RX beam is also directed to the CAS.

### Optical bench, B-side

The B-side (see right panel in Figure 5.12) contains the recombining beamsplitters for the TMI and the RFI. The local oscillator for both interferometers is derived from light that is delivered through the backlink fibre from the optical bench on the other MOSA. The TMI uses the light that is directed to the test mass via the test mass interface (TM-I/F). In front of the TM-I/F is a quarter-wave plate, converting the linearly polarised light coming from the A-side into circularly polarised light going to the test mass and back into the other linear polarisation when the light returns from the test mass. This allows a polarising beamsplitter to reflect the light going towards the test mass but to transmit the light returning from the test mass. After passing this polarising beamsplitter, the plane of polarisation of the light is rotated again by 90° by a half-wave plate to match the polarisation of the local oscillator. The returning light from the test mass and the local oscillator, both now at the same polarisation, are combined on a beamsplitter and read out by a quadrant photodetector (QPD) assembly. The B-side also houses redundant power monitors (PMONs) to monitor the power of the

→ THE EUROPEAN SPACE AGENCY

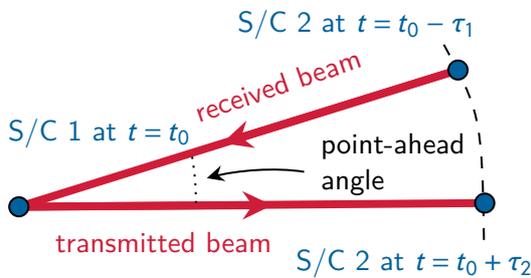

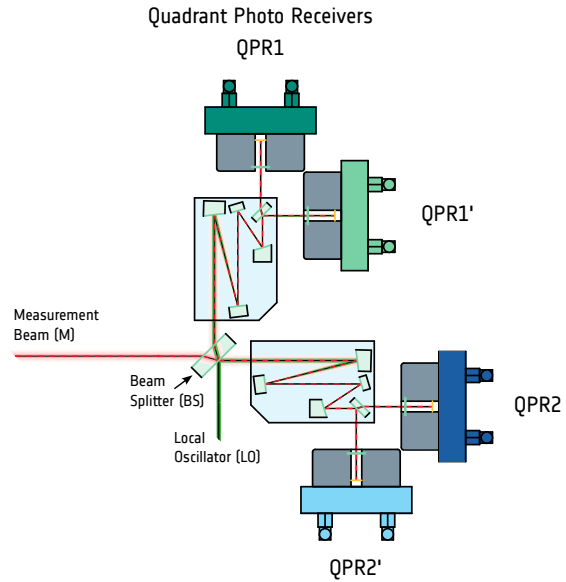

**Figure 5.13:** The point-ahead angle is determined by orbital dynamics and the light-travel-times $\tau_k$ between the respective spacecraft: the received beam is coming from the direction the distant spacecraft appeared at the time $t_0 - \tau_1$, whereas the transmit beam has to be sent to the direction where the distant spacecraft will be a time $t_0 + \tau_2$. The numerical values for the point-ahead angle are in the µrad range.

**Figure 5.14:** Redundancy scheme for the readout of each of the interferometers on the optical bench. Measurement beam (M) and local oscillator (LO) are combined on the beamsplitter (BS). Both outputs (1) and (2) are detected by a pair of QPRs (QPR1/QPR1' and QPR2/QPR2', respectively.)

light going to the test mass.

### Point-Ahead-Angle-Mechanism

The point-ahead angle mechanism (PAAM) is a mechanism that allows the transmit (TX) beam to be launched at an angle with the receive (RX) beam. The required angle is determined by orbital dynamics and appears due to the change of position of the spacecraft over the light-travel time $\tau$ (see Figure 5.13).

The TX beam steering angle needs to be adjusted at least once every few minutes with an accuracy of about 10 nrad to ensure the beam is received by the distant spacecraft. There will be one PAAM per optical bench, with two piezo stacks and capacitive sensors for redundancy and closed-loop control. The PAAM will be controlled via a dedicated control unit commanded by the on-board computer (OBC). As the PAAM is in the beam path of the transmit laser beam and hence part of the inter-satellite interferometer it needs to be mounted directly on the optical bench to ensure alignment stability.

### Fibre Switching Unit

The fibre switching unit (FSU) is a dedicated mechanism designed to switch between the optical fibres that deliver the prime and, respectively, redundant laser light to the optical bench. Each optical bench has two FSUs, one to switch between the prime and redundant optical bench feed fibres (coming from the laser system) and one to switch between the prime and redundant backlink fibre by rotating a $\lambda/2$-waveplate to change the polarisation of the light and two polarising beamsplitters to switch between the two polarisation-maintaining fibres in orthogonal orientation. Currently, the fibre switching unit (FSU) operational concept is to use it on the ground, during initial system setup in space, and thereafter only in case of any failure.

### *Quadrant Photo Receivers*

The quadrant photo receivers (QPRs) detect the optical beatnotes produced by the three interferometers on each optical bench, the ISI, the TMI, and the RFI. To minimise the sensitivity to amplitude modulations of the laser at the frequency of the beatnote and for redundancy reasons, each interferometer is read out by four QPRs (see Figure 5.14). Combining the signals of both outputs of the recombining beamsplitter minimises the sensitivity to amplitude modulations [434], doubling up the readout protects against failure of one of the QPRs.

The QPRs are mounted directly on the optical bench to ensure that the tight alignment requirements with respect to the optical beam can be maintained over the lifetime of the mission. However, due to the thermal load provided by the QPRs, they will be mounted on the backside of the optical bench to minimise thermally-induced distortions.

Each QPR features three main components: 1) an InGaAs quadrant photodetector (QPD) with four segments, serving to convert the optical signal into a photocurrent; 2) QPR-FEE (front-end electronics) with the main goal of amplifying each QPD segment's photocurrent and converting it into voltage [75], and 3) a mechanical enclosure providing the precise and stable positioning of each QPD with the optical beam, assuring the electromagnetic compatibility of the QPRs, and housing the QPD and the FEE in the limited volume allocated on the optical bench. The QPR voltage signals are delivered to the phase measurement subsystem (PMS) for subsequent phase measurement and auxiliary functions.

## 5.3.3 Phase Measurement System and Frequency Distribution System

The PMS, which, together with the interferometers is hosted on the optical bench, performs the high-fidelity measurements of relative motions (angular and longitudinal) of the spacecraft and free-falling test masses as well as the measurement of the reference laser. This includes the gravitational wave signals which arrive as phase fluctuations of the weak beam coming from the remote spacecraft, and subsequently are converted into phase fluctuations of the beatnote photocurrent in the range of 6–25 MHz (the exact values depend on the frequency plan, driven by the Doppler shifts and the laser RIN spectrum), detected with the quadrant photodetectors on the optical bench. In addition, the phasemeter needs to record auxiliary beatnotes with high fidelity for clock-transfer and correction and handle the inter-spacecraft absolute ranging and data transfer functionality [226]. All these functions have very tight requirements on high-resolution timing, in terms of timing noise as well as timestamp accuracy and consistency, both between the units aboard one spacecraft as well as between the three spacecraft in the constellation. Clock transfer and correction have been experimentally verified using a laboratory test-bed [374, 438]; ranging and data transfer have been demonstrated using a optical test-bed and a PMS prototype [246].

As the PMS needs a number of electrical signals in the radio frequency range to perform its functions, the current architecture foresees a frequency distribution system (FDS) integrated into the PMS. The FDS uses an ultra-stable oscillator (USO) to generate all frequencies and signals required on the spacecraft to allow high-resolution timing of phasemeter measurements, transfer of amplified ultra-stable oscillator (USO) noise via GHz sidebands to the other spacecraft, provision of a pilot tone signal to the PMS front to enable a correction of the analog-to-digital converter (ADC) sampling jitter, establishment of a single consistent high-resolution timer for all units on the spacecraft that attach timestamps to measurements, and providing a suitable clock frequency to all units on the spacecraft that require a synchronised clock. More details of this scheme are given in [225].

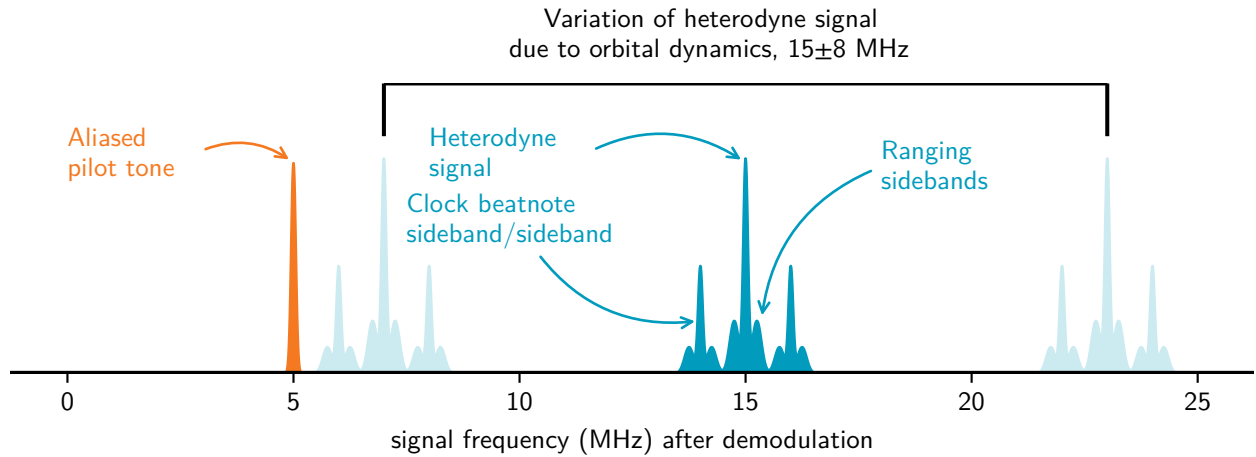

**Figure 5.15:** Sketch of the input spectrum for the inter-satellite interferometer phasemeter channels after digitisation.

### Phase Measurement System

After suitable pre-amplification and conditioning of the beatnote signals in the QPR, the phasemeter core digitises the signals with synchronised analog-to-digital converter (ADC), one per channel, at a sampling frequency of 80 MHz and a resolution of 14 bits.

Due to the inherent sampling jitter in each ADC, a pilot-tone correction scheme is implemented by adding a sinusoidal auxiliary tone at 75 MHz to each channel before digitisation, resulting in a 5 MHz tone after downsampling with 80 MHz. Since the pilot tone is well known, it can be used to measure the ADC timing jitter and correct for it in one of the first processing stages of the phasemeter raw data. Further details about the expected phasemeter performance and pilot tone correction can be found in [74].

Each photodiode channel is processed by a digital signal processor (DSP) implemented on field-programmable gate arrays (FPGAs). Processing depends on the channel's specific role within the optical metrology. The most demanding photodiode channels sensing the inter-satellite interferometer (ISI) have the highest complexity and are described in the following. A digitised signal from one photodiode segment (a "channel") contains a total of four tones (see Figure 5.15): the main interferometer signal carrier, two clock sideband-sideband beatnotes, and the aliased pilot-tone. In addition it contains spread-spectrum sidebands around the carrier containing the pseudo-random noise (PRN) ranging and data communication information.

The phase of each tone is tracked by an all-digital phase-locked loop (ADPLL), creating effectively a local, digital copy of the incoming signal frequency and phase by locking a numerically controlled oscillator (NCO) to the tone. The error signal for the tracking loop, referred to as Q-value, is generated by mixing (multiplying) the quadrature signal of the NCO with the incoming signal and low-pass filtering, generating a band-pass effect that enables tracking of each incoming tone individually without cross-talk.

The in-phase signal of the NCO is multiplied with the incoming tones as well to create a readout of the tone amplitude, referred to as I-value. The main readout of the all-digital phase-locked loop (ADPLL) is the frequency and the phase of the NCO, available in the so-called phase increment register (PIR) and the phase accumulation register (PAR). As the high frequency leads to a frequent wrap-around of the PAR, the PAR readout is only used when a suitable phase reference, or another channel, can be subtracted before decimation, such as the generation of differential wavefront sensing signals, where phase differences between different channels tracking the same overall frequencies have to be extracted. The PIR values are used for the main longitudinal phase readout, being converted back into phase by scaling and integrating them in post-processing.

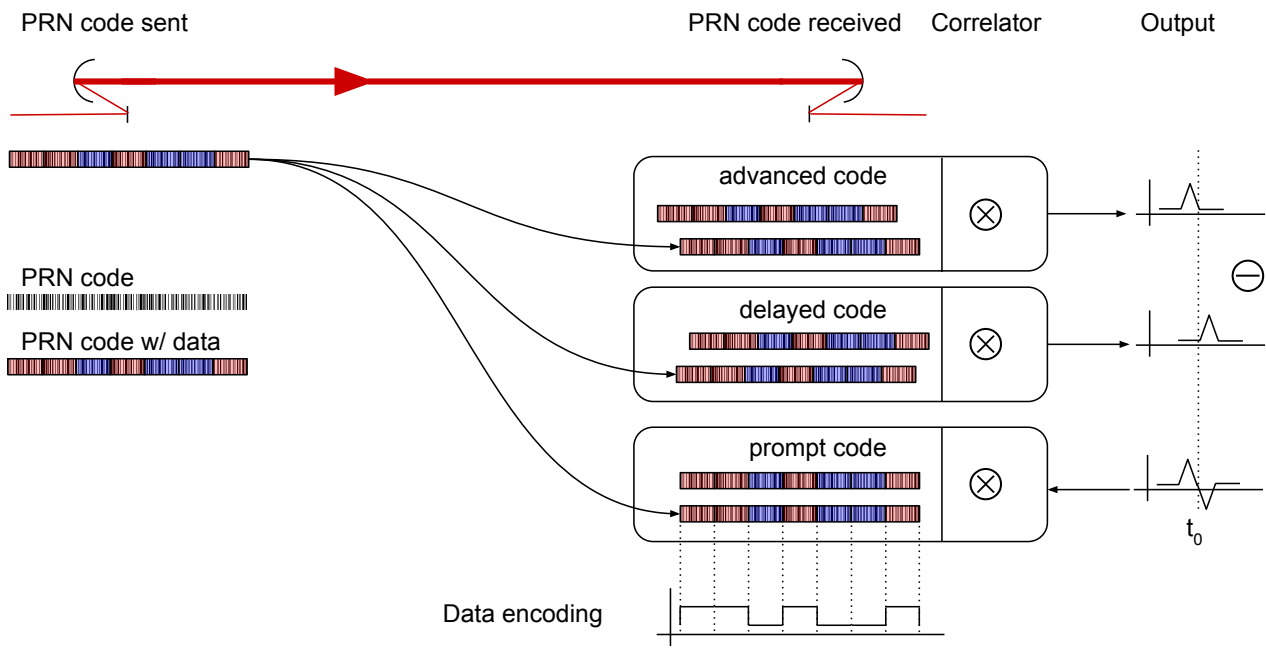

**Figure 5.16:** Schematic description of the ranging and data encoding using pseudo-random noise (PRN) codes. The received codes are correlated with slightly delayed and advanced copies of the PRN code present at the other spacecraft to establish the exact timing offset. The prompt copy is then correlated with the received code to extract the data code that is superimposed by flipping the phase of the PRN code, here symbolised by the red and blue parts of the code.

### Ranging data and inter-spacecraft communication

Extraction of the ranging signals is implemented via delay-locked loops (DLLs) that use the unfiltered Q output of the all-digital phase-locked loop measuring the phase of the inter-satellite interferometer as input. This implementation enables the DLL to extract information only from the PRN sidebands centred around the main carrier. The DLL contains copies of the known received PRN codes and establishes correlations between the local copy and the incoming signal by mixing them and integrating over given code segments. The DLL does this with a slightly delayed and a slightly advanced version of the code. By subtracting these two correlations an error signal is generated that is used to track the timing offset of the local PRN code, locking the local delay to the incoming one (see Figure 5.16). Similar to the ADPLL the readout signal here is given by the timeoffset register of the local PRN codes. Typically, a DLL is implemented for both codes present in each photocurrent, however, the readout of the local code is generally only useful for sanity checks or housekeeping measures. The PRN code time offset, both for the local copies controlled by the DLL and the copies being sent out to the far spacecraft require a fixed, constant time correlation to the local copy of the Spacecraft Elapsed Time (SCET) to ensure that all delays in the system can be properly accounted for in post-processing.

To extract the data encoded within the PRN codes the DLL also generates a punctual version of the local code and correlates it with the PRN sidebands. The result is then fed into a state machine that realises the specific decoding algorithm. Typically the PRN codes have a rate of about 1.25 MHz, with lengths of 1024 code bits, leading to a repetition rate of about 1.2 kHz, equivalent to an ambiguity-free range of 250 km. The data are encoded by switching the sign of the PRN codes 16 code bits per data bit, leading to a raw data rate in the order of about 78 ksamples/s. Additional encoding algorithms for bit error correction, redundancy and handshaking reduce this to an effective bit-rate optimised based on the final signal to noise ratio. The required ranging accuracy is about 30 m and has been demonstrated with a large margin [246]. Further post-processing of the data streams making use of the individual spacecraft orbits in LISA and the free-fall conditions of the test masses allow to increase the ranging accuracy from meters to a few centimeters [425, 426].

## 5.3.4    Lasers

LISA requires three bi-directional laser links between the spacecraft to record differential changes in test mass distances on the level of a few picometer via heterodyne interferometry. Laser phase noise couples into the phase measurements, requiring suppression of laser phase noise via stabilisation onto an optical reference combined with on-ground post-corrections using the Time-Delay Interferometry (TDI) algorithm (see Section 2.3 and Section 8.2.1). The amplitude fluctuation of the laser also needs to comply with strict requirements in the measurement band as not only does amplitude noise couple into phase noise to a certain extent, the position of the test mass is also probed by laser interferometry and hence susceptible to fluctuating radiation pressure.

In addition to the use for the interferometric measurement, the laser light transmitted between the spacecraft is also used for determining the *absolute* distance between the spacecraft with about 1 m accuracy ("ranging") and to transmit information between the spacecraft.

The laser system thus needs to provide highly phase-coherent, tuneable, and intensity-stable laser light for the interferometric measurement chain as well as high-fidelity phase modulation capabilities for ranging, local clock tone and data transfer. The total required output power of the laser system is 2 W at end-of-life to establish the shot-noise limited mission sensitivity; see Section 7.3 and the laser wavelength has been chosen to be 1064 nm. All requirements need to be guaranteed over the full mission duration, calling for an adequate redundancy scheme.

Each MOSA has a laser system associated with it that delivers light to the optical bench. Each laser system must be capable of operating in either of two modes: a *controller* mode, in which the laser frequency is locked to an optical cavity; and a *transponder* mode in which the laser frequency is offset phase-locked (with a tuneable frequency offset) to another optical signal through an interference generated on the optical bench. The tuneable offset allows the frequency of the transponder lasers to be individually tuned such that all beat notes throughout the constellation stay within the interferometric detection bandwidth, despite constantly changing Doppler shifts.

Each spacecraft contains a single main frequency stabilisation (MFS), to which either of the laser system aboard that spacecraft can be locked. Redundancy of the MFS is handled at the constellation level, as only one active MFS is needed at a time, thus double cold redundancy is ensured.

The laser development is currently ongoing under the responsibility of NASA [326, 441]

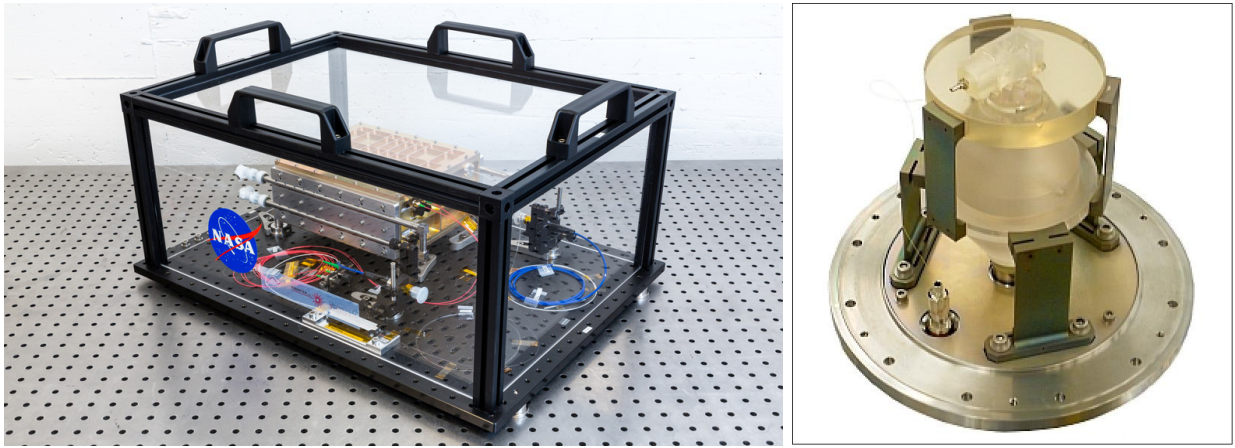

**Figure 5.17:** Left: TRL 5 Laser Demonstration Unit containing internally-redundant Main Oscillator Power Amplifier (MOPA) seed laser, phase modulator, and fiber power amplifier [326]. Image Credit: NASA/GSFC. Right: Optical cavity for the LRI on GRACE-FO [15, 397] providing direct heritage for the LISA main frequency stabilisation (MFS). Image Credit: NASA/JPL.

### The laser heads

Each laser system is composed of two identical laser heads, one active and one cold to provide redundancy. Each laser head consists of two separated units: the laser optical module (LOM), which in turn consists of optics and electro-optical elements for a functional laser transmitter; and the laser electrical module (LEM), which consists of electronics necessary to drive the laser head as well as interfaces with the spacecraft for electrical power, command and data handling, and instrument control functions including laser frequency control, optical power control, and laser phase modulation control [326].

The laser optical module consists of all optical components required for the generation and modulation of the laser light, i. e. a seed laser source, an amplifier, high-bandwidth phase modulator for clock and data transfer, and frequency and power tuning capabilities. The baseline design for the LOM is a MOPA at 1064 nm. The MOPA laser consists of a main oscillator laser source (the "seed", based on a Nd:YAG non-planar ring oscillator [248, 249]), a power amplifier for power scaling (a forward-pumped fiber amplifier with 10 μm double-clad Yb gain fiber), and a phase modulator for imprinting sidebands for clock noise transfer, ranging, and data information on the light [442]. A dedicated modulator inside the main frequency stabilisation provides phase modulation sidebands for the Pound-Drever-Hall [165, 345] locking technique used to stabilise the laser to the optical reference. The main optical output of the MOPA laser is a linearly polarised, continuous wave laser beam. This beam is sent to the optical bench via a single-mode polarization-maintaining fibre. A small fraction of the laser light is used for the frequency stabilisation at the optical reference, with the pickoff coming after the main oscillator.

The frequency of the respective transponder laser in each spacecraft is controlled such as to have an offset frequency to the incoming laser through a phase-locked loop. The frequency offsets are obtained from the constellation wide frequency plan which in turn is driven by the orbital dynamics of the constellation. The frequency plan is managed on-ground and uploaded regularly. The offsets are changing with a few Hz/s that switching direction at regular intervals, e. g. daily [224].

# 6 MISSION DESIGN

This chapter describes the mission design developed to fulfil the Science Objectives along with a brief overview of the spacecraft. Due to the competitive nature of the industrial studies run in Phase B1, only a generic overview is provided.

## 6.1 Mission Profile and Operational Orbit

The design of the LISA orbits is a critical aspect of the mission. The geometry of the LISA constellation determines the instrument's response to gravitational waves and its motion in the solar system barycentre induces frequency and amplitude modulations which aid in the identification and localisation of individual sources. The orbits also largely determine the spacecraft environment, especially the relevant thermal, gravitational, and radiation conditions. The LISA constellation will be realised through placing each of the three spacecraft in a heliocentric orbit at 1 AU with slightly different inclinations and arguments of periapsis, as first described by Faller and Bender [178]. Such constellations can be placed at varying distances either behind (trailing) or ahead (leading) of the Earth and are stable with only very limited station keeping. A smaller distance between the constellation and Earth simplifies communications and reduces transfer costs but results in larger disturbances across the constellation, caused by the Earth-Moon system. These disturbances result in secular changes to the constellation geometry which eventually limit the gravitational wave measurement. The proposed LISA configuration selects a distance that balances transfer and communication costs, while keeping a stable constellation geometry for a period of up to ten years to accommodate a potential extended mission. Earth-leading and Earth-trailing orbits provide identical science return, which allows the orbit type to be selected to optimise delta-v for a given launch window. Here, a single reference orbit is presented for the trailing configuration.

The LISA orbital design presents multiple challenges for the launcher capability as well as on the spacecraft themselves which we will address in the following sections.

### 6.1.1 Launch

LISA will be launched with an Ariane 64 in a direct escape trajectory towards the operational orbit. Launches are possible each day and all year-round. However, there are two months per year for which the delta-v demands are extraordinarily high compared to the other months. Should this become a mission design driver, the launch window could be reduced accordingly.

## 6.2 Transfer

LISA uses electric propulsion to transfer to the operational orbits. In contrast to LISA Pathfinder (LPF) and previous mission concept studies, there will be no dedicated propulsion module, as the thruster and the propellant tanks are integrated into the spacecraft.

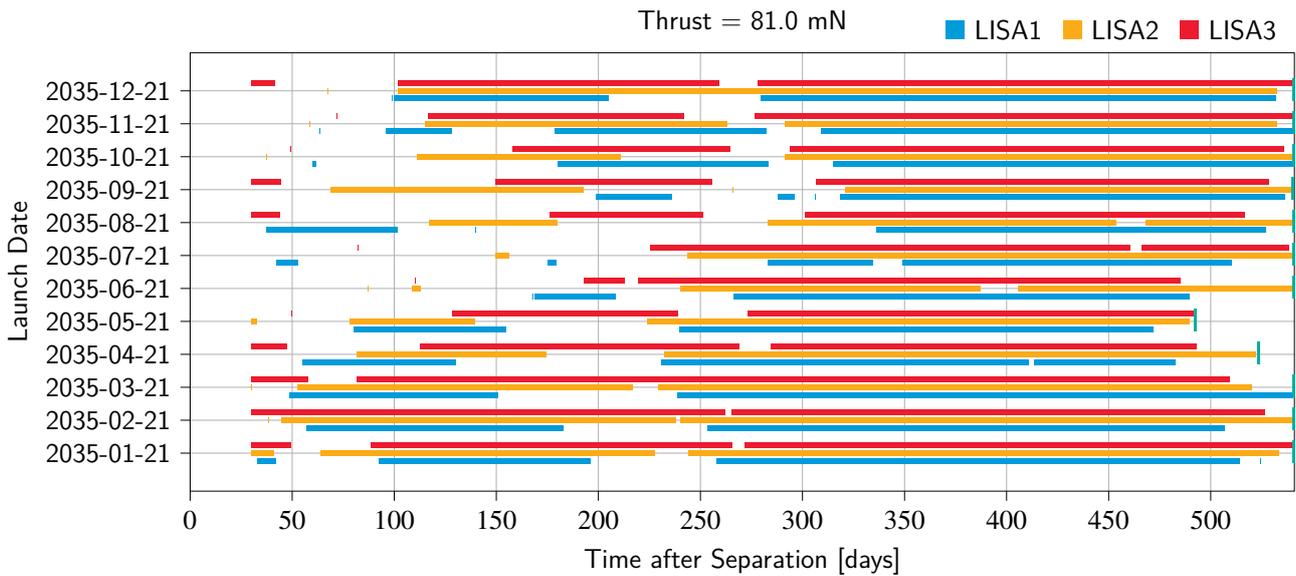

**Figure 6.1:** Time spent under thrust, so-called "thrust arcs" during the transfer for different launch dates and for all three LISA spacecraft. Early in the mission, there are many opportunities for commissioning activities.

Electric propulsion also implies a very different approach to operations during transfer. Many optimisation schemes are possible, varying parameters such as thrust levels, thrust times and directions. For the high mass of the LISA spacecraft, combined with using as small a thruster as possible, a significant amount of the transfer time is spent thrusting, as can be seen in Figure 6.1.

Nonetheless, during the transfer, commissioning and check-out routines can be performed, saving time once the constellation has reached its target orbit.

As the total delta-v required varies significantly depending on the launch date, and in order to not limit the available launch windows, the mission design is based on the worst case transfer case. If the worst two months per year were to be removed, the maximum delta-v would reduce by about 10–20 %. The delta-v also varies between the three spacecraft over the months, depending on the required inclination change. Maintaining identical spacecraft let us design for the worst case here as well.

## 6.2.1 Science Orbit

As described, the spacecraft are placed in heliocentric orbits such as to form a "cartwheel" constellation at a chosen distance behind or ahead of the Earth. With the additional perturbations, the evolution of the distance between the constellation and the Earth evolves over time, depending on the chosen Mean (Earth) Initial Displacement Angles (MIDAs). A negative value denotes a trailing configuration, a positive value a heading configuration. The smaller the MIDA, i.e. the closer the constellation is to Earth, the lower the required delta-v for transfer but the higher the initial perturbations from the Earth and thus the higher the perturbations on the constellation. Conversely, the further from Earth the MIDA, the higher the required delta-v, but the lower the perturbations. On both ends, the different limiting factors narrow the range of suitable angles to between 11 and 22°. The baseline orbit has a MIDA of 20°.

→ THE EUROPEAN SPACE AGENCY

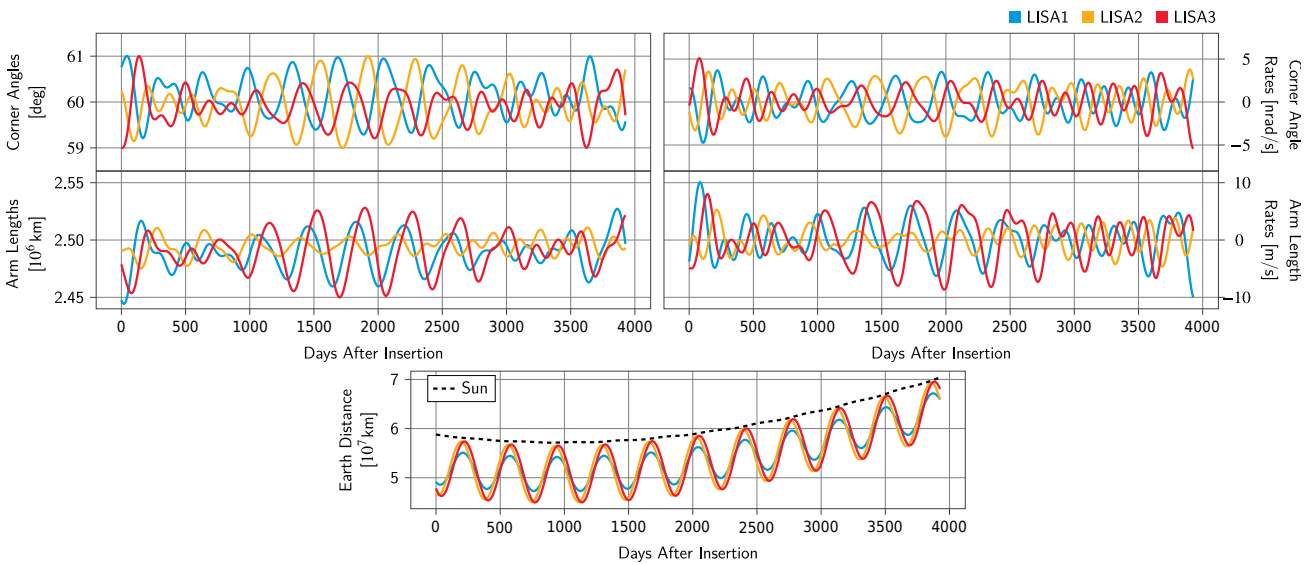

**Figure 6.2:** Overview of the orbital parameters for the baseline orbit.

The mean Earth distance is drifting over time. Since the spacecraft are bound to the dynamics of the Earth-Sun system, choosing a certain MIDA results in different drifting scenarios. The most interesting for LISA is a drift pattern that sees the constellation initially moving closer to the Earth before moving away again, optimising the average distance to Earth, as shown in the bottom graph in Figure 6.2.

Iterations between the capabilities of the instrument as described in Chapter 5 and the constraints on orbital dynamics lead to derived key performance requirements on the orbital design summarised in Table 6.1.

**Table 6.1:** Summary of the key performance requirements on the orbital design for LISA

| Parameter | Requirement |
|---|---|
| Arm-length rate of change | $<12\,\mathrm{m/s}$ |
| Corner Angle Variation | $<1.5°$ with $95\,\%$ confidence |
| Deterministic perturbing accelerations produced by spacecraft along the measurement direction | $<2\,\mathrm{nm/s^2}$ |
| Stochastic perturbing accelerations in any direction | $<1\,\mathrm{nm/s^2}$ |

Arm-length rate of change is a key parameter to ensure that the dynamic range of the beatnotes of the interferometric signals is within the capabilities of the acquisition systems. A good compromise was found in limiting the relative velocities to within $12\,\mathrm{m/s}$ (and thus the Doppler frequencies to within $\approx 12\,\mathrm{MHz}$). The range is limited by the bandwidth of the photoreceivers and the processing chain in the phasemeter.

Another key factor driven by obtainable technology is the variation of the inner angles of the constellation, i.e. the angle between two arms, also called corner angles. Changes here require active pointing of the laser beams and thus active mechanical parts that could impact the measurement. Also here a compromise is made by limiting the corner angle excursions to within a $1.5°$ half-cone.

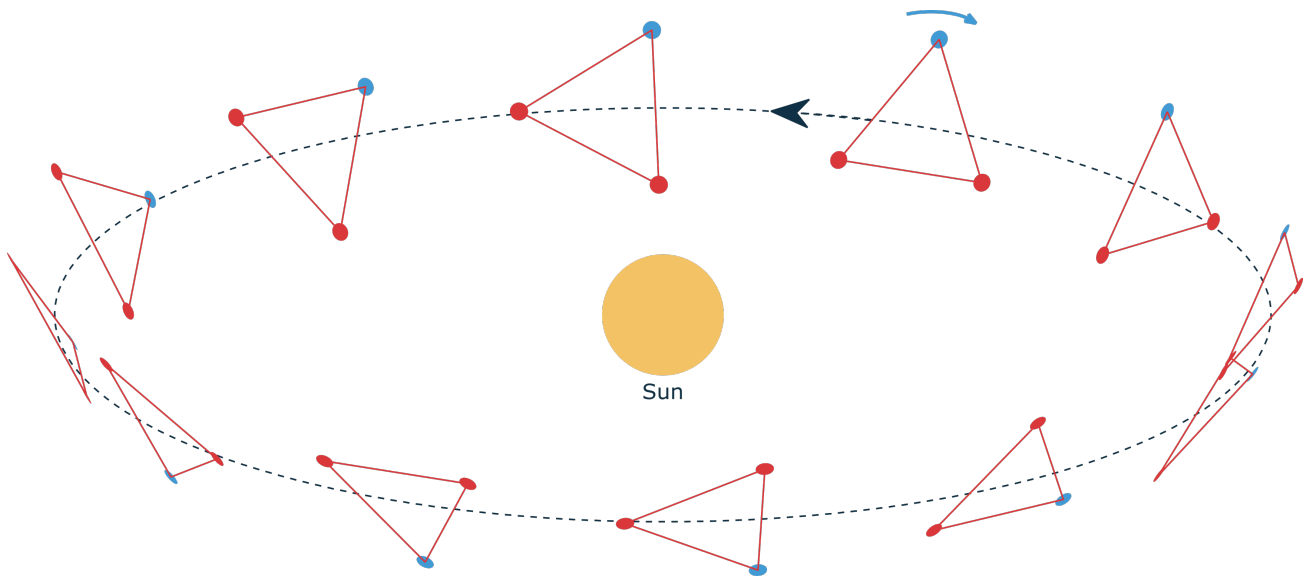

**Figure 6.3:** LISA orbits for the three individual spacecraft. The centre of mass of the constellation follows a heliocentric, circular orbit with no inclination. The orbits of the spacecraft have a small eccentricity $e = \eta/(2\sqrt{3})$ and an inclination of $\iota = \arcsin(\eta/2)$ where $\eta = L/1\,\mathrm{AU} \approx 0.017\,(L/2.5 \times 10^6\,\mathrm{km})$ is the armlength measured in AU. The $60°$ inclination of the constellation results in a backwards rotation of the constellation (blue arrow).

All of the parameters also depend on the injection accuracy, i.e. the deviation of the initial conditions from the nominal case, once the spacecraft are inserted into their final orbits. Monte-Carlo analyses with initial conditions drawn from the achievable injection accuracy were performed have shown that the orbits are compliant with the requirements even under the injection accuracy.

### 6.2.2 Decommissioning and Disposal

In compliance with applicable space debris mitigation guidelines, the mission will be passivated before the final switch-off command is sent. Passivation includes the de-pressurisation of all tanks, discharging of batteries and disconnecting of power lines in order to achieve a safe state of the spacecraft. No manoeuvres are planned, as the spacecraft are on orbits that are calculated to not approach the Earth within the next 100 years.

## 6.3 Constellation and Spacecraft Design

### 6.3.1 Spacecraft Overview

LISA will consist of three spacecraft that are nominally identical up to as-built deviations from the design. To accommodate all instrument units and spacecraft subsystems, a custom platform design was necessary. Each spacecraft has a mass of about 2500 kg and occupies more than 2/3 of the volume of a standard 20 ft shipping container. An overview of the key parameters is given in Table 6.2.

**Table 6.2:** Summary of the main constellation parameters.

| Spacecraft Overview | |
|---|---|
| Key Parameters | |
| Space Segment | 3 nearly identical spacecraft and launch adaptor |
| Launch Mass | 8212 kg |
| Power | 2300 W |
| Mission Lifetime | 6.25 years |
| Nominal Science Phase | 4.5 years |
| Constellation Orbit | |
| Orbit Type | Heliocentric 1 AU orbits |
| Armlength | 2.5 Mkm |
| Mean Initial Displacement Angle (MIDA) | ±20 degree leading or trailing the Earth |
| Subsystem Descriptions | |
| Power | Solar cells, main bus voltage 28V, small battery for launch and early operations |
| Communications | X-band up- and downlink to European Space TRACKing (ESTRACK) network, Steerable high-gain antenna, low gain antennas for safe mode, Data rate: 230 kbit/s, communication windows of 8 hours/day |
| Propulsion | 1.5 kW Electric propulsion for transfer |
| Attitude And Orbit Control System (AOCS) and Drag-Free Attitude Control System (DFACS) | 3-axis stabilised by cold gas thrusters. Drag-free attitude control via cold gas micro-Newton thrusters |

## 6.3.2      Design Drivers

### Mass

All three spacecraft will be launched on a single Ariane 64 launch vehicle, thereby limiting the mass and mass parameters available for the spacecraft design. With a total launcher capability of 8212 kg, each spacecraft should not exceed 2700 kg, before taking into account further reductions due to required launcher adaptors. Driven by LISA's powerful instruments and overall system complexity, the currently projected spacecraft mass is close to the available launcher performance. Albeit respecting all margins customary at this stage, first mass saving exercises have been run and strict limits been placed on the instruments.

### Environment

While in general the mission is exposed to a fairly benign interplanetary environment, there are several considerations to be taken into account. Radiation levels are low, with a total ionising dose of less than 40 krad over the duration of the mission, and with the dose already including the required design margin of factor two. For the opto-electronic components on LISA, proton fluxes and total non-ionising energy loss rates matter and have been assessed for the design.

Two key environmental considerations stem from LISA's high instrument sensitivity. First, at the low frequency end of LISA's measurement bandwidth, external thermal fluctuations dominate the stability. The Sun's total irradiance varies on timescales relevant to LISA and a dedicated model environment was developed based on available data. The solar cycle, while apparent at very low frequencies around 0.1 µHz, has limited impact on the LISA timescales, providing a good design point.

Secondly, micrometeoroids will impact LISA and affect the free-fall performance as well as the availability of the constellation for science [167, 398]. Only sparse data is available for LISA's orbits and the source models for micrometeoroid flux [209, 381, and references therein] vary in their predictions. A model with higher predicted cumulative fluence has been baselined for LISA, while other models could be used with adequate safety factors. The resulting flux on the spacecraft is directionally biased by the spacecraft motion and thus increased on forward-facing surfaces and reduced on trailing surfaces. From the assessments during the study phase, the impact of micrometeorites can be broadly assigned to three classes: those small enough to remain below the instrument sensitivity, those causing transient signals in the interferometer and lead to degraded performance for the duration of the hit, and those that lead to loss of science signal with varying recovery times. In the last class, recovery times vary depending on the magnitude of the exerted forces and torques from seconds to minutes to re-acquisition of the signal to more than a day for larger impacts that cause one spacecraft to lose attitude to the point where an autonomous recovery of the constellation is no longer possible and ground intervention is required. In total, micrometeorites are estimated to account for less than about 10 % downtime of the mission.

One further impact on LISA performance is the plasma environment and the effect of scintillations on the phase of the laser. This effect has been assessed and quantified in [245] and has been shown to be a source of considerable, but not limiting noise.

### Drag-Fee Attitude Control

The test masses on LISA are kept in free-fall by the Drag-Free Attitude Control Systems. This system is responsible for controlling the spacecraft in position and rotation, as well as actuate the optical assembly tracking mechanisms (OATMs) to steer the optical axis to follow the other spacecraft. As input, the DFACS receives the positional information of the test mass with respect to its housing from the Gravitational Reference System (GRS) and from the phase measurement subsystem (PMS) for the position along the sensitive axis. In addition, it receives Differential Wavefront Sensing (DWS) information from the phasemeter, providing angular information on the incoming laser beam from the remote spacecraft. A description of these measurements is provided in Chapter 5. These parameters are available for both Moving Optical Sub-Assemblies (MOSAs) and processed to calculate the required actuation forces and torques to make the spacecraft follow the test mass in its free-falling direction, actuate the test mass in the other directions and rotation, as well as actuate the OATM. The spacecraft thus follows the free-falling test masses in the two measurement axis, separated by the $\sim 60°$ corner angles. The major external force acting on the spacecraft is the solar radiation pressure, also to be compensated by the attitude control.

Based on the heritage acquired with LISA Pathfinder, Gaia [190], Euclid [274], and Microscope, LISA flies a cold-gas (nitrogen) micropropulsion system, providing up to $\sim 500\,\mu N$ of force per thruster. The propellant is stored at 320 bar in multiple tanks, placed strategically inside the spacecraft for gravitational balancing reasons. For control of the spacecraft, the system uses about 8 kg of cold gas per year on orbit.

In order to secure the availability of the cold gas system as flown on LISA Pathfinder, further lifetime qualification tests as well as production capability expansion have been initiated during the study phase.

During the transfer phase, and after separation from the launcher, additional systems ensure de-tumbling and the required coarser attitude control.

### Structure and Mechanisms

A central element of the LISA spacecraft is the LISA core assemblies (LCAs), containing the two MOSAs and providing protection against contamination and thermal fluctuations. The MOSAs are mounted inside the LCA via the OATMs to compensate the deviations in the corner angles by rotating the whole MOSA. During launch, launch locks secure the MOSAs against the vibrations and shocks produced by the rocket. These locks will be released after launch. These mechanisms are carefully designed to minimise the shocks onto the glass-ceramics inside, avoiding damage.

The MOSAs themselves provide the structure to which the telescope (Section 5.3.1), optical bench (Section 5.3.2), and Gravitational Reference System (GRS, Section 5.2) are mounted. Each element of the MOSA needs to be aligned on ground to within a few µm and µrad to minimise tilt-to-length (TTL) coupling. To help with this alignment, the beam alignment mechanism on the optical bench can be preset to compensate for misalignments during integration. Once on orbit, the structure provides the stability against introducing excessive tilts between its constituents, as well as pathlength changes due to mostly thermal loads.

The LCA is housed inside the custom-designed platform. Here, standard structural elements are used to build up the spacecraft, sized to accommodate all other units of the spacecraft, as well as propellant tanks. Since all three spacecraft are brought into space by a single launcher, separation mechanisms detach the spacecraft from the launcher and its adaptor. One more mechanism and associated launch lock are used for the high-gain antenna.

### Communication System

The data produced on board the constellation amounts to about 75 kbit/s, including science data, all required housekeeping, as well as an allocation for high-resolution science data to investigate anomalies or interesting events in the data. In order to enable low latency processing and the identification of merger events in due time, requirements dictate that the data is transmitted to Earth within 24 hours of it being measured by the constellation. In addition, to cover merger events, protected periods of up to 14 days are foreseen during which no operations shall be performed that impact science.

This poses multiple challenges on the design of the communication system. Obvious from the orbital evolution described in Section 6.2.1 is the increasing distance to Earth over time, up to $60 \times 10^6$ km after the nominal mission. Another factor is that in the spacecraft fixed frame, the Earth rotates around the spacecraft by about 1° per day in azimuth – 360° per year – and about ±4° in elevation. Available pass durations of the spacecraft over the ESA ESTRACK 35 m ground station antennas for deep space use are varying over the year between 8 and 12 hours per day. Combining this with the data rate produced on board and the requirements to download in 24 hours equates to a communication data rate of larger than 225 kbit/s.

This high data rate requires the use of a high-gain antenna, with a dish size of about 50 cm. This antenna is steered by an antenna pointing mechanism to cover the required ranges. Turning the antenna degrades the science performance during this operation and for a short amount of time (minutes) thereafter. Each spacecraft's antenna is covering a beamwidth of about 5 days – equating to 5° in azimuth – and the three spacecraft antennas are pre-positioned to cover about 15° in total by transitioning the communication from one to the next (see Figure 9.1), so the constellation can guarantee the protected period of 14 days (see Section 9.1.1) without having to rotate the antennas.

Since only one spacecraft is talking to Earth for ∼ 5 days at any instance, data from the remaining two spacecraft needs to be transferred to the currently communicating spacecraft. This is achieved by encoding this data on the laser links also used for the science measurement, effectively utilising the science instrument as a laser communication terminal (see 5.3.3.2 for details on the communication).

Low-gain antennas secure full-sky coverage for contingency cases such as safe modes and are driving the power required from the communication system in order to guarantee minimum data rates to operate the spacecraft. In such cases, each spacecraft can be addressed separately and does not require the laser links active.

### Power and Thermal

As the spacecraft are using electric propulsion to transfer to the operational orbit, power sizing is dominated by the transfer phase, leaving comfortable margins for the science phase. The instruments being off, but requiring heater power to keep them at their minimum non-operative temperatures, the transfer case is driven by the combination of heater power and electric propulsion. The phasemeter and laser systems have high power densities and thus need efficient radiators.

During science operations, thermal stability of the optical systems is paramount to limit pathlength changes through thermal expansion to below the picometer. While the glass ceramics used have very low coefficients of thermal expansion, stability requirements on the optical elements are still demanding at $5\,\mu K/\sqrt{Hz}$. This is achieved by placing the components in the centre of the spacecraft, as shielded as possible from external disturbances. Great care is taken to limit fluctuations from electronic equipment in proximity to the sensitive elements. The LCA is further protected from the spacecraft induced thermal noise by thermal shields.

### Time and Frequency

Timing requirements on LISA come from the Time-Delay Interferometry (TDI) processing which requires accurate timestamping across the constellation and to a common time reference frame to synthesise the virtual interferometer channels. This in turn places requirements on the timestamping of data on board of each spacecraft, based on a single source of truth for timing. An ultra-stable oscillator (USO) and timing system is providing the spacecraft clock, a pulse-per-second to resolve ambiguities, as well as a 50.3 MHz frequency reference to all units that require this.

In addition, the measurements of the interferometers are taken at 80 MHz to capture the heterodyne signals in the range from 5–25 MHz. Further frequencies are used in the GRS. Any spurious signals from other sources could interfere with these measurements and spoil the data. A frequency plan ensures that electromagnetic interference remains below susceptible level or avoids certain frequencies by design.

## 6.3.3      Integration

LISA presents a challenge from the integration standpoint. Most science missions have one spacecraft (exceptions e.g. Cluster [172]), some Earth Observation missions have two (GRACE-FO, NGGM [299], etc.). While not a production line like constellations (Galileo, GPS, others) or even mega-constellations (OneWeb, Starlink), LISA relies on small-series production of units and the spacecraft, presenting challenges for resources planning as well as facility availability. Figure 6.4 provides an overview of the required integration steps and the amount of units required.

Integration is planned with an optimised sequence, with slightly staggered integration activities, but a large overlap due to which all flight models will be on the main integration floor at some stage. While large available floor space is needed for this scheme, it opens opportunities for mitigating issues during integration by moving teams to different integration steps and between spacecraft. Nonetheless, items on the critical path have the ability to upset the carefully planned optimal sequence and endanger test facility slots. Sufficient margins are added to schedule for these items, as well as mitigations via alternative and late integration schemes.

In addition, it is a challenge especially for payload providers to manufacture three, six, or even twelve items within a short time-frame between the approval of requirements and the need date for integration on spacecraft level. This risk is mitigated by commensurate contingency planning, such as additional staff training, fast-response maintenance contracts, stockpiling, and additional facilities.

## 6.3.4      Verification

A sound, complete, and robust verification scheme is critical. At the same time, constraints in cost and available schedule influence the verification scheme. Margins have been allocated between the required science performance and the performance required by the TDI processing as well as the space segment. The flow-down philosophy is shown in Figure 6.5.

Verification of the mission against the applicable mission level performance requirement, specified in the noise in the TDI $A$ and $E$ channels, relies on a baseline reference implementation of the L0.5-L1 pipeline. Part of the performance has been allocated for the processing on ground with the remainder allocated to the noises produced by the constellation. From the reference pipeline, a set of analytical transfer functions for different noise categories has been derived that are used to determine the performance of the space segment once its data is processed on ground. This of course does not prohibit the development of better performing pipelines in the future, but secures the requirements baseline against changes.

Full verification of LISA's performance is not possible on ground nor on integrated spacecraft level. The test masses cannot be brought into sufficient free-fall, and the distance between spacecraft and the disturbance-free environment cannot be realised on ground. Thus the mission verification relies on a combination of analysis and test. To secure this approach, performance testing is performed on the lowest possible level, while higher integration levels verify functionality and key parameters to map against detailed models for verification by analysis. These parameters are chosen to secure performance based on the established model of the constellation.

### Lower Level Verification

A significant step in the verification programme is the Interferometric Detection Subsystem (IDS) Engineering Model test programme which aims to anchor the verification of requirements placed on unit level for the optical bench assembly, the phasemeter and auxiliary units to the higher level requirements of the IDS. As the IDS is responsible for the majority of functions and performance of the interferometers, this step is vital to secure the verification strategy described above. The test programme is run in France under the responsibility of CNES. At a later stage, one further full performance test programme is performed on the Qualification Model of the IDS. From then on, all IDS flight units will be individually tested and verified before being provided to the prime contractor for integration on the spacecraft.

Similarly, the GRSs will run a test programme in Italy on engineering model level, verifying that changes to the system from LPF due to e. g. obsolescence have not changed performance of the system. The GRS then follows a Qualification Model–Flight Model philosophy.

### Higher Level Verification

As the key performance tests are covered on the lower levels, the prime contractor will focus on the remaining performance parameters under its control, such as the correct alignment between telescope and optical bench to contain tilt-to-length couplings. After final integration onto the spacecraft a full set of functional tests is run together with tests confirming the performance via derived parameters and analysis of the validated model.

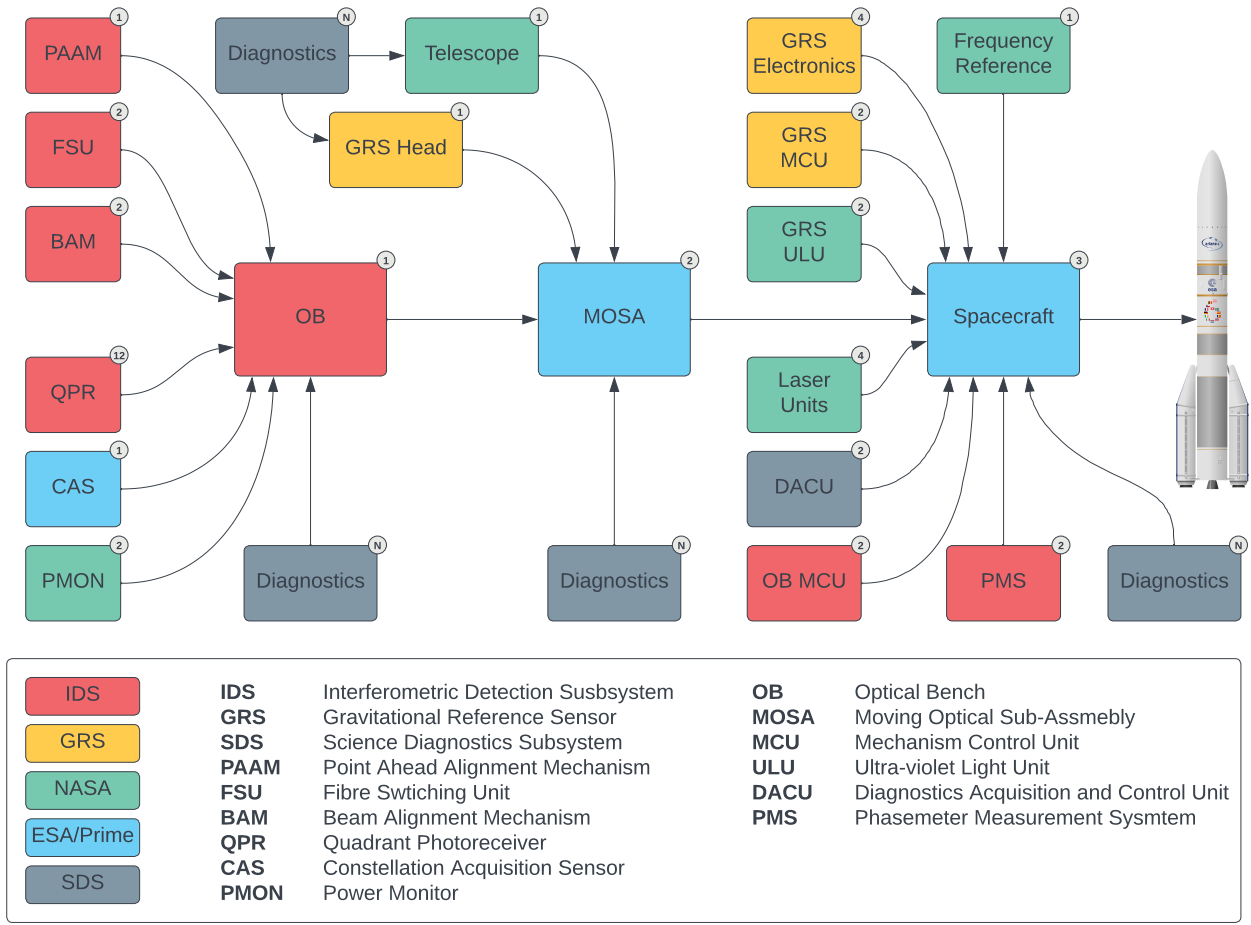

**Figure 6.4:** Overview of main integration sequence and number of integration steps. Each element is multiplied by its number and the number of the larger integration subsystem. In total this requires ~ 169 distinct "units" to be built, tested, delivered and integrated (not including diagnostics, platform units, harness, etc.)

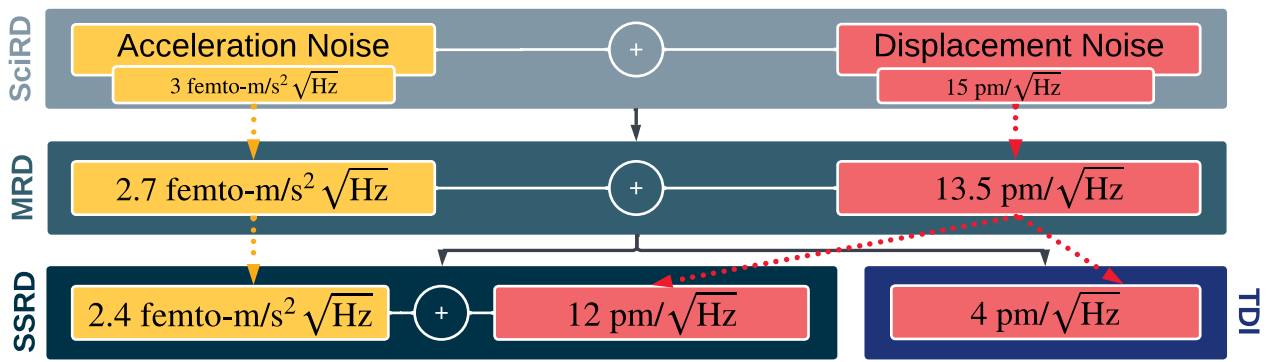

**Figure 6.5:** Flow down of converted Science Requirements (SciRD) to Mission Requirements (MRD) and further to the Space Segment (SSRD, applicable to the primes) as well as the ground segment for post processing noise (TDI).

# 7    MISSION PERFORMANCE

Chapter 4 introduced the top-level requirements for the LISA mission and their connection to the science objectives in Chapter 3. Here we present a more detailed description of the physical effects that limit LISA's sensitivity to gravitational wave (GW) sources and their relative importance. A similar analysis provides the foundation for the hierarchy of requirements documents that relate the top-level science and mission requirements to requirements on individual subsystems and components.

The standard approach to describing the sensitivity of a GW observatory without referring to a specific class of astrophysical signal is to convert the instrumental noise to an equivalent GW strain [61, 273]. This is typically expressed as a Power Spectral Density, $S_h(f)$, which is somewhat analogous to the "dark noise" for a Charge-Coupled Device (CCD) in the sense that it represents an instrumental foreground with which astrophysical or cosmological signals must compete.

## 7.1    Performance Modelling

As described in Chapter 2, LISA's sensitivity is determined by a few key factors: the arm-length of the constellation; the orientation of the constellation with respect to the line-of-sight to the source on the sky; non-inertial perturbations of the geodesic references; and noise in the interferometric measurement between these references. In order to compute $S_h(f)$, an observable where all of these quantities will be expressed must first be selected. Typical choices are the Time-Delay Interferometry (TDI) observables, particularly the pseudo-orthogonal combinations $A$ and $E$, which can be approximately treated as outputs from independent instruments. The transfer functions from each of the noise sources can then be used to compute the equivalent noise power in the observable. To express this as equivalent GW strain, it is necessary to divide by the response of that same observable to GW signals:

$$S_h(f) = \frac{T_{\text{acc}}(f) S_{\text{acc}}(f) + T_{\text{disp}}(f) S_{\text{disp}}(f)}{R_{\text{GW}}(f)} \tag{7.1}$$

where $S_{\text{acc}}(f)$ and $S_{\text{disp}}(f)$ are the spectral densities of the acceleration and displacement noise respectively; $T_{\text{acc}}(f)$ and $T_{\text{disp}}(f)$ are the transfer functions of the acceleration and displacement noises to a constellation observable such as a TDI variable; and $R_{\text{GW}}(f)$ is the response to GWs for that same observable. In general, $R_{\text{GW}}(f)$ is dependent on source location and orientation. For modelling the overall performance of the constellation and setting requirements, a location and orientation averaged response is used. In addition to noise generated at the instrument, there are some sources of noise associated with the on-ground processing required to remove the contributions of laser frequency noise from the displacement measurement. These noise sources, all of which are sub-dominant, are described in detail in Section 8.2.

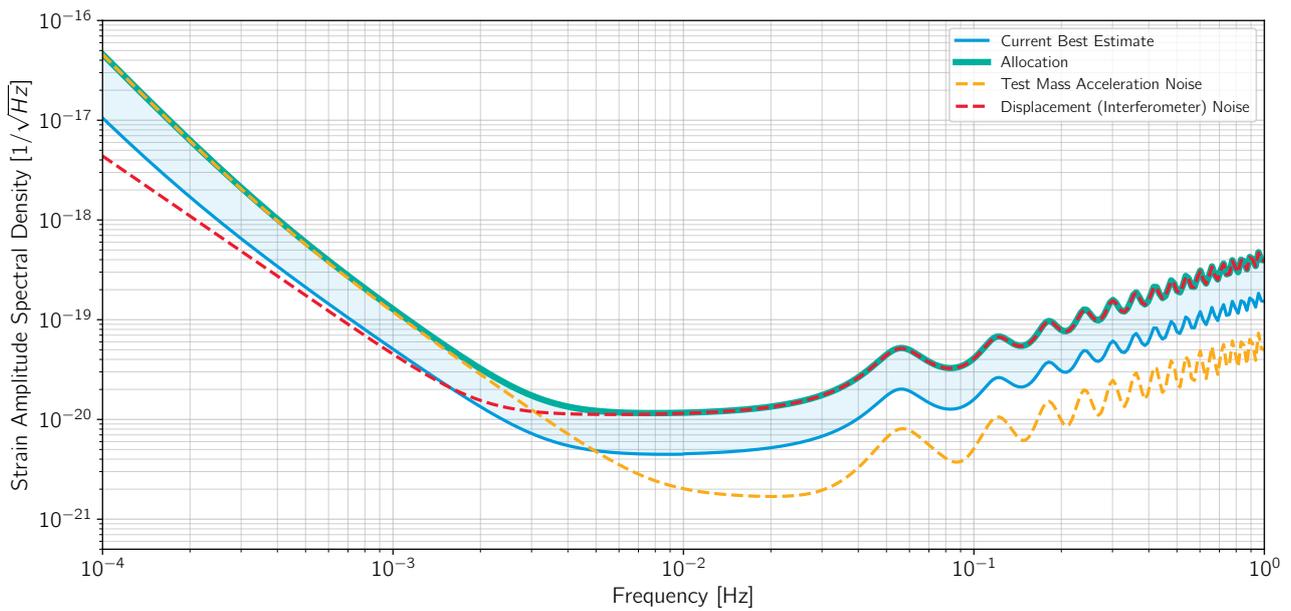

**Figure 7.1:** LISA's sensitivity to gravitational waves (GWs), expressed as equivalent GW strain amplitude spectral density and averaged over all source locations and orientations. The relative contributions from acceleration and displacement noise are shown by the yellow and red dashed lines, respectively.

Figure 7.1 shows the Amplitude Spectral Density $\sqrt{S_h(f)}$, computed using a sky and orientation averaged $R_{\mathrm{GW}}(f)$. The characteristic "wiggles" at higher frequencies are the result of the true zero response of the interferometric combinations for GWs with certain frequencies due to cancellation effects. The relative contributions of the acceleration and displacement are shown as well as the current best estimate and required values for each. Further discussion of the primary contributors to both the acceleration and displacement noise are described in Section 7.2 and Section 7.3 respectively. A detailed table of primary contributors to $S_h(f)$ can be found in Appendix A.

## 7.2    Sources of test mass acceleration noise

Test mass acceleration noise from stray forces competes directly with the gravitational wave tidal strain signal and limits the LISA sensitivity at all frequencies below roughly 4 mHz. While achieving high precision free-fall depends most critically on the local environment defined by the Gravitational Reference System (GRS, see Section 5.2) as a "test mass shield", it is truly a system level issue, depending on the spacecraft control as well as the magnetic, gravitational, electrostatic, and thermal environment on-board the spacecraft. The disturbances are all at least in part local, though some depend on coupling to external disturbances, most notably the noisy interplanetary magnetic field and cosmic ray charging.

Figure 7.2 shows some of the driving noises of test mass acceleration noise. At millihertz frequencies, the dominant noise is Brownian force noise from residual gas impacts, driving tight, ≈µPa, pressure requirements.

This "white noise" gives way at lower frequencies with a number of noise sources increasing in power as $f^{-1}$ and steeper. These include:

- gain fluctuations in the actuation voltages used to rotationally align the test mass to the local

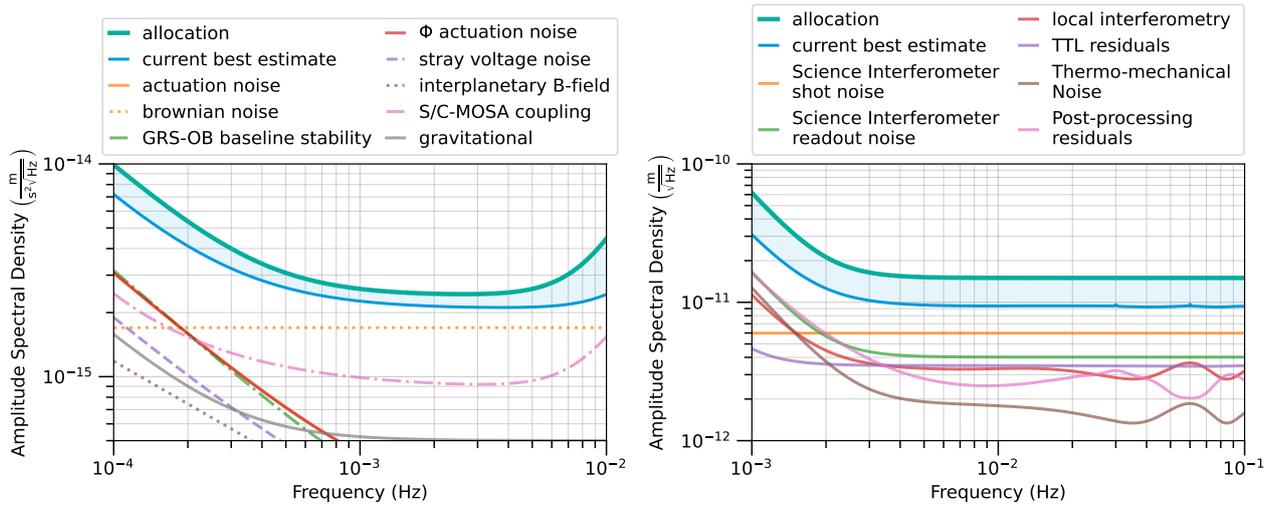

**Figure 7.2:** Left panel: Primary sources and groups of the test mass acceleration noise budget. Right Panel: Primary noise sources and noise groups of the interferometry budget.

interferometer. This drives the Front-End Electronics (FEE) actuation requirements and the precision of non-static spacecraft gravitational balancing.

- coupling to the spacecraft and Moving Optical Sub-Assembly (MOSA) motion via force gradients or *stiffness*. Driving factors here are the spacecraft control "jitter", translational and rotational, and thermally-driven relative motion between the Gravitational Reference System (GRS) and optical bench, as well as DC forces and force gradients.

- stray electrostatic forces, due to test mass charge, stray surface "patch" potentials, and low frequency FEE actuation noise.

- magnetic forces on the test mass bulk, most critically noise in the interplanetary field inducing a moment due to the finite test mass magnetic susceptibility and coupling to the local magnetic field gradient.

- various forces of thermal origin, coupling temperature and temperature gradient fluctuations into forces, which become relevant at and below the LISA 100 μHz lower band edge.

- fluctuations in the local gravitational field acting on the test mass, due to any motion or deformation in the spacecraft mass distribution.

The test mass acceleration noise budget [28] is both complicated and extremely difficult to verify on the ground, because of the extreme $\sim \mathrm{fm}/(\mathrm{s}^2\sqrt{\mathrm{Hz}})$ requirements, the long time scales and specific environment, in spite of decades of improving torsion pendulum measurements [29, 122, 366] aimed at quantifying specific force noise sources and setting upper limits surface force noise.

LISA Pathfinder however allowed a full flight test of free-falling LISA test masses, meeting and exceeding the LISA requirements [see Figure 5.3, 32, 41] across the entire LISA frequency band and in a representative spacecraft environment with the most critical GRS hardware and control techniques, in addition to allowing dedicated experimental campaigns to measure and understand specific limiting sources. As such, the LISA low frequency performance verification goes well beyond the important "bottom up" sum of individual identified noise terms, but additionally is anchored by a top-level global test, of heritage hardware, encompassing sources potentially outside the noise model.

## 7.3　　Sources of interferometer noise

Figure 7.2 shows the main contributing noise sources that limit the sensitivity of the interferometry on LISA. As described in Section 5.3, the final test mass to test mass measurement is composed by combining the science interferometer measurements (measurement of the phase/frequency of the received beam against a local beam) and the test mass interferometer (which compares two local beams, one of which was reflected off the test mass). Additionally, a reference interferometer (which measures common-mode noise of the two local lasers and associated system elements) is further used to reduce the impact of certain noise sources. For simplicity, the total noises in the local interferometers (test mass and reference) are shown as a single trace. The other contributions shown arise directly in the science (or long-arm) interferometer measurement. Some of these noises are further suppressed in on-ground processing, the residual effects of which are shown as a single trace. The primary noises that are unaffected by the on-ground processing are also shown.

As we see in Figure 7.2, one of the dominating noise sources is photon shot noise arising in the science interferometer measurement, where the low power received beam (a few hundred picowatts), which carries the gravitational wave information, is beat against a local oscillator beam. Stringent requirements are placed on the transmission and reflection of optics and coatings, as well as on the cleanliness of all optical components that impact the transmission from a laser on a local spacecraft to the photo-receivers on the far spacecraft. Most of the power is lost through the simple fact that Gaussian laser beams diverge as they travel through space, resulting in a wavefront that is a few km wide by the time it reaches the far spacecraft, and we only intercept a tiny fraction of that wavefront with our 30 cm telescope. Related to this, the overall performance of the electronics in the photo-receivers is critical in being able to measure the very low-level photo-currents produced by detecting the beat of the low-power beam against the local oscillator. Other local sources of optical path-length noise include the phase noise arising from the thermo-mechanics of the optical elements that comprise the optical system (telescope and interferometers). Control of this type of noise is done through careful design of the optical system and selection of components and coatings, as well as by placing stringent requirements on the thermal stability of the optical system.

The angular and lateral stability of the components of the optical system (including the test mass surface) couples with misalignments in the optical system to produce spurious path-length changes. This type of noise is carefully controlled by precise alignments of the components in the system (e. g., the alignment of the telescope to the optical bench), and by exquisite control of the spacecraft, MOSA and test mass to the optical beams through dedicated Drag-Free Attitude Control System (DFACS) control loops. In additional to the careful design and implementation of the space system, the angular jitters of the MOSA against the received beam, and the test mass against the MOSA are precisely measured to enable the DFACS control. These angular measurements can further be used on ground to estimate the misalignments in the system and to reduce (in post-processing) these impact of this so-called tilt-to-length (TTL) noise even further. Figure 7.2 shows the estimated TTL residuals assuming the system is built to the requirements and that the required level of suppression can be achieved on ground.

# 8     SCIENCE OBSERVATIONS

The operational concept for LISA's science phase is relatively simple: perform a time-resolved, all-sky survey of gravitational waves (GWs) sources in the millihertz band. During this survey, LISA will measure the combined effect on the LISA constellation of tens of thousands of GW sources of various types, enabling the broad suite of scientific investigations discussed in Chapter 3. This chapter describes the analysis approach to identifying, characterising, and cataloguing these sources. The emphasis is on the methodology and techniques employed in the analysis pipelines. Further details on the implementation, including the organisational structure and operational management, can be found in Chapter 9 and Chapter 10.

The LISA measurement concept, including the basic approach to data analysis, is introduced in Chapter 2, specifically in Section 2.3 and Section 2.4. Briefly, the measurements from the individual instrument subsystems on each LISA spacecraft (S/C) will be sent to the ground, where they will be combined to form data products that have sensitivity to GWs. Simultaneous "global-fit" search routines will identify and characterise source candidates within these data products. A catalogue of confirmed sources will be drawn from these candidates by applying thresholds, vetoes, and other criteria. Specialised pipelines optimised for low latency will target bright GW sources with potential electro-magnetic (EM) counterparts to enable the production of alerts for multi messenger astronomy. The entire process will be iterative and ongoing, producing regular updates to lists of sources, estimates of their measured parameters, and ancillary products such as estimates of the instrument parameters and noise performance.

## 8.1     Analysis Steps and Data Products

An overview of the LISA analysis flow, including the major processing steps and the resulting data product levels, is shown in Figure 8.1. The following sections in this Chapter describe each of these steps in more detail, discuss their challenges and describe the technical approaches that have been developed to address them.

Before describing the data processing steps, an overview of the different LISA data levels is given:

**L0 (Raw)**    Spacecraft telemetry after basic processing including: de-multiplexed and annotated with time correlation, quality flags and positioning data; anomalous, corrupted items corrected or removed; all measurements time-ordered and uniquely tagged. L0 data are not necessarily releasable to the public as they might contain information pertaining to third party intellectual property.

**L0.5**    L0 data which has been further processed and reformatted in the following way: de-packetised, de-compressed (if applicable), reformatted to the data exchange format of the Science Ground

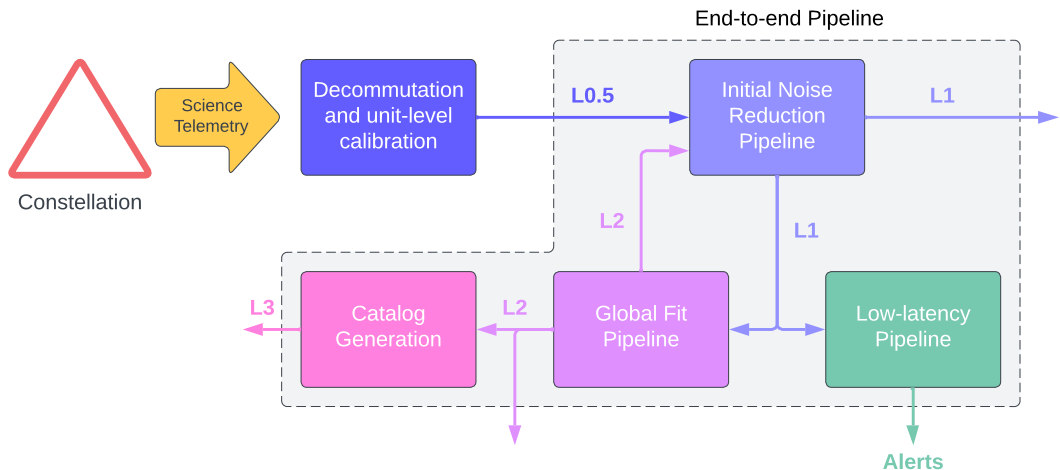

**Figure 8.1:** Schematic overview of the major steps in LISA data analysis and the corresponding data product levels.

System, clock data synchronised and converted to a common, non-Earth-bound astronomical timescale, binary units (analog-to-digital converter (ADC) counts), where applicable, converted to physical values. L0.5 data are fully releasable to the public.

**L1** Calibrated and noise corrected Time-Delay Interferometry (TDI) data streams. This is "analysable data" and the main input to the core downstream processing to isolate individual GW signals, i. e. major noise sources removed. L1 data constitutes the lowest-level data ready for scientific interpretation.

**L2** Output from the global fit pipelines, providing Probability Density Functions (PDFs) for identified sources. They include the reconstructed waveforms and detector signals for identified sources as well as regular updates on the critical parameters of transient sources, such as coalescence times and sky location. These are "analysed data". We expect several L2 products that vary depending on prior probabilities for source parameters and numbers of sources, on the methods used, and on the volume of the data engaged (the volume of the analysed data is increasing as LISA continues to observe).

**L3** Catalogue of GW source candidates with detection confidence, estimated astrophysical parameters and their strain time series $h(t)$. These are "consolidated data". L3 data also include the "residual" data stream, i. e. the L1 data stream with the contribution of identified sources removed. L3 data constitute the mission end products.

**Alerts** Specialised variants of L2/L3 data from dedicated low-latency pipelines to produce alerts for multi-messenger astronomy.

## 8.2  Generating Constellation Observables: From L0.5 to L1

The LISA "instrument" which measures GWs is the constellation itself. None of the individual payload components described in Chapter 5 is sensitive to GWs on its own; moreover, no single measurement local to one of the S/C or even single inter-S/C measurements can provide the required sensitivity. The first step in the LISA analysis is to combine data from the individual instrument elements in order to produce data products that contain the maximum information on the GWs and are minimally impacted by instrumental artefacts. The output of this step are L1 data products which serve as inputs to the subsequent search pipelines.

## 8.2.1     Laser frequency noise suppression with TDI

As described in Section 2.3, LISA's interferometric design results in the GW signal being combined with intrinsic noise in the laser's frequency that is orders of magnitude larger in the LISA band. Sensitivity to GWs is recovered by making multiple measurements at different points within the constellation, each of which includes different combinations of laser noise and GW signal. By combining these measurements with suitable time delays, it is possible to form a signal which suppresses the laser frequency noise but retains the GW sensitivity, albeit with a somewhat modified transfer function. This is the core of the Time-Delay Interferometry (TDI) algorithm that was developed more than two decades ago to support early versions of the LISA concept [200, 400, 403]. While the implementation is quite different, the effect of LISA with TDI is similar to that of the equal-arm Michelson interferometer that underlies ground-based GW designs: insensitive to fluctuations in the laser frequency while sensitive to differential arm motion caused by GWs.

TDI can be generalised [405] to accomodate constellation dynamics, i.e. the rotation of the constellation [377] and changing arm lengths [142, 404]; as well as suppressing spacecraft jitter by making use of the split-measurement setup described in Section 5.1. In its full form, the TDI algorithm takes in all interferometric measurements obtained from the constellation and can produce a number of different interferometric combinations, mimicking not only Michelson interferometers, but more complicated combinations that can be used if one or more laser links are lost [160]. More than 200 potential combinations have been studied [314], each with slightly different properties but the same general feature of suppressing the otherwise overwhelming laser frequency noise. A particularly interesting subclass are combinations which are also insensitive to GWs, at least at low frequencies. These so-called "null streams" will be useful in distinguishing between unmodelled GW signals and instrumental noise.

TDI requires accurate and precise estimates of the time-of-flight between the different LISA S/C. Crude approximations with errors of ~1 km can be derived from the orbital ephemeris. To reach the ~1 m accuracy required for TDI, an on-board ranging system is included as part of the LISA interferometric link. This system is described in more detail in Chapter 5. These on-board measurements can be combined with on-ground observations and an orbital dynamics model using a Kalman-like filter to further improve the ranging precision and disentangle ranging and clock information [356, 423]. As a risk mitigation, alternative approaches to TDI which include a search for the arm lengths [402] or non-TDI approaches based on principal component analysis [57] have demonstrated similar performance without requiring this a priori knowledge.

## 8.2.2     Clock noise suppression and timing synchronisation

After laser frequency noise, the second most significant noise source in the LISA measurement results from timing jitter in the onboard clocks used to time the measurements on each S/C. Each spacecraft hosts a single ultra-stable oscillator (USO) that drives the sampling of the phasemeter, which records the phase of the ~15 MHz signals resulting from the interference of the various optical beams. Jitter in this sampling clock degrades the accuracy of these phase measurements and is not suppressed by the baseline implementation of TDI. To address this, LISA employs a dedicated clock transfer system which allows the relative jitter between pairs of S/C clocks to be measured and sent to ground. This is accomplished by phase modulating the outgoing laser beams with a signal derived from the onboard clock. The phase difference of the two sidebands is measured using a dedicated phasemeter channel, which results in a measurement of the differential jitter between the S/C clocks [228]. This information can then be used in post-processing to correct for the effects of timing jitter in the TDI outputs. A generic clock noise-suppression algorithm applicable to any TDI combination has been developed and verified in simulations [221].

An additional step in the timing analysis for LISA is synchronisation which allows measurements made on board to be related to astrophysical events. This synchronisation will be implemented via a one-way clock synchronisation scheme similar to Gaia [255], where the communicating spacecraft sends its onboard time to Earth. This onboard time is then correlated with a well-defined global time scale with sub-millisecond accuracy. The resulting timing couples can be used to transfer measurements taken according to the spacecraft clocks to the Barycentric Celestial Reference System (BCRS) frame.

## 8.2.3 Optimising the Instrument Performance

In addition to laser frequency noise and clock jitter, which produce known noises in the LISA data stream that can be corrected with TDI and clock noise correction respectively, there may be opportunities to apply other corrections which improve the signal-to-noise ratio (SNR) for GW signals. An example of this is suppression of the so-called tilt-to-length (TTL) coupling, in which imperfections in the optical alignment between various components of the interferometer allow coupling of the angular jitter of the LISA Moving Optical Sub-Assemblies (MOSAs) into the longitudinal measurement. By using high-accuracy measurements of the jitter and estimates of the coupling constants, the effect of this angular jitter can be suppressed or removed. A 24-parameter tilt-to-length (TTL) coupling model has been developed [329], and these 24 parameters can either be recovered via calibration manoeuvres comparable to the LISA Pathfinder long cross-talk experiment [49], or by measuring the MOSA jitter via the Differential Wavefront Sensing (DWS) and fitting to the readout signals.

Other potential noise subtractions may also be applied during this step. It is likely that the precise recipe for optimal performance will be informed by residuals which are only available after bright sources have been identified and removed. In addition, evolution in the state of the instrument, either naturally or through intentional intervention by the mission team, will require updates to this optimisation process.

## 8.2.4 Technical Readiness of the Initial Data Processing

Since TDI applies to the full LISA constellation, there exists no method to perform a complete end-to-end test that precisely mimics all flight conditions. However, much of the technical risk can (and has) been retired through the use of numerical simulators and laboratory analogues. Numerical simulators allow the LISA constellation to be accurately modelled, including precise estimates of the orbital motion and the ~8 s light travel delays along the arms. The development of higher fidelity LISA simulators has paralleled the development of the TDI technique, gradually including more subtle effects [328, 334, 364, 413]. Current state-of-the-art LISA simulators such as LISANode or LISA Instrument [83] implement realistic orbits, a correct treatment of the various time frames (including a model for the onboard clocks), laser locking, representative frequency planning, and updated noise models [81, 82, 150, 220]. The left panel of Figure 8.2 shows the results of a recent simulation comparing the noise level measured by the inter-S/C interferometer, which is corrupted by laser noise, with a TDI channel that suppresses that noise.

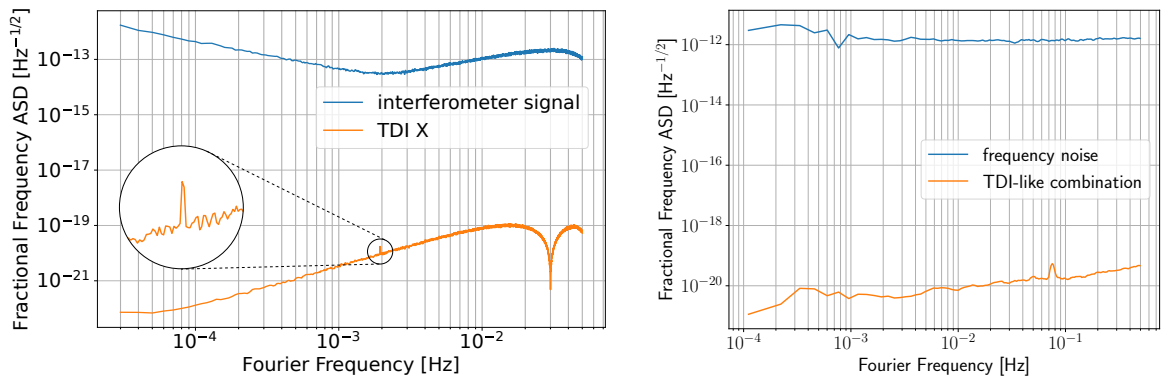

**Figure 8.2:** Numerical and experimental validation of TDI. Left panel: numerical simulation including realistic orbits, laser locking scheme, and clock transfer demonstrates suppression of frequency noise between individual long-arm measurement (top curve) and TDI combination (bottom curve). The estimated GW signal from the verification binary AM Canum Venaticorum was included in the simulation and is visible as the small peak around 2 mHz in the TDI signal. Right panel: results from the "Hexagon" experiment demonstrate over 8 orders of magnitude of frequency noise suppression in a system with independent optics, analogue chains, phasemeters, and clocks.

The concern with numerical simulations as a validation approach is that an unmodelled effect might limit the performance of TDI. Hardware analogues can be used to search for these "unknown unknowns" by making LISA-like measurements using representative hardware and measuring the noise suppression in TDI-like combinations. Such experiments verify some of the requirements that TDI places on the LISA instrument system, for example that the response of the phase measurement chain is linear at the part per billion level. Different approaches have been applied, including hardware-in-the-loop digital models, analogue electrical models, optical models with electronic delay lines, and fully-optical models [146, 210, 271, 272, 308, 449]. A recent example is the so-called "Hexagon" experiment in which three LISA-like optical interferences are generated in such a way that a specific linear combination should have zero phase noise. Current results from the Hexagon experiment are shown in the right panel of Figure 8.2 and demonstrate more than eight orders of magnitude of noise suppression in a system incorporating fully independent optical elements, phasemeters, and clocks [438].

Initial data processing is an active field of research, and although the performance according to the requirements is clear/demonstrated, current efforts concentrate on improving the realism of the simulations, on identifying the optimal channels to provide for scientific analysis in realistic setups [313, 315], on assessing the performance of various implementations for a L0.5-to-L1 pipeline under more and more realistic setups, and on identifying the preferred units or quantities to use as inputs of such a pipeline.

## 8.3     Source reconstruction: From L1 to L2

The second step in LISA's analysis pipeline is the identification and reconstruction of sources in the data stream. This step will use as input L1 data products, such as the TDI variables discussed above, as an input and produce a comprehensive statistical evidence for the entire set of candidate GW sources as L2 data. For most sources, the basic approach relies on the matched filtering technique, in which the data is compared against a source template derived from physical models of the expected systems called waveforms (see Section 8.3.1). Information from L2 data on GW sources is fed back to the L0.5 to L1 pipeline for extracting instrument characterisation, artefacts, noise level, and other optimisations.

Because the sources overlap both in time and in frequency, the search step must operate as a "global fit", simultaneously solving for the joint distribution of parameters for every source present in the data (see Section 8.3.2). From this joint distribution it will be possible to construct marginal distributions for individual events, and hence reconstruct the source waveforms, with uncertainties, accounting for confusion with other sources (see Section 8.4). It will also be possible to compute the correlation between pairs of events and to construct a distribution for the total signal component of the data and the residual after subtracting this total signal component. This residual will be valuable when searching for unknown sources.

## 8.3.1 Waveforms: current status and prospective

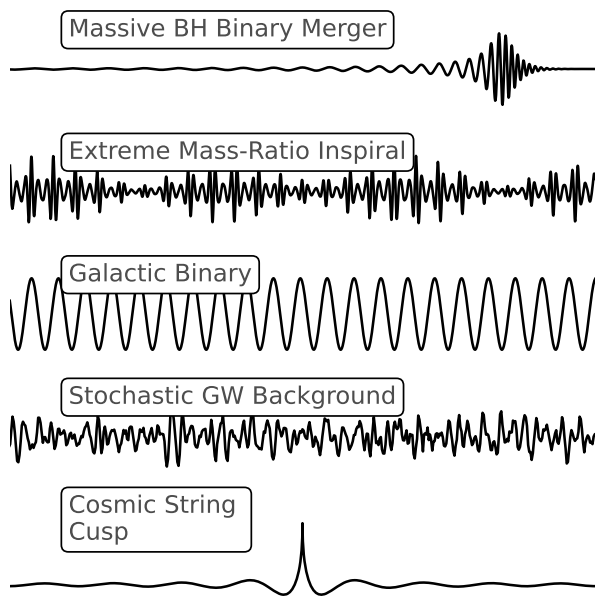

**Figure 8.3:** Shapes of the waveforms corresponding to the GW emission of (from top to bottom): Massive BH binary mergers; Extreme-mass-ratio inspirals; a single Galactic binary; a typical stochastic process; and a cosmic string cusp.

The accuracy of the waveform model that links source parameters to observed data directly limits the precision of LISA's source reconstruction. Waveform models are on a development path to reach the accuracy needed to control the systematic errors. Methods already exist to fold uncertainties into source reconstruction [311]. Most GW sources in the LISA data are expected to be compact objects in a binary configuration. When isolated, and far from coalescence, such systems are characterised by a set of 17 physical parameters. These are the masses and (3D) angular momenta of the two binary components, the eccentricity and inclination of the orbit of the binary and parameters characterising the sky location, distance and orientation of the system relative to the Solar System. For Galactic binary sources, the model can be simplified as the signals will be approximately monochromatic, with perhaps one and at most two frequency derivatives detectable, the effect of the spins being too small to be detected. For some systems there could be additional effects due to the environment, such as tidal effects, magnetic effects and planets around Galactic binaries or accretion disc effects in extreme mass-ratio inspirals (EMRIs). Models including such effects will have additional parameters. Stochastic backgrounds, another type of sources, are characterised by a variable number of parameters that describe the overall amplitude, and spectral shape.

Typical GW signals that we expect to find in the LISA data stream are illustrated in Figure 8.3. Waveform modelling for LISA has a strong foundation in the development of such models for ground-based observations, which has been used to detect GWs from binaries and to infer their properties since 2015. Three main methods are used to compute waveforms from first principles, covering different parts of the parameter space (see Figure 8.4).

Numerical Relativity (NR) solves the field equations directly, building on the 2005 breakthrough [62,

117, 348] with high order finite differencing or spectral methods [236, 387]. Due to computational cost, routine simulations are currently only feasible for bodies of comparable masses (up to a mass ratio of $q = 1/20$) with tens of GW cycles prior to merger. Further out from the merger, a post-Newtonian (PN) expansion [98] in the orbital velocity is used, while for extreme mass ratios, the expansion in the mass ratio is used to solve Einstein equations [68].

Current comparable-mass waveforms already allow for a robust detection of weak signals in ground-based observations, where the space of signals is described to sufficient accuracy by existing models. With some extension in their parameter-space coverage, these models are expected to be accurate and complete enough to identify and estimate the parameters of most of the anticipated LISA signals. However, at current accuracy levels, waveform errors would limit the accuracy of inference of astrophysical sources and populations, as well as tests of General Theory of Relativity (GR). The continuation of current waveform development programmes is required to avoid this.

In addition, several new LISA-specific challenges arise: generally, eccentric orbits may be more common in sources in the LISA band, e.g. Galactic binaries, compared to ground-based detectors. For them, current PN waveforms are also only adequate when the binary is detached (i.e. their stellar atmospheres are not interacting) and tidal interaction is sufficiently small; otherwise mass transfer and tidal effects need to be modelled for astrophysical interpretation, even if the GW signal is close to monochromatic. For some massive Black Hole binaries (MBHBs), their very high SNR will require comparable-mass models with one to two orders of magnitude greater accuracy across the full parameter space of possible signals. Meanwhile, EMRIs and intermediate mass-ratio inspirals (IMRIs) require completely new waveform models from the Gravitational Self-Force (GSF) programme due to their extreme mass ratios. Here, we will summarise the state-of-art in modelling GWs for the problematic MBHBs, EMRIs and IMRIs, as well as outline the plans for improving the waveform accuracy.

### *Merging massive Black Hole binaries*

LISA will observe MBHBs throughout the history of the Universe and a few of these signals could have very high SNR, potentially in the thousands (about two orders of magnitude louder than any detected signal to date). These high SNRs present a challenge in the global fit as the MBHB waveform needs to be exquisitely known to extract the maximal information (i.e. with negligible parameter bias), and to minimise residuals, thus ensuring the other events are extracted without contamination.

Three principal programmes have been established to synthesise models that cover the whole space of MBHB signals, using NR data sets and perturbative approaches such as PN and GSF. These are the effective one-body (EOB) formalism [114, 205, 352, 353], which describes the GW signal in terms of Hamiltonian orbital dynamics; "surrogate models", which interpolate numerical data sets [414]; and IMRPhenom [174, 346], which models the GW signal directly in terms of piecewise closed form expressions.

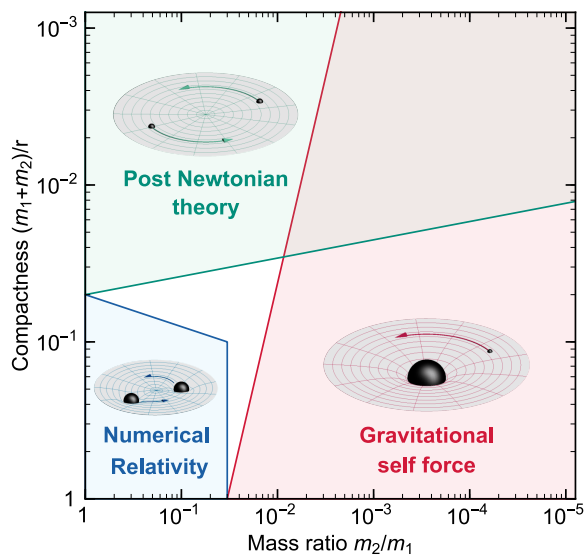

**Figure 8.4:** Parameter space of compact binaries and the relevant modelling techniques. In the top-right area (green+red), both GSF and PN are valid [264].

These programmes complement each other and are already continuously updated to meet the growing accuracy requirements of ground-based detectors. Models produced by all three approaches are routinely used for parameter estimation of detected signals [9], and, with continued development, are expected to eventually meet the requirements of LISA and third-generation ground-based detectors [264, 351]. Building on the experience gained with modelling and data analysis for current GW detectors, the LISA Consortium waveform working group and work package team will develop procedures to review waveform models, and monitor and forecast their accuracy, e.g. by extending [351] to the latest models and adding analysis techniques such as described in [148].

Ultimately, the effectiveness of waveform models is limited by the accuracy and parameter-space coverage of the NR simulations that inform them. Both of these factors are currently limited by computational cost. For MBHBs of comparable mass, current NR codes [264] can produce the required accuracy to extract the waveform of a MBHB for reasonable SNRs [181] for up to 20 GW cycles including ringdown. However, for the highest anticipated SNRs and more extreme mass-ratios, more accurate NR calculations will be needed. To date, $\sim 10^4$ waveforms for Black Hole binaries are publicly available via NR catalogues [111, 222, 244].

The NR community will have to explore the region of more extreme mass ratios and spins and develop codes that can evolve such systems more efficiently. In this sense, recent results indicating that the GSF approach exhibits surprising accuracy in the comparable-mass regime [427, 452] can help reducing the number of NR simulations required at extreme mass ratios.

We envision that the iterative parameter estimation procedures to deal with overlapping signals may benefit from a hierarchy of models, ranging from fast and reasonably accurate across the whole parameter space to more accurate, slower models for specific regions.

### Extreme- and intermediate-mass-ratio inspirals

Very-extreme-mass-ratio systems introduce unique challenges because of their disparate time and length scales, and their long, complex waveforms. Generic EMRIs are approximately tri-periodic, with fundamental frequencies corresponding to the small mass' azimuthal motion, the precession of its periapsis, and the precession of its orbital plane. The emitted GWs contain many harmonics of these three frequencies, which slowly evolve due to radiation reaction over the signals' $\sim 10^5$ observable cycles [233, 344], and which generically pass through transient resonances during their evolution [185, 344].

Effectual "kludge" models have been developed that capture this waveform morphology over short segments of an evolution [134]. Such models, or even more minimal ones [424], should suffice for the detection of most EMRI signals, with an acceptable loss of SNR [135]. However, precise parameter estimation will require accurate models that remain in phase with a signal over the full observable evolution [134, 135, 424]. Accurate modelling is based on the GSF formalism of Black Hole perturbation theory [68, 344], which leverages the extreme mass ratio of the binary. The binary's spacetime metric is expressed as a series expansion in powers of the small mass ratio ($q \equiv m_2/m_1$ with $m_2 \ll m_1$), comprising the background metric of the primary plus perturbations due to the secondary. These perturbations contain the emitted gravitational radiation, and they exert a "self-force" on the secondary which drives its inspiral.

→ THE EUROPEAN SPACE AGENCY

Using a multi-timescale method, all expensive computations are performed as an offline step on the parameter space of slowly evolving variables, and waveforms are then generated by taking these precomputed data as inputs for a rapid evolution through the space [233, 252, 287, 305, 344, 451]. This allows online waveform generation that is efficient enough for data analysis pipelines [136, 252].

It is expected that fairly low-order GSF approximations should suffice for all EMRIs and for most if not all IMRIs. This expectation is based on the fact that in the mass-ratio expansion, the waveform phasing error is of order of $O(1)$ for leading-order "adiabatic" models and $O(q)$ for next-to-leading-order one-post-adiabatic (1PA) models [233, 344].

Like for MBHBs, a major challenge for EMRIs and IMRIs is coverage of the high-dimensional parameter space. Models at 1PA already exist for the (astrophysically-unrealistic) non-spinning, circular systems [427], but theoretical and computational obstacles remain to extend these models to spinning, eccentric and inclined systems. Waveforms must also incorporate transient resonances [90, 107, 185, 214, 317, 363, 382, 450] and the secondary's spin [168, 301, 337, 379]. Treatments of these are steadily advancing, and they represent a small additional effort relative to baseline 1PA computations.

On the other hand, work to develop search and inference strategies suitable for EMRI signals is required. The impact of modelling error on EMRI detection efficiency is currently being studied [300], both with and without search techniques that account for such error (e. g., semi-coherent filtering [191]). Current tests employ kludge models [135] to generate both the simulated signal and the analysis templates [52, 55, 424], and more work is needed to determine how well these models stand up to more realistic simulated data.

### Environmental effects, beyond-*GR* and other waveforms

MBHBs, EMRIs, and IMRIs all involve Massive Black Holes (MBHs), embedded in a rich enviroment of surrounding bodies and gas, that can modify the waveform. In the case of MBHBs, these are expected to have a negligible impact on the waveform [69]. In many models of Galactic cores, particularly quiescent ones, the impact on EMRIs and heavy IMRIs is also expected to be negligible during the observable, late stages of an inspiral. Waveform models in these cases should not require modification from the baseline case of an isolated binary.

Probing the external environment is nonetheless possible, as described in Section 3.3, requires waveform models including these impacts. The influence of a nearby body can be included through the addition of a tidal field [107, 340], the dominant effect of which is to introduce tidal resonances [107, 214]. The effects of an accretion disk (which can be significant for Active Galactic Nuclei - AGN) or of a dark matter cusp can be included through the addition of an effective torque or force on the secondary [383].

Much of this discussion carries over to tests of the GR and searches for new fundamental physics using EMRIs. In most viable candidate theories beyond GR, corrections due new physics are proportional to powers of the mass ratio, and the corrections to the primary, in particular are also small [297]. Large classes of theories can then be modelled simultaneously through modular additions to baseline GSF models; the addition of a scalar charge on the secondary, for example. It is also conceivable that the primary is not a Kerr Black Hole. In that case the baseline GSF model can be also modified in a modular way, by adding small modifications to the gravitational multipole moments away from

Kerr.

Tests of GR with MBHBs, and detection of possible exotic objects that may appear similar to MBHBs, have large similarities with current methods used in ground-based observing [6]. Such tests have steadily improved since the first detection of GWs, but will be carried out at much greater precision with LISA observations due to the much larger SNRs. They include theory-agnostic tests such as parameterised tests of deviations from GR [444] and consistency tests [6, 198]. In addition, the development of PN inspiral models [86, 87] and Numerical Relativity simulations for specific alternative theories [330], or for exotic objects such as boson stars [331], are active fields. It is expected that inspiral-merger-ringdown waveform models for such exotic signals will become available over the next decade, and can be used in parallel with theory-agnostic tests.

The detection of bursts from cusps and kinks on cosmic strings (see Section 3.8) also requires waveform models. Recent NR simulations of a collapsing loop  [50] have paved the way and are expected to be followed by simulations of more complicated loops with cusp/kink features.

As modifications to baseline waveform models are constructed, it is crucial to investigate possible degeneracies between environmental and beyond-GR effects, and to mitigate them in the cases when they can limit LISA's ability to constrain astrophysical models and fundamental theories.

## 8.3.2    Global Fit

The LISA data are signal-dominated, containing tens of thousands of resolvable astrophysical sources overlapping in both time and frequency and stochastic signals of astrophysical and/or cosmological nature. The signal-rich data are coupled with the presence of gaps and glitches in the data together with a slowly-varying noise floor.

The high source density precludes analysis strategies currently employed by ground-based interfero-metric detectors where searches for different source classes are conducted independently and where signal-free data are used to empirically measure the confidence of detection candidates. Instead, LISA analyses depend on the simultaneous fitting of complete astrophysical, cosmological, and instrument models to the observed data. The need for this so-called "global fit" was identified as the primary challenge to the LISA analysis early in the mission formulation. Because LISA source catalogues will contain a mixture of continuous sources (e. g. ultra-compact binaries in the Milky Way) and transients (e. g. MBH mergers) that will correlate with one another in the data streams, each iteration of the global analysis uses all of the available data. Science inferences for all sources will improve throughout the mission as newly acquired data increase the integrated signal-to-noise ratio for persistent signals and improve foreground subtraction for transients.

## 8.3.3    Addressing Instrumental Imperfections

Beyond Gaussian stationary noise, LISA data are likely to be affected by other noise artefacts. Some of them are predictable, and others cannot be precisely anticipated. We classify them into several types guided by the reasoning based on their nature, the way they appear in the data and the ways they can be addressed by the data analysis.

The first type is referred to as **gaps**, which designates the loss of data, either due to the absence of telemetry or to the parts of data being highly contaminated by noise, to the extent that they cannot be used anymore in the data analysis. Current projection of mission operations and duty cycle provide scenarios for gap patterns. During the TDI process, gaps will enlarge by about 100 seconds because of interpolation and delays.

The strategy to deal with data gaps is two-fold. The most straightforward method is to apply a smooth tapering window in the time domain, allowing for a progressive transition at the gap edges before any processing in the frequency or time-frequency domain is performed. Windowing data gaps reduces the noise frequency leakage and should in principle be sufficient in most cases, but its efficiency depends on the type of GW sources and the gap pattern. For continuous sources like Galactic binary (GB), time windowing is acceptable for high enough frequencies and relatively rare gaps [58]. However, for frequent and short gaps, leakage becomes dominant at low frequency, and more sophisticated methods such as statistical imputation [58, 101] are necessary. The gap mitigation strategy will be different for transient sources like MBHB. In this case, the main impact parameter is the locations of the gaps relative to the merger time, which carries the bulk of the SNR. Gaps are usually harmless for both detection and parameter estimation when far from the merger [159], however, the closer to the merger time, the higher the impact. In situations where gaps can be scheduled, this feature shall drive the planning of mission operations.

The second type of artefacts are called **glitches**, which refers to the signals in the data that have a non-astrophysical origin. They can have various known or unknown causes, but they feature characteristic transient signal shape and can appear at different places inside the instrument. Projections of at least some kinds of LISA glitch populations can be inferred from LISA Pathfinder (LPF) observations [42].

The planned strategy to handle glitches includes detection, identification and mitigation of their impact on the data analysis and science extraction. The crudest mitigation we envision is to mask the data stretches affected by the spurious events, before treating them as data gaps. A more refined mitigation is to use a phenomenological model to fit their signal as part of the global fit. Glitch detection can be done through matched filtering techniques, tracking anomalous excess of power using wavelets or other mathematical functions, as done in [56]. Identifying glitches as originating from instrumental effects is a more difficult task, though. In particular, they must be distinguished from GW bursts. Instrument characterisation efforts using different TDI combinations, in particular the null channel will help to define them. Another possible way to tackle this problem is through Bayesian model comparison, where we compare the evidences for two models: one where we assume that the glitch comes from an astrophysical event with a particular sky location to be determined; another where the glitch is supposed to come from an instrumental effect assuming an injection point, like the kick of one test-mass. A proof of concept of this approach was carried out in [360], building on a wavelet analysis used in LIGO/Virgo [141]. For the glitches comparable in duration with merging MBHBs we can use consistency tests as a veto similar to those developed in the ground-based community [17].

The third type of artefacts take the form of **spectral lines**. These artefacts are continuous nearly sinusoidal processes that appear as vertical lines rising about the noise spectrum, and their characteristic frequency can oscillate as a function of time. They can originate from electronic control systems or mechanical processes, as observed during LPF operations [38].

The biggest challenge spectral lines pose is their relative similarity with continuous GW source signals. Spectral lines are present in ground-based GW detectors and existing mitigation techniques can be adapted. For example, one approach is to treat spectral lines as features of the instrumental noise power spectral density and fit them in a Bayesian framework using Lorentzian shapes [282]. This strategy naturally integrates into the global fit pipelines. Note that the orbital motion of the LISA spacecraft imprints a precisely predictable phase modulation to continuous GW signals, whereas modulations in noise spectral lines are likely to be stochastic. This difference will make narrow-band noise and signal features in the data robustly distinguishable.

The fourth type of artefact is **non-stationary noise**. It includes the change of the noise spectral density with time or Gaussian noise bursts. In addition, this generic category also includes the possible non-Gaussianity other than the 3 first types of artefacts.

Continuous noise non-stationarities are expected to arise in the phasemeter measurements. However, the noise Power Spectral Density (PSD) is likely to evolve slowly, making the process locally stationary. Accounting for this variability is particularly important to detect and characterise continuous GW sources, because fixed-PSD analyses could fail in detecting faint sources or derive incorrect parameter uncertainties. Hence, non-stationarity will be incorporated within the noise covariance estimation component of the global fit. One way to achieve this is to use time-frequency decomposition such as wavelets, which have already been introduced in GW data analysis [140].

Finally, **imperfect calibration** of the output strain data and lack of knowledge of the instrumental noise (see Section 8.3.3) could also hinder signal reconstruction. Both types of uncertainty will be included within the global fit through the inclusion of additional fitting parameters, so no bias should arise, but reconstruction uncertainties will be larger than the ideal case. Calibration of LISA is expected to be good enough that signal reconstruction will only be mildly affected [371]. Noise uncertainties should have a limited impact on resolvable source reconstruction and on the characterisation of modelled stochastic backgrounds [59]. However, this may be limiting for unmodelled background reconstruction.

## 8.4    Construction of the final results: From L2 to L3

The global fit pipeline produces an unwieldy probabilistic collection of candidate detections with correlated source parameters among signals overlapping in time and frequency. These results need to be consolidated into easy-to-use source catalogues which will serve as the entry point for the vast majority of LISA end users, and as the input to many of the science investigations planned for the data. Going from the global fit output to a discrete source catalogue is a lossy data compression step requiring tools, documentation, and support to enable users to work back to the lower level data products should their research interests be sensitive to the details of the processing, source subtraction, and subjective choices necessarily made in the construction of catalogues. This will be coordinated through the LISA Science Team (LST) as described in Chapter 9.

Thus the final data products must strike a balance between being an easy-to-use and interpret source catalogue, while having a powerful and flexible interface to the precise details of the upstream analyses. Figure 8.5 shows the joint sky maps of all catalogue sources from one of the LISA Data Challenges (LDC2a, see Section 8.6 for more details) analyses for the GBs (left) and MBHBs (right). Viewing the GB sky map in Galactic coordinates clearly shows the Galactic disk and bulge structures in the reconstructed source positions. Here, the MBHB map features an ecliptic coordinate option to facilitate multimessenger observations.

A complication to the construction of source catalogues comes from the anticipated parallel and independent efforts for developing and operating global fit pipelines. Multiple pipelines are critical for LISA given that it is a unique instrument operating in a never-before-observed band of the GW spectrum targeting sources that, with a few exceptions, have never been directly detected. Consolidating the input from multiple global fit pipelines into a single self-consistent catalogue is a non-trivial but vitally important challenge. For loud astrophysical sources the procedures for combining results are straightforward so long as different pipelines are using consistent assumptions about the data and underlying astrophysical models. For weaker signals near the detection threshold, combining catalogues from different pipelines has an inherent complexity due to the fully probabilistic nature of a global fit. Catalogues with different confidence levels attached to sources are anticipated, in which each source is denoted by the pipeline used in its recovery. In principle, this class of problem exists in other missions, such as Gaia, but more development is needed to arrive at a procedure for producing self-consistent consolidated L3 data products.

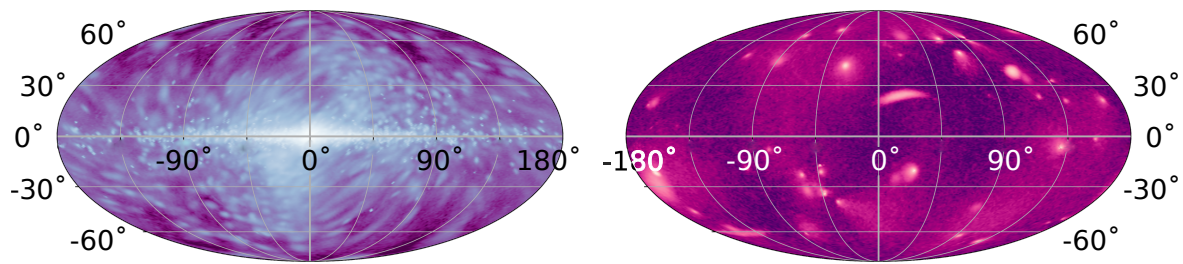

**Figure 8.5:** Joint sky maps from global analysis of LDC2a for Galactic ultra-compact binaries (left) and massive Black Hole mergers (right). The Ultra-Compact Binaries (UCB) map is in Galactic coordinates and clearly shows the disk and bulge structure of the Milky Way.

Early exploratory efforts at catalogue consolidation use the results from the LDC2a global fit analyses. For example, ~4300 GBs signals were detected with a high confidence in both global fit pipelines. Another ~1600 GBs were identified in one of the two pipelines at lower confidence (see Figure 8.9).

The user interface to source catalogues will include sophisticated database query capabilities so that end users can select individual source or sources based on type, event time, sky location, parameter ranges, etc. Similarly interactive tools for constructing customised data sets for further study are also necessary for science exploitation. Examples include constructing residual data sets containing only individual sources or source types to enable re-analysis using different models and/or assumptions. The most extreme case would be residuals after all catalogue entries have been removed in order to search for sources or source types not included in the global fit.

## 8.5       Low Latency Analysis: Alerts

Global fit processing will occur periodically after sufficient collection of new data. Time-critical science, most notably the sharing of information with the multimessenger observing community, cannot be delayed by the cadence and processing time of the global fit. Instead there will be separate Low Latency Alert Pipelines (LLAPs) that are operating in real time to detect and improve astrophysical inferences from high-value transients, which will autonomously release those results through an alert system similar to what has been established for other transient astronomy missions/observatories. The L1 data will be analysed continuously by the LLAP as it becomes available with the requirement that processing be completed within one hour from when the L0 data are communicated to the ground stations.

The main objective of the LLAP is to identify and localise new astrophysical transient signals and to refine the localisation of ongoing transients as new data are acquired. The main target of these pipelines are coalescing MBHBs which are expected to be detectable days to months before merger, with the source localisation rapidly improving during the late stages of the signal (see Figure 9.2). The LLAP processing will also be sensitive to unanticipated transients, i. e. GW "bursts".

To do so, the first stage of LLAP is to remove the astrophysical foreground sources already identified from the global fit processing. Searches for new sources, and characterisation of ongoing transients, will then be conducted on the residuals after the global fit subtraction. These cleaned data will also be returned to the Science Operation Centre (SOC) for instrument characterisation, as the GW signals present in the data will conceal the instrument noise.

The outputs of the LLAP processing will include alerts to the multimessenger observing community for new transients that have entered the LISA measurement band; predictions for the merger time of MBHBs which will be used by the external community to plan multimessenger observing campaigns as well as the SOC to schedule *protected periods* for the data to ensure that there are no planned disruptions to the LISA observations during the merger phase of the binary; and regular updates to the localisation of long-lived transients as new data help refine the estimates.

The LLAP processing is optimised for computational speed and has a larger tolerance for reduced detection efficiency and/or accuracy in parameter estimation than the global fit. The prioritisation on latency will govern algorithm design choices that are different from the global fit, including liberal use of various approximations (e. g., waveform models, instrument response calculations, treatment of instrument noise and astrophysical foregrounds, etc.) or alternative methods from the traditional matched filtering approach ubiquitous in GW astronomy (e. g., machine learning methods).

While LLAP prototypes are at an earlier stage of development than the global fit pipelines, a proof-of-concept demonstration of the low-latency search and characterisation step was performed on the "blind" version of LDC2a data. This "blind" data set uses a different statistical realisation of the same astrophysical population as the training LDC2a data used for the global fit demonstrations, but with hidden information about the number of sources and their parameters, as well as the instrumental noise level. This analysis treated the UCBs signals as a stationary noise source and whitened the data using a noise model that includes spectral lines adapted from a pipeline used for LIGO, Virgo, and KAGRA (LVK) analyses. The whitened data were searched in month-long segments and event candidates were characterised using a single computing node in a few hours of processing time. The pipeline successfully identified all 6 simulated signals, and the true sky position of the source is within 90 % credible interval for all sources and within 50 % credible interval for 2 sources (out of 6). The sky uncertainty (90 % credible region) varies between 90 and few thousands square degrees. While significant development is needed to create an end-to-end low latency pipeline, this proof-of-concept demonstration shows that the required accuracy and latency are achievable with present-day algorithmic and computational technologies.

## 8.6    Demonstrating LISA Analysis with the LISA Data Challenges

The need for, and difficulty of, developing global fit algorithms inspired a coordinated LISA consortium effort to prototype data analysis algorithms well in advance of mission operations through the LISA Data Challenges (LDCs).

The LDCs have provided data and analysed results for a series of challenges, beginning with simulations focusing on individual source types (LDC1) aiming at building the data analysis tools and verifying/optimising their performance. The second challenge (LDC2) was split between two major objectives: (i) prototyping the global fit using the simulated data with $3 \times 10^7$ Galactic binary, dozens of MBHBs, and stationary instrumental noise with unknown a priori level (LDC2a); (ii) testing robustness of the single-source data analysis algorithms to the presence of data gaps, noise transients (or "glitches"), and non-stationary noise (LDC2b).

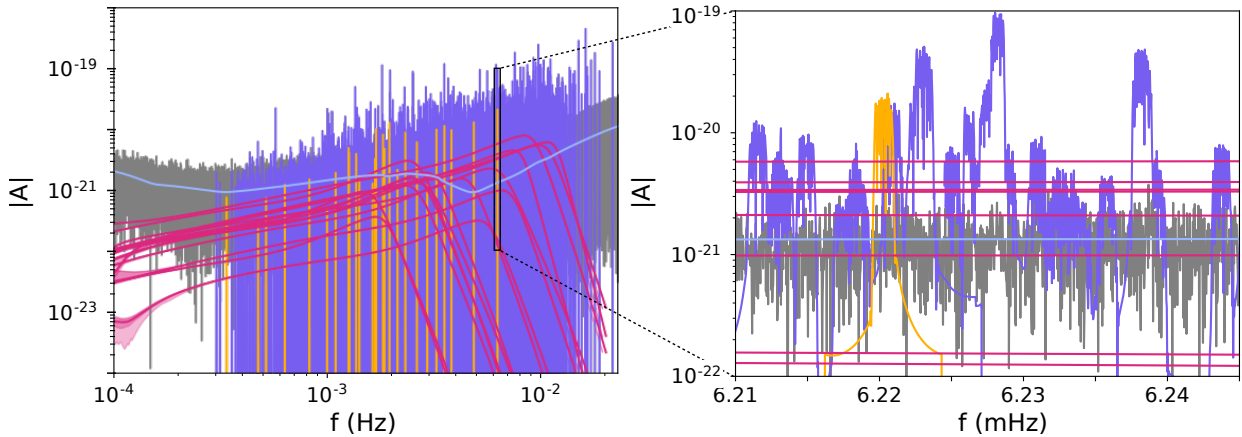

**Figure 8.6:** Results from global fit analysis of LDC2a data showing simultaneous reconstructions of the resolvable Galactic binary (purple), verification binaries (orange), massive Black Hole mergers (magenta), residuals (grey), and noise model (blue). The right panel focuses on the frequency interval containing HM Cancri. Both plots display the amplitude spectral density of the TDI A Channel.

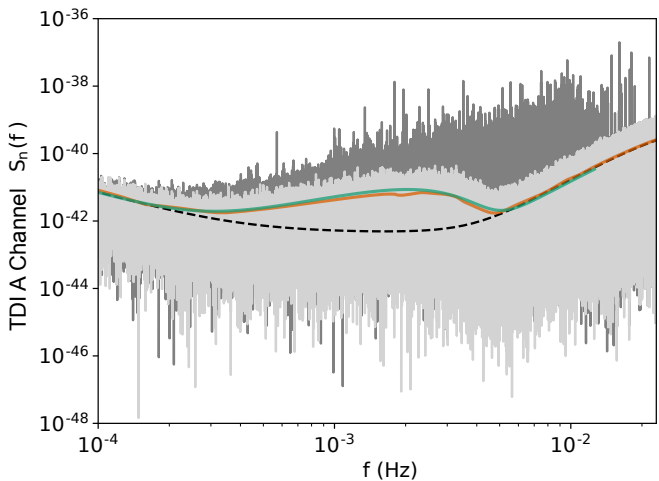

**Figure 8.7:** Power spectrum of the TDI A channel LDC2a data and noise models. The dark grey line shows the original data. Light grey is the residual after removal of detected signals. Green and orange show the noise level inferred by the global fit pipelines. The dashed black line shows the theoretical estimate including instrumental noise. The difference between the inferred and theoretical spectrum between $2 \times 10^{-4}$ Hz and $6 \times 10^{-3}$ Hz is due to the unresolved Galactic foreground.

Two independent search pipelines were constructed and applied to the LDC2a dataset to demonstrate the feasibility of the global fit. Both pipelines use similar underlying philosophies of forward modelling the LISA response using a parameterised model of the full set of detectable sources, and exploring the large parameter space using stochastic sampling algorithms. The pipelines leverage the relatively mild correlations between different source types to efficiently sample the large parameter space in sequential "blocks" composed of smaller parameter subsets where correlations are more likely. The scheme will also work when different parameter blocks are strongly correlated but with lower efficiency.

A graphical summary of the global fit analysis in the frequency domain is given in the left panel of Figure 8.6. The recovered UCBs are shown in purple, verification binaries (i. e. LISA sources already known through electromagnetic observations) in orange, merging MBHBs in magenta and the estimated noise level is blue. The detected sources are plotted together with the residual in grey. For the verification binaries the global fit pipeline uses the electromagnetically-observed sky position and orbital period to conduct a targeted analysis to (i) provide the best possible constraints on the binary parameters by incorporating what is already known about the sources, or (ii) provide upper limits on the GW amplitude for sources below LISA's detection threshold.

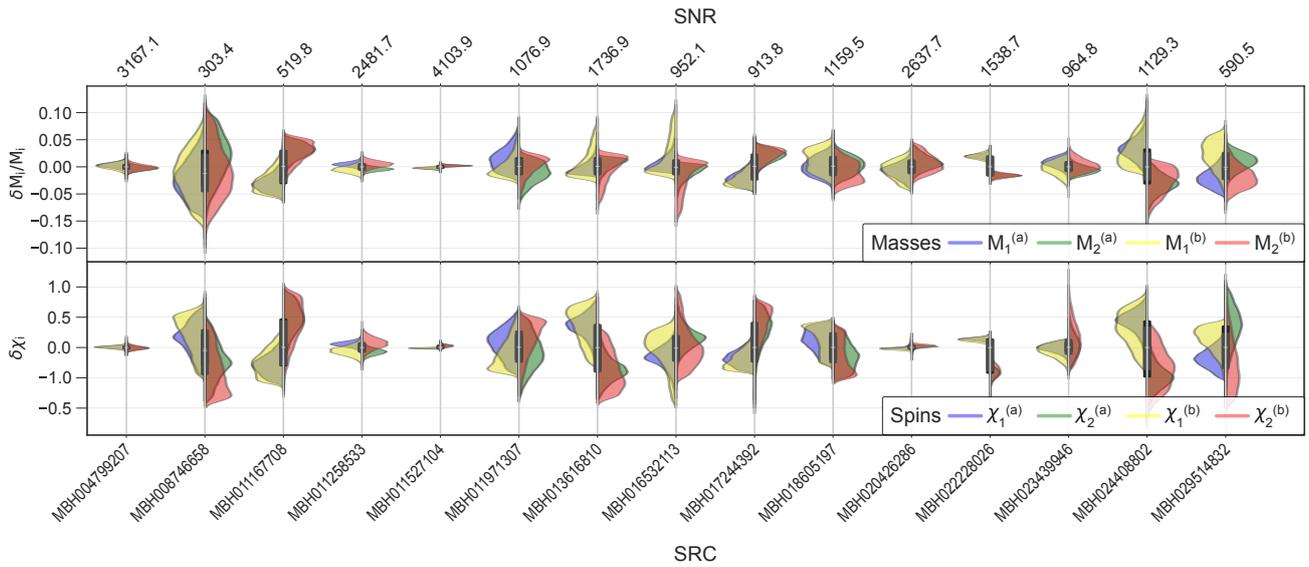

**Figure 8.8:** Results from global fit analysis of LDC2a data: estimation of masses (fractional errors are given in the upper panel) and spins (absolute error in the lower panel) of MBHBs. The left part of all violin plots corresponds to the primary object (heavier MBH). Zero corresponds to the true (injected) value.

The right panel shows a typical residual (in grey) after subtracting the identified sources while zooming in on a narrow band segment containing one of the loudest known verification binaries, HM Cnc (orange), to show how it and other UCBs (purple) are densely packed into the data, with the MBHBs (magenta) sweeping through the frequency band on their way to merger.

Despite the similarity in the overall strategy, the implementation and tools used for the two prototype global fit algorithms are independently developed. Both pipelines have demonstrated consistent results: (i) identifying all 15 MBHBs and their parameters correctly (see Figure 8.8) (ii) unambiguously recovering (with correlation >0.9 against the injected catalogue) ~5000 UCB sources (see Figure 8.9), and (iii) inferring the

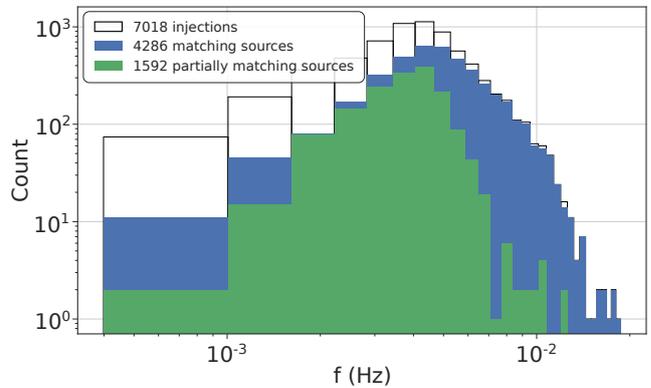

**Figure 8.9:** Identified GBs in the simulate LISA (LDC2a) data. Source reliably identified by both pipelines (blue), sources found by individual pipelines but with cross-pipeline correlation <0.9 (green), and (white) are all sources above SNR > 8 present in the simulated data.

instrumental noise at levels similar to the theoretical predictions when excluding the region of the Galactic foreground ($2 \times 10^{-4}$–$6 \times 10^{-3}$ Hz) as shown in Figure 8.7

At the current stage of the global fit pipeline development, efforts have focused on the astrophysical populations expected to be the dominant contributors to the LISA data, namely the most numerous (UCBs) and the loudest (MBHBs) sources.

Other astrophysical sources, most notably EMRIs, stellar-mass Black Hole binaries (sBHBs) and stochastic gravitational-wave backgrounds (SGWBs), must be incorporated into the global fit architecture but current theoretical predictions suggest that those sources will be subdominant in the amplitude and in the event rate. A parallel study is ongoing to develop algorithms for detecting these sources (LDC1b) which will be incorporated into the next generation data simulations and algorithm development cycles. For example, a recent proof-of-concept study exploring the simultaneous fit of instrument noise and stochastic GW signals has shown that, under certain assumptions, it is possible to disentangle power-law SGWBs from instrumental noise without relying on a fixed model for the noise PSD's shape [59]. Conversely, other studies suggest that agnostic detection of SGWBs with a parameterised noise model is possible, provided that the model is accurate [336].

In addition to using data simulation campaigns to expand the astrophysical capabilities of prototype LISA analysis pipelines, the LDCs have produced the LDC2b datasets which contain simplified astrophysical populations (one MBHBs merger or only the verification binaries) but include more realistic observing conditions including gaps in the time series data, noise transients ("glitches"), and non-stationary instrument noise due to the astrophysical foreground. The preliminary analysis of these datasets has shown that glitches could be detected through power excess identification and then discarded in the analysis, avoiding biases in the source parameter estimation. However, the ability to tell whether the power excesses come from astrophysical or instrumental origin is under assessment. Besides, the gap patterns simulated in the LDC2b were found not to have significant impact on the verification binary parameter estimation, provided that an optimally smooth time-domain windowing is applied to mitigate noise power leakage effects. The characterisation of MBHBs is also mildly affected by LDC2b gaps, as data interruptions arise sufficiently far away from the merger time. However, glitches or gaps can induce an apparent bias in the parameter estimation. A carefully tuned time windowing or more sophisticated techniques can be required to mitigate this effect.

# 9 GROUND SEGMENT AND OPERATIONS

## 9.1 Mission Operations Centre and Ground Stations

The ground segment and operations infrastructure for the Mission Operations Centre (MOC) of the LISA mission will be set up by ESA at the European Space Operations Centre (ESOC) in Darmstadt, Germany. The role of the LISA MOC is expected to be fairly standard, and the MOC will be based on extensions of the ground segment infrastructure existing at the time of LISA, customised to meet the mission specific requirements thereby maximising the sharing and reuse of facilities and tools made available from other ESA Science missions.

The MOC is responsible for all operations of the LISA mission during all mission phases covering both nominal and contingency operations. This includes the following tasks: planning, scheduling and execution of the ground station contacts; performing uplink of the satellite and payload telecommands and receiving telemetry through the ground stations and communications network; monitoring the spacecraft platform health and safety; monitoring the payload safety and reacting to contingencies and anomalies according to procedures provided by the LISA Science Operation Centre (SOC) to restore the constellation to normal science mode; generation and uplink of the master schedule using the inputs from the SOC and adding the necessary platform commanding; validation of the master schedule against the mission resources (power, data storage); performing uplinks of payload on-board software updates as generated, validated and delivered by the Instrument Teams via the SOC; providing flight dynamics support, including performing trajectory and attitude analysis and preparation commanding sequences for input to the master schedule updates related to all orbit and attitude manoeuvres during the Transfer Phase; producing and providing ancillary data to the SOC (orbit files, pointing information, housekeeping telemetry, etc.); and supporting the Science Ground Segment (SGS) on all aspects concerning spacecraft operations.

### 9.1.1 Mission Operations

The LISA mission is planned with a short 3 day Launch and Early Operations Phase (LEOP), included in the 1 month Near Earth Commissioning Phase (NECP). While underway towards the operational orbit the transfer activities may run in parallel to the commissioning phase. Flight operations teams will be optimised throughout this period to ensure full MOC team and ground station coverage during the critical stages of the mission. After the orbit insertion the mission transitions into the Science Commissioning and Calibration Phase (SCCP) during which ground station pass durations and shift team sizes are gradually further adapted to achieve those as planned for routine science operation.

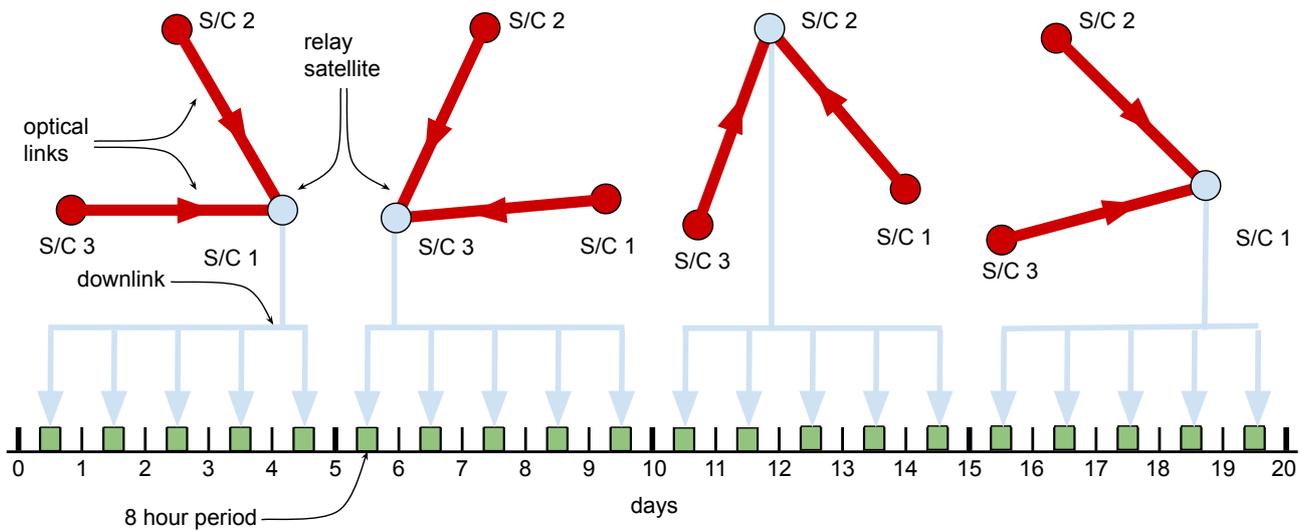

**Figure 9.1:** Communication with the constellation is routed through one of the spacecraft ("relay") for 5 days, establishing a link for 8 hours a day. During these 5 days, the other spacecraft communicate to the relay spacecraft via optical communication, using the existing laser link. After 5 days, the next spacecraft serves as relay, handing off to the third spacecraft after another 5 days, completing the cycle after 15 days.

During the nominal science phase, operations of LISA are foreseen to be highly automated. MOC will communicate with one satellite for 5 days with a ground contact of around 8 hours per day. During the ground contact the satellite will serve as relay for the other two satellites; after the 5 days, the communication will be handed over to another satellite and recurrently the communication token is passed every 5 days (see Figure 9.1). The mission timelines are uplinked periodically (e.g. weekly) and the pass durations are primarily used for downlink of the science data and collection of tracking data needed for orbit determination. In the event of ground anomalies LISA is foreseen to continue nominal operations without the loss of science data for a period of up to 14 consecutive days and in all post LEOP mission phases the spacecraft will be able to survive without ground contact. No station keeping manoeuvres are expected during the nominal science phase.

In addition to the nominal science phase, LISA has two special modes of operations that are designed to maximise the science return from transient sources, such as Massive Black Hole (MBH) mergers. The *protected periods* are periods of uninterrupted data taking that can last up to 14 days. For those protected periods, any planned interruption of the nominal science phase through e.g. antenna rotation or fast discharge of any of the test masses is rescheduled to occur either before or after the protected period. To allow for sufficient planning, such protected periods will have to be requested at least two days before their start.

The other special mode of operation affects the download of data from the constellation. In case of a transient event that can be well enough localised in the sky to warrant alerting other detectors, e.g. in-orbit telescopes or large ground-based facilities, it can be beneficial to have the data available outside the nominal daily 8 hour communication window. During such a *low-latency period*, data on board the constellation will be downloaded in a quasi-real time mode if the nominal communication schedule allows. Should the low-latency period occur outside the nominal communication window, i.e. in the 16 hours between passes, data will be downloaded depending on the availability of ground stations.

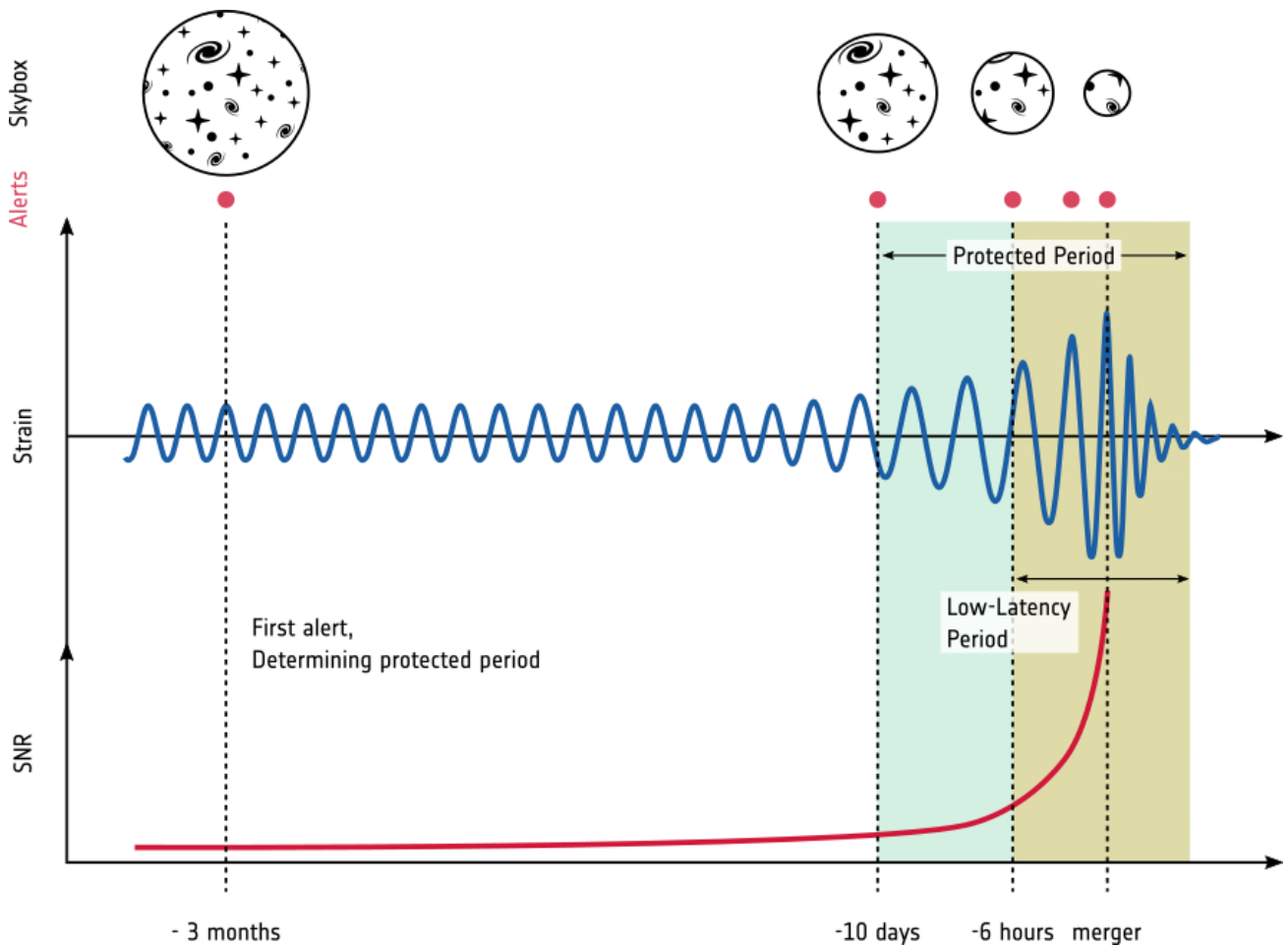

**Figure 9.2:** Schematic representation of the protected period and the low latency. The protected period can be requested at the time of the first alert, when the approximate time of the merger of a massive Black Hole binary is known. A few hours before the merger the accumulated signal-to-noise ratio (SNR) increases significantly and the uncertainty region for the sky position ("skybox") shrinks significantly to the point where other observatories can start observations. The low-latency period allows to monitor the continuously shrinking skybox and to update the alerts.

With this scheme total latency between data generation in the constellation and providing updates to the critical parameters of the transient source, i.e. typical sky position, of less than 1 hour can be achieved. The number of those events during the operational lifetime of LISA is estimated to be small. See Figure 9.2 for a schematic overview of the protected periods and low-latency periods.

## 9.2 Science Operations Centre

The Science Operation Centre (SOC) will be located at the European Space Astronomy Centre (ESAC), Madrid, Spain and its functions are detailed in the following section. SOC, MOC, and the ground stations form the Mission Operations Ground segment and will be fully funded by ESA for the full duration of the mission. The SOC is also part of the Science Ground Segment (SGS) whose other main components are the Distributed Data Processing Centre (DDPC) provided by the ESA Member States and the NASA Science Ground Segment (NSGS). The SOC is MOC's only interface with the SGS. Payload commanding and calibration activities onboard are prepared by the SOC and sent to MOC for upload to the constellation. On the downlink side, telemetry from the constellation together with auxiliary data (see below) is transferred from MOC to SOC. The subsequent sections describe the assumed responsibilities and the setup of the SOC from a functional and organisational viewpoint.

It is assumed that the SOC will have the following roles and main responsibilities: lead the overall validation of the Science Ground Segment (SGS); design, development, procurement, integration, validation and maintenance of the software and computational resources under the SOC responsibility; design, development, and maintenance of the LISA Parameter database; design, development, operations, and maintenance of the LISA public archive; perform the long term planning and scheduling of the science operations of the observatory; storage of all levels of data products (L0 to L3) and all auxiliary mission data and their dissemination within the SGS; maintenance and updating of the instrument calibration data; reception of the raw telemetry from the MOC; processing of the Raw telemetry to L0 data products in a near-real-time; design, development (with support by the DDPC and NSGS), and operations of a near-real-time pipeline to produce validated L0.5 data products from the L0 data; design, development (with support by the DDPC and NSGS), and operations of a near-real-time pipeline to produce validated L1 data products from the L0.5 data; development (with support by the DDPC and NSGS) of the capacity to produce L2 data products from the L1 data; management, through the Project Scientists and the LISA Science Team (LST), the generation of the L3 source catalogue; management of periodic data releases of the L1 to L3 data products via the LISA public archive; design, development (with support by the DDPC), and operations of the Quick-look data processing systems; integration and operations of the Low Latency Alert Pipeline (LLAP) provided by the DDPC; and issue and update alerts for transient signals.

## 9.2.1     Functional breakdown

Figure 9.3 depicts a top-level overview of the SOC with its main interfaces and operational elements. The design and procurement of all systems, including scientific algorithms and tools, is assumed to be done in close collaboration with the LISA Consortium (see Chapter 10).

### *Main Data Repository*

The Main Data Repository (MDR) acts as the central repository and hub for all data produced throughout the SGS. Due to the relatively small amount of total data ($< 100$ TB), it is assumed that the MDR can be mirrored at the Distributed Data Processing Centre (DDPC) and the NASA Science Ground Segment (NSGS). Data mirroring is a standard feature and will provide a convenient, transparent, and reliable replication service and data transfer mechanism among all SGS sites. The MDR is to be regarded as a working data repository of all data (L0.5 to L3, calibration, auxiliary) needed by the SGS to create the final mission products.

### *Near-Real-Time Processing*

The Near-Real-Time Pipeline (NRT) is assumed to be a sequence of three main modules (see also Figure 9.3): the initial data treatment, during which the augmented telemetry is converted to L0.5 data (see description of L0.5 in Section 8.2); the L1 data pipeline, that turns L0.5 into L1 data through application of the Time-Delay Interferometry (TDI) algorithms (see Section 8.2 for details); and a quick look system to assess the scientific health of the payload and quality of the produced L0.5 and L1 data in the form of diagnostic plots and reports. The NRT runs directly upon data ingestion with minimal human intervention.

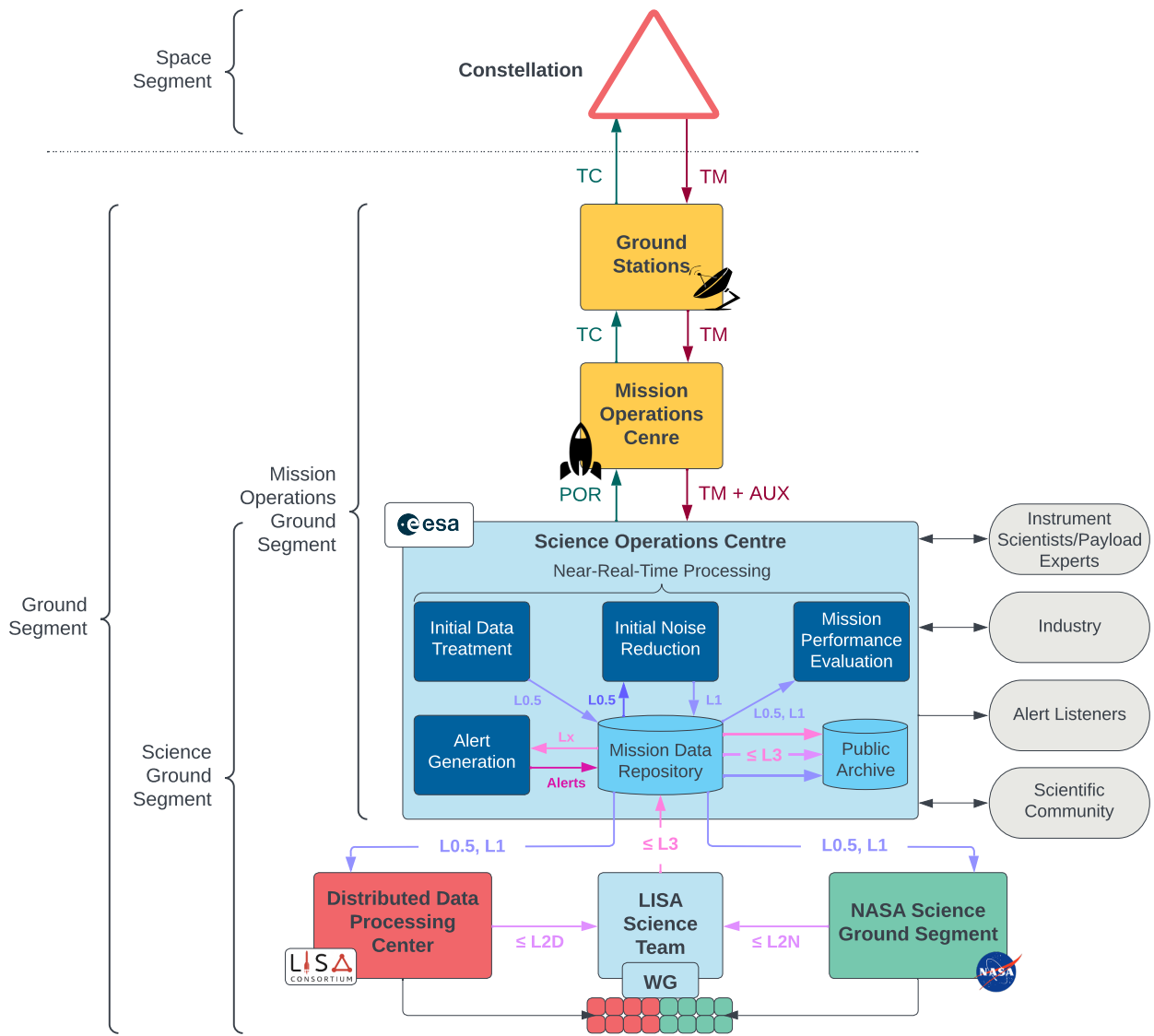

**Figure 9.3:** Top-level overview of LISA's main constituents, their interfaces, and consumers of the mission's final data products.

### Alert Generation

Alerts will be generated with the Low Latency Alert Pipeline (LLAP) which is described in Section 8.5. The LLAP will run in the SOC to issue rapid alerts with the needed source parameters to a network of Event Listeners. Requests for extended ground station coverage will have been made when the event has been originally forecast and validated to allow for low-latency periods (see Section 9.1.1).

### Public Archive

The public archive will be the primary repository for the final mission products and serve such data to the scientific community. Data sets will be released during LISA's lifetime as controlled data releases instead of a continuous stream of data. The data releases will predominantly contain a catalogue of gravitational wave (GW) sources of increasing size and accuracy, including intermediate data products. The final mission data release is will contain updated data products released earlier, including all software and models used in the construction of the data products. The SOC will be responsible for the procurement and operation of the public archive but parts will be delegated to the ESAC Science Data Centre (ESDC) group at European Space Astronomy Centre (ESAC). The SOC will be involved in all preparatory activities leading to the data releases in close collaboration with ESDC and the Consortium. Data within the Science Ground Segment (SGS) are exchanged not via the public archive but by means of the Main Data Repository (MDR) and associated transfer mechanism. The public archive will only serve the mission products for the general public.

### External Interfaces

The three main interfaces of the SOC are with the MOC, the DDPC, and the NSGS as described in the following sections. Others are:

**LISA Instrument Scientists and Payload Experts Group (LISPEG)** The LISA Instrument Scientists and Payload Experts Group (LISPEG) will be co-chaired by the SOC and the Performance & Operations (P&O) group. This is expected to be a group of instrument experts from the LISA Consortium and NASA staff with detailed in-depth knowledge of the payload and its operations. The LISPEG's mainly interact with the Instrument Operations Team within the SOC and jointly recommend payload operations requests. Functionally the LISPEG might be regarded as members of the SOC. In any case, SOC will need to be in full control of operational requests and deliverables (e. g. calibration files) coming from this group.

**Industry** This is the team of industrial experts from the prime- and sub-contractors. It is expected that they will be supporting the Commissioning period and be available to support anomaly investigations during the mission. They will surely mostly interact with the MOC but likely also have an interface with the SOC.

**Event Listener** This is a group of humans or software agents which receive the alerts from the LLAP to organise contemporaneous electro-magnetic observations with astronomical facilities. The format of the data is expected to follow then-existing standard coordination mechanisms and protocols of Multi-Messenger astronomy.

**Scientific Community** The general public interested in the LISA data releases. Their main interface is given by the available facilities in the public archive to access the data products. A small user help desk within the SOC will support the community with questions and requests.

→ THE EUROPEAN SPACE AGENCY

## 9.3      Distributed Data Processing Centre

The DDPC is an entity that extracts source data (L2 Data) from the noise corrected data (L1 data). It also develops some of the software that will run at SOC and provide an offline follow-up of the alerts. It is under the responsibility of the ESA member states and is designed, developed and operated by the ESA member states in close collaboration with the SOC and the LISA Consortium. The DDPC provides dedicated infrastructure and is managing the LISA data processing using several physical computing centres or Data Computing Centres (DCCs) located within ESA member states. In terms of hardware (processing, storage, etc.) it can be seen as a distributed centre. The main DCC is providing the primary interface to the SOC serving a coordinating role for the other centres. During operations, the DDPC encompasses a team of operators including operational scientists, personnel for system monitoring and management of the infrastructure.

At a high level, the DDPC provides several functions in service to the mission and to the realisation of its science objectives: software, computational resources, and operational pipelines, to conduct the core analysis – beginning from L1 to extract content about GW sources and produce the data (L2); supporting the consolidation of the L3 data products together with other L2 providers under the leadership of ESA; software pipeline components to be incorporated in the SOC as needed for L0.5→L1 processing, and for processing of information for quick alerts (follow-up of candidates); infrastructure, tools and processes for LISA Consortium expert teams to develop modular models and operationally represented prior and posterior knowledge, statistical analysis systems, using simulated data; infrastructure and tools for LISA Consortium experts to challenge, monitor, verify and redevelop the models and tools provided once confronted with real data; infrastructure and tools enabling disseminating data to the DCC; and quality control tools and supervision software to support development and operations.

## 9.4      NASA Science Ground Segment

The NASA Science Ground Segment (NSGS) comprises NASA's contributions to data analysis and support for science investigations in the LISA mission. The NSGS supports the scientific goals of the NASA LISA Project and fulfils NASA's commitments to ESA in areas of science and data analysis as specified in the NASA-ESA Memorandum of Understanding (MoU). The primary objectives of the NSGS are the following: Deliver a suite of data products from the LISA mission which enables the US and global scientific communities to conduct scientific investigations as identified in the ESA LISA Science Requirements Document; Coordinate with SGS efforts in Europe to perform cross-checks of pipeline outputs and to develop a single consolidated catalogue for joint publication on ESA and NASA archives; Provide the US research community with tools and extended data products to enable science investigations beyond those identified in the ESA Science Requirements Document; and facilitate the broadest possible participation in LISA science by providing access to data products, tools, and expertise

While the specific architecture of the NSGS will be finalised in later phases of the NASA project, two key functional elements are anticipated. A *NASA Data Processing Center* will be responsible for the development and operation of all analysis pipelines required by the NASA project. The Data Processing Center will ingest low-level data products and information from the SOC and run one or more end-to-end data pipelines to produce a suite of data products up to and including L2. While the design, implementation, and operation of these pipelines will be managed by NASA, it is anticipated that these activities are closely coordinated with ESA and the DDPC so as to take maximum advantage of the parallel pipeline architecture. Most importantly, the output of the DDPC and NSGS global fit pipelines will be consolidated into a single joint source catalogue for distribution on ESA and NASA archives (see Section 9.5 and Section 10.7.2 for management and publication details). Members of the NASA Data Processing teams will also contribute to project science activities such as performance and operations and preparation of data releases.

A *NASA LISA Science Center* will provide public access to released data products, including consolidated catalogues jointly developed with European partners. In addition, the NASA LISA Science Center will provide user support services to the LISA user community and organise a Guest Investigator program to provide grant support for conducting scientific investigations with LISA data products.

## 9.5      Routine Science Operations Concept

During science operations, telemetry data from the constellation is downlinked to the ESTRACK ground stations and transferred to the MOC via public internet lines. From the MOC, telemetry and auxiliary data are transferred to SOC to a staging area "inbox". From there, the data will be ingested into a read-only, long-term telemetry archive and, likewise flow into the NRT system consisting of three main modules (see Section 9.2.1.2). Each of them will write to and read from a MDR which also acts as a central data distribution system.

From the MDR, L0.5 and L1 data will be sent to the two ground segment entities, the DDPC and the NSGS. Both entities will perform global fits to the L1 data, obtain individual sets of L2 data and deliver those back to ESA. There is no direct exchange of data between DDPC and the NSGS but all transfers go through SOC (hub-spokes data dissemination topology).

A L2/L3 coordination team, chaired and coordinated by the ESA Project Scientist will include experts from NASA, ESA and the Consortium and is responsible to consolidate all L2 datasets to a final L3 catalogue. After final approval from the LISA Science Team, the SOC will provide the data releases to the community.

# 10   MANAGEMENT

## 10.1   Mission Organisation and Management

The LISA mission is implemented as an ESA mission in partnership with NASA. ESA maintains overarching responsibility for the mission and the management of partners and contractors. The customer of the mission is the scientific community to support research on and with the gravitational wave environment of our Universe, represented by a LISA Science Team (LST) interfacing with the ESA and NASA LISA Project Scientists.

As is common in ESA's Science Programme, the LISA mission is supported by contributions from the member states. The approach to the scientific exploitation of the mission will be described in the Science Management Plan (SMP), to be approved by ESA's Science Programme Committee (SPC). The legal framework of the contributions and partnership agreements will be captured in the Multi-Lateral Agreement (MLA) between ESA and its member states, and via a Memorandum of Understanding (MoU) with NASA. For the MLA, an oversight committee, called the steering committee, will be established with representatives from all signatories of the MLA. The overall mission organisation is shown in Figure 10.1.

LISA consists of three main mission segments: the space segment consisting of the constellation of three spacecraft, the launch segment consisting of the launcher and launch services, and the ground segment, consisting of the mission operations and science operations. Partners and Member states are involved in multiple elements of the LISA product tree.

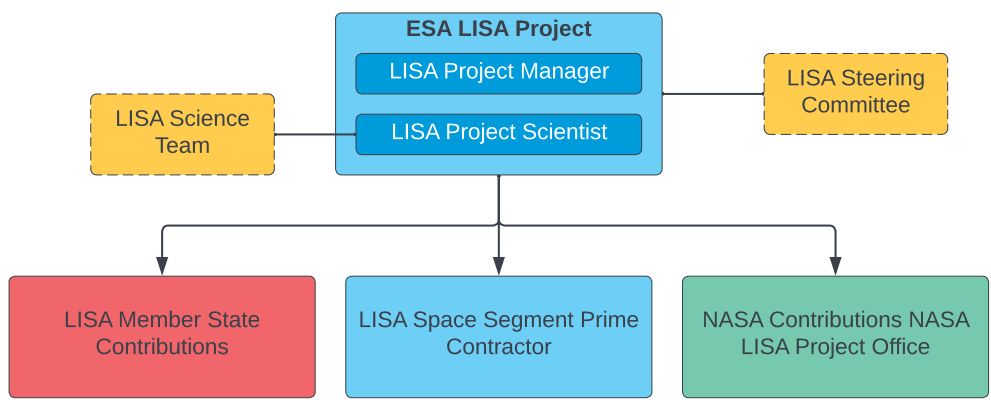

**Figure 10.1:** The overall organisation of the LISA mission.

During the implementation phase (B2/C/D/E1), the mission will be led by the ESA LISA Project Manager, located at the European Space Research and Technology Centre (ESTEC), responsible for the delivery of the constellation into orbit on scope, on time, and on budget.

During operations, the mission will be led by the ESA LISA Mission Manager, located at the European Space Astronomy Centre (ESAC), responsible for the successful operation of the mission until its deactivation, disposal, and the conclusion of post-operations. The LISA Mission Archive, containing the scientific data of the mission will be maintained by ESA also after the conclusion of the mission.

## 10.2    ESA Member State Contributions

The ESA member states envisage to support the mission providing hardware and services. The hardware contributions will be delivered from the member states to ESA for further delivery to the prime contractor as customer-furnished items (CFIs). Four major hardware contributions are identified. The member states further contribute to the ESA's mission oversight by providing expert support for performance and operations planning, as well as the for the development of data processing pipelines.

The member state hardware contributions are listed here:

**Gravitational Reference System (GRS)** – provided under leadership of Italy with contributions from Switzerland on the GRS Front-End Electronics (FEE) and from NASA via ESA on the Charge Management Device;

**Interferometric Detection Subsystem (IDS)** – provided under leadership of Germany with contributions from Germany on the point-ahead angle mechanisms, Germany and Denmark on the phase measurement subsystem (PMS), the United Kingdom on the Optical Bench, France on the assembly, integration, and testing activities and facilities, Belgium on the beam alignment mechanisms and the quadrant photo receiver (QPR) Front-End Electronics, the Netherlands on the point-ahead angle mechanisms, the quadrant photo receivers, and the IDS Mechanisms Control Unit (IDS-MCU), the Czech Republic on the Fibre-Switching Unit Actuators;

**Science Diagnostics Subsystem (SDS)** – provided by Spain;

**Ground Support Equipment Optical Test Systems (OTS)** – provided by France to ESA for use by the prime contractor during the integration and test phase.

Under French leadership, the member states undertake the organisation of the L1 to L2 data processing at the Distributed Data Processing Centre (DDPC). This structure also supports ESA in the generation of L3 data. Under German leadership, the member states support the Agency's independent LISA Mission performance budget and performance monitoring function. For this purpose, dedicated personnel will be assigned, covering the main functions of optical metrology, free-fall, and data processing.

## 10.3    NASA Partnership

NASA's contributions to LISA will include major hardware elements; a science ground segment implementing analysis pipelines; contributions to collaborative efforts on systems engineering, science operations, and analysis; and support for science investigations. These contributions will be managed by a NASA LISA Project Office at the Goddard Space Flight Center (GSFC) operating under the authority of the NASA Astrophysics Division and Science Mission Directorate.

The governing document for the ESA-NASA partnership will be a Memorandum of Understanding while the working-level relationship will be defined and maintained by a Joint Project Implementation Plan (JPIP). NASA's hardware deliverables, including telescopes, laser systems, and charge management devices, will be delivered to ESA for subsequent delivery to ESA's prime contractor as "customer-furnished item". NASA's contributions to performance monitoring will be described in the JPIP, those to data processing will be described in the Science Operations Concept Document and the NASA Science Implementation Plan. These are expected to include an independent processing chain resulting in L2 data and contribution to a consolidation process that will yield a single L3 catalogue for publication on the ESA and NASA archives. Representatives from NASA's LISA Project Office will be embedded in ESA Instrument Operations Team to facilitate the operation of LISA as a gravitational wave instrument. Coordination of science activities will be made through the ESA's Science Working Team, which will include NASA-selected representatives in proportion to US contributions.

## 10.4 Industrial Organisation

After the mission adoption ESA will issue a single Invitation to Tender (ITT) for the full scope of activities of the implementation phase (B2/C/D/E1). The resulting contract with the LISA prime contractor will cover, amongst others, the management, system engineering, and management of sub-contractors of the space segment activities, thus the full LISA constellation.

## 10.5 The LISA Consortium

The LISA Consortium is an open, grass roots organisation that originated in the LISA community. It put in a proposal to the ESA call for the Cosmic Vision L3 mission in 2017 that has led to the current LISA mission. The LISA Consortium has contributed significantly to the development of the mission, in bringing together experts on hardware, data processing and gravitational wave (GW) science and played an important role in the division of mission contributions from the member states and the US in close collaboration with ESA. After adoption the LISA Consortium will evolve to adapt to the formation of the ESA-created LST and the LISA project. The Consortium will remain the open organisation for the scientific community; it will have science interest groups that complement the LST and its working groups, it will provide pathways for early career scientists to join the LISA community, and it will promote LISA and its science to the public. The LISA Consortium will also provide a pool of scientific expertise that can be drawn on as needed to support the implementation of the Science Ground Segment and the Performance & Operations activities. There will be close collaboration between the LISA Consortium and the LST and as well as the LISA project, e.g. in outreach.

## 10.6 Operations Management

ESA's European Space Operations Centre (ESOC) implements the Mission Operations Centre (MOC), operates the spacecraft, and delivers telemetry and attitude data to the Science Ground Segment via the ESA Science Operation Centre (SOC). ESA's ESAC implements the SOC, which acts as the central node for the science mission planning, and is the single interface to the MOC. The SOC performs an initial quality check of the raw data (Level 0), creates Level 0.5 data (see Section 8.2 for description of data levels), and runs the Level 0.5 to Level 1 data processing pipelines developed in conjunction with the DDPC and the NASA Science Ground Segment (NSGS). The SOC is also responsible for the development and operations of the Main Data Repository (MDR) and the public archive, and for issuing and updating alerts. The LISA Science Team (LST) will be asked to review and endorse top-level requirements (in all areas of the mission) that impact science return, and to monitor all aspects of the subsequent implementation. Details of the LST composition and responsibilities are reported in Section 10.7.1.

## 10.7        Science Management

This section summarises the contents of the SMP. The SMP will be the top-level science management document for the mission, and will require approval by the SPC in the context of mission adoption. It is emphasised that this subsection is subject to change pending approval of the SMP.

### 10.7.1        The LISA Science Team and Project Scientist

After the mission adoption, ESA will appoint a LISA Science Team (LST), co-chaired by the ESA and NASA Project Scientists. The tasks and composition of the LST will be detailed in the SMP. NASA will nominate a number of members to the LST commensurate with the NASA contribution to the mission.

The LST supports the LISA Project Scientists in monitoring the correct implementation of the scientific objectives of the mission and in maximising its scientific return. The LST acts as a focus for the interest of the scientific community in LISA and links to the LISA Consortium. In general, members of the LST are expected to monitor the development and operations of the mission and give advice on all aspects which affect its scientific performance. They perform specific scientific tasks as required to discharge their responsibilities during development and operation.

### 10.7.2        Data rights and protected periods

An initial verification period after the start of the nominal science operations phase will be used to validate the instrument performance, data processing pipelines and verify mission data products (the exact length of this period is detailed in the SMP). Before the launch of the mission, the LST will establish a number of Science Topical Panels (STPs) which will focus on one specific science question during the Early Release Science Time (ERST). Each STP will be led by a chair and up to two co-chairs. Chairs and co-chairs will be selected through a coordinated call issued by ESA and NASA, and will constitute a working group of the LST. A commensurate number of chairs and co-chairs will be selected from scientists from US institutions. Membership of the STPs is open to the scientific community. STP members will be encouraged to publish science interpretation papers related directly to their STP science theme using only data acquired during the ERST period. At the end of the ERST, LISA data products from Level 0.5 to Level 3 will be published via the public archive. In addition to the data products, a set of instrument performance and science validation papers will be published by the instrument teams, DDPC, NSGS and SOC. Scientist that have contributed significantly to the LISA mission get the right to sign all these papers.

Subsequent data releases will occur on a regular schedule thereafter. Each data release will contain the Level 0.5 to Level 2 data products covering the full science operations phase, as well as the updated Level 3 catalogue of identified sources with associated instrument papers. Data products obtained from the Low Latency Alert Pipeline (transient sources) are made publicly available immediately through alerts to enable rapid follow-up observations.

# 11 COMMUNICATIONS AND OUTREACH

Space missions are inspirational to general audiences, and the LISA Mission is no exception. Many of the unique aspects of LISA will prove inspiring and interesting to general audiences, from the technology required to send laser beams millions of kilometres, to the data analysis challenges, and the astrophysical and cosmological observations. As LISA is the first mission of its kind, it will be important to prepare tailored resources to help inform the general public about LISA.

ESA is responsible for planning and coordinating press release activities relating to the LISA mission. This will start with standard news coverage and set of merchandise (such as pins, stickers, etc.) based on the mission logo. ESA will also be responsible for developing the communication plan and its implementation for the launch. In addition, press coverage and releases will be done by ESA's international partners such as the LISA Consortium and NASA in close coordination with ESA.

The LISA Consortium will focus on the scientific rewards of the LISA mission in its outreach work, and also inform their audiences about mission milestones, research highlights, including in the field of multi-messenger astronomy, technological innovations and general progress in close cooperation with ESA, NASA and industry. To lay the groundwork for an effective outreach program even before adoption, the Consortium is implementing a comprehensive communication strategy. This includes relaunching a new website, expanding all social media activities, establishing information flow procedures both within the consortium and with partners.

In order to prepare the scientific community for LISA data, a series of "Masterclass" workshops are planned. Those workshops will function within the "train the trainer" concept, where scientists from ESA member states can apply to being trained in a workshop on LISA data and data processing tools and return to their home countries to train their own community with support from ESA. This concept was successfully applied in the JWST mission and will maximize the scientific turnout for all member states interested in participating in LISA science.

# A    PERFORMANCE MODEL

Chapter 7 presented an overview of the formalism used to quantify LISA's sensitivity to gravitational wave (GW) signals and the major contributions to the instrumental noise. As part of the mission systems engineering process, a more comprehensive and detailed model of LISA's mission performance has been developed and will continue to be refined as the mission progresses in its development [230]. Table A.1 and Table A.2 present a subset of the entries in the current version of this performance model. Each entry is expressed as the strain spectral density of an equivalent GW signal and is evaluated at three reference frequencies spanning the measurement band. The root-sum-squared total of the three major sub-categories test mass (TM) Acceleration Noise, Interferometer displacement noise, and post-processing errors) are shown as major rows.

**Table A.1:** List of primary instrumental noise contributions for LISA, expressed as equivalent gravitational wave strain spectral density at three reference frequencies across the LISA band. Noise sources are grouped into three categories: disturbances on the test masses, noise in the interferometric displacement measurement, and ground processing errors.

| Noise Term [$10^{-21}\,\mathrm{Hz}^{-.5}$] | 0.1 mHz | 1 mHz | 10 mHz |
|---|---|---|---|
| Total | 2.86e+04 | 90.45 | 8.8 |
| TM Acceleration Noise | 2.81e+04 | 87.84 | 0.96 |
| Brownian Noise | 6.27e+03 | 63.51 | 0.65 |
| TM-SC/MOSA coupling Force Noise | 8.86e+03 | 36.79 | 0.59 |
| Actuation Noise | 1.14e+04 | 25.54 | 0.25 |
| Magnetic Noise | 1.16e+04 | 22.34 | 0.12 |
| Gravitational Noise | 1.09e+04 | 21.39 | 0.19 |
| Temperature Force Noise | 1.45e+04 | 18.66 | 0.12 |
| Stray Electrostatics Noise | 8.45e+03 | 14.29 | 0.088 |
| Laser Pressure Noise | 1.8e+03 | 12.98 | 0.13 |

**Table A.2:** Summary of primary instrumental noise contributions continued

| Optical Metrology Noises | 1.5e+03 | 17.42 | 8.4 |
|---|---|---|---|
| Inter-Satellite IFO PMS | 582 | 6.08 | 1.6 |
| Inter-Satellite IFO Shot Noise | 4.84 | 4.85 | 5.3 |
| Inter-Satellite IFO StrayLight | 434 | 4.54 | 1.2 |
| Inter-Satellite IFO QPR Chain Noise | 165 | 3.79 | 3.7 |
| Unmodelled Far Field WFE Non linear TTL | 290 | 3.03 | 0.8 |
| Non Correctable TTL | 190 | 1.99 | 0.52 |
| Inter-Satellite IFO Plasma Scintillation | 43.1 | 1.84 | 0.085 |
| Inter-Satellite IFO 1f RIN Noise | 0.258 | 0.26 | 0.28 |
| Inter-Satellite IFO 2f RIN Noise | 0.00186 | 0.00 | 0.002 |
| Inter-Satellite IFO Telescope Spacer Stability | 579 | 6.05 | 1.5 |
| Total BAM Rx and Tx Noise | 99.6 | 1.18 | 0.29 |
| OB Baseplate Thermoelastic | 57.9 | 0.59 | 0.081 |
| Mirror Thermoelastic | 57.9 | 0.59 | 0.081 |
| Test Mass Thermoelastic | 50.6 | 0.53 | 0.13 |
| Unmodelled Local Oscillator TTL | 29 | 0.30 | 0.04 |
| GRS Optical Window Thermoelastic | 20.5 | 0.21 | 0.027 |
| Component Thermoelastic | 9.94 | 0.10 | 0.014 |
| PAAM Piston | 9.42 | 0.10 | 0.013 |
| Test Mass Waveplates Thermoelastic | 5.3 | 0.05 | 0.0069 |
| Inter-Satellite IFO Waveplates on Rx and Tx Thermoelastic | 3.53 | 0.04 | 0.0046 |
| Inter-Satellite IFO Waveplates Thermoelastic on Rx only | 1.77 | 0.02 | 0.0025 |
| Reference Noise | 648 | 7.62 | 3.3 |
| Thermo-mechanical noise | 526 | 5.50 | 1.5 |
| TTL Short Term Drift Contribution to ISI TTL Residual Noise | 142 | 2.57 | 2.1 |
| TTL Coefficient Contribution to ISI TTL Residual Noise | 80.9 | 1.26 | 0.99 |
| TDI and Post-Processing Noises | 5.07e+03 | 12.68 | 2.4 |
| Residual Timing Jitter | 5.04e+03 | 12.36 | 1.8 |
| TTL Coefficient Error Contribution to ISI TTL Residual Noise | 101 | 1.81 | 1.5 |
| Residual In-Band Laser Noise | 491 | 1.58 | 0.42 |
| Residual MOSA Longitudinal Jitter | 148 | 1.54 | 0.41 |

→ THE EUROPEAN SPACE AGENCY

# Acronyms

**1PA**    one-post-adiabatic
**ADC**    analog-to-digital converter
**ADPLL**    all-digital phase-locked loop
**AGN**    Active Galactic Nuclei
**AOCS**    Attitude And Orbit Control System
**ASD**    Amplitude Spectral Density
**ATM**    augmented telemetry
**AU**    Astronomical Unit
**Au**    gold
**BAM**    beam alignment mechanism
**BCRS**    Barycentric Celestial Reference System
**BH**    Black Hole
**BHB**    Black Hole binary
**BSM**    Physics beyond the Standard Model
**CAS**    Constellation Acquisition Sensor
**CBE**    current best estimate
**CCD**    Charge-Coupled Device
**CFI**    customer-furnished item
**CMB**    Cosmic Microwave Background
**CMD**    Charge Management Decive
**CoM**    Center of Mass
**CPTA**    Chinese Pulsar Timing Array
**CS**    Cosmic String
**CTE**    thermal expansion coefficient
**CVM**    caging and venting mechanism
**DCC**    Data Computing Centre
**DDPC**    Distributed Data Processing Centre
**DFACS**    Drag-Free Attitude Control System
**DLL**    delay-locked loop
**DM**    Dark Matter
**DOF**    degree of freedom
**DR**    data release
**DSP**    digital signal processor
**DWS**    Differential Wavefront Sensing

**ECSS**    European Cooperation for Space Standardization
**EH**    electrode housing
**EM**    electro-magnetic
**EMRI**    extreme mass-ratio inspiral
**EOB**    effective one-body
**EPTA**    European Pulsar Timing Array
**ERST**    Early Release Science Time
**ESA**    European Space Agency
**ESAC**    European Space Astronomy Centre
**ESDC**    ESAC Science Data Centre
**ESOC**    European Space Operations Centre
**ESTEC**    European Space Research and Technology Centre
**ESTRACK**    European Space TRACKing
**EW**    Electroweak
**FDS**    frequency distribution system
**FEE**    Front-End Electronics
**FOPT**    First-Order Phase Transition
**FPAG**    Fundamental Physics Advisory Group
**FPGA**    field-programmable gate array
**FSU**    fibre switching unit
**GB**    Galactic binary
**GPRM**    grabbing, positioning, and release mechanism
**GR**    General Theory of Relativity
**GRS**    Gravitational Reference System
**GRSH**    Gravitational Reference Sensor Head
**GSF**    Gravitational Self-Force
**GSFC**    Goddard Space Flight Center
**GUT**    Grand Unified Theory
**GW**    gravitational wave
**Hg**    mercury
**IDS**    Interferometric Detection Subsystem
**IMBH**    Intermediate-Mass Black Hole
**IMR**    Inspiral-Merger-Ringdown
**IMRI**    intermediate mass-ratio inspiral
**InGaAs**    Indium-Gallium Arsenide
**InPTA**    Indian pulsar Timing Array
**ISI**    inter-satellite interferometer
**ISUK**    Inertial Sensor UV Kit
**ITT**    Invitation to Tender
**JILA**    Joint Institute For Laboratory Astrophysics
**JPIP**    Joint Project Implementation Plan
**KAGRA**    Kamioka Gravitational Wave Detector

| | |
|---|---|
| **L0** | Level 0 data |
| **L0.5** | Level 0.5 data |
| **L1** | Level 1 data |
| **L2** | Level 2 data |
| **L3** | Level 3 data |
| **LAGOS** | Laser Antenna for Gravitational-Radiation Observation in Space |
| **LCA** | LISA core assembly |
| **LDC** | LISA Data Challenge |
| **LEM** | laser electrical module |
| **LEOP** | Launch and Early Operations Phase |
| **LH** | laser head |
| **LHC** | Large Hadron Collider |
| **LIGO** | Laser Interferometer Gravitational-Wave Observatory |
| **LISA** | Laser Interferometer Space Antenna |
| **LISPEG** | LISA Instrument Scientists and Payload Experts Group |
| **LLAP** | Low Latency Alert Pipeline |
| **LO** | local oscillator |
| **LOM** | laser optical module |
| **LPF** | LISA Pathfinder |
| **LRI** | Laser-Ranging Interferometer |
| **LS** | laser system |
| **LST** | LISA Science Team |
| **LVK** | LIGO, Virgo, and KAGRA |
| **MBH** | Massive Black Hole |
| **MBHB** | massive Black Hole binary |
| **MCMC** | Markov-Chain Monte Carlo |
| **MDR** | Main Data Repository |
| **MFS** | main frequency stabilisation |
| **MIDA** | Mean (Earth) Initial Displacement Angle |
| **MLA** | Multi-Lateral Agreement |
| **MO** | main oscillator |
| **Mo** | molybdenum |
| **MOC** | Mission Operations Centre |
| **MOPA** | Main Oscillator Power Amplifier |
| **MOSA** | Moving Optical Sub-Assembly |
| **MoU** | Memorandum of Understanding |
| **MRD** | Mission Requirement Document |
| **NANOGrav** | North American Nanohertz Observatory for Gravitational Waves |
| **NCO** | numerically controlled oscillator |
| **NECP** | Near Earth Commissioning Phase |
| **NR** | Numerical Relativity |
| **NRT** | Near-Real-Time Pipeline |
| **NS** | Neutron Star |
| **NSGS** | NASA Science Ground Segment |
| **NSP** | nominal science phase |
| **OATM** | optical assembly tracking mechanism |
| **OB** | optical bench |
| **OBC** | on-board computer |
| **OTS** | Optical Test Systems |
| **PA** | power amplifier |
| **PAA** | point-ahead angle |
| **PAAM** | point-ahead angle mechanism |
| **PAR** | phase accumulation register |
| **PARX** | public archive |
| **PBH** | Primordial Black Hole |
| **PDF** | Probability Density Function |
| **PIR** | phase increment register |
| **PMF** | polarization-maintaining fibre |
| **PMON** | power monitor |
| **PMS** | phase measurement subsystem |
| **PN** | post-Newtonian |
| **PPE** | Parameterized Post-Einsteinian |
| **PPTA** | Parkes Pulsar Timing Array |
| **PRN** | pseudo-random noise |
| **PSD** | Power Spectral Density |
| **PT** | Phase Transition |
| **Pt** | platinum |
| **PTA** | Pulsar Timing Array |
| **QPD** | quadrant photodetector |
| **QPR** | quadrant photo receiver |
| **RF** | radio frequency |
| **RFI** | reference interferometer |
| **RIN** | relative intensity noise |
| **RX** | receive |
| **S/C** | spacecraft |
| **sBH** | stellar-mass Black Hole |
| **sBHB** | stellar-mass Black Hole binary |
| **SCCP** | Science Commissioning and Calibration Phase |
| **SCET** | Spacecraft Elapsed Time |
| **SciRD** | Science Requirements Document |
| **SDS** | Science Diagnostics Subsystem |
| **SGS** | Science Ground Segment |
| **SGWB** | stochastic gravitational-wave background |
| **SI** | Science investigation |
| **SMP** | Science Management Plan |
| **SNR** | signal-to-noise ratio |
| **SO** | Science Objective |
| **SOC** | Science Operation Centre |
| **SPC** | Science Programme Committee |
| **SSB** | solar system barycentre |
| **SSRD** | Space Segment Requirements Document |
| **STP** | Science Topical Panel |
| **TDI** | Time-Delay Interferometry |

| | | | | |
|---|---|---|---|---|
| **Ti** | titanium | | **USO** | ultra-stable oscillator |
| **TM** | telemetry | | **VB** | verification binary |
| **TM** | test mass | | **VC** | vacuum chamber |
| **TM-I/F** | test mass interface | | **Virgo** | Virgo |
| **TMI** | test mass interferometer | | **VLBI** | Very-long-baseline interferometry |
| **TRL** | Technology Readiness Level | | **VMS** | Very Massive Star |
| **TTL** | tilt-to-length | | **W** | tungsten |
| **TX** | transmit | | **WD** | White Dwarf |
| **UCB** | Ultra-Compact Binaries | | **XMRI** | extremely mass-ratio inspiral |
| **ULU** | UV light unit | | **Yb** | ytterbium |

# Glossary of terms

**ΛCDM** The ΛCDM is a parameterisation of the Big Bang cosmological model in which the universe contains three major components: first, a cosmological constant denoted by Λ associated with dark energy; second, the postulated cold dark matter (abbreviated CDM); and third, ordinary matter. It is frequently referred to as the standard model of Big Bang cosmology. 58, 59

**Alerts** Specialised variants of / data from dedicated low-latency pipelines to produce alerts for multi-messenger astronomy.

**Amplitude Spectral Density** The Amplitude Spectral Density (ASD) is an alternative name for the square-root of the Power Spectral Density (PSD) 19, 20, 108, *see also* Power Spectral Density

**apoapse** The furthest point of a body to the centre object on its orbit around that centre object. Analogous to apogee and aphel *see also* periapse

**Ariane 64** Ariane 64 is a variant of the future Ariane 6 launcher family developed by ArianeGroup. It will be able to lift up to 11 500 kg to a geosynchronous transfer orbit (GTO) and up to 21 500 hg to low earth orbit. 96, 101

**Athena** ESA's Advanced Telescope for High Energy Astrophysics (Athena) mission is a future large mission of ESA target to observe the hot and energetic Universe. 41, 51, 58

**audio carrier frequencies** Carrier frequencies in the audio range. Chiefly used in the control of the test mass in the Gravitational Reference System where the typical range is between 10 Hz and 1 kHz 83

**Cassegrain** The term describes a telescope architecture characterised by a parabolic primary mirror (M1) and a hyperbolic secondary mirror (M2). The design is named after Laurent Cassegrain, a 17$^{th}$ century french priest and scientist. 87

**CHAMP** The CHAMP (Challenging Minisatellite Payload) satellite was a German earth observation mission relevant for the geosciences and atmospheric physics. Its payload made possible the simultaneous determination of Earth's gravity and magnetic fields. 78

**Chinese Pulsar Timing Array** The Chinese Pulsar Timing Array (CPTA) is the youngest member of the PTA collaboration family. The CPTA can make use of observations with the recent Five Hundred Metre Spherical Telescope (FAST), which began full operations in 2020.

**chirp mass** The chirp mass $M_c$ of a binary system determines the leading-order orbital evolution of the system as a result of energy loss from emitting gravitational waves. It is defined as

$$M_c = \frac{(m_1 m_2)^{3/5}}{(m_1 + m_2)^{1/5}}$$

where $m_1$ and $m_2$ are the masses of the two constituents of the binary system. The chirp mass in the observer frame is denoted $\mathcal{M}$ and is related to the chirp mass in the source frame $M_c$ through the redshift $z$

$$\mathcal{M} = M_c(1 + z)$$

27, 29, 31, 33, 37, 39, *see also* symmetric mass ratio

**Cluster** Cluster is an ESA mission to study small-scale structures of the magnetosphere and its environment in three dimensions. Cluster is constituted of four identical spacecraft that fly in a tetrahedral configuration. The separation distances between the spacecraft can be varied between 600 km and 20 000 km, according to the key scientific regions. 104

**CNES** The Centre national d'études spatiales (CNES) is the French space agency. 105, 149

**cosmic noon** A period between 1–4 Gyr after the Big Bang during which most of the stars and galaxies in the Universe formed. 35, 38, 39

**cosmic reionisation** A period in the evolution of the Universe at redshifts $30 \gtrsim z \gtrsim 6$, about 100 Myr to 1 Gyr after the Big Bang during which the first stars (re)-ionised the neutral interstellar medium [433]. 35, 37–39

**delta-v** delta-v (also denoted $\Delta v$) is a measure of the impulse per unit of spacecraft mass that is needed to perform a an in-space orbital manoeuvre. It is a scalar that has the units of speed. 96, 97 **dispersion** Used here in the context of describing the statistics of posterior parameter distributions and taking the meaning of the upper $c_+$ and lower limits $c_-$ of the 90 % confidence interval around the central value $\mu$ (taken to be the expectation value or the median, depending on context). Denoted by

$$\mu_{c_-}^{c_+}$$

27, 37, 57

**ELT** The Extremely Large Telescope (ELT) is a 39 m diameter telescope under construction on the Cerro Armazones mountain at an altitude of about 3046 m in the central part of Chile's Atacama Desert about 20 km from Cerro Paranal, home of ESO's Very Large Telescope (VLT). 58

**ESTRACK** ESTRACK is a global system of ground stations providing links between satellites in orbit and ESOC, the European Space Operations Centre, Darmstadt, Germany. The core Estrack network comprises seven stations in seven countries. 103, 151

**Euclid** Euclid is an ESA mission designed to explore the evolution of the dark Universe. It will make a 3D-map of the Universe (with time as the third dimension) by observing billions of galaxies out to 10 billion light-years, across more than a third of the sky. Euclid has been launched on 1 July 2023. 16, 102

**European Pulsar Timing Array** The European Pulsar Timing Array (EPTA) is a multinational European collaboration of pulsar astronomers with the aim is to increase the precision and quality of pulsar science results by combining the efforts and resources of the various member institutions and telescopes.

**extreme mass-ratio inspiral** Extreme mass-ratio inspirals (EMRIs) are binary systems with a mass ratio of $q < 10^{-4}$ 15, see also intermediate mass-ratio inspiral & extremely mass-ratio inspiral

**extremely mass-ratio inspiral** Extremely mass-ratio inspirals (XMRIs) are binary systems with a mass ratio of $q < 10^{-6}$ see also extreme mass-ratio inspiral & intermediate mass-ratio inspiral

**Gaia** Gaia is an ESA mission to chart a three-dimensional map of the Milky Way, revealing the composition, formation and evolution of the Galaxy. Gaia provides unprecedented positional and radial velocity measurements with the accuracies needed to produce a stereoscopic and kinematic census of about one billion stars in the Milky Way. 20, 102

**GOCE** The Gravity Field and Steady-State Ocean Circulation Explorer (GOCE) is measuring Earth's gravity field and model the geoid with extremely high accuracy. It was launched on 17 March 2009. Owing to its sleek shape, GOCE is often cited as one of ESA's most elegant space probes. The mission ended on 11 November 2013 after a planned destructive re-entry into the atmosphere. 78

**GRACE** NASA's GRACE twin satellites, launched 17 March 2002, are making detailed measurements of Earth's gravity field changes and revolutionising investigations about Earth's water reservoirs over land, ice and oceans, as well as earthquakes and crustal deformations. 78

**GRACE-FO** The Gravity Recovery and Climate Experiment Follow-On (GRACE-FO) mission is a partnership between NASA and the German Research Centre for Geosciences (GFZ). GRACE-FO is a successor to the original GRACE mission, which orbited Earth from 2002–2017. 23, 95, 104, 149

**GW150914** GW150914 is the designation of the first observation of a binary Black Hole merger [5], demonstrating both the existence of binary stellar-mass Black Hole systems and the fact that such mergers could occur within the current age of the universe. The event was located at a distance of about $440^{+160}_{-180}$ Mpc and involved two Black Holes with masses of $35^{+5}_{-3}$ $M_\odot$ and $30^{+3}_{-4}$ $M_\odot$, respectively, resulting in a post-merger Black Hole of $62^{+4}_{-3}$ $M_\odot$. 51, 56, 57, 71, 72

**GW170817** GW170817 was a gravitational wave signal observed by the LIGO and Virgo detectors on 17 August 2017, originating from the shell elliptical galaxy NGC 4993 [3]. The signal was produced by the last minutes of a binary pair of neutron stars' inspiral process, ending with a merger, resulting in the gamma ray burst GRB 170817A 1.7 s later. It is the first gravitational wave (GW) observation that has been confirmed by non-gravitational means. 57

**GW190521** GW190521 is the designation of an observation of a binary Black Hole merger [8] at a distance of 5.3 Gpc with masses of $85^{+21}_{-14}$ $M_\odot$ and $66^{+17}_{-18}$ $M_\odot$, respectively, resulting in a $142^{+28}_{-16}$ $M_\odot$ post-merger Black Hole. The primary falls in the mass gap predicted by (pulsational) pair-instability supernova theory, in the approximate range 65–120 $M_\odot$, the final mass of the merger classifies it as an Intermediate-Mass Black Hole (IMBH). 26, 49, 50, 71

**Indian pulsar Timing Array** The Indian Pulsar Timing Array Experiment (InPTA) is a pulsar timing experiment searching for low frequency nanohertz Gravitational Waves in operation since 2016. InPTA is a Indo-Japanese collaboration which uses the upgraded Giant Meterwave Radio Telescope (uGMRT) for monitoring a sample of nearby millisecond pulsars for this purpose.

**intermediate mass-ratio inspiral** Intermediate mass-ratio inspirals (IMRIs) are binary systems with a mass ratio of $1 \times 10^{-4} < q < 10^{-2}$. *see also* extreme mass-ratio inspiral & extremely mass-ratio inspiral

**Intermediate-Mass Black Hole** Intermediate-Mass Black Holes bridge the gap between the high-mass end of the stellar-mass Black Holes and the low-mass end of Massive Black Holes with a mass range of $10^2\,M_\odot$ to $10^5\,M_\odot$. They are likely products of run-away mergers in dense stellar clusters. , *see also* stellar-mass Black Hole & Massive Black Hole

**JWST** The James Webb Space Telescope (Webb) is the next great space science observatory following Hubble, designed to answer outstanding questions about the Universe and to make breakthrough discoveries in all fields of astronomy. Webb will see farther into our origins: from the formation of stars and planets, to the birth of the first galaxies in the early Universe. Webb is an international partnership between NASA, ESA and CSA. The telescope launched on 25 December 2021 on an Ariane 5 from Europe's Spaceport in French Guiana. 35, 140

**KAGRA** The Kamioka Gravitational Wave Detector (KAGRA) is a gravitational wave detector situated underground in the Kamioka mine in Japan. KAGRA employs cryogenic optics to minimize the thermal noise and improve sensitivity. , 143, *see also* LIGO & Virgo

**Kerr metric** The Kerr metric describes the geometry of empty spacetime around a rotating uncharged axially symmetric Black Hole. 52–55, 119, 120

**Level 0 data** Spacecraft telemetry after basic processing including: De-multiplexed and annotated with time correlation, quality flags and positioning data; Anomalous, corrupted items corrected or removed; All measurements time-ordered and uniquely tagged. L0 data are not necessarily releasable to the public as they might contain proprietary information. , 16

**Level 0.5 data** L0 data which has been further processed and reformatted in the following way: De-packetised, de-compressed (if applicable), Reformatted to the data exchange format of the Science Ground System, Clock data synchronised and converted to a common, non-Earth-bound astronomical timescale, Binary units (analog-to-digital converter (ADC) counts), where applicable, converted to physical values. L0.5 data are fully releasable to the public. , 16

**Level 1 data** Calibrated and noise corrected Time-Delay Interferometry (TDI) data streams. This is "analysable data" and the main input to the core downstream processing to isolate individual GW signals, i. e. major noise sources removed. L1 data constitutes the minimum viable science data. , 16

**Level 2 data** Output from a global fit pipelines, providing Probability Density Functions (PDFs) for identified sources. They include the reconstructed waveforms and detector signals for identified sources as well as regular updates on the critical parameters of transient sources, such as coalescence times and sky location. This is "analysed data". We expect several L2 products that vary depending on prior probabilities for source parameters and numbers of sources, on the methods used and on the volume of the data engaged (the volume of the analysed data is increasing as LISA observes for longer). , 16

**Level 3 data** Catalogue of GW source candidates with detection confidence, estimated astrophysical parameters and their strain time series $h(t)$. This is "consolidated data". L3 data also include the "residual" data stream, i. e. the L1 data stream with the contribution of identified sources removed. L3 data constitute the mission end products. , 17

**LIGO** The Laser Interferometer Gravitational-Wave Observatory (LIGO) was designed to open the field of gravitational-wave astrophysics through the direct detection of gravitational waves predicted by Einstein's General Theory of Relativity. LIGO's multi-kilometre-scale gravitational wave detectors use laser interferometry to measure the minute ripples in space-time caused by passing gravitational waves from cataclysmic cosmic events such as colliding neutron stars or Black Holes, or by supernovae. LIGO consists of two widely-separated interferometers within the United States—one in Hanford, Washington and the other in Livingston, Louisiana—operated in unison to detect gravitational waves. , 4, 15, 18, 144, *see also* KAGRA & Virgo

**LISA Data Challenge** The LISA Data Challenges (LDCs) are an open, collaborative effort by LISA Consortium's LDC working group to tackle unsolved problems in LISA data analysis, while developing software tools that will form the basis of the future LISA data system. They are comprised of a collection of simulated LISA data with increasing difficult levels (i.e. noise sources added) aimed at testing LISA data analysis tools.

**LRI** The Laser Ranging Interferometer is a technology demonstration that uses laser interferometry instead of microwaves to measure fluctuations in the separation between the two GRACE-FO spacecraft. 95

**LSST** The 10-year Legacy Survey of Space and Time (LSST) will deliver a 500 petabyte set of images and data products that will address some of the most pressing questions about the structure and evolution of the universe and the objects in it. LSST is the main survey conducted on the Vera Rubin Observatory *see also* Vera Rubin Observatory

**Massive Black Hole** Massive Black holes reside at the centres of galaxies with a mass range of $10^5\,M_\odot$ to $10^9\,M_\odot$ or even larger. , 15, *see also* stellar-mass Black Hole & Intermediate-Mass Black Hole

**Mean (Earth) Initial Displacement Angle** The mean initial angle subtended between the barycentre of the constellation and the Earth. The apex of the MIDA is at the Sun. As the orbit of the constellation's barycentre is circular, in contrast to the Earth's elliptical orbit, the actual angle between the constellation's barycentre and the Earth varies depending on the initial data. However, the orbital dynamics of the constellation are predominantly driven by the mean angle, assuming a circular Earth's orbit.

**Microscope** MICROSCOPE is a mission jointly undertaken by the Observatoire de la Cote d'Azur, CNES, and ZARM aimed at testing the equivalence principle with an unprecedented sensitivity of about $10^{-15}$. 102

**multimessenger sources** Multimessenger sources are sources that have the potential to be detected simultaneously (or near simultaneously) in both the electro-magnetic (EM) and GW domain. 28

**Nancy Grace Roman Space Telescope** The Nancy Grace Roman Space Telescope is NASA's planned space telescope with wide field of view and a method to block starlight using a coronograph. 41

**NGGM** ESA's Next Generation Gravity Mission (NGGM) is a candidate Mission of Opportunity for ESA-NASA cooperation in the frame of the MAss change and Geosciences International Constellation (MAGIC) . The mission aims at enabling long-term monitoring of the temporal variations of Earth's gravity field at relatively high temporal (down to 3 days) and increased spatial resolutions (up to 100 km) at longer time. 104

**no-hair theorem** The no-hair theorem states that all stationary Black Hole solutions of the Einstein–Maxwell equations of gravitation and electromagnetism in general relativity can be completely characterised by only three independent externally observable classical parameters: mass, electric charge, and angular momentum. 16, 54, 55

**NANOGrav** The North American Nanohertz Observatory for Gravitational Waves (NANOGrav) is an international collaboration dedicated to exploring the low-frequency gravitational wave universe through radio pulsar timing. Its telescopes currently comprise the Green Bank Telescope (GBT), the Karl G. Jansky Very Large Array (VLA), and the Canadian Hydrogen Intensity Mapping Experiment (CHIME). The William E. Gordon Telescope at the Arecibo Observatory was contributing until its collapse in December 2020. , 14, 144

**periapse** The closest point of a body to the centre object on its orbit around that centre object. Analogous to perigee and perihelion 44, *see also* apoapse

**post-Newtonian** The post-Newtonian (PN) approximation of Einstein's field equation is an expansion of the energy tensor and the metric in terms of $1/c$. The approximation holds well for sources with weak self-gravity and small velocities.

**Power Spectral Density** The Power Spectral Density (PSD) of a quantity $x(t)$ can be defined via the Fourier transformation $\tilde{x}(f)$ of $x(t)$.

$$S_x(f) \propto |\tilde{x}(f)|^2$$

The exact proportionality factors depend on the definition of the Fourier transform used, but can always be inferred from the relationship that gives the PSD its name and that motivates the units (here denoted by $[\cdot]$) for the PSD.

$$\mathrm{var}\, x(t) = \int_0^\infty S_x(f)\, \mathrm{d}f$$

$$[S_x(f)] = \frac{[x]^2}{\mathrm{Hz}}$$

And the Amplitude Spectral Density (ASD):

$$\left[\sqrt{S_x(f)}\right] = \frac{[x]}{\sqrt{\mathrm{Hz}}}$$

see also 19, 107, 150

**Project phases**  The full definition of the mission phases are given in the relevant European Cooperation for Space Standardization (ECSS) document [350], here we give an abbreviated version

**Phase 0**  Mission analysis. Develop the preliminary technical requirements specification; identify possible mission concepts.

**Phase A**  Feasibility. Elaborate possible system and operations concepts and system architectures; identify critical technologies and propose pre-development activities.

**Phase B**  Preliminary definition. Often split in two sub-phases B1 and B2. Establish a preliminary design definition for the selected system concept, elaborate the baseline cost at completion.

**Phase C**  Detailed definition. Completion of the detailed design definition of the system at all levels; production and development testing of engineering models; production, development testing and pre-qualification of selected critical elements and components.

**Phase D**  Qualification and production. Complete qualification testing and associated verification; complete manufacturing, assembly and testing of flight hardware/software and associated ground support hardware/software

**Phase E**  Operations/utilisation. Launch and early orbital operations; on-orbit science operations; science ground segment operations

**Phase F**  Disposal.

4, 96, 137

**Pulsar Timing Array**  A Pulsar Timing Array is a set of Galactic pulsars that is monitored and analysed to search for correlated signatures in the pulse arrival times on Earth. As such, they are Galactic-sized gravitational wave detectors. , 15

**Schwarzschild radius**  The Schwarzschild radius or the gravitational radius is a physical parameter in the Schwarzschild solution to Einstein's field equations that corresponds to the radius defining the event horizon (boundary beyond which events cannot affect an observer) of a non-rotating Black Hole. It is a characteristic radius associated with any quantity of mass ($M$) and is given by:

$$r_s = \frac{2GM}{c^2}$$

where $G$ is the Gravitational constant, and $c$ is the speed of light.  47

**Sgr A\***  Sagittarius A\* (SgrA\*) is the supermassive Black Hole at the Galactic Centre of the Milky Way. It is located near the border of the constellations Sagittarius and Scorpius, about 5.6° south of the ecliptic, visually close to the Butterfly Cluster and Lambda Scorpii. 43, 44

**signal-to-noise ratio**  The signal-to-noise ratio (SNR) $\rho$ of a signal $h(t)$ is defined [51] via the Fourier transform of the signal $\tilde{h}(f)$ as

$$\rho^2 = 4 \int_0^{f_{\mathrm{max}}} \frac{|\tilde{h}(f)|^2}{S_n(f)}\, \mathrm{d}f$$

$$= 8T \int_0^{f_{\mathrm{max}}} \frac{S_h(f)}{S_n(f)}\, \mathrm{d}f$$

where $S_n(f)$ denotes the Power Spectral Density of the noise and $T$ the measurement time. 15, 121, see also Power Spectral Density

**SKA**  The SKA Observatory (SKAO) is a next-generation radio astronomy-driven Big Data facility that will revolutionise our understanding of the Universe and the laws of fundamental physics. The SKAO is composed of respectively hundreds of dishes and thousands of antennas, covering the frequency range from 0.05–24 GHz.  41, 51, 58

**stellar-mass Black Hole**  Stellar-mass Black holes are linked to the life-cycle of massive stars, are widespread in all galaxies, and have a mass range of $1\,M_\odot$ to $10^2\,M_\odot$. , 16, see also Massive Black Hole & Intermediate-Mass Black Hole

**symmetric mass ratio**  The symmetric combination of the two masses $m_1$ and $m_2$ in a binary system

$$\nu \equiv \frac{m_1 m_2}{(m_1 + m_2)^2}$$

27, see also chirp mass

→ THE EUROPEAN SPACE AGENCY

**TRL** The technology readiness levels (TRLs) are a method for understanding the technical maturity of a technology. They are indicated by a numerical scale $(0 - 9)$; for space missions, TRL 5 has the meaning of "critical functions of the element are identified and verified in a relevant environment", TRL 6 of "critical functions of the element are verified and design of the element is achieved" [394]. 84, 95

**ULE** ULE® is a is a titania-silicate glass produced by Corning, providing a very low CTE of $(0 \pm 30) \times 10^{-9}\,\mathrm{K}^{-1}$. 88

**Vera Rubin Observatory** The Vera Rubin Observatory conduct the 10-year Legacy Survey of Space and Time (LSST). LSST will deliver a 500 petabyte set of images and data products that will address some of the most pressing questions about the structure and evolution of the universe and the objects in it. 20, 41, 51, 58, 149, see also LSST

**Virgo** Virgo is the gravitational wave detector of the the European Gravitational Observatory (EGO), located in the Italy, near Pisa in the Commune of Cascina. Virgo is a laser interferometer with two perpendicular, 3 km-long arms: its purpose is detecting gravitational waves from astrophysical sources. , 4, 15, 18, see also LIGO & KAGRA

**Viton** Viton® is a brand name of DuPont for its synthetic rubber and fluoropolymer elastomer, commonly used in O-rings and other moulded or extruded products. 80

**VLBI** Very-long-baseline interferometry is a type of astronomical interferometry used in radio astronomy. In VLBI a signal from an astronomical radio source, such as a quasar, is collected at multiple radio telescopes on Earth or in space. The distance between the radio telescopes is then calculated using the time difference between the arrivals of the radio signal at different telescopes. This allows observations of an object that are made simultaneously by many radio telescopes to be combined, emulating a telescope with a size equal to the maximum separation between the telescopes. 145

**X-band** Frequency band used for satellite communications. ESTRACK uses the frequency band between 7.145 GHz and 8.500 GHz 100

**ZARM** The Center of Applied Space Technology and Microgravity (ZARM) is an institute of the faculty of Production Engineering at the University of Bremen. The main facility of ZARM is the Bremen Drop Tower, designed for short-term experiments under high-quality microgravity conditions. 149

**Zerodur** Zerodur® is a lithium-aluminosilicate glass-ceramic produced by Schott AG, providing a very low CTE of $(0.05 \pm 0.10) \times 10^{-6}\,\mathrm{K}^{-1}$. 3, 88

# Bibliography

This bibliography is also available as a ADS library

→ THE EUROPEAN SPACE AGENCY